\documentclass[11pt, a4paper, oneside]{Thesis} 

\graphicspath{{Pictures/}} 

\usepackage[square,sort,compress,numbers]{natbib}
\usepackage{cite}
\usepackage{mathtools}
\renewcommand{\thefootnote}{\arabic{footnote}}
\usepackage{pifont}
\usepackage{cancel}
\usepackage{perpage}
\usepackage{amsmath,latexsym} 
\usepackage{float}
\newcommand{\gc}[1]{\widehat{#1}}
\title{\ttitle} 
\DeclareUnicodeCharacter{2212}{-}
\DeclareUnicodeCharacter{2212}{+}
\newcommand\blfootnote[1]{%
     \begingroup
     \renewcommand\thefootnote{}\footnote{#1}%
     \addtocounter{footnote}{-1}%
      \endgroup
    }
\begin{document}

\setstretch{1.3} 

\fancyhead{} 
\rhead{\thepage} 
\lhead{} 

%

\thesistitle{Exploring Wormholes in Modified Theories of Gravity}
\documenttype{THESIS}
\supervisor{Prof. Pradyumn Kumar Sahoo}
\supervisorposition{Professor}
\supervisorinstitute{BITS-Pilani, Hyderabad Campus}
\examiner{}
\degree{Ph.D. Research Scholar}
\coursecode{DOCTOR OF PHILOSOPHY}
\coursename{Thesis}
\authors{\textbf{ZINNAT HASSAN}}
\IDNumber{2019PHXF0463H}
\addresses{}
\subject{}
\keywords{}
\university{\texorpdfstring{\href{http://www.bits-pilani.ac.in/} 
                {Birla Institute of Technology and Science, Pilani}} 
                {Birla Institute of Technology and Science, Pilani}}
\UNIVERSITY{\texorpdfstring{\href{http://www.bits-pilani.ac.in/} 
                {BIRLA INSTITUTE OF TECHNOLOGY AND SCIENCE, PILANI}} 
                {BIRLA INSTITUTE OF TECHNOLOGY AND SCIENCE, PILANI}}



\department{\texorpdfstring{\href{http://www.bits-pilani.ac.in/pilani/Mathematics/Mathematics} 
                {Mathematics}} 
                {Mathematics}}
\DEPARTMENT{\texorpdfstring{\href{http://www.bits-pilani.ac.in/pilani/Mathematics/Mathematics} 
                {Mathematics}} 
                {Mathematics}}
\group{\texorpdfstring{\href{Research Group Web Site URL Here (include http://)}
                {Research Group Name}} 
                {Research Group Name}}
\GROUP{\texorpdfstring{\href{Research Group Web Site URL Here (include http://)}
                {RESEARCH GROUP NAME (IN BLOCK CAPITALS)}}
                {RESEARCH GROUP NAME (IN BLOCK CAPITALS)}}
\faculty{\texorpdfstring{\href{Faculty Web Site URL Here (include http://)}
                {Faculty Name}}
                {Faculty Name}}
\FACULTY{\texorpdfstring{\href{Faculty Web Site URL Here (include http://)}
                {FACULTY NAME (IN BLOCK CAPITALS)}}
                {FACULTY NAME (IN BLOCK CAPITALS)}}

\maketitle

\clearpage
\setstretch{1.3} 

\pagestyle{empty} 
\pagenumbering{gobble}

\addtocontents{toc}{\vspace{2em}} 
\frontmatter 
\Certificate
\Declaration
\begin{acknowledgements}
I would like to express my sincere thanks and gratitude to my supervisor, \textbf{Prof. Pradyumn Kumar Sahoo}, Professor, Department of Mathematics, BITS-Pilani, Hyderabad Campus, for his unwavering support, guidance, and vast experience throughout my Ph.D. journey. I am deeply grateful for his endless patience and faith in me, which inspired me to develop a deeper understanding of the subject and fostered my aptitude for scientific research.

I sincerely thank the members of my Doctoral Advisory Committee (DAC), \textbf{Prof. Bivudutta Mishra} and \textbf{Prof. Sashideep Gutti}, for their valuable suggestions and constant encouragement, which significantly contributed to the improvement of my research work.

It is my privilege to thank the Head of the Department, the DRC convener, all faculty members, and the staff of the Department of Mathematics for their support throughout this Ph.D. journey.

I owe a deep sense of gratitude to all my co-authors for their insightful suggestions, discussions, encouragement, and collaboration. I am especially grateful to \textbf{Dr. Ghulam Mustafa} for his guidance and unwavering support.

I gratefully acknowledge \textbf{BITS-Pilani, Hyderabad Campus} for providing the necessary facilities, and the \textbf{Department of Science and Technology (DST), Government of India, New Delhi}, for supporting my research through the \textbf{INSPIRE Fellowship} (File No. DST/INSPIRE Fellowship/2019/IF190911).

I would also like to thank my friends and colleagues—\textbf{Nur Salam}, \textbf{Dr. Sanjay Mandal}, \textbf{Gaurav Gadbail}, \textbf{Moreshwar Tayde}, \textbf{Sayantan Ghosh}, and \textbf{Dheeraj Singh Rana}—for their friendship and support. A special note of thanks goes to my best friend, \textbf{Mr. Nasibur Rahman}, who has always stood by me through every situation. I am also thankful to \textbf{Dr. Pinkimani Goswami, USTM} for consistently motivating me in my research pursuits.

I extend my heartfelt gratitude to my parents, \textbf{Mr. Mozibar Rahman} and \textbf{Mrs. Zoshefa Begum}, my siblings, and my in-laws for their unwavering love and support throughout my journey. I am especially thankful to my beloved wife, \textbf{Mrs. Zubina Tasrin}, whose constant love, understanding, and encouragement have brought immense joy and strength into my life.

I also thank my nieces, \textbf{Zerifa Zahan} and \textbf{Mehek Hassan}, for filling our lives with happiness and laughter. Lastly, I fondly remember my late grandparents, \textbf{Mr. Abdus Sarawar} and \textbf{Mrs. Latiful Begum}, whose love and care I continue to cherish.

\vspace{0.5 cm}
Zinnat Hassan,\\
ID: 2019PHXF0463H.
\end{acknowledgements}
\begin{abstract}
In this thesis, we investigate traversable wormhole spacetimes within different modified theories of gravity. The matter source at the wormhole throat is acknowledged to be an anisotropic energy-momentum tensor. Before discussing the investigated problems, the preliminaries are discussed in chapter-\ref{Chapter1}. In chapter-\ref{Chapter1}, we discuss the background of wormhole geometry, some important concepts, general relativity, and the modified theories of gravity. In chapters-(\ref{Chapter2}-\ref{Chapter4}), we investigate wormhole solutions in various scenarios in the context of $f(Q)$ gravity, while wormhole solutions in the galactic halos within $4$D Einstein Gauss-Bonnet (EGB) gravity are discussed in chapter-\ref{Chapter5}.\\
Chapter-\ref{Chapter2} aims to discuss wormhole solutions under a linear barotropic EoS, anisotropic EoS, and some specific shape functions in $f(Q)$ gravity. We discuss the viability of shape functions and the stability analysis of the wormhole solutions for each case. \\
Chapter-\ref{Chapter3} discusses the influence of GUP on Casimir wormholes with three different redshift functions. Two famous Generalized Uncertainty Principle (GUP) models such as  Kempf, Mangano, and Mann (KMM) and Detournay, Gabriel, and Spinde (DGS) models, are considered and discussed Casimir wormholes under different forms of $f(Q)$ gravity. We analyzed the obtained wormhole solutions with energy conditions, especially null energy conditions at the wormhole’s throat, and encountered that some arbitrary quantity disrespects the classical energy conditions at the wormhole throat of radius $r_0$. Later, the volume integral quantifier is also discussed to calculate the amount of exotic matter required near the wormhole throat.\\
Chapter-\ref{Chapter4} explored the possibility of traversable wormhole formation in the dark matter halo regions. We obtain the exact wormhole solutions based on the Bose-Einstein Condensate (BEC), Navarro-Frenk-White (NFW), and Pseudo-Isothermal (PI) matter density profiles. We present a novel wormhole solution supported by these dark matters using the density profile and rotational velocity along with the modified field equations to calculate the redshift and shape functions of the wormholes. 
We also examine the shadow and deflection of light by wormholes in the presence of various dark matter models.\\
In chapter-\ref{Chapter5}, we analyze wormhole formations in $4$D EGB gravity based on three different choices of dark matter profiles, such as the NFW, Universal Rotation Curve (URC), and Scalar Field Dark Matter (SFDM) profiles. We imposed the Karmarkar condition to find the exact solutions for the shape functions under two non-constant redshift functions. We discussed the energy conditions for each dark matter profile and noticed the influence of the Gauss bonnet coefficient in violating energy conditions. Further, some physical features of wormholes, viz. complexity factor, active gravitational mass, and total gravitational energy, have been explored.
\end{abstract} 

\Dedicatory{\bf \begin{LARGE}
Dedicated to
\end{LARGE} 
\\
\vspace{0.2cm}
\it My Parents and My Wife\\}




\lhead{\emph{Contents}} 
\tableofcontents 
\addtocontents{toc}{\vspace{1em}}
\addtocontents{toc}{\vspace{1em}}
\lhead{\emph{List of Tables}}
\listoftables 
\addtocontents{toc}{\vspace{1em}}
\lhead{\emph{List of Figures}}
\listoffigures 
\addtocontents{toc}{\vspace{1em}}



\lhead{\emph{List of symbols and Abbreviations}}
\listofsymbols{ll}{
$g_{\alpha\beta}:$ \,\,\,\,\,\,\,\,\, Metric tensor\\
$g:$ \,\,\,\,\,\,\,\,\,\,\,\,\,\,\, Determinant of $g_{\alpha\beta}$\\
$\tilde{\Gamma}^{\alpha}_{\;\:\beta\gamma}:$ \,\,\,\,\,\, General affine connection\\
${\Gamma}^{\alpha}_{\;\:\beta\gamma}:$ \,\,\,\,\,\, Levi-Civita connection\\
$K^{\alpha}_{\;\:\beta\gamma}:$ \,\,\,\,\, Contorsion tensor\\
$L^{\alpha}_{\;\:\beta\gamma}:$ \,\,\,\,\,\, Disformation tensor\\
$\nabla_{\gamma}$: \,\,\,\,\,\,\,\,\,\,\, Covariant derivative\\
${R}^{\alpha}{}_{\beta\gamma\sigma}:$ \,\,\, Riemann tensor \\
$R_{\mu\nu}:$\,\,\,\,\,\,\,\,\,\, Ricci tensor \\
$R:$ \,\,\,\,\, \,\,\,\,\,\,\,  Ricci scalar \\
$T_{\mu\nu}:$ \,\,\,\,\,\,\,\,\, Stress-energy tensor\\
$P^\alpha_{\;\beta\gamma}:$ \,\,\,\,\,\,\,\, Non-metricity conjugate\\
$Q_{\alpha\beta\gamma}: $\,\,\,\,\,\,\, Non-metricity tensor\\
GR:\,\,\,\,\, \,\,\,\,\,\, General Relativity\\
TEGR:\,\,\,\, Teleparallel Equivalent of General Relativity\\
STEGR:\,\,\, Symmetric Teleparallel Equivalent of General Relativity\\
$\Lambda$CDM:\,\,\,\, $\Lambda$ Cold Dark Matter\\
WH:\,\,\,\,\,\,\,\,\,\,\, Wormhole\\
EoS:\,\,\,\,\,\,\,\,\,\,\,\, Equation of State\\
4D:\,\,\,\,\, \,\,\,\,\,\,\, Four Dimension\\
EGB:\,\,\,\,\, \,\, Einstein Gauss Bonnet\\
WIMP:\,\,\,\, Weakly Interacting Massive Particles\\
CMB:\,\,\,\,\,\,\,\, Cosmic Microwave Background\\
NFW:\,\,\,\,\, \, Navarro, Frenk, and White\\
BEC:\,\,\,\,\, \,\, Bose-Einstein Condensate\\
PI:\,\,\,\,\, \,\,\,\,\,\,\,\, Psudo-Isothermal\\
MOND:\,\,\,\, Modified Newtonian Dynamics\\
URC:\,\,\,\,\, \,\, Universal Rotational Curve\\
WEP:\,\,\,\,\, \, Weak Equivalence Principle\\
EEP:\,\,\,\,\, \, Einstein Equivalence Principle\\
SEP:\,\,\,\,\, \, Strong Equivalence Principle\\
PPN:\,\,\,\,\, \, Parameterized Post-Newtonian\\
GB:\,\,\,\,\, \,\,\,\, Gauss Bonnet\\
ZTF:\,\,\,\,\, \,\,\,\, Zero Tidal Force\\

}

\addtocontents{toc}{\vspace{2em}}

%
%


\clearpage 





\mainmatter 

\pagestyle{fancy} 


\chapter{Introduction} 
\label{Chapter1}

\lhead{Chapter 1. \emph{Introduction}} 

This thesis titled {\bf Exploring Wormholes in Modified Theories of Gravity} has been focused on investigating wormhole geometries in modified theories of gravity. Before going to the investigated problems, the present chapter discusses the history, mathematical notations, basic elements, and fundamental theories of gravity.
\section{Motivations}
Wormholes are theoretical constructs providing shortcuts across the fabric of space-time, allowing transportation across different locations \cite{a1,a2,a3}.
A crucial aspect allowing such phenomena is the ``throat", a minimal radius that differentiates it from black holes \cite{a4}, which removes the event horizon from the geometric structure. These theoretical concepts require dependence on a well-established and experimentally validated theoretical framework for examination. The most comprehensive framework used to explain gravitational phenomena, including wormholes, is Einstein's General Relativity (GR) theory,  proposed in 1915 \cite{a5}.\\
It can be claimed that the concept of wormholes remains theoretical, with an agreement suggesting that these phenomena might emerge in areas characterized by extreme gravitational fields \cite{a2}. However, there is a lack of experimental proof either supporting or disproving their existence. Despite the absence of concrete evidence, it is crucial to note that this should continue ongoing research in the field. Perhaps wormholes, if they can be engineered at all, could only be artificially created through the efforts of an extremely technologically advanced civilization. The question is, do those structures allow in the GR formalism? This line of inquiry is essentially trying to validate if wormholes are possible and probe further into the foundations upon which GR is built. Since GR predicts wormholes as possibilities, a further case that substantiates the need to consider these phenomena is the consistency of this theory that has been well corroborated by myriad tests and is adopted in general \cite{a6,a7,a8,a9,a10,a11,a12,a13}. Otherwise, if GR implies that the wormhole configuration is not a physical solution to merge black holes, this would have serious consequences for its entire formulation as a gravitational model.\\
The idea was quite simple: while GR is regarded as the most successful gravitational theory in the field of physics, it is crucial to remember that despite its significant achievements, GR remains a scientific theory. This property of scientific theories is part of what makes science a ``self-correcting" process: as new experimental or theoretical information becomes available, it can either corroborate predictions made by the theory and further support its validity or challenge them and require alternative explanations to be developed. This suggests that it may be time to refine or discard the theory. In particular, relevant in the case of GR, where various phenomena, such as the Universe's acceleration \cite{a14}, the rotation curves of galaxies \cite{a14}, and GR's resistance to quantization attempts \cite{a15}, suggest that the theory may need to be reformulated. This motivates a more detailed exploration of theories beyond GR. See Ref. \cite{a16} for a review of the theories and what implications they have on cosmology. In the context of wormhole geometry, it has been demonstrated that traversable wormholes must violate the null energy conditions in the throat region \cite{a17}, implying the necessity for exotic matter to maintain such structures. On the other hand, these exotic matters have never been observed in any known astrophysical systems. This raises the possibility of modified gravity theories in which modifications in the gravitational sector may satisfy this requirement \cite{a18,a19,a20,a21,a22}.\\
Further, modifying a theory can be one of the most useful techniques for gaining more profound knowledge. To do so, two approaches can be employed. One can either question its foundational principles and construct a different, consistent theory or explore whether the theory can be considered as part of a more general class of theories. In the latter approach, GR is treated as the starting point, and theories are chosen to investigate how far and in which direction one can deviate from it \cite{a16}. These modified gravity theories must be simple and manageable, deviating from GR in only one aspect, and their viability is discussed when applied to specific gravitational or cosmological phenomena. Overall, this trial-and-error approach to modifying gravity should be seen as ``thought experiments" to probe the foundations of GR.\\
One significant area of exploration within these modified gravity frameworks involves the conceptualization and mathematical modeling of traversable wormholes without invoking unobserved exotic matter by proposing modifications to gravity's behavior. This endeavor aims to illuminate the possible physics of wormholes and deepen our understanding of gravity's nature and scope under different theoretical modifications, such as in the contexts of $f(Q)$ gravity and $4$D EGB theory. This approach underscores a methodical exploration of gravity's theoretical boundaries, fostering a deeper dive into our Universe's known and unknown aspects.
\section{Historical overview of wormholes}

\subsection{Non-traversable wormholes}
\begin{figure}
    \centering
    \includegraphics[scale=0.8]{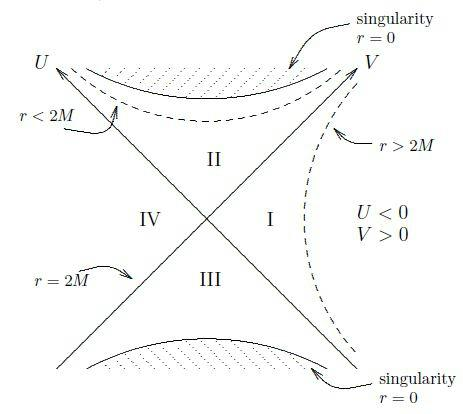}
    \caption{A Kruskal diagram with its 4 separate regions. This figure is adapted from Ref. \cite{aa5}.}
    \label{fig1ch1}
\end{figure}

The initial conceptual instance of a wormhole that cannot be passed through is derived from the Schwarzschild solution of Einstein's field equations within the framework of GR. In 1916, Ludwig Flamm \cite{aa1} discovered an alternative solution to the Schwarzschild black hole, now referred to as a white hole. This secondary solution describes a different region of space-time but is mathematically linked through a channel in space-time \cite{aa1}. Albert Einstein and Nathan Rosen further investigated this concept in 1935 through an article that primarily aimed at developing a theory of matter and electricity that avoids field singularities by solely employing the general relativity tensors $g_{\mu\nu}$ and the $\phi_\mu$ of the Maxwell theory. This exploration led to what is now known as the Einstein-Rosen bridge \cite{aa2}. In order to comprehend this structure mathematically and topologically, we must first examine the metric of the solution to Einstein's field equations and the corresponding coordinate systems.\\
The spherically symmetric with no electric and magnetic charge of the Schwarzschild metric to Einstein’s field equations can be read as
\begin{equation}
ds^2 = -\left(1 - \frac{2M}{r}\right) dt^2 + \frac{1}{\left(1 - \frac{2M}{r}\right)} dr^2 + r^2(d\theta^2 + \sin^2\theta d\varphi^2),
\end{equation}
where the radial coordinate $r>2M$, $0\leq\theta\leq\pi$, and $0\leq\varphi\leq 2\pi$. An initial look may suggest a singularity at $r = 2M$; however, this is merely a result of the chosen coordinate system, which creates an illusion of a singular point that, in reality, does not exist. A shift towards a different coordinate system, notably the Eddington–Finkelstein (EF) coordinates, proves beneficial in overcoming this limitation. This setup differentiates between black holes, using ingoing EF coordinates, and white holes, employing outgoing EF coordinates \cite{aa4}. Specifically, for black holes, transforming from $t$ to $t = v- r^k$ where $r^k$ is defined by the relation $dr^k/dr = \left(1 - \frac{2GM}{r}\right)^{-1}$, modified the metric as
\begin{equation}
ds^2 = -\left(1 - \frac{2M}{r}\right) dv^2 + 2dvdr + r^2(d\theta^2 + \sin^2\theta d\varphi^2).
\end{equation}
For white holes, the change in co-ordinates $t = u+r^k$, resulting in a revised metric
\begin{equation}
ds^2 = -\left(1 - \frac{2M}{r}\right) du^2 - 2dudr + r^2(d\theta^2 + \sin^2\theta d\varphi^2).
\end{equation}
Nonetheless, the mathematical framework linking black holes and white holes draws from a common coordinate system known as the Kruskal-Szekeres (KS) coordinates. The KS coordinates, represented by $(U, V, \theta, \varphi)$, follow the definitions $U=-e^{-u/4M}$ and $V=e^{v/4M}$. Initially, the metric only applies to $U < 0$ and $V > 0$; however, through analytical extension, we can derive the maximally expanded Schwarzschild solution
\begin{equation}
ds^2 = -\frac{32 M^3}{r} e^{-r/2M} dU dV + r^2 (d\theta^2 + sin^2\theta d\varphi^2),
\end{equation}
with $-\infty <U,\,V<\infty$ and $``r"$ linked to $U$ and $V$ through
\begin{equation}
UV = -\frac{(r - 2M)}{2M} e^{r/2M}.
\end{equation}
This leads to the rise of the Kruskal diagram (see Fig. \ref{fig1ch1}), dividing the entire space-time into four distinct quadrants based on the sign of $U$ and $V$. Region I corresponds to the part of space where $r > 2M$ in Schwarzschild coordinates. Both region I and region II are surrounded by ingoing EF coordinates, which are essential in describing black hole geometry. Meanwhile, region I and region III are covered by outgoing EF coordinates, which are suitable for white holes. Region IV, a new region, is isometric to the region I under the transformation $(U, V) \rightarrow (-U, -V)$. The singularity located at $r = 0$ is described by $UV = 1$, while the $r = 2M$ boundary corresponds to $UV = 0$, indicating that either $U = 0$ or $V = 0$. In this diagram, each point represents a 2-sphere with radius $r$. Another way to interpret the diagram is to see it as showing the causal structure of radial motion for fixed polar angles $\theta$ and $\varphi$.\\
In region I, the ratio $\frac{U}{V} = e^{-t/2M}$ holds, meaning that constant-time slices in Schwarzschild coordinates are straight lines passing through the origin on the Kruskal diagram. These hypersurfaces are partially within region I and region IV. To better visualize this geometry, it helps to shift to isotropic coordinates $(t, \rho, \theta, \varphi)$, where $\rho$ represents the new radial coordinate
\begin{equation}
r = \left(1 +\frac{M}{2\rho}\right)^2 \rho = \rho + M + \frac{M^2}{4\rho^2}.
\end{equation}
The metric in isotropic coordinates becomes
\begin{equation}
ds^2 =-\left(\frac{1 -\frac{M}{2\rho}}{1 +\frac{M}{2\rho}}\right)^2 dt^2+\left(1 +\frac{M}{2\rho}\right)^4 \left(d\rho^2+\rho^2 d\Omega^2\right).
\end{equation}
For a given $r$, there are two possible solutions for $\rho$. These two values of $\rho$ are related by the isometry $\rho \rightarrow M^2/4\rho$, which has a fixed point at a 2-sphere with a radius of $2M$. This isometry connects regions I and IV, similar to the $(U, V) \rightarrow (-U, -V)$ transformation. For region I, $\rho > M/2$, while for region IV, $M/2 > \rho > 0$. Isotropic coordinates, therefore, only cover regions I and IV as $\rho$ becomes problematic at $r < 2M$ (see Ref. \cite{aa5}).\\
When considering the geometry for $t = \text{constant}$, when we close $\rho = M/2$ from either side, the radius of the 2-sphere at each point decreases to a minimum of $r = 2M$ at $\rho=M/2$, called the minimal 2-sphere. There are two asymptotically flat regions as $\rho \rightarrow \infty$ and $\rho \rightarrow 0$, which are connected by a ``throat" with a minimum radius equal to that of the minimal 2-sphere, i.e., $2M$. This structure is mathematically known as the \textit{Einstein-Rosen bridge}.
\subsection{Traversable wormholes}
The year 1973 marked a significant milestone in analyzing traversable wormholes, with pioneering work conducted separately by Homer G. Ellis \cite{aa6} and Kirill A. Bronnikov \cite{aa7}, whose findings uncovered that traversable wormholes could exist within the framework of GR. The model presented by Ellis \cite{aa6}, often cited as the initial complete representation of such a wormhole, showcases a static, spherically symmetric solution to the Einstein vacuum field equations. This is achieved by integrating a scalar field $\Phi$ that is minimally connected to the space-time geometry but with a reversed coupling direction (negative rather than positive) \cite{aa6}
\begin{equation}
R_{\mu\nu}-\frac{1}{2}R g_{\mu\nu}=2\left(\Phi_{,\mu}\Phi_{,\nu}+\frac{1}{2}\Phi^{,\sigma}\Phi_{,\sigma}g_{\mu\nu}\right).
\end{equation}
The above mathematical description provided outlines the metric as follows
\begin{eqnarray}
ds^2 &=& -dt^2 + (d\rho - f(\rho)dt)^2 + r^2(\rho)(d\theta^2 + \sin^2\theta d\varphi^2),\\
&=&  -[1 - f^2(\rho)]dT^2 + \frac{1}{1 - f^2(\rho)} d\rho^2 + r^2(\rho)(d\theta^2 + \sin^2\theta d\varphi^2),
\end{eqnarray}
where 
\begin{equation}
  dT = dt + \frac{f(\rho)}{1 - f^2(\rho)} d\rho.
\end{equation}
The functions \(f(\rho)\) and \(r(\rho)\) must be determined from the field equations. The range for the coordinates is given by
\begin{equation}\nonumber
-\infty<t<\infty; \quad -\infty<\rho<\infty; \quad 0<\theta<\pi; \quad -\pi<\varphi<\pi.
\end{equation}
The behavior of solutions is influenced by two parameters, $m$ and $n$, which must adhere to the condition $0 \leq m < n$ but are otherwise unrestricted. The expressions for the functions are demonstrated as follows
\begin{equation}
f(\rho)=-\sqrt{1 - e^{-\left(\frac{2m}{n}\right)\tau(\rho)}},
\end{equation}
\begin{equation}
r(\rho)=\sqrt{(\rho - m)^2 + a^2}e^{\left(\frac{m}{n}\right)\tau(\rho)},
\end{equation}
where \(\tau(\rho) = \frac{n}{a}\left[\frac{\pi}{2}-\tan^{-1}\left(\frac{\rho-m}{a}\right)\right]\) and \(a = \sqrt{n^2 - m^2}\) \cite{aa6}.
The movement of particles through this wormhole is similar to the movement of liquid on turbulent draining curves. Depending on the direction of the initial velocity of each particle, we can obtain different trajectories \cite{aa6}. For more details about Ellis's \textit{drainhole} theory, see the Ref. \cite{aa6}.\\
In 1988, K. Thorne, M. Morris, and U. Yertsever \cite{a17,MorrisThrone} made significant advancements in the field by publishing two influential papers that introduced the concept of the Morris-Thorne (MT) wormhole. They identified nine key characteristics for a stable, traversable wormhole, some of which are essential for its existence, while others were chosen for the sake of simplifying calculations:
\begin{itemize}
    \item[1.] To make the calculations easier, the metric should be both spherically symmetric and time-independent. 
The MT wormhole metric is expressed as
\begin{equation}\label{1ch1}
    ds^2 = -e^{2\phi(r)}dt^2 + \frac{dr^2}{\left(1 - \frac{b(r)}{r}\right)} + r^2d\Omega^2
\end{equation}
where $r$ is the radial distance and $d\Omega^2 = d\theta^2+\sin^2\theta d\varphi^2$. $b(r)$ defines the spatial structure of the wormhole and is known as the ``shape function", while $\phi(r)$, governing the gravitational redshift, is called the ``redshift function".
   \item[2.] The solution must satisfy the Einstein field equations everywhere.
   \item[3.] The wormhole must feature a throat that connects two distinct, asymptotically flat regions of space-time.
   \item[4.] To allow for two-way travel, the wormhole cannot contain horizons, meaning that $\phi(r)$ must remain finite everywhere.
   \item[5.] The tidal gravitational forces experienced by travelers must be bearably small.
   \item[6.] A traveler should be able to pass through the wormhole within a finite and reasonably small proper time.
   \item[7.] The matter and energy fields responsible for the space-time curvature must have a stress-energy tensor that aligns with physical expectations.
   \item[8.] The solution should be stable under small perturbations.
   \item[9.] The wormhole should be feasible to construct, implying that the required mass should be significantly smaller than the total mass of the Universe, and the formation time should be much less than the Universe’s age.
\end{itemize}
The authors categorized the first four conditions as ``basic wormhole criteria", while the fifth and sixth conditions related to making the wormhole traversable for human travelers, were termed ``usability criteria".\\
To simplify the calculation of the Einstein tensor components, we switch to a different coordinate system that makes the process more straightforward. In this new system, we adopt a set of orthonormal vectors that represent the ``proper reference frame of observers who remain stationary at fixed coordinates \((r, \theta, \varphi)\)":
\begin{equation}
e_t = e^{-\phi}e_{t0}, \quad e_r = \left(1 - \frac{b}{r}\right)^{-\frac{1}{2}} e_{r_0}, \quad e_\theta = \frac{1}{r} e_{\theta_0}, \quad e_\varphi = \frac{1}{r \sin \theta} e_{\varphi_0},
\end{equation}
where $e_{t_0}, e_{r_0}, e_{\theta_0}, e_{\varphi_0}$ are the initial orthonormal basis vectors given by
\begin{equation}
e_{t_0} = \frac{\partial}{\partial t}, \quad e_{r_0} = \frac{\partial}{\partial r}, \quad e_{\theta_0} = \frac{\partial}{\partial \theta}, \quad e_{\varphi_0} = \frac{\partial}{\partial \varphi}.
\end{equation}
In this newly adopted basis, the metric coefficients reduce to their standard form, as seen in special relativity
\[
g_{\mu \nu} = e_\mu \cdot e_\nu = \eta_{\mu \nu} =
\begin{pmatrix}
-1 & 0 & 0 & 0 \\
0 & 1 & 0 & 0 \\
0 & 0 & 1 & 0 \\
0 & 0 & 0 & 1
\end{pmatrix}.
\]
This representation simplifies the calculations by aligning the metric coefficients with those of flat space-time. Now, by calculating the Riemann tensor \(R^\mu_{\nu\rho\sigma}\), the Ricci tensor \(R_{\mu\nu}\), and the Ricci scalar \(R\), and solving the Einstein field equations, we arrive at the following results
\begin{equation}
G_{tt} = \frac{b'(r)}{r^2},
\end{equation}
\begin{equation}
G_{rr} = -\frac{b}{r^3} + 2\left(1 - \frac{b}{r}\right) \frac{\phi'(r)}{r},
\end{equation}
\begin{equation}
G_{\theta\theta} = \left(1 - \frac{b}{r}\right)\left[\phi''(r) - \frac{b''(r)}{2} - \frac{b}{2r(r - b)} \phi'(r) + (\phi'(r))^2 + \frac{\phi'(r)}{r} - \frac{b'(r)}{2r^2(r - b)}\right] = G_{\varphi\varphi}.
\end{equation}
Next, we express the components of the stress-energy tensor as follows
\begin{equation}
T_{tt} = \rho(r), \quad T_{rr} = P_r(r), \quad T_{\theta\theta} = T_{\varphi\varphi} = P_t(r).
\end{equation}
Here, $\rho(r)$ represents the energy density, $P_r(r)$ is the radial pressure, and $P_t(r)$ is the tangential pressure. We can also denote the radial pressure in terms of the radial tension, i.e., $P_r(r)=-\tau(r)$. Using the equation \(G_{\mu\nu} = 8\pi T_{\mu\nu}\), we find
\begin{equation}\label{p1}
\rho(r) = \frac{b'(r)}{8\pi r^2},
\end{equation}
\begin{equation}\label{p2}
\tau(r)=\frac{1}{8\pi} \left[\frac{b}{r^3} - 2\left(1 - \frac{b}{r}\right) \frac{\phi'(r)}{r}\right],
\end{equation}
\begin{equation}\label{p3}
P_t(r) = \frac{1}{8\pi} \left(1 - \frac{b}{r}\right)\left[\phi''(r) - \frac{b''(r)}{2} - \frac{b}{2r(r - b)} \phi'(r) + (\phi'(r))^2 + \frac{\phi'(r)}{r} - \frac{b'(r)}{2r^2(r - b)}\right].
\end{equation}
By selecting appropriate forms for $b(r)$ and $\phi(r)$, according to the constraints outlined earlier, we can solve for $\rho(r)$ and $\tau(r)$. The most significant conditions arise at the wormhole's throat, where the radial tension must exceed the energy density, i.e., $\tau_0 > \rho_0$. Materials that exhibit this behavior are referred to as ``exotic". This situation introduces challenges, especially when considering measurements by an observer traveling through the wormhole close to the speed of light. Such an observer would perceive a negative energy density. Additionally, a stationary observer may also observe a negative energy density $\rho_0 < 0$, which naturally follows from this scenario.
The following subsection will explore specific conditions where this phenomenon occurs.
\section{Energy conditions}
To gain a deeper understanding of the matter inside the wormhole, Morris and Thorne \cite{a17} introduced the dimensionless function $\xi= \frac{(\tau-\rho)}{|\rho|}$. By applying Eqs. \eqref{p1} and \eqref{p2}, we can derive the expression
\begin{equation}\label{p4}
\xi=\frac{(\tau-\rho)}{|\rho|}=\frac{b/r-b'-r(1-b/r)\phi'}{|b'|}.
\end{equation}
When we combine Eq. \eqref{p4} with the flaring-out condition $\frac{d^2r}{dz^2}=\frac{b-r b'}{b^2}>0$, the exoticity function $\xi(r)$ becomes
\begin{equation}
\xi= \frac{2b^2}{r|b'|} \frac{d^2r}{dz^2}-\frac{r(1-b/r)\phi'}{|b'|}.
\end{equation}
Given that $\rho$, and therefore, $b'$ are finite, and considering that $(1-b/r)\phi'\rightarrow 0$ at the throat, we obtain the relation
\begin{equation}
\xi(r_0)=\frac{\tau_0-\rho_0}{|\rho_0|} > 0.
\end{equation}
The condition $\tau_0>\rho_0$ presents a significant challenge, as it indicates that the radial tension at the throat must surpass the energy density. Morris and Thorne, therefore, classified the matter meeting this requirement as ``exotic matter". Further analysis will show that this type of matter violates the null energy condition, as well as all other energy conditions.\\
Let us explore the energy conditions, especially when the energy-momentum tensor is diagonal, represented as
\begin{equation}
T_{\mu}^{\nu} = \text{diag}(-\rho, P_1, P_2, P_3).
\end{equation}
Here, \(P_i\) denotes the pressure components. When all pressures are equal (\(p_1 = p_2 = p_3\)), the tensor simplifies to that of a perfect fluid. Although we generally expect classical forms of matter to adhere to these energy conditions, certain quantum fields, such as those involved in the Casimir effect, may violate them.
\begin{itemize}
  \item[\ding{114}] \textbf{Null Energy Condition (NEC):} The NEC dictates that for any null vector \(k^\mu\)
  \begin{equation}
   T_{\mu\nu}k^\mu k^\nu \geq 0.   
  \end{equation}
  In the context of the diagonal energy-momentum tensor, this condition simplifies to 
  \begin{equation}
     \rho + P_i \geq 0,\quad \forall i.
  \end{equation} 
Essentially, this implies that the energy density observed by any null observer with a four-velocity \(k^a\) must be non-negative.
\end{itemize}
\begin{itemize}
    \item[\ding{114}] \textbf{Week Energy Condition (WEC):} The WEC requires that for any timelike vector \(U^\mu\), the inequality 
    \begin{equation}
    T_{\mu\nu}U^\mu U^\nu \geq 0
    \end{equation}
holds. This means the energy density observed by any timelike observer must be positive, which translates to
\begin{equation}
  \rho \geq 0 \quad \text{and} \quad \rho + P_i \geq 0 \quad \forall i.
\end{equation}
Notably, the WEC includes the NEC as a special case.
\end{itemize}
\begin{itemize}
    \item[\ding{114}] \textbf{Strong Energy Condition (SEC):} The SEC states that for any timelike vector \(U^\mu\), the inequality
    \begin{equation}
    \left(T_{\mu\nu} - \frac{T}{2}g_{\mu\nu}\right)U^\mu U^\nu \geq 0
    \end{equation}
    must hold, where \(T\) is the trace of the energy-momentum tensor. For the diagonal tensor, this condition is expressed as 
    \begin{equation}
    \rho + P_i \geq 0 \quad \text{and}\quad \rho + \sum_i P_i \geq 0.
    \end{equation}
The SEC is closely tied to maintaining the attractive nature of gravity, implying the NEC but not necessarily the WEC.
\end{itemize}
\begin{itemize}
    \item[\ding{114}] \textbf{Dominant Energy Condition (DEC):} The DEC asserts that for any timelike vector \(U^\mu\), the conditions
    \begin{equation}
   T_{\mu\nu}U^\mu U^\nu \geq 0 \quad \text{and}\quad T_{\mu\nu}U^\nu
    \end{equation}
    being non-spacelike must be met. These requirements ensure that the locally observed energy density is positive and that the energy flux is either timelike or null. For a diagonal tensor, this leads to
    \begin{equation}
      \rho \geq 0 \quad \text{and}\quad P_i \in [-\rho, +\rho], 
    \end{equation}
ensuring the speed of sound remains below the speed of light, which is often associated with the stability of the system.
\end{itemize}
In summary, we examine four key energy conditions, i.e., NEC, WEC, SEC, and DEC, under a diagonal energy-momentum tensor framework. These conditions are respectively given by
\begin{itemize}
    \item NEC: \(\rho + P_r \geq 0\), \(\rho + P_t \geq 0\).
    \item WEC: \(\rho \geq 0\), \(\rho + P_r \geq 0\), \(\rho + P_t \geq 0\).
    \item SEC: \(\rho + P_r \geq 0\), \(\rho + P_t \geq 0\), \(\rho + P_r + 2 P_t \geq 0\).
    \item DEC: \(\rho \geq |P_r|\), \(\rho \geq |P_t|\).
\end{itemize}
By applying the Einstein field equations at the wormhole throat and considering the finite nature of the redshift function, the flaring-out condition leads to $(\rho + P_r)|_{r_0} < 0$, which violates the NEC and, by extension, all pointwise energy conditions. While classical matter is generally expected to satisfy these conditions, quantum fields like those involved in the Casimir effect are known to violate them. Thus, the flaring-out condition implies a violation of the NEC at the throat, suggesting that the wormhole's geometry requires support from exotic matter. Negative energy densities are not strictly necessary, but negative pressures at the throat, such as $P_r(r_0) = -1/(8\pi r^2_0)$, are essential to maintain the throat. While challenging the NEC may seem doubtful from a classical perspective, experimental evidence of quantum effects violating this condition supports the possibility of such violations, although to a minimal degree. Whether these violations can be significant enough to sustain a traversable wormhole remains uncertain. To address this challenge, various approaches have been explored, including evolving wormhole space-times \cite{ec1,ec2,ec3}, rotating solutions \cite{ec4}, thin-shell wormholes using the cut-and-paste method \cite{ec5,ec6,ec7}, and modifications to gravity theories \cite{a19,a20,ec11,ec12}.
\section{The Casimir effect}
Casimir force is a very tiny attractive force arising between uncharged metals due to quantum fluctuations in the electromagnetic field in a vacuum. This phenomenon was first predicted by the Dutch physicist Hendrick Casimir in 1948 \cite{Casimir1}. According to quantum field theory, a vacuum contains particles (photons) that all the time come and go in fluctuating \cite{Casimir2}. These fluctuating particles generate an attractive force. The quantum Casimir effect is typically discussed for two nearly touching parallel plates in the topic of quantum field theory (since this leads to a nonzero null density between them). Casimir proposed that in calculating the vacuum energy, only virtual photons with wavelengths that fit an integer number of times between the plates should be considered \cite{Casimir1}. This reduces energy density between the plates as they are brought closer together, implying a small attractive force. This can also be explained by the fact that the limited space between the plates allows only smaller exotic particles to exist. The difference in particle types and quantities between the exterior and interior of the plates leads to a pressure differential, creating a force that attracts the plates together. This phenomenon is known as the Casimir force.\\
The Casimir force, which acts between uncharged conducting surfaces, was described by Schwinger as one of the most counterintuitive results of quantum electrodynamics. This force per unit area on parallel flat plates with separation $d$ is given by \cite{Casimir1}
\begin{equation}
    \frac{F(d)}{A} = \frac{\pi^2 \hbar c}{240 d^4},
\end{equation}
where $\hbar$ is the reduced plank constant. The above relation can be derived by examining the electromagnetic mode structure between the plates in comparison to free space and assigning a zero-point energy of $\frac{1}{2}\hbar\kappa$ (here $\kappa$ denotes the frequency) to each electromagnetic mode (photon). The modification of the mode structure due to the plates alters the total energy density as a function of separation $d$, resulting in an attractive force. This prediction was significant as it was one of the earliest indications of a physical effect arising directly from zero-point fluctuations, contemporary with, independent of Bethe's work on the Lamb shift \cite{ben}.\\
Despite the fact that the Casimir force was predicted more than fifty years ago, it is only in the last eight years that precise measurements (within a few percent) have only been achieved. The fundamental constants involved in this equation are $\hbar$ and $c$; notably, the electron charge $e$ is absent. This absence indicates that the electromagnetic field does not couple to matter in the conventional sense. Instead, the plates impose boundary conditions on the field, making their microscopic properties irrelevant within the perfect conductivity limit. The constant $c$ converts the electromagnetic mode wavelength to a frequency, while $\hbar$ converts this frequency to an energy.\\
The term ``Casimir effect" encompasses several long-range interactions, including the retarded Van-der-Waals interaction between atoms or molecules, the Casimir–Polder interaction between an atom and a material surface, and the attraction between bulk material bodies. The latter, specifically, is the Casimir force, which depends only on the bulk properties of the materials involved. It is widely recognized that the concept of system boundaries affecting physical properties has significant implications, ranging from condensed matter physics to quantum chromodynamics. For a more general case of this argument, see \cite{Casimir4,Casimir5}. More focused reviews are available in \cite{Casimir6,Casimir7} (compiled for Dr. Casimir's 80th birthday) and in \cite{Casimir8} (for his 90th birthday). Also, a comprehensive study can be found in \cite{Casimir9}. \\
In 1996, Steven Lamoreaux performed an experiment that measured the Casimir force with a 5\% uncertainty compared to the theoretical predictions \cite{Casimir10}. The Casimir force is influenced by both bosons and fermions, which have a repulsive effect. While all types of particles contribute to the force, only the contribution from photons is detectable. According to theory, the vacuum's lowest energy state, known as zero-point energy, becomes infinite when considering all photon modes. The original derivation of the Casimir force involves complex mathematical situations where differences in infinities cancel each other out. There are still unresolved issues and puzzles related to this effect, especially when integrating it with quantum gravity theories. Nevertheless, it is expected that solutions to these issues will arise from the development of a comprehensive theory of quantum gravity.\\
In 2005, Jaffe \cite{Casimir11} clarified that zero-point fluctuations described by quantum field theory cannot be observed in laboratory experiments through the vacuum value of the stress tensor (vacuum energy density). 
Jaffe also demonstrated that the Casimir force can be calculated without referencing zero-point energy \cite{Casimir11}, indicating that zero-point energy might be a useful mathematical construct rather than a physical necessity for deriving measurable results. 

\subsection{Derivation of Casimir force}
Let $k_x,\,k_y,$ and $k_z$ represent the wave numbers in the \( x, y, \) and \( z \) directions, respectively. Consider two plates parallel to the $x-y$ plane, separated by a distance $a$. The space between these plates can be defined by $0 \le x \le \sqrt{A}$, $0 \le y \le \sqrt{A}$, and $0 \le z \le a$, where the plates have an area $A$. Using periodic boundary conditions, we have
\begin{equation}
k_x = \frac{2\pi n_x}{\sqrt{A}} \Rightarrow dn_x = \frac{\sqrt{A}}{2\pi} dk_x,\quad k_y = \frac{2\pi n_y}{\sqrt{A}} \Rightarrow dn_y = \frac{\sqrt{A}}{2\pi} dk_y, \quad k_z = \frac{n_z \pi}{a},
\end{equation}
with \( (n_x, n_y, n_z) \in \mathbb{Z} \). The frequency of the wave is given by $\omega_n = v|\vec{k}| = v \sqrt{k_x^2 + k_y^2 + k_z^2}$. In a vacuum, the speed of an electromagnetic wave is $c$, so $\omega_{n_z} = c \sqrt{k_x^2 + k_y^2 + k_z^2}$. The vacuum energy is the sum of all possible modes. The zero-point (ground state) energy associated with the $n_z$th mode is
\begin{equation}
E_{n_z} = \frac{\hbar \omega_{n_z}}{2}.
\end{equation}
The total energy of all combined modes is then the sum over all $n_z$
\begin{equation}
E = \sum_{n_z=1}^{\infty} \frac{\hbar \omega_{n_z}}{2}.
\end{equation}
For simplicity, let $n \equiv n_z$. The expectation value of the energy over the entire area of the plates is found by integrating over all possible values of $n_x, n_y$, and all possible modes
\begin{equation}\label{1ch3a1}
\langle E \rangle = \frac{\hbar}{2} \int \int \sum_{n=1}^{\infty} \omega_n \, dn_x \, dn_y = \frac{A \hbar}{8\pi^2} \int \int \sum_{n=1}^{\infty} \omega_n \, dk_x \, dk_y.
\end{equation}

This expression clearly diverges, indicating an infinite sum. By applying zeta regularization, we can determine a finite energy per unit area by defining a quantity $\langle E(s) \rangle$, which approaches Eq. \eqref{1ch3a1} as $s\rightarrow 0$ (here the notation $s$ is used traditionally in the study of the Riemann zeta function).
\begin{eqnarray}
\frac{\langle E(s) \rangle}{A} &=& \frac{\hbar}{8\pi^2} \int \int \sum_{n=1}^{\infty} \omega_n |\omega_n|^{-s} \, dk_x \, dk_y \\
&=& \frac{\hbar}{8\pi^2} \sum_{n=1}^{\infty} \int \int \omega_n |\omega_n|^{-s} \, dk_x \, dk_y.
\end{eqnarray}
Simplifying this expression (using Mathematica to integrate over $dk_x$ and $dk_y$) yields
\begin{equation}
\frac{\langle E(s) \rangle}{A} = \frac{\hbar c^{1-s} \pi^{2-s}}{2a^{3-s} (3 - s)} \sum_{n=1}^{\infty} |n|^{3-s}.
\end{equation}
The sum diverges around the value of $s$ approaching zero. However, if we assume that the damping of high-frequency excitations, corresponding to the analytical continuation of the Riemann zeta function to $s = 0$, makes sense physically somehow, then we have
\begin{equation}
\frac{\langle E \rangle}{A} = \lim_{s \to 0} \frac{\langle E(s) \rangle}{A} = -\frac{\hbar c \pi^2}{6 a^3} \zeta(-3).
\end{equation}
Also, it is known that $\zeta(-3)=\frac{1}{120}$ and hence the above expression gives
\begin{equation}\label{cf1}
\frac{\langle E \rangle}{A} = -\frac{\hbar c \pi^2}{720 a^3}.
\end{equation}
The Casimir force per unit area between two parallel plates in a vacuum is then
\begin{equation}
\frac{F_c}{A} = -\frac{d}{da} \frac{\langle E \rangle}{A} = -\frac{\hbar c \pi^2}{240 a^4}.
\end{equation}
This result, derived using analytical continuation, shows that the infinities were resolved to obtain a finite, experimentally verifiable solution. The negative sign indicates that the force is attractive, and the presence of \( \hbar \) signifies its quantum nature. Also, it is important to mention here that, in the original derivation, Casimir used Euler-Maclaurin summation with a regularizing function to handle non-convergent sums \cite{Casimir1}.

\section{Generalized uncertainity principle}
In this section, we will briefly explore the concepts and motivations behind the GUP in the context of quantum gravity. GUP is a phenomenological approach that introduces an absolute minimal length into the theory. Various methods of connecting quantum mechanics and gravity suggest the necessity of a minimal length \cite{gup1,gup2,gup3,gup4,gup5}. According to the Heisenberg uncertainty principle ($\Delta x \Delta p \geq \hbar/2$), quantum mechanics establishes a relationship between the uncertainties in position and momentum, $\Delta x \sim \frac{\text{Const.}}{\Delta p}$. From a gravitational perspective, probing smaller distances requires higher center-of-mass energies/momenta. At sufficiently high energy/momentum, a micro-black hole forms, with its event horizon size approximated by the Schwarzschild radius, $r_{\text{Sch}} = \frac{2G\Delta E}{c^2}$, where $\Delta E$ replaces the traditional black hole mass, $M$. By setting $c = 1$, replacing $r_{\text{Sch}}\) with $\Delta x$ and $\Delta E$ with $\Delta p$, the relation becomes $\Delta x = 2G\Delta p$, indicating a linear connection between $\Delta x$ and $\Delta p$. Since we set $c = 1$, hence mass, energy, and momentum are interchangeable. Combining this linear relation from gravity with the inverse relationship from quantum mechanics yields $\Delta x \sim \hbar/\Delta p + G\Delta p$ (ignoring factors of 2). The interplay between these terms results in a minimum value for $\Delta x$ at $\Delta p_m \sim \sqrt{\hbar/G}$, leading to a minimal length $\Delta x_m \sim \sqrt{\hbar G}$.\\
It has been shown that the minimal $\Delta x$ derived from the GUP provides a potential method to avoid the point singularities in specific GR solutions, such as black hole space-times. If distances smaller than $\Delta x_m$ cannot be resolved, this may help bypass the singularities predicted by GR. This is one of the aspirations for theories of quantum gravity—to eliminate the singularities of classical GR. Another approach to avoiding these singularities involves non-commutative geometry \cite{gup9}, where coordinates do not commute with each other, for example, $[X, Y] \neq 0$ or $[Y, Z] \neq 0$.\\
A significant advantage of the phenomenological GUP approach to quantum gravity is its potential for experimental validation. It allows for tests to determine whether a minimal distance resolution, $\Delta x_m$, exists and, if so, to measure its scale. Some tests of the GUP scenario depend on astrophysical phenomena. For instance, Ref. \cite{gup10} proposed testing minimal lengths based on the dispersion of high-energy photons from short gamma-ray bursts. The idea was that a minimal distance scale, in theory, would modify the standard energy-momentum relation of special relativity, $E^2 = p^2 c^2 + m^2 c^4$. This revised relationship would result in an energy-dependent speed of light in a vacuum, causing photons of different energies to disperse as they travel long distances. In 2009, the Fermi gamma-ray satellite detected high-energy photons from a distant gamma-ray burst. Examining this data using the approach from \cite{gup10}, the Fermi satellite observations placed restrictions on deviations from $E^2 = p^2 c^2 + m^2 c^4$ due to a minimal distance scale. Surprisingly, if these deviations were linear in energy (i.e., $p^2 c^2 = E^2 [1 + \zeta(E/E_{\text{QG}}) + \ldots]$, where $E_{\text{QG}}$ is the quantum gravity scale and $\zeta$ is a parameter of order 1), the observations of \cite{gup11} implied that $E_{\text{QG}} > E_{\text{Planck}}$, signifying no deviation up to the Planck energy scale. This translates to a quantum gravity length scale $l_{\text{QG}} < l_{\text{Planck}}$, conflicting with the expectation that signs of quantum gravity would appear before reaching the Planck scale.
\subsection{A brief discussion on GUP models}
We now briefly review some foundational aspects of GUP models, focusing on how modified operators influence the existence of a minimal length. The uncertainty relationship between two physical quantities is closely related to the commutation relationship between the operators representing these quantities. Generally, for two operators \(\hat{A}\) and \(\hat{B}\), the relationship between the uncertainties and the commutator is given by
\begin{equation}
\Delta A \Delta B \geq \frac{1}{2i} \langle [\hat{A}, \hat{B}] \rangle,
\end{equation}
where the uncertainties are defined as \(\Delta A = \sqrt{\langle \hat{A}^2 \rangle - \langle \hat{A} \rangle^2}\) and similarly for \(\langle \hat{B} \rangle\). For standard position and momentum operators in position space, \(\hat{x} = x\) and \(\hat{p} = -i\hbar \partial_x\), the usual commutator \([\hat{x}, \hat{p}] = i\hbar\) leads to the standard uncertainty principle
\begin{equation}
\Delta x \Delta p \geq \frac{\hbar}{2}.
\end{equation}
Alternatively, in momentum space, the operators are \(\hat{x} = i\hbar \partial_p\) and \(\hat{p} = p\). To derive a GUP characterized by \(\Delta x \sim \frac{\hbar}{\Delta p} + \lambda \Delta p\), Ref. \cite{gup20} proposed the following modified commutator
\begin{equation}
[\hat{x}, \hat{p}] = i\hbar(1 + \lambda \hat{p}^2).
\end{equation}
Here, the parameter $\lambda$ replaces $G$ as a phenomenological constant that characterizes the scale at which quantum gravity effects become significant.
Various experiments have been proposed, like seeing the modified dispersion relation of a photon via gamma-ray burst \cite{gup10}; there is a phenomenological discussion in \cite{dasexp} on how the various experimental methods can probe the fundamental length scale of $\lambda$ like Lamb shift $\lambda<10^{36}$, Landau level $\lambda<10^{56}$, tunneling $\lambda<10^{21}$. For various other experiments, one can see \cite{gupmain,sabinegup} and the reference therein. Corrections of the Casimir effect due to GUP are also well known and are discussed in \cite{gupcasimir}. The application of minimal time scale and GUP has already been successfully used in solving the Wheeler-Dewitt equation for the Universe's evolution \cite{3}. In recent years, GUP has also been used to find dispersion relation during Hawking radiation of Schwarzschild de Sitter black holes \cite{4}, and one can also get a limit of minimal length scale already knowing the blackhole evaporation formula using semiclassical quantum gravity.\\
Note that position and momentum are no longer conjugate variables in the classical sense, meaning we cannot interpret the position eigenspace as the actual physical position due to the altered position-momentum relationship. To still refer to position, we can project states onto the maximally localized state, also known as the ``quasi-position representation", as discussed in \cite{gup20}.\\
There are two primary methods to obtain the physical position states or ``maximally localized states". The first method, proposed by Kempf, Mangano, and Mann \cite{gup20}, is the canonical approach. However, Detournay, Gabriel, and Spindel \cite{gup21} later demonstrated that the canonical approach does not encompass all squeezed states, which represent the ``maximally localized states". They employed a variational program to identify these states.\\
According to \cite{gupcasimir}, in n-dimensional space, the GUP can be defined as
\begin{equation}
 [x_i, p_j] = i\hbar[f(p^2)\delta_{ij} + g(p^2)p_ip_j].   
\end{equation}
These forms are dictated by spherical symmetry, and the functions \(f\) and \(g\) are not arbitrary, as discussed in \cite{gup20}.
The quantum state can be expressed as
\begin{equation}\label{1ch3b1}
\psi_r = \frac{1}{(\sqrt{2\pi\hbar})^3}\Omega(p)\exp\left({-\frac{i}{\hbar}[k(p) \cdot r - \hbar \omega(p)t]}\right),
\end{equation}
where \(\omega(p)\) denotes the dispersion relation, \(\Omega\) is the measure, and \(k\) is the wave vector.\\
We will discuss two popular GUP methods below: the KMM approach \cite{gup20}, which uses squeezed states, and the DGS approach \cite{gup21}, which employs the variational principle and is more general.
Further details on the field-theoretic formulations of GUP, including issues of ultraviolet divergence, can be found in \cite{gupcasimir,gup23}.\\
\textbf{KMM model:} The specific form of maximally localized states depends on the number of dimensions and the model considered. One approach, proposed in \cite{gup20} known as KMM model, involves selecting the generic functions \( f(\hat{p}^2) \) and \( g(\hat{p}^2) \) as given in \cite{gup23}
\begin{equation}\label{1ch3b2}
f(\hat{p}^2) = \frac{\lambda \hat{p}^2}{\sqrt{1+2\lambda \hat{p}^2} - 1}, \quad g(\hat{p}^2) = \lambda.
\end{equation}
Removing the hat notation for simplicity, the KMM construction of maximally localized states provides the following functions for Eq. \eqref{1ch3b1}
\begin{equation}
    \kappa_i(p)=\left(\frac{\sqrt{1+2\lambda p^2}-1}{\lambda p^2}\right)p_i,
\end{equation}
\begin{equation}
    \omega(p)=\frac{pc}{\hbar}\left(\frac{\sqrt{1+2\lambda p^2}-1}{\lambda p^2}\right),
\end{equation}
\begin{equation}
    \Omega(p)=\left(\frac{\sqrt{1+2\lambda p^2}-1}{\lambda p^2}\right)^{\frac{\delta}{2}},
\end{equation}
where $n$ denote the number of space-time dimensions, and $\delta=1+\sqrt{1+\frac{n}{2}}$ represents the KMM approach. Now we could determine the identity operator from the scalar product of maximally localized states
\begin{equation}
    \int \frac{d^n p}{\sqrt{1+2\lambda p^2}}\left(\frac{\sqrt{1+2\lambda p^2}-1}{\lambda p^2}\right)^{(n+\delta)}\vert p\rangle\langle p \vert=1.
 \end{equation}

\textbf{DGS model:}
As defined earlier, various maximally localized states may correspond to a given choice of generic functions \eqref{1ch3b2}. The DGS \cite{gup21} maximally localized forms are provided by Eq. \eqref{1ch3b1} with
\begin{equation}
    \kappa_i(p)=\left(\frac{\sqrt{1+2\lambda p^2}-1}{\lambda p^2}\right)p_i,
\end{equation}
\begin{equation}
    \omega(p)=\frac{pc}{\hbar}\left(\frac{\sqrt{1+2\lambda p^2}-1}{\lambda p^2}\right),
\end{equation}
\begin{equation}
\Omega(p)=\left[\Gamma\left(\frac{3}{2}\right)\left(\frac{2\sqrt{2}}{\pi\sqrt{\lambda}}\right)^{\frac{1}{2}}\right]\left[\frac{1}{p}\frac{\lambda p^2}{\sqrt{1+2\lambda p^2}-1}\right]^{\frac{1}{2}}\textbf{J}_{\frac{1}{2}} \left[\frac{\pi \sqrt{\lambda}}{\sqrt{2}}\left(\frac{\sqrt{1+2\lambda p^2}-1}{\lambda p^2}\right)p\right],
\end{equation}
where $\textbf{J}_{\frac{1}{2}}$ is the Bessel function of the first kind. Now by solving the above expression, one can obtain
\begin{equation}
\Omega(p)=\frac{\sqrt{2}}{\pi}\frac{\sqrt{\lambda}p}{\sqrt{1+2\lambda p^2}-1} \text{sin}\left[\frac{\sqrt{2}\pi(\sqrt{1+2\lambda p^2}-1)}{2\sqrt{\lambda}p}\right].
\end{equation}
The modified identity operator for the momentum eigenstates $\vert p\rangle$ for this case
\begin{equation}
    \int \frac{d^n p}{\sqrt{1+2\lambda p^2}}\left(\frac{\sqrt{1+2\lambda p^2}-1}{\lambda p^2}\right)^{n}\vert p\rangle\langle p \vert=1.
 \end{equation}
\section{Dark matter}
Our current understanding contains merely about 5\% of the Universe. The Lambda Cold Dark Matter ($\Lambda$CDM)  model \cite{dm1}, the most successful cosmological model to date, indicates that only 5\% of the Universe's mass-energy is baryonic matter, which we observe as stars, gas, and dust. The remaining mass-energy is predominantly composed of two `dark' components: dark energy (68\%) and dark matter (27\%) \cite{dm2,dm3}.
Dark matter is assumed to be non-baryonic (not composed of protons or neutrons), cold (having non-relativistic particle velocities at the epoch of radiation-matter equality), and collisionless (interacting only via gravity); for example, see Refs. \cite{dm4,dm5,dm6}. This is what makes dark matter so elusive and difficult to study. \\
Weakly Interacting Massive Particles (WIMPs) were one of the first leading candidates for dark matter particles \cite{dm8,dm9,dm10}. 
These particles have masses around $100$ $GeV$ and interact via forces similar in strength to the weak nuclear force. WIMPs were thought to have been created thermally in the early Universe, and particle physics theories predicted the existence of such particles. If WIMPs were produced this way, their current abundance would require a self-annihilation cross-section consistent with predictions, a concept known as the WIMPs Miracle \cite{dm11}. This made WIMPs a highly promising candidate for dark matter.
Despite extensive searches, no direct evidence of WIMPs has been found, leading to a decline in confidence regarding their role as dark matter \cite{dm12,dm13}. This has opened the door to other potential candidates, such as axions. Axions were initially proposed to resolve the strong charge-parity problem in quantum chromodynamics \cite{dm14,dm15}. They are supposed to interact weakly with standard model particles and remain non-relativistic (cold), fitting within a mass range of $10^{-5} - 10^{-2}$ $eV$. These characteristics position axions as a compelling dark matter candidate, attracting significant interest from cosmologists \cite{dm16}. Beyond WIMPs and axions, many other dark matter particles have been proposed. For comprehensive reviews of different candidates, see some Refs. \cite{dm17,dm18}.\\
Numerous experiments have been conducted to directly or indirectly detect dark matter particles (see Refs. \cite{dm19,dm20,dm21,dm22}). These experiments are often developed to match specific particle hypotheses, making the search similar to finding a needle in a cosmological haystack. To date, no dark matter particle has been definitively detected \cite{dm12}. In the absence of direct detection, we study the 5\% of the Universe's matter that we can observe to infer the properties of dark matter. By analyzing the effects of dark matter on visible, baryonic matter, we can limit its potential properties and improve the search for dark matter particles \cite{dm16}.
\subsection{Evidence of dark matter}
\begin{figure}[h]
    \centering
    \includegraphics[scale=0.6]{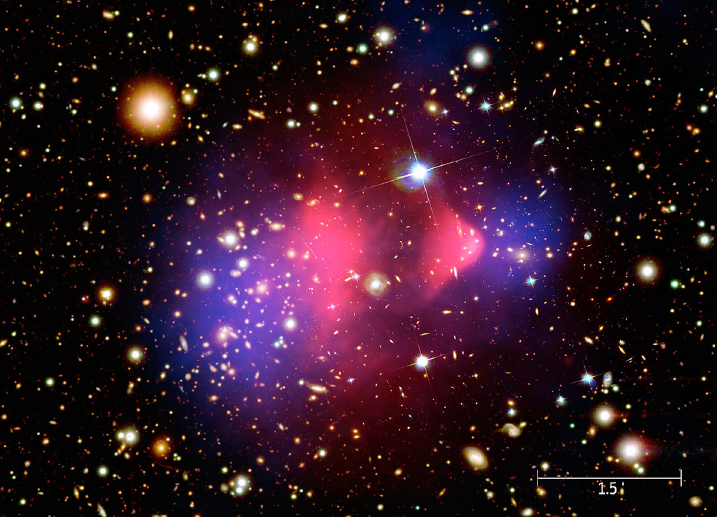}
    \caption{An image of the Bullet Cluster, showing the collision of two galaxy clusters. Image Credit: X-ray: NASA/CXC/CfA/M.Markevitch et al.; Optical: NASA/STScI;
Magellan/U.Arizona/D.Clowe et al.; Lensing Map: NASA/STScI; ESO WFI; Magellan/U.Arizona/D.Clowe et al.}
    \label{fig:in1}
\end{figure}

The term ``dark matter" is most commonly linked to Zwicky's work in 1933 \cite{dm26} (English version available in \cite{dm27}) and his follow-up research in 1937 \cite{dm28}. Zwicky studied eight galaxies within the Coma cluster and, assuming virial equilibrium, found their observed velocity dispersions to be around 1000 km/s, much higher than the theoretically predicted value of about 80 km/s. This indicated that the Coma cluster needed to be at least 400 times more massive than what was observed. Zwicky proposed that this discrepancy was due to non-luminous or dark matter. His hypothesis was supported by a similar study of the Virgo cluster \cite{dm29}.\\
Despite Zwicky's findings, the astronomical community did not seriously consider dark matter until the 1970s when galaxy rotation curve studies introduced by Rubin and Ford \cite{dm30} demonstrated strong evidence for missing mass within galaxies. They used improved spectroscopy technology to conduct detailed spectroscopic observations of the Andromeda galaxy (M31) using the 21 cm hydrogen emission line \cite{dm30}. Later, this work was followed by numerous other spectroscopic observations of galaxies (see Refs. \cite{dm31,dm32,dm33,dm34,dm35,dm36}).
These observations consistently revealed that galaxy rotation curves flattened out at large radii, maintaining a constant circular velocity instead of the expected decrease predicted by the Keplerian model. This unexpected flattening indicated the presence of additional, unseen mass. These findings showed that galaxies are embedded in dark matter halos that extend far beyond their visible components, supporting the concept that dark matter is abundant throughout the Universe, not just in galaxy clusters.\\
Since its initial discovery, additional evidence supporting the existence of dark matter has accumulated. Gravitational lensing, for instance, has played a pivotal role by revealing the necessity for additional mass within various cosmic structures. This phenomenon occurs when the gravitational field of a foreground object bends light rays from a background object \cite{dm37,dm38}. Analyzing these lensed images allows astronomers to infer the mass distribution of the foreground object. Gravitational lensing has proven to be a helpful method for studying and providing evidence of the presence of dark matter. A notable instance is a cosmic event involving the merger of two galaxy clusters, famously referred to as the Bullet cluster, depicted in Fig. \ref{fig:in1}. In this figure, the hot gas within the clusters, or the baryonic matter, is visualized in a pink region, as observed through X-ray telescopic analysis \cite{dm39,dm40}, whereas the dark matter component is shown in blue, identified through the technique of weak lensing \cite{dm41,dm42}. The distinct separation between these regions underscores dark matter's collisionless nature, contrasting with the interacting baryonic matter. Similar studies of other galaxy cluster collisions (e.g., \cite{dm43,dm44}) have reaffirmed these findings. \\
Evidence for dark matter extends beyond individual systems to larger cosmological scales. Particularly influential are cosmic abundance calculations derived from observations of the Cosmic Microwave Background (CMB). The CMB, discovered in \cite{dm45,dm46}, represents radiation from the early Universe, now cooled to approximately $2.7\,K$. Detailed mapping of the CMB, notably by COBE \cite{dm47,dm48,dm49}, WMAP \cite{dm50c,dm51c}, and most recently, Planck \cite{dm2,dm3}, reveals tiny temperature fluctuations that signify density variations in the primordial Universe. These variations seeded the formation of cosmic structures. By examining these temperature fluctuations across different angular scales, astronomers can probe the cosmological composition, including the contribution of dark matter. These observations broadly verify predictions from the $\Lambda$CDM model \cite{dm2,dm3}. Combining CMB data with detailed studies of galaxy abundances and clustering \cite{dm52c,dm53c,dm54c} provides further constraints on the properties of dark matter, guiding our understanding of its role in the cosmos.
\subsection{Dark matter halos}
Dark matter halos are dense concentrations of dark matter that have undergone a process known as virialization \cite{dm55c}. This approach typically results in regions with densities several hundred times greater than the average matter density of the Universe, which often defines the boundaries of these halos. One common definition is the virial boundary \cite{dm56c}, which generally depends on cosmology and evolves with cosmic time.
According to the hierarchical model of structure formation, the majority of the Universe's mass resides within these halos, making them fundamental for studying the distribution of matter on large scales. Halos serve as tracers of the underlying density field, and their clustering patterns provide insights into cosmology and the formation of cosmic structures. However, it is important to note that halos indicate a bias in their clustering behavior compared to dark matter itself, and understanding this bias is crucial for accurately interpreting clustering statistics.\\
Defining the edge of a dark matter halo is complex, so sizes are often discussed in terms of the virial radius, $r_{200}$, the radius within which the average density is $200$ times the critical density of the Universe at redshift $z = 0$. The virial mass, $M_{200}$, enclosed within this radius is defined as
\begin{equation}
M_{200} = (4/3) \pi 200 \rho_{\text{crit}} r_{200}^3,
\end{equation}
where $\rho_{\text{crit}}$ is the critical density. The virial mass relates to the halo's concentration through the mass-concentration relation \cite{dm57c,dm58c,dm59c}.
\subsection{Some dark matter profiles}
Different types of dark matter density profiles have been proposed in the literature. In this section, we shall briefly discuss some of the most common dark matter density profiles.
\subsubsection{Navarro-Frenk-White (NFW) profile}
The internal structure of a dark matter halo is often described by its density profile, $\rho(r)$. In 1991, Dubinski and Carlberg \cite{dm50} demonstrated that the density profiles of dark matter halos in N-body simulations could be effectively modeled employing a broken power law known as the Hernquist profile \cite{dm51}
\begin{equation}
\rho(r) = \rho_s \left(\frac{r}{r_s}\right)^{-\gamma} \left(1 + \frac{r}{r_s}\right)^{\gamma - \beta},
\end{equation}
where $\rho_s$ represents the effective density, and $r_s$ denotes the scale radius. In this profile, $\gamma$ represents the slope of the density near the halo's center ($\gamma=1$), while $\beta$ describes the slope in the outer regions ($\beta=4$). Later, Navarro, Frenk, and White \cite{dm52,dm53} proposed a universal profile for halos of varying masses, which slightly adjusts the outer slope to $\beta=3$, leading to the widely-used NFW profile
\begin{equation}\label{4aaa1}
\rho(r) = \rho_s \left(\frac{r}{r_s}\right)^{-1} \left(1 + \frac{r}{r_s}\right)^{-2}.
\end{equation}
Advances in simulation resolution have revealed that this profile is not universally applicable, as $\gamma$ can gradually change with radius \cite{dm54,dm55,dm56}. An alternative model is the three-parameter Einasto profile \cite{dm57}, which has been found to match simulation results better \cite{dm54}. Despite this, the NFW profile remains a useful and commonly applied model, particularly for smaller halos like dwarf galaxies, due to its relative simplicity \cite{dm58c}.

\subsubsection{Bose-Einstein Condensate (BEC) profile}
Some idea suggests that dark matter may be a collection of ultra-light scalar particles with mass at the order of magnitude of \( 10^{-22}\) eV, which would form BEC, first introduced in \cite{dm60,dm61} and further developed in \cite{dm62}. The wave-like properties of these particles are believed to prevent gravitational collapse, forming stable halo cores and significantly reducing small-scale power. The work in \cite{dm63} also studies the behavior of BEC in both atomic vapors and galactic dark matter. A cosmological model was also proposed where bosonic dark matter slowly condenses into dark energy \cite{dm64}, resulting in the Universe's accelerated expansion and the rapid formation of black holes from the smallest fluctuations, which then act as seeds for the first galaxies. Additionally, interactions mediated by scalar particles among baryons within a Bose-Einstein condensate, extending beyond the Compton wavelength, have been explored in \cite{dm65}. If the Universe's dark matter is indeed in such a condensate state, interactions between baryonic matter and dark matter could potentially leave detectable anomalies in the CMB peak structure. In a medium made of scalar particles with non-zero mass, scalar-mediated interactions between nucleons become infinite when the medium transitions to a BEC phase. This phenomenon has been studied in astrophysical contexts \cite{dm66}, including its impact on the equilibrium of degenerate stars.\\
In a quantum system consisting of $N$ interacting condensed bosons, most of the particles occupy the same single-particle quantum state. Directly calculating the ground state of such a system, especially when dealing with a large number of particles, is often impractical due to the high computational demands. However, certain approximation methods can simplify this complex process. One effective approach is the mean-field theory, which simplifies the problem by isolating the condensate's contribution to the bosonic field operator. Additionally, in a medium made of scalar particles with non-zero mass, the range of Van-der-Waals type scalar interactions among nucleons becomes theoretically infinite when the medium transitions to a Bose-Einstein condensate phase. The many-body Hamiltonian that describes these interacting bosons, confined by an external potential $V_{\text{ext}}$, can be expressed in the framework of second quantization as \cite{dm67}
\begin{equation}\label{31}
\hat{H}=\int d\Vec{r}\hat{\Psi}^{\dag}(\Vec{r})\left[-\frac{\hbar}{2m}\bigtriangledown^2+V_{rot}\Vec{r}+V_{ext}\Vec{r}\right]\hat{\Psi}(\Vec{r})
+\frac{1}{2}\int d\Vec{r} d\Vec{r}^{'}\hat{\Psi}^{\dag}(\Vec{r})\hat{\Psi}^{\dag}\Vec{r}^{'}V(\Vec{r}-\Vec{r}^{'})\hat{\Psi}(\Vec{r})\hat{\Psi}(\Vec{r}^{'}),
\end{equation}
where $\hat{\Psi}(\Vec{r})$ and $\hat{\Psi}^{\dag}(\Vec{r})$ are boson field operators that responsible for annihilating and creating a particle at position $\hat{r}$ respectively, and the term $V(\Vec{r}-\Vec{r}')$ is the two-body interatomic potential between particles \cite{dm68}. Also, $V_{rot}(\Vec{r})$ represents the potential corresponding to the rotation of the condensate, defined as
\begin{equation}
V_{rot}(\Vec{r}) = f_{\text{rot}}(t) \frac{m\omega^2}{2} \Vec{r}^2,
\end{equation}
where the parameter $\omega$ represents the angular velocity of the condensate, whereas $f_{rot}(t)$ is a function that accounts for any time-dependent variations of the rotational potential.
Now, to find the radius $r_s$ of the BEC, we begin by assuming that $V_{\text{ext}}(\Vec{r})$ corresponds to the gravitational potential $V$, which satisfies the Poisson equation \cite{dm67}
\begin{equation}
\nabla^2 V = 4\pi G \rho_m,
\end{equation}
with $\rho_m = m\rho$ being the mass density in the BEC. By focusing on the first approximation and neglecting the rotation of the BEC (i.e., $V_{\text{rot}} = 0$), one finds the radius $r_s$ of the BEC as follows \cite{dm67}
\begin{equation}
r_s = \pi \sqrt{\frac{\hbar^2 \varepsilon}{G m^3}}.
\end{equation}
Here, the scattering length $\varepsilon$ is associated with the scattering cross-section of the particles in the condensate. Various studies in the literature have estimated the mass and scattering length of condensate dark matter particles. For instance, astrophysical observations of the Bullet Cluster suggest a scattering length $\varepsilon$ on the order of $10^{-7} \mathrm{fm}$, with the mass of the dark matter particle being roughly in the range of micro-electron volts ($\mu eV$) \cite{dm68}.\\
Now, the density distribution of the dark matter BEC can then be expressed as \cite{dm69}
\begin{equation}\label{becd1}
\rho(r) = \rho_s \frac{\sin(kr)}{kr},
\end{equation}
where $\rho_s$ is the central density of the condensate and $k=\frac{\pi}{r_s}$.
\subsubsection{Pseudo Isothermal (PI) profile}
There is an important class of dark matter models associated with modified gravity, such as Modified Newtonian Dynamics (MOND) \cite{Begeman}. In the MOND model, the dark matter density profile is described by the PI profile
\begin{equation}\label{4aa1}
\rho_\text{PI}=\frac{\rho_s}{1+\left(\frac{r}{r_s}\right)^2},
\end{equation}
where $\rho_s$ is the central dark matter density and $r_s$ is the scale radius. Unlike the NFW profile, which predicts a sharp density increase (or ``cusp") at the center, the PI profile features a flat central region that better matches the observed rotation curves of low surface brightness and dwarf galaxies. This makes the PI profile especially effective in cases where the NFW model falls short of accurately describing galaxy dynamics \cite{Begeman}. Many studies have highlighted the success of the PI profile in matching observed galactic data. For example, de Blok et al. \cite{Frenk4} compared the PI and NFW profiles and found that the PI model's flat core is a better fit for the observed density profiles of low surface brightness galaxies, addressing the ``core-cusp problem". Likewise, Gentile et al. \cite{pi1} analyzed spiral galaxy rotation curves and found that the PI profile consistently fits observed data across various radii, further confirming its applicability. The PI profile has also been effective in modeling dwarf galaxies, as demonstrated by Oh et al. \cite{pi2}, who used it to accurately represent the dark matter distribution in galaxies from the THINGS survey without requiring additional adjustments. In the context of wormhole geometry, the PI profile has been examined in \cite{pi3}, and it was found that the dark matter density for an axisymmetric traversable wormhole is similar to that of a black hole spike. However, the dark matter density varies with the wormhole's spin in the opposite direction. This versatility emphasizes the potential of the PI profile in both observational and theoretical physics.
\subsubsection{Universal Rotation Curve (URC) profile}
The URC model posits that the density of dark matter within a galaxy diminishes as one moves further from the galaxy's center, in contrast to the visible matter, which is primarily focused near the center. This model derives from observations indicating that the rotation curves of galaxies do not decrease with distance from the center as would be expected if only visible matter were present. This implies the existence of an additional, invisible mass component that accounts for the high rotation speeds of stars and gas in the outer regions of galaxies. The URC dark matter density profile is widely accepted because it applies across the galactic halo's central and outer regions.\\
The URC dark matter density profile is given by \cite{dm70}
\begin{equation}\label{urc1}
\rho(r) = \frac{\rho_s\, r_s^3}{(r + r_s)(r^2 + r_s^2)},
\end{equation}
where $r_s$ and $\rho_s$ are the core radius and central density. For instance, in the case of the Milky Way galaxy, $r_s = 9.11$ kpc and $\rho_s = 5\times 10^{-24}\,g/cm^3$, respectively.\\
The above URC density profile reproduces the isothermal profile
\begin{equation}
\rho(r)=\frac{\rho_s}{1+\frac{r^2}{r_c^2}},
\end{equation}
in the inner regions ($r \ll r_s$) and a finite central density $\rho_0$. Here, $r_s$ is identical to the core radius $r_c$ of the isothermal fit formula. At larger radii, the mass distribution of the isothermal profile would diverge proportionally with $r$. However, Eq. \eqref{urc1} results in mass profiles that diverge logarithmically as $r$ increases, aligning with the predictions from cosmological CDM models, such as those proposed by Navarro et al. \cite{dm71}.
Another critical aspect of this model is the rotation curve, which is a key indicator supporting the presence of dark matter in galaxies. Persic et al. \cite{dm72} analyzed using $H_\alpha$ data, supplemented by certain radio rotation curves, and observed a wide range of rotation curves across different galaxies. Their study revealed that rotation curves could be effectively modeled for various luminosities and galaxy types, including spirals, low-surface-brightness ellipticals, and dwarf-irregular galaxies. As a result, they introduced the concept of a ``universal rotation curve" to reflect the widespread applicability of this phenomenon, replacing the traditional term ``rotation curves". Overall, the URC profile provides a valuable framework for analyzing the distribution of dark matter in galaxies and understanding its role in their dynamics. It is important to note that the shapes of rotation curves can vary from one galaxy to another, and researchers use different equations and models to account for these variations when interpreting their data.
\section{Basis of theoretical understanding}
In this section, we revisit some core principles that form the foundation of GR. Any alternative theories of gravity, by necessity, must align with at least some of these foundational concepts. Our main goal here is to familiarize the reader with the essential ideas and principles that GR encompasses.
\subsection{A theory of principle}
It has been noted \cite{gr1} that what is commonly called ``experimental tests of GR", such as the bending of light, the precession of Mercury's orbit, and the gravitational redshift seen in distant cosmic sources like galaxies, are fundamentally tests of underlying principles rather than direct examinations of the theory itself. This perspective provides a foundation for developing viable gravity theories. GR, as a well-established theory, aligns closely with this foundational framework. Dicke \cite{gr2} suggested two basic assumptions:
\begin{itemize}
    \item All physical events occur within a 4-dimensional manifold known as space-time.
    \item The equations governing physical laws should not depend on the choice of coordinates—this is the Principle of Covariance.
\end{itemize}
This conceptual approach is often referred to as the Dicke framework. Regarding the first assumption, it is important to clarify that it does not inherently imply the existence of additional structures like a metric or affine connection within space-time. The second assumption deserves careful consideration, as some researchers, like Wald \cite{gr3}, offer alternative interpretations of covariance, distinguishing between\\
\textbf{General Covariance:} There are no preferred vector fields or bases of vector fields related solely to the structure of space that appear in any physical law.\\
\textbf{Special Covariance:} If there is a family of observers, denoted by $\mathcal{O}$, and another family, $\mathcal{O}'$, derived from $\mathcal{O}$ through an isometric transformation, then if $\mathcal{O}$ measures a physical field, $\mathcal{O}'$ should be able to perform the exact measurement, ensuring consistency across both sets of observers.\\
Dicke further added two additional criteria: gravity should be associated with one or more fields of a tensorial nature (such as scalar, vector, or tensor fields), and the field equations should be derivable from an invariant action using a stationary action principle.
\subsection{The principle of equivalence}
The principle of equivalence is one of the fundamental principles in GR and gravitational theories. This principle is said to have played a key role in guiding Einstein during the development of his theory of gravity \cite{gr4}, and it has been a topic of significant debate ever since. It should also be noted that even in Newtonian Mechanics, a form of this principle existed, which appeared at the start of Newton's \textit{Principia} (see for example \cite{gr2}, page 13 and Figure 2.1). This is the principle of inertia, which states that in a Newtonian perspective, inertial mass and passive gravitational mass are equal: $m_I = m_p$ \footnote{The inertial mass is that present in Newton's third law $\vec{F}=m_I\vec{a}$ and the passive gravitational
mass is the one in Newton’s gravitational law $\vec{F}=m_p\vec{g}$.}. Put more straightforwardly, the principle states that all objects in free-fall experience the same acceleration regardless of their composition or mass.\\
Current interpretations and variations of this principle can be found in the literature \cite{gr1}.\\
\textit{Weak Equivalence Principle} (WEP): The path taken by an uncharged test particle\footnote{For a related definition, check Ref. \cite{gr2}.}, regardless of its initial conditions, is unaffected by its internal structure or composition.\\
\textit{Einstein Equivalence Principle} (EEP): A natural extension of the statement in WEP, this principle asserts that the outcomes of non-gravitational experiments are independent of both the velocity of the experiment (Local Lorentz Invariance - LLI) and the position in space-time (Local Position Invariance - LPI).\\
\textit{Strong Equivalence Principle} (SEP): This is considered a generalization of the WEP with regard to self-gravitating bodies, which must satisfy WEP along with any local experiment remaining unchanged over transformations under LLI and LPI. \\
The Schiff conjecture is that ``any complete, self-consistent theory of gravity must have the EEP in addition to the WEP” (refer to \cite{gr2}, Section 2.5).\\
Note that this is only one view of the equivalence principle. A second but still physically equally valid interpretation is that there is an accelerated reference frame where ``the very nature of gravity vanishes, and the local mechanical motion of other bodies will follow their paths as if unperturbed by gravity"\footnote{Except for possible tidal effects due to inhomogeneities in the gravitation field.} \cite{gr2}, a similar example can be found in \cite{gr5}. Additionally, due to variations in terminology, what we refer to as the WEP is sometimes labeled as Galileo's principle \cite{gr55}.\\
However, there are certain limitations when using these equivalence principles to develop modified theories of gravity. Three key issues are worth noting: the first is the connection between the SEP and GR. It has been argued that SEP holds exclusively for GR, though no definitive proof has been provided. The second issue involves defining `test particles', precisely determining how small a particle must be to disregard its gravitational field, which likely depends on the specific theory. The third concern is that some theories may appear to either comply with or violate a given equivalence principle based on the variables chosen for the description; this is particularly evident in scalar-tensor theories and the distinction between the Jordan and Einstein frames.
\subsection{Metric postulates}
The principles summarized above form the foundation upon which a theory must be constructed. But what happens if we are presented with such a theory, either in the form of an action or field equations, and we wish to determine whether it adheres to certain principles? What mathematical expressions do these principles take? Conversely, when a principle is established, how does it manifest mathematically, and what constraints does it impose on the theory? Thorne and Will \cite{gr6} addressed these questions in 1971 when they introduced what are now known as the metric postulates. They proposed the following:
\begin{itemize}
    \item[1.]  In space-time, a non-degenerate second-rank tensor is known as the metric tensor, denoted as \(g_{\mu\nu}\).
    \item[2.] If \(T_{\mu\nu}\) represents the stress-energy tensor associated with non-gravitational matter fields, and if \(\nabla_{\mu}\) is the covariant derivative derived from the Levi-Civita connection associated with the metric, then \(\nabla_{\mu} T^{\mu\nu} = 0\).
\end{itemize}
Theories that satisfy these metric postulates are referred to as metric theories. Two key points regarding these postulates are worth noting: first, geodesic motion can be derived from the second postulate \cite{gr7}; second, the definition of \(T_{\mu\nu}\), as well as the concept of non-gravitational fields, lacks precision and clarity \cite{gr8}.\\
Despite these ambiguities, the metric postulates have proven to be extremely useful. They serve as a foundation for the parameterized post-Newtonian (PPN) expansion, a framework that has been widely used to test and constrain alternative theories of gravity. However, they also have significant limitations. As noted by Thorne and Will \cite{gr6}, any metric theory can be reinterpreted in a way that appears to contradict the metric postulates. 
Before concluding this subsection, it is essential to highlight that the principles discussed here represent just one possible set among many. For example, \cite{otr1} offers an alternative set of principles.
\section{Space-time: A geometric perspective}
In GR, the fundamental idea is to interpret geometry as a way to describe space-time, specifically focusing on differential geometry. At the core of differential geometry, this is the concept of a manifold, which can be viewed as a mathematical space that, on a small scale, resembles the familiar Euclidean space $\mathbb{R}^n$. Essentially, space-time is represented as a four-dimensional manifold where a symmetric, non-degenerate metric $g_{\alpha\beta}$ is defined; meaning the determinant of $g$ is non-zero. Additionally, there is a concept known as the connection, represented as $\tilde{\Gamma}^\alpha_{\beta\gamma}$ (for further information on the connection, see Ref. \cite{sp1}), which describes how vectors are transported along a curve while maintaining their orientation consistent with the underlying space. This connection is essential in defining a derivative that is suitable for curved space-time, known as the covariant derivative, represented as $\nabla$. For a tensor $T^\alpha_\beta$, the covariant derivative is given by
\begin{equation}
\nabla_\gamma T^\alpha_\beta = \partial_\gamma T^\alpha_\beta + \tilde{\Gamma}^\gamma_{dc} T^\delta_\beta - \tilde{\Gamma}^\delta_{\beta\gamma} T^\alpha_\delta.
\end{equation}
In the framework of GR, the connection $\tilde{\Gamma}^\gamma_{\alpha\beta}$ is related to the metric $g_{\alpha\beta}$ and is known as the Levi-Civita connection. The Levi-Civita connection is derived to ensure that the connection is torsion-free and compatible with the metric.
Furthermore, the energy-momentum tensor $T_{\alpha\beta}$ and the covariant derivative $\nabla_\gamma$, which arises from the Levi-Civita connection, must satisfy the condition $\nabla_\gamma T^{\alpha\beta} = 0$. To ensure this condition holds, the connection must be symmetric with respect to its lower indices, i.e.,
\begin{equation} \Gamma^\alpha_{\beta\gamma} = \Gamma^\alpha_{\gamma\beta}. 
\end{equation}
Also, the metric must be preserved under covariant derivative, which means \(\nabla_\gamma g_{\alpha\beta} = 0\). With these conditions, the connection takes the form of the Levi-Civita connection, expressed as
\begin{equation}\label{cr1}
\Gamma^\alpha_{\beta\gamma} = \frac{1}{2} g^{\alpha\delta} \left( \partial_\beta g_{\gamma\delta} + \partial_\gamma g_{\delta\beta} - \partial_\delta g_{\beta\gamma} \right).
\end{equation}
Furthermore, the Riemann tensor represents the curvature of the manifold, reflecting the curvature of space-time. This tensor can be computed from the connection as follows
\begin{eqnarray}
R^\alpha_{\beta\gamma\delta} = \partial_\gamma \Gamma^\alpha_{\beta\delta} - \partial_\delta \Gamma^\alpha_{\beta\gamma} + \Gamma^\tau_{\beta\delta} \Gamma^a\alpha_{\tau\gamma} - \Gamma^\tau_{\beta\gamma} \Gamma^\alpha_{\tau\delta}.
\end{eqnarray}
The Ricci tensor is obtained by contracting the Riemann tensor's first and third indices as follows
\begin{equation}
R_{\alpha\beta} = R^\gamma_{\alpha\gamma\beta} = -R^\gamma_{\alpha\beta\gamma}.
\end{equation}
At last, the Ricci scalar, which describes the curvature, is obtained by contracting the Ricci tensor with the metric. It can be read as
\begin{equation}
R = g^{\alpha\beta} R_{\alpha\beta}. 
\end{equation}

\section{General Relativity}
Now, to explain the dynamics of the gravitational field, we can draw a close analogy to the Poisson equation, $\nabla^2 \Psi = 4\pi\rho$, which governs the gravitational potential in Newtonian gravity. In the empty space of the Universe, the equations of GR simplify to $R_{\mu\nu} = 0$. It is important to note that since the Ricci tensor involves the metric as a second-order differential expression, the equation $R_{\mu\nu} = 0$ is similar to Laplace's equation, $\nabla^2 \Psi = 0$.\\
We need to consider additional factors to extend this analogy to situations involving matter. In GR, the metric is the fundamental field, and any other fields are considered ``matter fields" that influence the gravitational field. Therefore, gravity is exclusively associated with the metric's second-rank tensor field. Accordingly, the field equation should have a left side that depends only on the metric and a right side that represents the influence of all other fields. The energy-momentum tensor \(T_{\mu\nu}\) acts as the source of the gravitational field, similar to mass density in the Poisson equation. For a detailed discussion, one can refer to Ref. \cite{sp2}, and for challenges related to defining this quantity, see Refs. \cite{sp3,gr8}. Also, to maintain the analogy with the Poisson equation, the field equation in GR must also be a second-order differential equation.\\
To summarize, GR is based on the following key assumptions:
\begin{itemize}
    \item The connection must be symmetric in its lower indices, meaning \(\Gamma^\mu_{\nu\gamma} = \Gamma^\mu_{\gamma\nu}\).
    \item The metric must be covariantly conserved, i.e., \(\nabla_\gamma g_{\mu\nu} = 0\).
    \item The gravitational field is determined solely by the metric; no other fields contribute to this interaction.
    \item The field equations must be second-order partial differential equations.
    \item The field equations must be covariant.
\end{itemize}
Using these assumptions, we can follow Einstein's original derivation to arrive at the field equations of GR
\begin{equation}\label{9}
G_{\mu\nu}\equiv R_{\mu\nu}-\frac{1}{2}g_{\mu\nu}R=\frac{8\pi G}{c^4} T_{\mu\nu}.
\end{equation}
Here, $G_{\mu\nu}$ represents the Einstein tensor, and $T_{\mu\nu}$ describes the energy-momentum component of the Universe. 
It is worthy of mentioning here that the Einstein equation can be derived using the variation principle from varying the Einstein-Hilbert action
\begin{equation}\label{gr10}
S=\frac{1}{2\kappa}\int d^4x\sqrt{-g} R+\int d^4x\sqrt{-g} \mathcal{L}_m,
\end{equation}
where $\kappa\equiv 8\pi G/c^4$ and $\mathcal{L}_m$ is the matter Lagrangian. In the above action, the first term is the gravitational part, and the second term is the matter part.
\section{The geometrical trinity of gravity}
One of the key characteristics of metric-affine geometry is that when a metric is present, the components $\tilde{\Gamma}^{\alpha}_{\beta\gamma}$ of a general affine connection can be distinctly decomposed into three terms as follows
\begin{equation}\label{gt1}
\tilde{\Gamma}^{\alpha}_{\;\:\beta\gamma}={\Gamma}^{\alpha}_{\;\:\beta\gamma} + K^{\alpha}_{\;\:\beta\gamma} + L^{\alpha}_{\;\:\beta\gamma},
\end{equation}
where the component ${\Gamma}^{\alpha}_{\beta\gamma}$ represents the Levi-Civita connection
\begin{equation}
{\Gamma}^{\alpha}_{\;\:\beta\gamma}=\frac{1}{2} g^{\alpha\sigma} (\partial_{\beta} g_{\sigma\gamma} + \partial_{\gamma} g_{\sigma\beta} - \partial_{\sigma} g_{\beta\gamma}).
\end{equation}
$K^{\alpha}_{\;\:\beta\gamma}$ is the contortion tensor which is given by
\begin{equation}
K^{\alpha}_{\;\:\beta\gamma}=\frac{1}{2}{T^{\alpha}}_{\beta\gamma}+T_{(\beta}{}^{\alpha}{}_{\gamma)},
\end{equation}
with the torsion tensor ${T^{\alpha}}_{\beta\gamma}\equiv 2{\tilde{\Gamma}^{\alpha}}_{\;\:[\beta\gamma]}$.\\
$L^{\alpha}_{\;\:\beta\gamma}$ known as the disformation tensor
\begin{equation}
L^{\alpha}_{\;\:\beta\gamma}=\frac{1}{2} g^{\alpha\sigma}\left(-Q_{\beta\sigma\gamma}-Q_{\gamma\sigma\mu}+Q_{\sigma\beta\gamma}\right),
\end{equation}
where $Q_{\alpha\beta\gamma}$ is the non-metricity tensor.
The components ${\Gamma}{}^{\alpha}_{\beta\gamma}$, $K^{\alpha}_{\beta\gamma}$, and $L^{\alpha}_{\beta\gamma}$ are responsible for quantifying curvature, torsion, and non-metricity, respectively. Note that the above decomposition \eqref{gt1} of the affine connection co-efficient $\tilde{\Gamma}^{\alpha}_{\;\:\beta\gamma}$ for each metric affine geometry implies an interesting consequence. 
The curvature tensor \(\gc{R}^{\alpha}{}_{\beta\gamma\sigma}\) can also be allow similar decomposition
\begin{equation}\label{eq:gennaffcurvdec}
\gc{R}^{\alpha}{}_{\beta\gamma\sigma} = R^{\alpha}{}_{\beta\gamma\sigma} +\nabla_{\gamma}\gc{D}^{\alpha}{}_{\beta\sigma} - \nabla_{\sigma}\gc{D}^{\alpha}{}_{\beta\gamma} + \gc{D}^{\alpha}{}_{\tau\gamma}\gc{D}^{\tau}{}_{\beta\sigma} - \gc{D}^{\alpha}{}_{\tau\sigma}\gc{D}^{\tau}{}_{\beta\gamma}\,,
\end{equation}
where the distortion tensor $\gc{D}^{\alpha}{}_{\beta\gamma}$ is defined as
\begin{equation}
\gc{D}^{\alpha}{}_{\beta\gamma} = \tilde{\Gamma}^{\alpha}_{\;\:\beta\gamma}-\Gamma^{\alpha}_{\;\:\beta\gamma} = {K}^{\alpha}{}_{\beta\gamma} + {L}^{\alpha}{}_{\beta\gamma}\,.
\end{equation}
Two noteworthy cases arise when the curvature tensor vanishes \(\gc{R}^{\alpha}{}_{\beta\gamma\sigma} \equiv 0\). The first scenario involves the metric teleparallel connection $\Gamma^{\alpha}_{\;\:\beta\gamma} \equiv \tilde{\Gamma}^{\alpha}_{\;\:\beta\gamma}$, allowing the Riemann tensor of the Levi-Civita connection to be expressed as
\begin{equation}
{R}^{\alpha}{}_{\beta\gamma\sigma} = K^{\alpha}{}_{\tau\sigma}K^{\tau}{}_{\beta\gamma} - K^{\alpha}{}_{\tau\gamma}K^{\tau}{}_{\beta\sigma} + {\nabla}_{\sigma}K^{\alpha}{}_{\beta\gamma} - {\nabla}_{\gamma}K^{\alpha}{}_{\beta\sigma}\,,
\end{equation}
involving the contortion tensor. The second case considers the symmetric teleparallel connection $\Gamma^{\alpha}_{\;\:\beta\gamma} \equiv \tilde{\Gamma}^{\alpha}_{\;\:\beta\gamma}$, where the Riemann tensor is formulated using the disformation tensor
\begin{equation}
{R}^{\alpha}{}_{\beta\gamma\sigma} = {L}^{\alpha}{}_{\tau\sigma}{L}^{\tau}{}_{\beta\gamma} - {L}^{\alpha}{}_{\tau\gamma}{L}^{\tau}{}_{\beta\sigma} + {\nabla}_{\sigma}{L}^{\alpha}{}_{\beta\gamma} - {\nabla}_{\gamma}{L}^{\alpha}{}_{\beta\sigma}\,.
\end{equation}
From these decompositions, the Ricci scalar in the metric teleparallel framework can be read as
\begin{equation}\label{gt2}
{R} = K^{\alpha}{}_{\gamma\beta}K^{\gamma\beta}{}_{\alpha} - K^{\alpha}{}_{\gamma\alpha}K^{\gamma\beta}{}_{\beta} - 2{\nabla}_{\alpha}K^{\alpha\beta}{}_{\beta} = -T + 2{\nabla}_{\alpha}T_{\beta}{}^{\beta\alpha}\,,
\end{equation}
while in the symmetric teleparallel scenario, it becomes
\begin{equation}\label{gt3}
{R} = {L}^{\alpha\beta\gamma}{L}_{\gamma\alpha\beta} - {L}^{\alpha}{}_{\alpha\gamma}{L}^{\gamma\beta}{}_{\beta} + {\nabla}_{\beta}{L}_{\alpha}{}^{\alpha\beta} - {\nabla}_{\alpha}{L}^{\alpha\beta}{}_{\beta} = -{Q} + {\nabla}_{\beta}{Q}_{\alpha}{}^{\alpha\beta} - {\nabla}_{\alpha}{Q}^{\alpha\beta}{}_{\beta}\,.
\end{equation}
The Ricci tensor plays a crucial role in the gravitational component of the Einstein-Hilbert action in GR
\begin{equation}\label{gt4}
S_{EH} = \frac{1}{2\kappa} \int d^4x \sqrt{-g} \, R.
\end{equation}
Utilizing a metric or symmetric teleparallel geometry allows for the transformation of the Einstein-Hilbert action into alternative forms. Both Ricci scalar expressions \eqref{gt2} and \eqref{gt3} contain total divergence terms, which contribute to boundary terms when inserted into the Einstein-Hilbert action \eqref{gt4}. Omitting these boundaries and matter terms leads to the following actions
\begin{equation}
S_{TEGR} = -\frac{1}{2\kappa} \int d^4x \sqrt{-g} \, T,
\end{equation}
for the Teleparallel Equivalent of General Relativity (TEGR), and
\begin{equation}
S_{STEGR} = -\frac{1}{2\kappa} \int d^4x \sqrt{-g} \, Q,
\end{equation}
for the Symmetric Teleparallel Equivalent of General Relativity (STEGR) \cite{sp5}. These formulations, collectively known as the \textit{geometric trinity of gravity} \cite{sp6}, yield dynamics identical to those of general relativity, highlighting their equivalence in treating the metric as the sole fundamental field variable.

\section{Understanding the need for modifications in GR}
It is known that our Universe is believed to be homogeneous and isotropic on cosmological scales and is described by the following flat FLRW metric
\begin{equation}
ds^2=-dt^2+a^2(t)d\Bar{x}^2=-a^2(\eta)\left[d\eta^2-d\Bar{x}^2\right],
\end{equation}
where $\Bar{x}$ is the $3-$space. $t$ and $\eta$ represent the cosmic time and conformal time, respectively.
The $\Lambda$CDM model, regarded as the standard framework of cosmology, aligns well with most of the observational data \cite{sp7}. This model assumes that GR governs gravitational behavior at cosmological scales. As suggested by its name, $\Lambda$CDM posits that dark matter and dark energy, represented by the cosmological constant $\Lambda$, dominate the Universe's total energy composition.\\
However, the success of this standard model comes with significant trade-offs, as ordinary atomic matter constitutes less than 5\% of the Universe. Furthermore, the true nature of dark matter and dark energy remains to be discovered, with their presence inferred solely through gravitational effects, as neither has been directly detected. The $\Lambda$CDM model also faces numerous theoretical and observational challenges, some of which are highlighted below:
\begin{itemize}
    \item \textbf{Fine-tuning problem:} Observations estimate that the current energy density of the cosmological constant is around $\rho_\Lambda \sim 10^{-47} GeV$ \cite{sp8,Padmanabhan/2003}. If we assume that this constant arises from vacuum energy density, there is a significant discrepancy between this value and the vacuum energy density predicted by quantum field theory, which is about $\rho_{vac}\sim 10^{74}$ GeV \cite{sp9}. In simple terms, the observed value of $\rho_\Lambda$ clashes with expected energy scales and suggests a need for extreme fine-tuning.
    \item \textbf{Coincidence problem:} Cosmological observations indicate that the present value of the dark energy density parameter is approximately $\Omega_{\Lambda}^{(0)}\sim 0.7$, which is of the same order as the current matter-energy density parameter $\Omega_m^{(0)}\sim 0.3$. This is puzzling because, while matter-energy density $\rho_m$ changes over time, the dark energy density $\rho_\Lambda$ remains constant, yet they coincide closely at the present epoch. This apparent coincidence suggests that the parameters of the early Universe would need fine-tuning, leading to what is known as the cosmological coincidence problem \cite{sp10}. It is noteworthy that this issue arises in many other dark energy models as well \cite{sp11}.
    \item \textbf{Tensions between early and late Universe observations:} While the $\Lambda$CDM model generally aligns with cosmological observations, including those from the CMB and late-time acceleration, recent data reveals tensions within this framework \cite{sp12}. Notably, the value of the Hubble constant $(H_0)$ derived from late Universe observations shows a $5\sigma$ tension compared to the $H_0$ value obtained from CMB observations \cite{sp13}. Estimates of $H_0$ from early Universe data yield higher values than those derived from local Universe observations. With improving observational precision, it has been suggested that resolving this $H_0$ tension may require looking beyond the $\Lambda$CDM model. Another significant cosmological tension involves observations of large-scale structures and corresponding parameters derived from CMB data. This $3\sigma$ tension is observed in the constraints on the matter density parameter $\Omega_m$ and the amplitude $\sigma_s$ of matter fluctuations estimated from local Universe measurements versus CMB observations \cite{sp14,sp15}.
\end{itemize}
Several modified theories of gravity have been proposed to deal with these challenges \cite{a16,sp16}. These alternatives include modified gravity models involving the modification of geometry \cite{sp17} as well as scalar-tensor theories \cite{sp10}. Research has demonstrated that these models can account for the observed late-time evolution of the Universe. However, based on current cosmological observations, most of these models are not preferred over the $\Lambda$CDM model. Therefore, it is crucial to develop alternative models that more closely match observational data.
\section{Modified gravity theories: representative examples}
Modified gravity takes a more unconventional approach by modifying the geometric side of Einstein's equations rather than modifying the stress-energy tensor. This means that changes are made to the structure of gravity itself, resulting in field equations that differ from those of GR. Below are some ways GR can be modified:
\begin{itemize}
    \item \textit{Higher-order field equations:} Einstein's equations involve derivatives only up to second order, but some theories introduce higher-order derivatives in the field equations. In contrast, these theories are often mathematically complex and can introduce Ostrogradski's instability \cite{mg1}, which leads to an unstable Hamiltonian. However, certain theories, like $f(R)$ gravity \cite{mg2}, manage to avoid this instability despite having fourth-order field equations.
    \item \textit{Higher dimensions:} Theories with higher dimensions extend beyond the familiar 3+1 space-time of GR by introducing additional spatial or temporal dimensions. Although our observable Universe seems to be four-dimensional, higher-dimensional models are valuable at high energies. Examples include the Kaluza-Klein theory \cite{mg3}, which introduces a five-dimensional space-time, and the Dvali-Gabadadze-Porrati theory \cite{mg4}, which describes our Universe as a four-dimensional brane within a five-dimensional Minkowski space-time.
    \item \textit{Torsion and non-Metricity:} These models expand the Einstein-Hilbert action by including torsion and non-metricity. The Lagrangian is constructed using scalars like $T$ (derived from the torsion tensor) \cite{mg5} or $Q$ (from the non-metricity tensor) \cite{sp6}. While these theories with $T$ and $Q$ can reproduce GR \cite{sp6}, modified versions like $L = f(T)$ or $L = f(Q)$ transform gravity functions.
    \item \textit{Additional fields:} Some theories modify gravity by incorporating extra fields, such as scalars $\phi(t, x)$, vectors $v_\mu(t, x)$, or tensors $A_{\mu\gamma}(t, x)$, alongside the usual metric tensor of GR. Introducing one or more such fields leads to various models, including Einstein-Aether theory \cite{mg7} (with vector fields), quintessence \cite{mg8} (with scalar fields), and Bigravity \cite{mg9} (tensor fields). Among these, quintessence is the simplest and illustrates the blurred lines between dark energy and modified gravity since its effects can mimic those of a dynamic dark energy fluid.
\end{itemize}
The approaches outlined above are some of the most commonly used methods for modifying gravity. We will briefly discuss some modified theories in the following subsections.

\subsection{The $f(R)$ gravity}
We first study a simple extension of GR that can be achieved by replacing the Ricci scalar $R$ in the Einstein-Hilbert action \eqref{gr10} with a general function of $R$, resulting in $f(R)$ gravity theories. The action for $f(R)$ gravity is given by
\begin{equation}\label{fr1}
S_{f(R)} = \frac{1}{2\kappa} \int d^4x \sqrt{-g} f(R) + \int d^4x \sqrt{-g} \mathcal{L}_m.
\end{equation}
Varying this action leads to the field equation
\begin{equation}\label{fr2}
F R_{\mu\nu} - \frac{1}{2} f(R)g_{\mu\nu} - \nabla_\mu \nabla_\nu F + g_{\mu\nu} \Box F = \kappa T_{\mu\nu},
\end{equation}
where  $F \equiv \partial f/\partial R$. It is known that the Bianchi identities ensure that $\nabla_\mu T_{\mu\nu}^{(M)} = 0$. The trace of this field equation results in
\begin{equation}
R F(R) + 3 \Box F(R) - 2 f(R) = \kappa T.
\end{equation}
Since $\Box F(R)$ does not vanish for arbitrary $f(R)$, this equation implies a propagating scalar degree of freedom that does not exist in GR. Two notable features make $f(R)$ gravity an appealing alternative to GR:
\begin{itemize}
    \item $f(R)$ functions are flexible enough to include higher-order Ricci scalar terms, capturing high-energy modifications to GR, yet the resulting equations of motion remain solvable.
    \item $f(R)$ theories avoid the Ostrogradsky instability \cite{sp18}.
\end{itemize}
The field equations \eqref{fr2} involve higher-derivative terms. Hence, it is challenging to find exact solutions. However, conformal rescaling of the metric can transform the action \eqref{fr1} into the Einstein-Hilbert action with a minimally coupled scalar field. To prevent instabilities and ghost-like behavior, any viable $f(R)$ theory must satisfy the following conditions \cite{sp19}
\begin{equation}
\frac{\partial f}{\partial R} > 0, \quad \frac{\partial^2 f}{\partial R^2} > 0, \quad \text{for} R \geq R_0 > 0,
\end{equation}
where $R_0$ is the current value of $R$. The first successful $f(R)$ gravity model was the inflationary model proposed by Starobinsky \cite{sp20}
\begin{equation}
f(R) = R + \frac{R^2}{6M^2},
\end{equation}
where $M$ is a model parameter. This model has two key features:
\begin{itemize}
    \item When the $R^2$ term dominates, it leads to an exact de Sitter expansion without the need for an additional scalar field.
    \item When the linear $R$ term dominates, it provides a natural exit from inflation.
\end{itemize}
An approximate solution for this model is
\begin{equation}
H \approx H_i - \frac{M^2}{6}(t - t_i); \quad a \approx a_i e^{[H_i(t - t_i) - (M^2/12)(t - t_i)^2]},
\end{equation}
where $H_i$ is the initial Hubble parameter, and $a_i$ is the scale factor at the onset of inflation. The Starobinsky model has been shown to be consistent with the Planck-2018 data \cite{sp21}.\\
Viable $f(R)$ gravity models have been developed for low-energy scales, specifically to describe the late-time acceleration of the Universe \cite{sp19,sp22}. Some of these models include
\begin{equation}
f(R) = R - \mu R_c \frac{\left(R/R_c\right)^{2n}}{\left(R/R_c\right)^{2n} + 1},
\end{equation}
and
\begin{equation}
f(R) = R - \mu R_c \left[1-\left(1+\frac{R^2}{R_c^2}\right)^{-n}\right],
\end{equation}
where $n,\,\mu,\,R_c > 0$ \cite{sp19,sp22}.
Interestingly, the Birkhoff theorem does not apply to $f(R)$ theories \cite{sp23}. Recent studies have shown that an infinite number of exact static spherically symmetric vacuum solutions exist for certain classes of $f(R)$ gravity. For instance, in GR, the zero-spin limit ($J \to 0$) of the Kerr black hole reduces uniquely to the Schwarzschild solution. However, if $f(R)$ gravity allows numerous spherically symmetric vacuum solutions, this suggests that the no-hair theorem, which states that black holes are fully characterized by mass, charge, and spin, may not hold in $f(R)$ theories.

\subsection{Symmetric teleparallel gravity: equivalence to GR and beyond}
In this subsection, we explore the theory of symmetric teleparallelism by first introducing the concept of non-metricity, which plays a crucial role in developing alternative models of gravity. The geometry is typically defined by a Lorentzian metric $g$ and a linear connection $\tilde{\Gamma}$. The metric $g$ is expressed using co-frame 1-forms as
\begin{equation}
g = g_{\alpha\beta}dx^\alpha \otimes dx^\beta = \eta_{ab}\vartheta^a \otimes \vartheta^b,
\end{equation}
 where $\eta$ represents the Minkowski metric, and $\{\vartheta^a\}$ is the dual co-frame of a general frame $\{e_a\}$, with the relation $\vartheta^b(e_a) = \delta^b_a$. The full connection 1-form $\tilde{\Gamma}^a_b$ relates to the covariant derivative $D(\tilde{\Gamma})$, and it can be uniquely decomposed as \cite{fq0}
\begin{equation}\label{fdd}
\tilde{\Gamma}^a_{\;\:b} = \omega^a_{\;\:b} + K^a_{\;\:b} + L^a_{\;\:b},
\end{equation}
where $\omega^a_{\;\:b}$ represents the Levi-Civita connection 1-form, $K^a_{\;\:b}$ is the contorsion 1-form, and $L^a_{\;\:b}$ is the deformation 1-form.
In component form, the above expression \eqref{fdd} can be read as \eqref{gt1}.\\
The non-metricity 1-form associated with this connection is given by
\begin{equation}
Q_{ab} = \frac{1}{2}D(\tilde{\Gamma})\eta_{ab} = \tilde{\Gamma}_{(ab)} = -A^c_{\;\:b}\eta_{ac} - A^c_{\;\:a}\eta_{cb},
\end{equation}
where $A^a_{\;\:b} = K^a_{\;\:b} + L^a_{\;\:b}$, and the condition $D(\omega)\eta_{ab} = 0$ is utilized. When  $K^a_{\;\:b}$ vanishes, the non-metricity in components can be written as
\begin{equation}\label{ga2}
Q_{\alpha\beta\gamma} = \nabla_\alpha g_{\beta\gamma} = -L^\rho_{\alpha\beta}g_{\rho\gamma} - L^\rho_{\alpha\gamma}g_{\rho\beta},
\end{equation}
or equivalently,
\begin{equation}
    L^\alpha_{\beta\gamma}=\frac12Q^\alpha_{\beta\gamma}-Q_{(\beta\gamma)}^{\quad\;\alpha},
\end{equation}
Here, $\nabla_\alpha$ represents the components of $D(\tilde{\Gamma})$. The torsion 2-form and curvature 2-form are given by
\begin{equation}
T^a = D(\tilde{\Gamma})\vartheta^a = d\vartheta^a+\Gamma^a_{\;\:b}\wedge \vartheta^b=A^a_{\;\:b}\wedge \vartheta^b,
\end{equation}
and
\begin{equation}
R^a_{\;\:b}(\tilde{\Gamma}) = D(\tilde{\Gamma})\tilde{\Gamma}^a_{\;\:b} = d\tilde{\Gamma}^a_{\;\:b} + \tilde{\Gamma}^a_{\;\:c} \wedge \tilde{\Gamma}^c_{\;\:b},    
\end{equation}
respectively. If $\tilde{\Gamma}^a_{\;\:b} = \omega^a_{\;\:b}$, the Riemannian curvature becomes
\begin{equation}
R^a_{\;\:b}(\omega) = d\omega^a_{\;\:b}+\omega^a_{\;\:c}\wedge\omega^c_{\;\:b}.
\end{equation}
Thus, the total curvature is
\begin{equation}\label{ga11}
R^a_{\;\:b}(\tilde{\Gamma})=R^a_{\;\:b}(\omega)+D(\omega)A^a_{\;\:b}+A^a_{\;\:c}\wedge A^c_{\;\:b}.
\end{equation}
Defining $h^{ab...} = \ast(\vartheta^a \wedge \vartheta^b \wedge \dots)$, the curvature 4-form can be written as
\begin{eqnarray}
        R(\tilde{\Gamma})&=& R^a_{\;\:b}\wedge h^b_{\;\:a}\\
        &=& R(\omega)+D(\omega)A^a_{\;\:b}\wedge h^b_{\;\:a}+A^a_{\;\:c}\wedge A^c_{\;\:b}\wedge h^b_{\;\:a}\\
        &=& R(\omega)+d(A^a_{\;\:b}\wedge h^b_{\;\:a})-A^a_{\;\:b}\wedge D(\omega)h^b_{\;\:a}+A^a_{\;\:c}\wedge A^c_{\;\:b}\wedge h^b_{\;\:a}.
\end{eqnarray}
This formulation shows that $R(\omega)$ is equivalent to the Hilbert-Einstein Lagrangian 4-form in GR. If we want to describe gravity solely with non-metricity, we impose conditions that the connection must be torsionless and curvatureless (i.e., $T^a(\tilde{\Gamma}) = 0$ and $R^a_{\;\:b}(\tilde{\Gamma}) = 0$).
By making these assumptions, the Einstein-Hilbert Lagrangian simplifies to \cite{sp5,fq000,fq1}
\begin{equation}
\mathcal{L}_{EH} = \frac{1}{2\kappa}R(\omega) =-d(\frac1{2\kappa}L^a_{\;\:b}\wedge h^b_{\;\:a})+\frac1{2\kappa}L^a_{\;\:c}\wedge L^c_{\;\:b}\wedge h^b_{\;\:a},
\end{equation}
with the condition $D(\omega)h^b_{\;\:a} = 0$. If we discard the exact form and define the non-metricity 4-form $Q$, we arrive at the Lagrangian for STEGR
\begin{equation}\label{fq1234}
\mathcal{L}_{STEGR} = \frac1{2\kappa}L^a_{\;\:c}\wedge L^c_{\;\:b}\wedge h^b_{\;\:a}= \frac{1}{2\kappa}Q.
\end{equation}
In components, we can rewrite as
\begin{equation}\label{fq123}
\mathcal{L}_{STEGR}=-\frac{\sqrt{-g}}{2\kappa}g^{\mu\nu} \left(L^\alpha_{\beta\nu}L^\beta_{\mu\alpha} - L^\beta_{\alpha\beta}L^\alpha_{\mu\nu}\right)d^4x = \frac{\sqrt{-g}}{2\kappa}Qd^4x,
\end{equation}
where the non-metricity scalar $Q$ is defined as
\begin{equation}
Q \equiv -g^{\mu\nu}\left(L^\alpha_{\beta\nu}L^\beta_{\mu\alpha} - L^\beta_{\alpha\beta}L^\alpha_{\mu\nu}\right).
\end{equation}
By introducing the non-metricity conjugate $P^\alpha_{\;\beta\gamma}$, defined as
\begin{equation}
P^\alpha_{\;\beta\gamma}=-\frac12L^\alpha_{\beta\gamma}+\frac{1}{4}\left( Q^\alpha-\tilde{Q}^\alpha \right)g_{\beta\gamma}-\frac14\delta^\alpha_{(\beta}Q_{\gamma)},
\end{equation}
where $Q_\alpha=Q_{\alpha\beta}^{\quad\beta}$ and $\tilde{Q}_\alpha=Q^\beta_{\;\alpha\beta}$ are the two independent traces of the non-metricity tensor, we find that
\begin{equation}\label{ga3}
Q = -g^{\mu\nu}\left(L^\alpha_{\beta\nu}L^\beta_{\mu\alpha} - L^\beta_{\alpha\beta}L^\alpha_{\mu\nu}\right) = -P^{\alpha\beta\gamma}Q_{\alpha\beta\gamma}.
\end{equation}
Taking the variation of Eq. \eqref{fq1234} with respect to the metric $g$, one can obtain \cite{fq000,fq11}
\begin{equation}
 \left[ D(\omega)\left( L^{ab}-Q^{ab} \right)+L^a_{\;c}\wedge L^{cb} \right]\wedge h_{kab}=2\kappa\tau_k,
\end{equation}
where $\tau_k$ represents the energy-momentum 3-form. In component form, the above equation becomes
\begin{equation}
2\bar{\nabla}_\alpha P^\alpha_{\;\mu\nu} + \left(L^\alpha_{\beta\nu}L^\beta_{\mu\alpha} - L^\beta_{\alpha\beta}L^\alpha_{\mu\nu}\right) + \frac{1}{2}g_{\mu\nu}Q = \kappa T_{\mu\nu},
\end{equation}
where $\bar{\nabla}_\alpha$ denotes the component of $D(\omega)$.\\
In $f(Q)$ gravity, we replace $Q$ in the Lagrangian \eqref{fq123} with an arbitrary function $f(Q)$ and consider the gravitational action
\begin{equation}
S = \frac{1}{2\kappa}\int\sqrt{-g}f(Q)d^4x.
\end{equation}
and hence the field equation for $f(Q)$ gravity becomes
\begin{equation}\label{ga4}
2\bar{\nabla}_\alpha f_Q P^\alpha_{\;\mu\nu} + f_Q\left(L^\alpha_{\beta\nu}L^\beta_{\mu\alpha} - L^\beta_{\alpha\beta}L^\alpha_{\mu\nu}\right) + \frac{1}{2}g_{\mu\nu}f = \kappa T_{\mu\nu},
\end{equation}
where $f_Q = \frac{df(Q)}{dQ}$.
In a more unified form, utilizing $\nabla_\alpha$ instead of $\bar{\nabla}_\alpha$, the above equation \eqref{ga4} can be rewritten as
\begin{equation}\label{ga44}
    -\frac2{\sqrt{-g}}\nabla_\alpha \left( \sqrt{-g}f_QP^\alpha_{\;\mu\nu} \right)+f_Q \left( P_\nu^{\;\alpha\beta}Q_{\mu\alpha\beta}-2P^{\alpha\beta}_{\;\:\;\:\mu}Q_{\alpha\beta\nu} \right)+\frac12g_{\mu\nu}f= \kappa T_{\mu\nu}.
\end{equation}
Additionally, varying the action with respect to the connection gives
\begin{equation}\label{FQ6}
\nabla_{\mu}\nabla_{\gamma}\left( \sqrt{-g}f_Q {P^{\gamma}}_{\mu\nu}\right)=0\,.
\end{equation}
Further, several extensions of $f(Q)$ have been proposed, such as $f(Q,T)$ gravity \cite{fqt}, $f(Q,C)$ gravity (where $C$ denotes the boundary term) \cite{fq2}, etc.
\subsection{$4$D Einstein-Gauss-Bonnet gravity}
Here, we shall briefly explore the background of the Einstein-Gauss-Bonnet gravity within four-dimensional space-time. Previously, Einstein–Gauss-Bonnet, the extension of GR, was long considered a trivial case in the context of four-dimensional space-time. This perspective shifted in 2020 when Glavan and Lin \cite{egb1} introduced a re-scaling of the coupling constant, which made it possible for the effects of Einstein–Gauss-Bonnet gravity to manifest even in four dimensions. The resulting theories, known as ``$4$D Einstein–Gauss–Bonnet" ($4\text{D}\,\text{EGB}$) gravity, exhibit several intriguing properties.\\
\textit{Lovelock's condition:} 
Based on Lovelock’s theorem, as referenced in \cite{egb2,egb3}, it is established that Einstein’s GR stands as the unique gravitational theory provided certain criteria are satisfied for rank-2 tensors \( A_{\mu\nu} \):
\begin{itemize}
    \item[(i)] $A^{\mu\nu}$ is symmetric, i.e., $A^{\mu\nu}=A^{\nu\mu}$.
    \item[(ii)] $A^{\mu\nu}$ is associated with the metric tensor $g_{\rho\sigma}$ and its first two derivatives, i.e., $A^{\mu\nu}=A^{\mu\nu}\,(g_{\rho\sigma},g_{\rho\sigma,\tau},g_{\rho\sigma,\tau\chi})$.
    \item[(iii)] $A_{\mu\nu}$ is divergence free, i.e., $\nabla_\nu A^{\mu\nu} = 0$.
\end{itemize}
Glavan and Lin \cite{egb1} presented a method to bypass the conditions of Lovelock’s theorem and presented a model that respects all the assumptions (\textit{i-iii}) while showing modified dynamics.
Lovelock’s field equations can be derived from the Lagrangian density
\begin{equation}\label{gb1}
\mathcal{L} = \sqrt{-g} \sum_{j} \alpha_j R_j,
\end{equation}
where
\begin{equation}\label{gb2}
R_j \equiv \frac{1}{2^j} \delta^{\mu_1\nu_1...\mu_j\nu_j}_{\alpha_1\beta_1...\alpha_j\beta_j} \prod_{i=1}^{j} R^{\alpha_i\beta_i}_{\mu_i\nu_i}\quad \text{and}\quad \delta^{\mu_1\nu_1...\mu_j\nu_j}_{\alpha_1\beta_1...\alpha_j\beta_j} \equiv j! \delta^{\mu_1}_{[\alpha_1} \delta^{\nu_1}_{\beta_1} ... \delta^{\mu_j}_{\alpha_j} \delta^{\nu_j}_{\beta_j]}.
\end{equation}
Here, $g$ is the determinant of the metric, $\delta^\mu_\nu$ is the Kronecker delta, $R^\mu_{\nu\rho\sigma}$ are the components of the Riemann tensor, and $\alpha_j$ are arbitrary constants. The square brackets denote anti-symmetrization.\\
The tensor $A_{\mu\nu}$, which satisfies conditions (\textit{i-iii}), can be derived by integrating the Lagrangian density from Eq. \eqref{gb1} over a region $\Omega$ in D-dimensional space-time to form an action $S$. Varying this action with respect to the inverse metric $ g^{\mu\nu}$ yields
\begin{equation}\label{gb3}
\delta S = \delta \int_\Omega d^D x \, L = \int_\Omega d^D x \, \sqrt{-g} A_{\mu\nu} \delta g^{\mu\nu} + \int_{\partial\Omega} d^{D-1}x \, \sqrt{h} B,
\end{equation}
where $h$ is the determinant of the induced metric on the boundary $\partial\Omega$, and
\begin{equation}\label{gb4}
A^\mu_\nu = -\sum_{j} \frac{\alpha_j}{2j+1} \delta^{\mu \rho_1 \sigma_1...\rho_j \sigma_j}_{\nu \alpha_1 \beta_1...\alpha_j \beta_j} \prod_{i=1}^{j} R^{\alpha_i \beta_i}_{\rho_i \sigma_i}.
\end{equation}
For further details on the derivation and the boundary term $B$, one can check the Refs. \cite{egb2,egb4}.\\
It is evident from Eq. \eqref{gb4} that the summation will stop when $2j + 1 > D$, where $D$ represents the dimension of space-time. This arises from the definition of \(\delta^{\mu_1 \nu_1 ... \mu_j \nu_j}_{\alpha_1 \beta_1 ... \alpha_j \beta_j}\) in Eq. \eqref{gb2}, where the number of available index values must exceed the number of lower indices for the expression to be non-zero.
In $D=1$ or $2$ dimensions, only one term exists in the Lovelock tensor $A^\mu_\nu$, which takes the form $\sim (\text{Riemann})^0$, i.e., a constant. Whereas in $D=3$ and $4$, two terms are possible: a constant term and a term proportional to $\sim (\text{Riemann})^1$.\\
As we move to higher dimensions, specifically when $D > 4$, Einstein’s equations are no longer the most general set of field equations that satisfy conditions (\textit{i-iii}). In $D=5$ or $6$, the tensor $A^\mu_\nu$ can include three terms, with the highest order term being $\sim (\text{Riemann})^2$, which corresponds to $j = 2$ in the sums of Eqs. \eqref{gb1} and \eqref{gb4}. This results in the Lagrangian density
\begin{equation}\label{gb6}
\mathcal{L} = \sqrt{-g} [s_0 + s_1 R + s_2 \mathcal{G}],
\end{equation}
where 
\begin{equation}\label{gb7}
\mathcal{G} = R^2 - 4 R_{\mu\nu} R^{\mu\nu} + R_{\mu\nu\rho\sigma} R^{\mu\nu\rho\sigma},
\end{equation}
is known as the \textit{Gauss-Bonnet term}. Extremizing the action associated with this Lagrangian leads to the Lanczos tensor \cite{egb6,egb7}
\begin{equation}\label{gb8}
A^\mu_\nu = -\frac{1}{2} s_0 \delta^\mu_\nu + s_1 \left(R^\mu_\nu - \frac{1}{2} \delta^\mu_\nu R\right) + s_2 \left(2 R^{\mu\alpha\rho\sigma} R^\nu_{\alpha\rho\sigma} - 4 R^{\rho\sigma} R^\mu_{\ \rho \nu \sigma} - 4 R^{\mu\rho} R^\nu_{\rho} + 2 R R^\mu_\nu - \frac{1}{2} \delta^\mu_\nu \mathcal{G}\right).
\end{equation}
The second part of the above Eq. \eqref{gb8} vanishes identically in $D\leq 4$. This outcome is also consistent with the Chern theorem applied to the action derived by integrating Eq. \eqref{gb6} over the space-time manifold \cite{egb5}.\\
To address this issue, Glavan and Lin suggested re-scaling the Gauss-Bonnet (GB) constant $s_2$ by the factor \cite{egb1}
\begin{equation}
s_2 \rightarrow \frac{s_2}{D-4}.
\end{equation}
This modification implies that terms in the Lanczos tensor \eqref{gb8} involving $s_2$ could remain finite and non-zero. Their idea was that introducing a divergence into $s_2$ could counterbalance the vanishing behavior of the additional terms in Eq. \eqref{gb8} as the dimension $D\rightarrow 4$. If this were true, then the GB term could directly impact the 4D theory of gravity.\\
Now, let us begin with the standard Einstein–Gauss–Bonnet action, focusing solely on the gravitational sector and ignoring matter contributions
\begin{equation}
S_{GB} = \frac{1}{16\pi G} \int d^D x \sqrt{-g} \left(-2\Lambda + R + \hat{s} \mathcal{G}\right),
\end{equation}
where $\hat{s}$ is a constant, and $D$ represents the number of space-time dimensions, which remains unspecified at this point. Varying the action with respect to the metric leads to the field equations
\begin{equation}
G_{\mu\nu} + \Lambda g_{\mu\nu} = \hat{s} H_{\mu\nu},
\end{equation}
where $H_{\mu\nu}$ is the GB tensor, and it can be read as
\begin{equation}
H_{\mu\nu} = 15\delta_{\mu[\nu} R_{\rho\sigma} R^{\rho\sigma} R_{\alpha\beta]} = -2 \left( R R_{\mu\nu} - 2 R_{\mu\alpha\nu\beta} R^{\alpha\beta} + R_{\mu\alpha\beta\sigma} R^{\,\,\alpha\beta\sigma}_{\nu} - 2 R_{\mu\alpha} R^{\,\,\alpha}_{\nu} - \frac{1}{4} g_{\mu\nu} \mathcal{G} \right).
\end{equation}
Since the right-hand side is antisymmetrized over five indices, it automatically vanishes for dimensions $D < 5$. This is the conventional setup of the Einstein–Gauss-Bonnet theory. However, a novel approach proposed in \cite{egb1} suggests that the vanishing of $H_{\mu\nu}$ in four dimensions can be bypassed by rescaling the GB coupling constant as follows
\begin{equation}\label{gb77}
\hat{s} = \frac{s}{D - 4},
\end{equation}
where $s$ is a new finite constant, and this limit is considered as $D \to 4$. This approach appears viable when examining the trace of the field equations, which includes a GB term expressed as
\begin{equation}
g^{\mu\nu} H_{\mu\nu} = \frac{1}{2} (D - 4) \mathcal{G}.
\end{equation}
Here, the factor ($D - 4$) could be precisely canceled by the proposed rescaling of $\hat{s}$, resulting in a non-zero contribution to the trace as $D \to 4$ \footnote{Note that $\mathcal{G}$ itself is not required to vanish in the four-dimensional limit.}
\begin{equation}
\hat{s} g^{\mu\nu} H_{\mu\nu} =\frac{s}{\cancel{(D-4)}}\frac{1}{2}\cancel{(D-4)}\mathcal{G}=
\frac{s}{2} \mathcal{G}.
\end{equation}
In \cite{egb1}, it has been further argued that this non-zero contribution may not be limited to the trace but could influence the entire theory. They point out that when the field equations are expressed in differential form, they take the form
\begin{equation}
\epsilon_{a_{D}} = \sum_{p=0}^{D/2-1} s_p (D - 2p)\epsilon_{a_1...a_D} R^{a_1, a_2} \wedge \ldots \wedge R^{a_{2p-1}, a_{2p}} \wedge e^{a_{2p+1}} \wedge \ldots \wedge e^{a_{D-1}} = 0,
\end{equation}
where $e^a$ denotes the vielbein. The factor ($D - 4$) surfaces in the field equations when $p = 2$ for the GB term. While intriguing, the expected results do not emerge straightforwardly. Nevertheless, some $D$-dimensional space-times behave consistently under the proposed rescaling when taken to the limit $D \to 4$.
Thus, in the $D$ dimensions, the EGB gravity action incorporates a rescaled GB coupling  to restore dimensional regularization results in the expression \cite{egb1}
\begin{equation}\label{gb}
\mathcal{I}=\frac{1}{16 \pi}\int d^D\,x\sqrt{-g}\left(R+\frac{s}{D-4} \mathcal{G}\right)+\mathcal{S}_{matter},
\end{equation}
where $s$ is a finite non-vanishing GB constant in $D = 4$.
By varying the action \eqref{gb} with respect to the metric tensor, one can find the field equations
\begin{equation}\label{2a3}
G_{\mu\nu}+\frac{s}{D -4}H_{\mu\nu}=8\pi T_{\mu\nu},
\end{equation}
\section{Conclusions}
In this chapter, we have provided a comprehensive introduction, beginning with a detailed exploration of the historical development of wormhole geometry. We then discussed the Casimir effect and studied the significance of the GUP in relation to the Casimir force. Additionally, we discussed the concept of dark matter, highlighting several widely accepted dark matter profiles. Fundamental concepts of GR were reviewed, along with their limitations, leading to a discussion on various modified theories of gravity. This chapter concludes with an overview of these modifications. In the upcoming chapters, we will address some problems in wormhole geometry by applying the theoretical foundations laid out in this section.

\chapter{Traversable wormhole geometries in $f(Q)$ gravity} 
\label{Chapter2} 

\lhead{Chapter 2. \emph{Traversable wormhole geometries in $f(Q)$ gravity}} 
\blfootnote{*The work in this chapter is covered by the following publication:\\
\textit{Traversable wormhole geometries in $f(Q)$ gravity}, Fortschritte der Physik, \textbf{69}, 2100023 (2021).}
This chapter presents a detailed discussion of wormhole geometry in the context of modified $f(Q)$ gravity under various scenarios like the effect equation of state (EoS). The detailed study of the work is outlined as follows:
\begin{itemize}
    \item We have constructed the field equations for static and spherically symmetric Morris-Throne wormhole metric in $f(Q)$ gravity.
    \item We discuss wormhole geometries for two functional form of $f(Q)$ gravity such as linear $f(Q)=\alpha Q$ and non-linear $f(Q)=a\,Q^2+B$ models.
    \item We obtain wormhole solutions by assuming some cases, namely barotropic EoS, anisotropic EoS, and specific shape functions. 
    \item We discussed the viability of shape functions and the stability analysis of the wormhole solutions for each case and found that the NEC violates each wormhole model, which concluded that our outcomes are realistic and stable.
\end{itemize}
\section{Introduction} 
Over the years, significant growth has been witnessed in the extensions of GR \cite{Laurentis} involving torsion-based gravity \cite{Krssak}. Nevertheless, in 1999, the non-metricity theory came to light after the proposal of the so-called symmetric teleparallel gravity \cite{sp5,Conroy}. In this modified gravity, both curvature and torsion are set to zero; hence, gravitation is linked to the non-metricity tensor and affiliated to the nonmetricity scalar $Q$. Jimenez et al. have generalized this theory, which has acquired significant attention from researchers, namely $f(Q)$ gravity \cite{fq1}, where the gravitational field is expressed by the non-metricity scalar $Q$ only. This theory has successfully encountered various background and perturbation observational data such as the Supernovae type Ia (SNIa), CMB, Redshift Space Distortion, Baryonic Acoustic Oscillations (BAO), etc., \cite{Soudi,Banos,Salzano,Koivisto11}, and this conflict demonstrates that the $f(Q)$ gravity could challenge the $\Lambda$CDM model \cite{Anagnostopoulos}. Moreover, we could see the growing interest of $f(Q)$ gravity in the field of astrophysical objects as well. Black holes in $f(Q)$ gravity have been investigated in \cite{Fell}. Further, the static and spherically symmetric solutions under anisotropic fluid for $f(Q)$ gravity have been discussed by Wang et al. in \cite{Wang2}. Recently, a class of static spherically symmetric solutions in $f(Q)$ gravity have been investigated in \cite{Calza}.\\
This chapter focused on exploring the wormhole geometries in symmetric teleparallel gravity or $f(Q)$ gravity. The wormholes are supported by exotic matter, and that is an entirely unsolved problem. This issue motivates us to study wormhole geometries through modified theories where curvature explains the wormhole and retains standard matter. 
Here, we have studied three types of wormhole geometries by considering (i) a relation between the radial and lateral pressure, (ii) the phantom energy EoS, and (iii) specific shape functions for $b(r)$. We have calculated and discussed the properties of $b(r)$ for three cases. To do the stability analysis of the wormhole solutions, we have tested the energy conditions. Besides this, we have measured the exotic matter for all wormhole solutions.\\
The plan of this chapter is layered as follows: 
in section \ref{ch1sec3}, we discussed the framework of the traversable wormhole geometries in $f(Q)$ gravity. We also discussed the energy conditions and three types of wormhole solutions. In section \ref{ch1sec5}, we construct the motion equations for two different forms of $f(Q)$. Then, we discuss three wormhole solutions for the linear form of $f(Q)$ and two wormhole solutions for the non-linear form of $f(Q)$ in sections \ref{ch1sec6} and \ref{ch1sec7}, respectively. In section \ref{ch1sec9}, we discussed the volume integral quantifier to measure the exotic matter. Finally, we conclude our outcomes in section \ref{ch1sec10}.

\section{Wormhole field equations in $f(Q)$ gravity}\label{ch1sec3}
For the present interest, let us consider the matter is described by an anisotropic stress-energy tensor of the form
\begin{equation}
\label{3ch1}
T_{\mu}^{\nu}=\left(\rho+P_t\right)u_{\mu}\,u^{\nu}+P_t\,\delta_{\mu}^{\nu}+\left(P_r-P_t\right)v_{\mu}\,v^{\nu}
\end{equation}
where $u_{\mu}$ is the four-velocity and $v_{\mu}$ the unitary space-like vector in the radial direction.\\
The non-metricity scalar $Q$ for the wormhole metric in \eqref{1ch1} takes the form
\begin{equation}
\label{4ch1}
Q=-\frac{2}{r}\left(1-\frac{b(r)}{r}\right)\left(2\phi^{'}(r)+\frac{1}{r}\right).
\end{equation}
Now, the field equations for the metric \eqref{1ch1} under anisotropic matter \eqref{3ch1} within the framework of modified symmetric teleparallel gravity can be derived as
\begin{equation}
\label{5ch1}
\left[\frac{1}{r}\left(-\frac{1}{r}+\frac{rb^{'}+b}{r^2}-2\phi^{'}\left(1-\frac{b}{r}\right)\right)\right]f_Q \\
-\frac{2}{r}\left(1-\frac{b}{r}\right)f_{QQ}Q^{'}-\frac{f}{2}=-\rho,
\end{equation}
\begin{equation}
\label{6ch1}
\left[\frac{2}{r}\left(1-\frac{b}{r}\right)\left(2\phi^{'}+\frac{1}{r}\right)-\frac{1}{r^2}\right]f_Q+\frac{f}{2}=-P_r,
\end{equation}
\begin{multline}
\label{7ch1}
\left[\frac{1}{r}\left(\left(1-\frac{b}{r}\right)\left(\frac{1}{r}+\phi^{'}\left(3+r\phi^{'}\right)+r\phi^{''}\right)-\frac{rb^{'}-b}{2r^2} \left(1+r\phi^{'}\right)\right)\right]f_Q \\
+\frac{1}{r}\left(1-\frac{b}{r}\right)\left(1+r\phi^{'}\right)f_{QQ}Q^{'}
+\frac{f}{2}=-P_t,
\end{multline}
and
\begin{equation}\label{8ch1}
\frac{\cot{\theta}}{2}f_{QQ}Q^{'}=0,
\end{equation}
where ${'}$ represents $\frac{d}{dr}$.
\section{Wormhole solutions in different forms of $f(Q)$}\label{ch1sec5}
In this study, we will consider two different forms of $f(Q)$: (i) linear form of $f(Q)$, and (ii) non-linear form of $f(Q)$. By using these forms of $f(Q)$'s, we will get different field equations, and subsequently, we shall discuss by assuming particular models. Note that, in this chapter, we consider the constant redshift function, i.e., $\phi'(r)=0$.
\subsection{Field equations in $f(Q)=\alpha Q$}
To proceed further, we have presumed the linear functional form  of $Q$ as
\begin{equation}\label{b}
f(Q)=\alpha Q,
\end{equation}
 where `$\alpha$' is a constant, which is the teleparallel gravitational term. The linear form of $f(Q)$ recovers the STEGR, which helps us to compare our wormhole solutions to their fundamental level. Further, the redshift function $\phi(r)$ must be finite and non-vanishing at the throat $r_0$. So one can consider $\phi(r)=constant$ to achieve the de Sitter and anti-de Sitter asymptotic behavior. Therefore the field equations in \eqref{5ch1}-\eqref{7ch1} reads
\begin{equation}
\label{9ch1}
-\frac{b^{'}(r)}{r^2}\alpha=\rho,
\end{equation}
\begin{equation}
\label{10ch1}
\frac{b(r)}{r^3}\alpha=P_r,
\end{equation}
\begin{equation}
\label{11ch1}
\left(\frac{b^{'}(r)}{2\,r^2}-\frac{b(r)}{2\,r^3}\right)\alpha=P_t.
\end{equation}
Now, in the next section, we are going to discuss three special cases of wormhole solutions for our study.
\subsection{Field equations in $f(Q)=a\,Q^2+B$}
In this study, we have assumed a particular power-law form of $f(Q)$, i.e., $f(Q)=a\,Q^2+B$, where $a$ and $B$ are constants. Researchers have already shown that the power-law model can oblige the regular thermal extending history, including the cold dark matter-dominated stage and the radiation. Therefore using the model we developed the field equations \eqref{5ch1}-\eqref{7ch1} as follows
\begin{equation}
\label{12ch1}
\frac{2 a \left( b(r)-r\right) \left(11 b(r)-(6\,b^{'}(r)+7)\,r\right)}{r^6}+\frac{B}{2}=\rho,
\end{equation}
\begin{equation}
\label{13ch1}
\frac{2 a \left(3 (b(r))^2-4\,r\, b(r)+r^2\right)}{r^6}-\frac{B}{2}=P_r,
\end{equation}
\begin{equation}
\label{14ch1}
-\frac{6\,a \left(b(r)-r\right) \left(2 b(r)-(b^{'}(r)+1) r\right)}{r^6}-\frac{B}{2}=P_t.
\end{equation}

\section{Wormhole models with linear $f(Q)$ model}\label{ch1sec6}
In this section, we shall discuss three special cases of wormhole solutions by using the Eqs. (\ref{9ch1}-\ref{11ch1}).
\subsubsection{Wormhole (WH1) solution with $P_t=mP_r$}
\begin{figure}[h]
\centering
\includegraphics[width=8.5 cm]{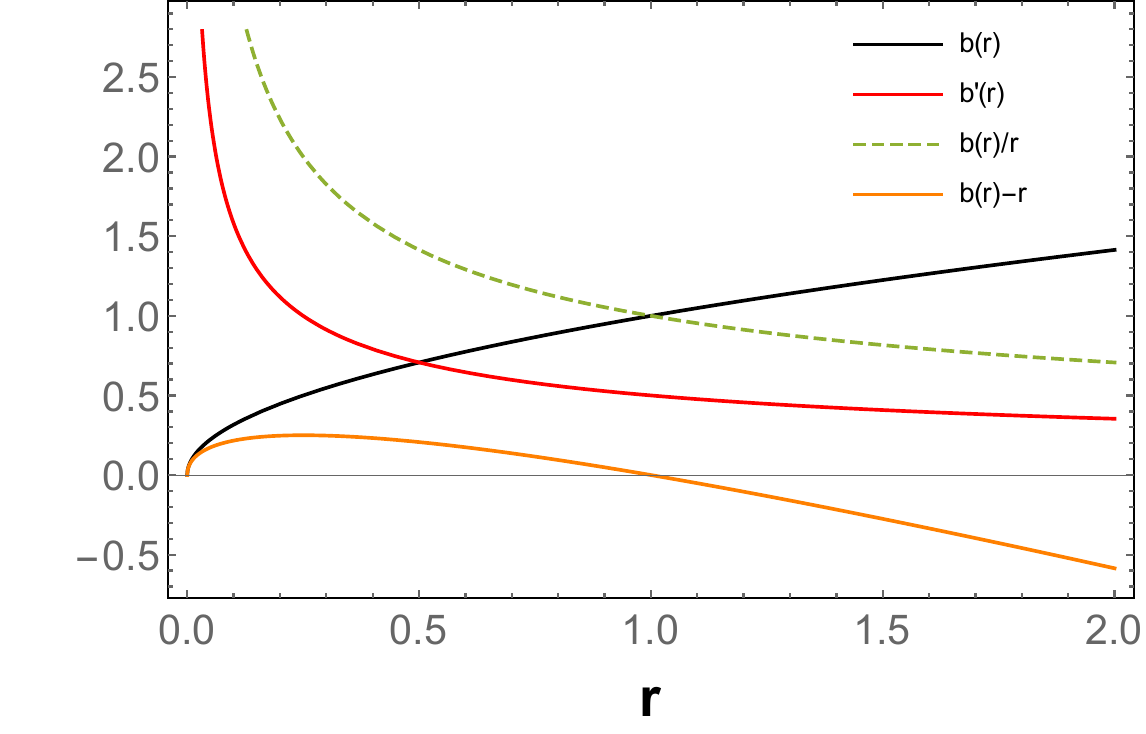}
\caption{The plot of shape function $b(r)$, flaring out condition $b'(r)<1$, throat condition $b(r)-r<0$, and asymptotically flatness condition $\frac{b(r)}{r}\rightarrow0$ as $r\rightarrow\infty$ with varying $r$, for $m=-0.25$ (for WH1).}
\label{1chf1}
\end{figure}

In first model, we assume the pressures $P_r$ and $P_t$ are related as (for more details see Ref. \cite{Moraes/2017})
\begin{equation}
\label{15ch1}
P_t=mP_r,
\end{equation}
where $m$ is an arbitrary constant.
By using equations \eqref{10ch1} and \eqref{11ch1} in equation \eqref{15ch1}, one can obtain
\begin{equation}
\label{16ch1}
b(r)=kr^{(1+2m)},
\end{equation}
where $k$ is an integrating constant.
Without loss of generality, we consider $k=1$. For $m<0$, Eq. \eqref{16ch1} retained the asymptotically flatness condition i.e. $b(r)/r\rightarrow 0$ for $r\rightarrow \infty$. In Fig. \ref{1chf1}, we depict the quantities $b(r), b(r)/r, b(r)-r$ and $b^{'}(r)$ with varying the radial component $r$, for $m=-0.25$. One can clearly see that $b(r)-r$ cuts the $r$-axis at $r_0=1$ in Fig. \ref{1chf1}. Note that, for a stable wormhole, the shape function $b(r)$ needs to obey flaring, throat, and asymptotically conditions. Profiles in Fig. \ref{1chf1} show that the shape function satisfied all the conditions that are required for a stable traversable wormhole.
Now using equation \eqref{15ch1}, we can rewrite the energy density $\rho$, radial pressure $P_r$ and tangential pressure $P_t$ from \eqref{9ch1}-\eqref{11ch1} as follows
\begin{equation}
\label{17ch1}
\rho=-\alpha (1+2m)\,r^{2(m-1)},
\end{equation}
\begin{equation}
\label{18ch1}
P_r=\alpha\,r^{2(m-1)},
\end{equation}
and
\begin{equation}
\label{19ch1}
P_t=m\,\alpha\,r^{2(m-1)}.
\end{equation}
Now, from the above Eqs. \eqref{17ch1}-\eqref{19ch1}, we can obtain the expressions for NEC
\begin{equation}
\label{20ch1}
\rho+P_r=-2\,m\,\alpha\,r^{2(m-1)},
\end{equation}
\begin{equation}
\label{21ch1}
\rho+P_t=-(m+1)\,\alpha\,r^{2(m-1)},
\end{equation}
As we know, energy conditions are the best geometrical tool to test the cosmological models' self-stability. So, we adopted this technique to test our models. Moreover, we have presumed the linear functional form of $f(Q)$. Hence, our models will retain the standard energy conditions.
\begin{figure}[h]
\centering
    \includegraphics[width=8.5 cm]{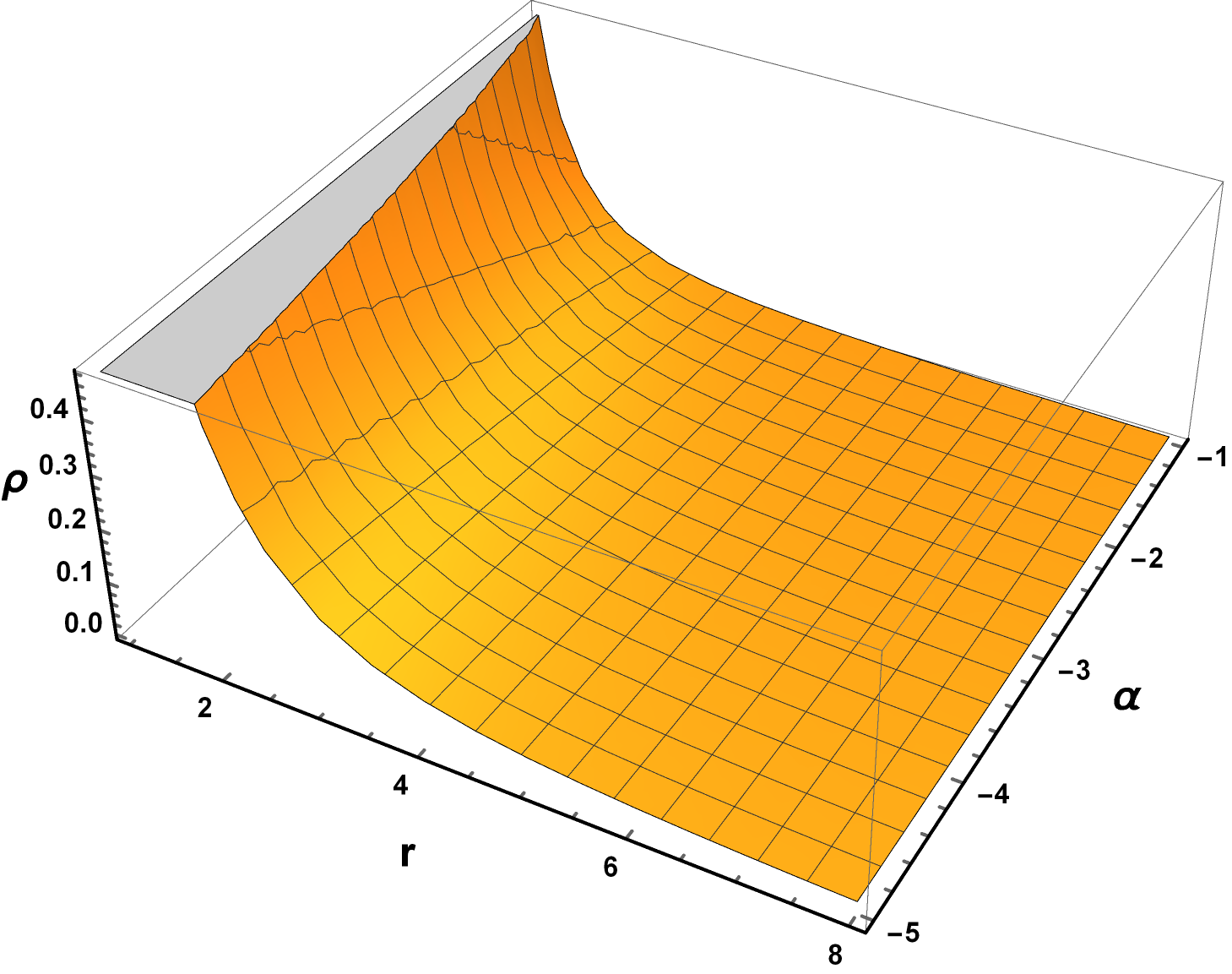}
    \caption{The plot of energy density, $\rho(r)$ w.r.t. $r$  and $\alpha$ with $m=-0.25$ for WH1.}
    \label{1chf2}
\end{figure}
Figs. \ref{1chf2} and \ref{1chf3}, depicts the behavior of the energy conditions. It can be seen from Fig. \ref{1chf2} that the energy density, $\rho$ is positive throughout the space-time. From Fig. \ref{1chf3}, one can observe that NEC for the radial pressure violates, i.e., $\rho+P_r< 0$, whereas NEC for the lateral pressure, i.e., $\rho+P_t\geq 0$ obeys. Also, DEC is satisfied, i.e., $\rho-P_r\geq 0$ and $\rho-P_t\geq 0$ satisfied. These profiles of energy conditions aligned with the properties of exotic matter which is responsible for a traversable wormhole. From equations \eqref{17ch1}-\eqref{19ch1}, the SEC yields $\rho+P_r+2P_t=0$. This similar result was obtained in \cite{Elizalde/2019}.
\begin{figure}[h]
\centering
  \includegraphics[width=6.5 cm]{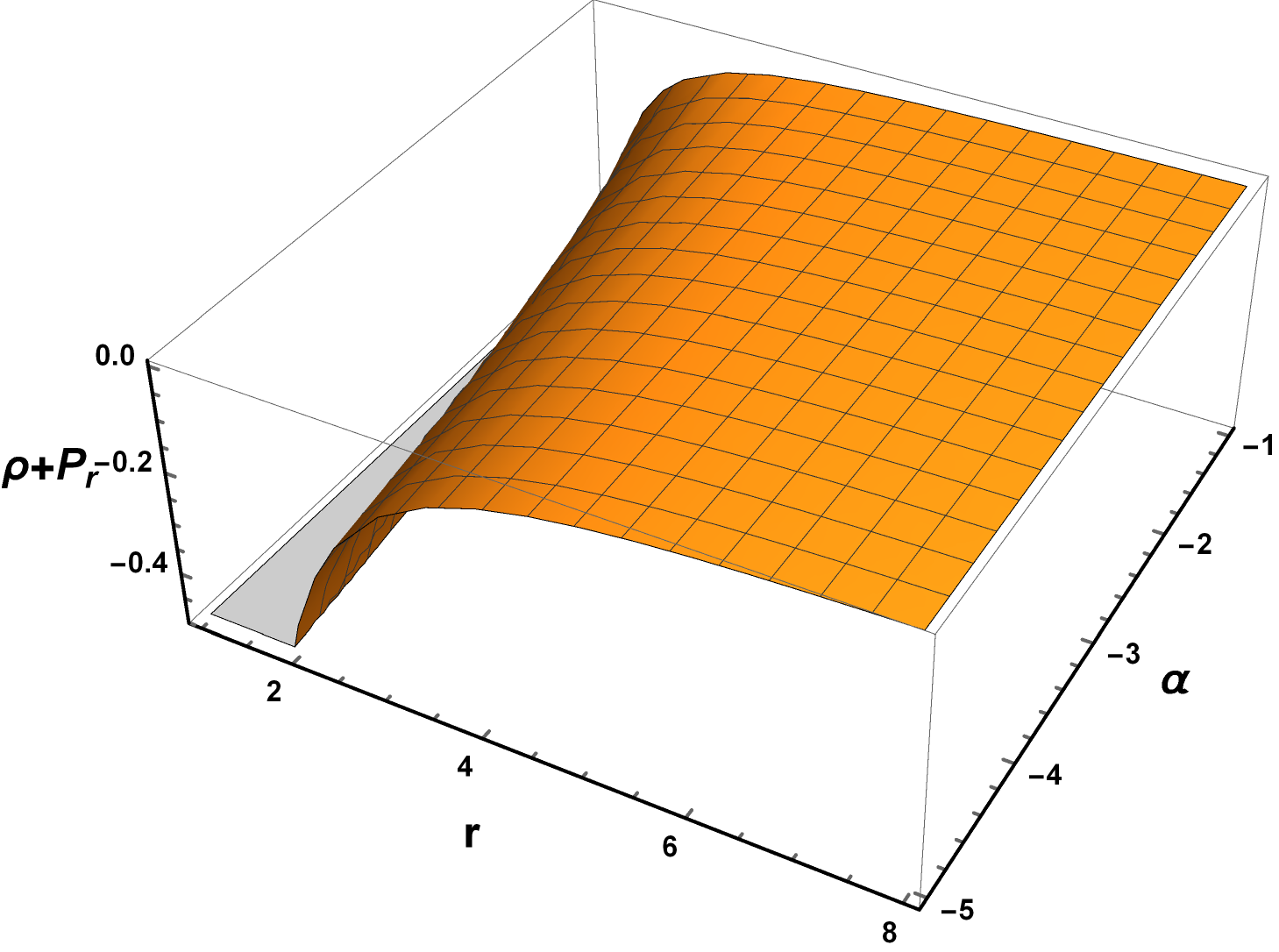}
  \includegraphics[width=6.5 cm]{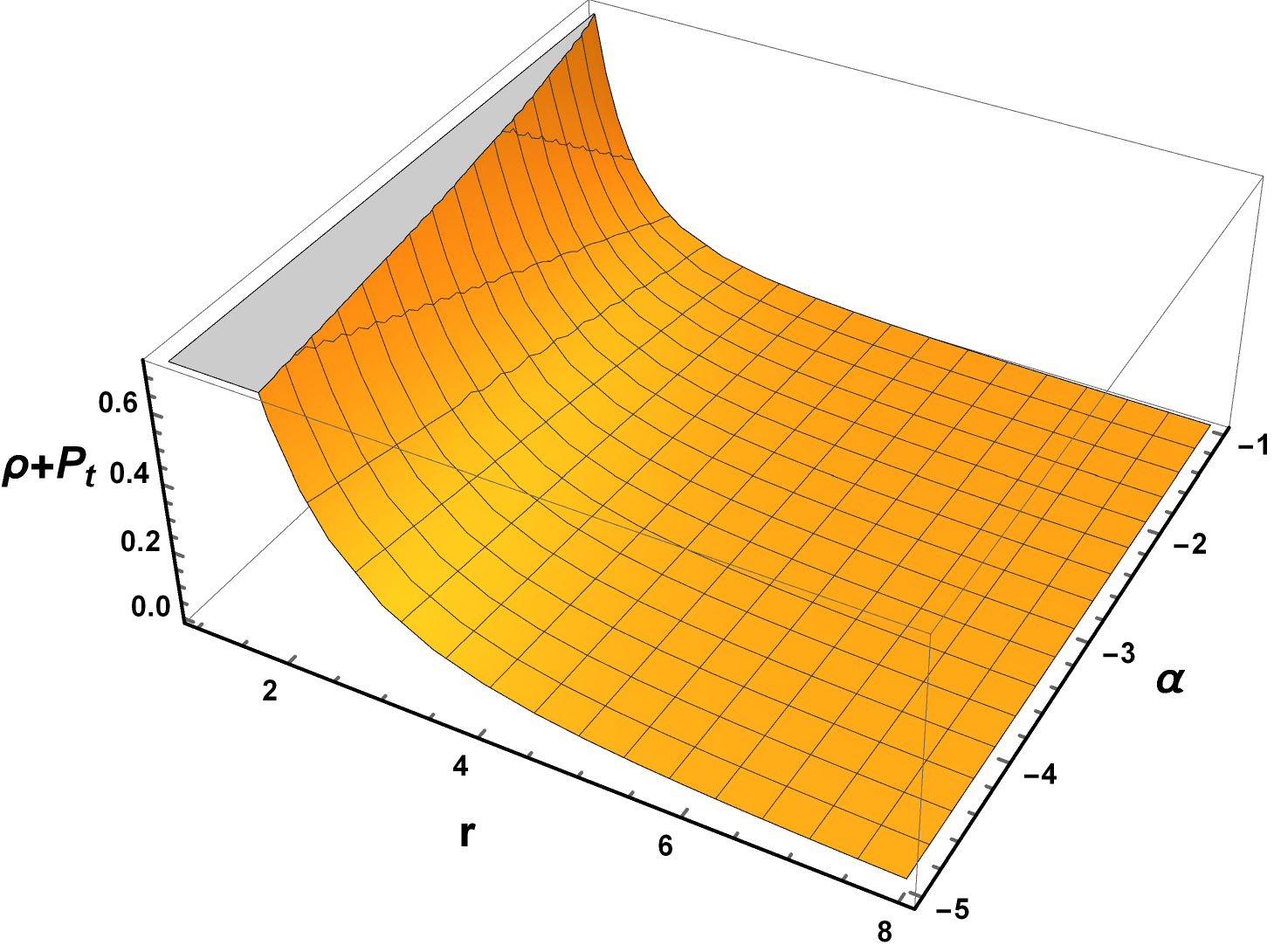}
  \includegraphics[width=6.5 cm]{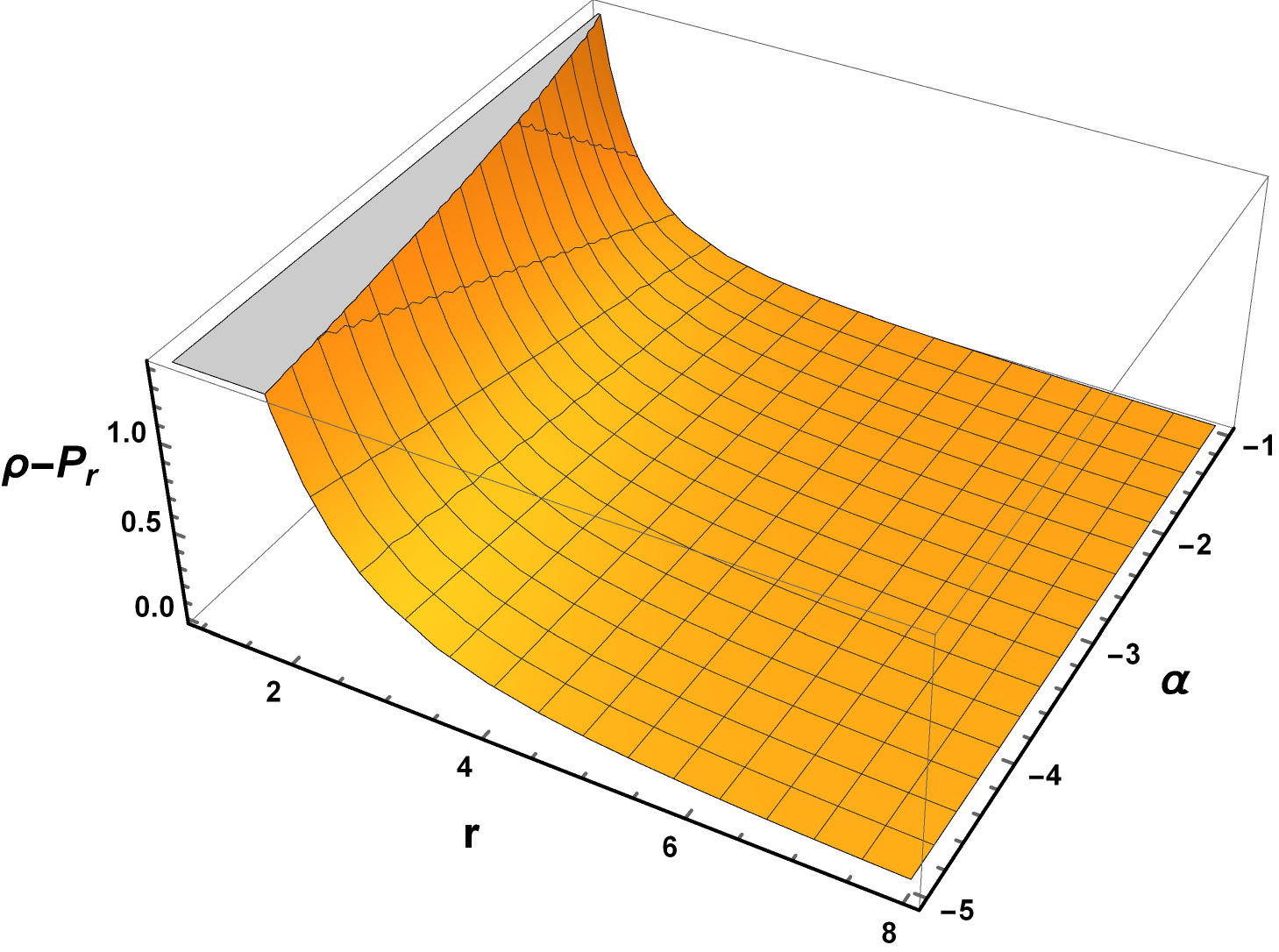}
  \includegraphics[width=6.5 cm]{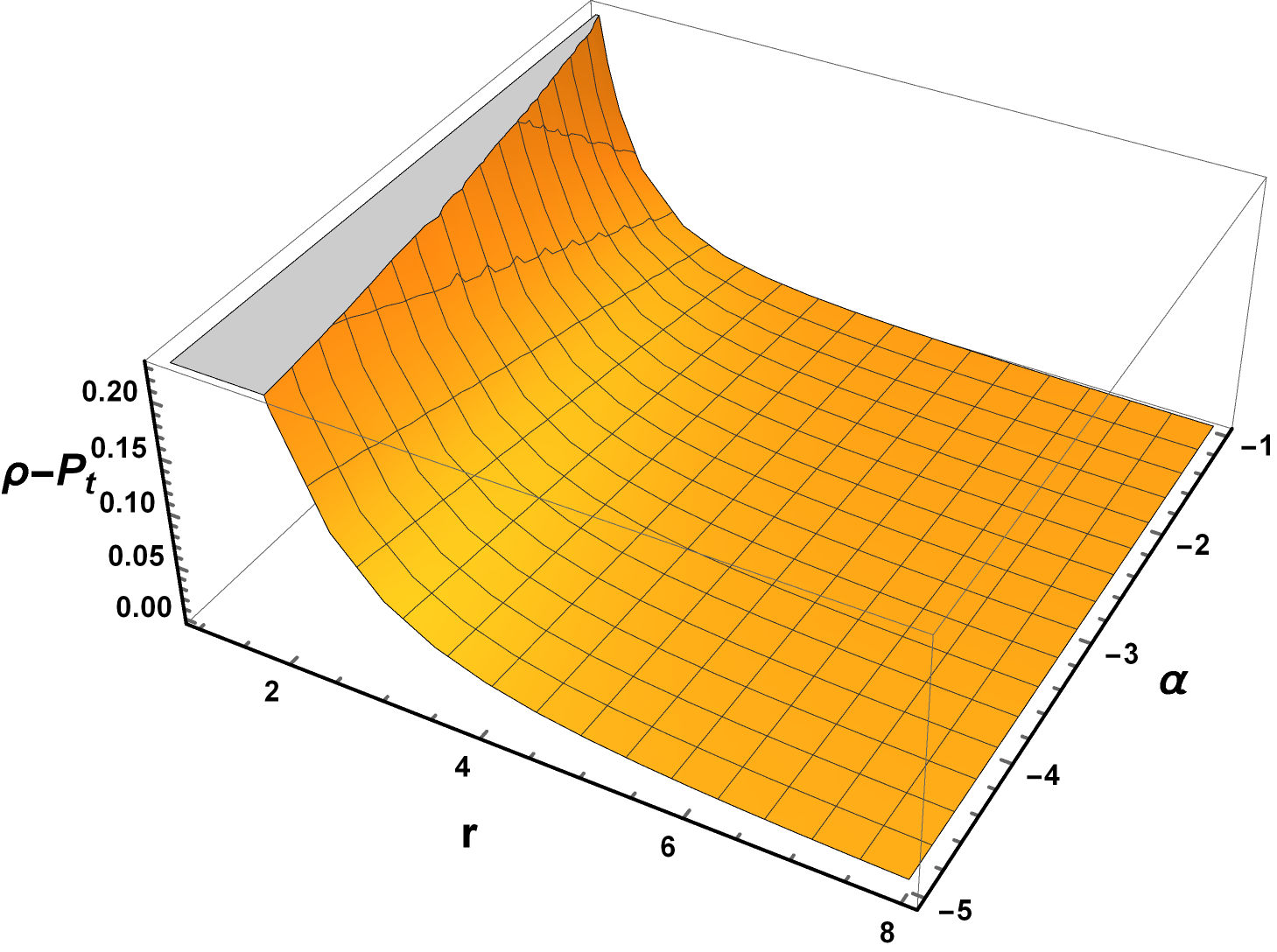}
\caption{The plots of WEC, NEC and DEC w.r.t. $r$  and $\alpha$ with $m=-0.25$ for WH1.}
\label{1chf3}
\end{figure}
\subsubsection{Wormhole (WH2) solution with $P_r=\omega\rho$}
\begin{figure}[h]
\centering
	\includegraphics[width=8.5 cm]{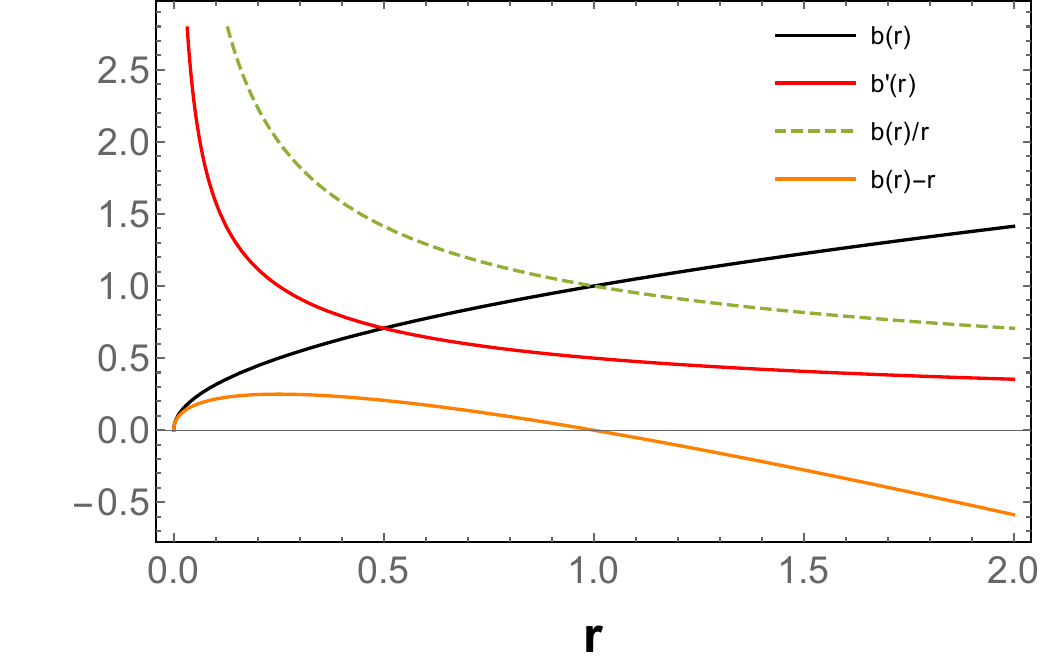}
	\caption{The plot of shape function $b(r)$,  flaring out condition $b'(r)<1$, throat condition $b(r)-r<0$, and asymptotically flatness condition $\frac{b(r)}{r}\rightarrow0$ as $r\rightarrow\infty$ for $c=1, \omega=-2$ for WH2. }
	\label{1chf4}
\end{figure}

Next, we assume a relation between $\rho$ and $P_r$ as a linear EoS such as \cite{kimet/2018,Lobo/2005,Francisco/2005,Mauricio/2013}.\\
\begin{equation}
\label{24ch1}
P_r=\omega\rho,
\end{equation}
where $\omega$ is an EoS parameter.\\
 Therefore, using Eqs. \eqref{9ch1} and \eqref{10ch1} in Eq. \eqref{24ch1}, one can obtain the following shape function
\begin{equation}
\label{25ch1}
b(r)=cr^{-\frac{1}{\omega}},
\end{equation}
where $c$ is an integrating constant.\\
Note that, to satisfy the asymptotic flatness condition, i.e., $\frac{b(r)}{r}\rightarrow 0$ as $r\rightarrow \infty$, \,$\omega$ should be less than -1, i.e., $\omega<-1$.\\
The profiles of the necessary conditions for a shape function, i.e., throat condition, flare-out condition, and asymptotically flatness conditions, are depicted in Fig. \ref{1chf4}. From Fig. \ref{1chf4}, one can see that the shape function $b(r)$ is in the increasing direction as $r$ increases. For $r>r_0$, $b(r)-r<0$, which represents the consistent throat condition for wormholes, and from the same Fig. \ref{1chf4}, it is clear that $b(r)-r$ cuts the $r$-axis at $r_0=1$. Besides, the flaring out condition satisfied at $r_0$ i.e. $b^{'}(r_0)=b^{'}(1)\approx0.667<1$. Asymptotically flatness condition is also satisfied, i.e. $\frac{b(r)}{r}\rightarrow 0$ as $r\rightarrow \infty$ satisfied. Therefore, from Fig. \ref{1chf4}, one can notice that the shape function satisfies all the required conditions for a traversable wormhole.
Now, we discuss the traversable wormhole space-time supported by the phantom energy with the presence of exotic matter. For phantom energy EoS, $P_r=\omega\rho$ with $\omega<-1$ and using the shape function \eqref{25ch1} in field equations \eqref{9ch1}-\eqref{11ch1}, we can obtain the relations among stress-energy tensor components
\begin{equation}
\label{26ch1}
\rho=\frac{c\,\alpha}{\omega}\,r^{-\left(\frac{1}{\omega}+3\right)},
\end{equation}
\begin{equation}
\label{27ch1}
\rho+P_r=c\,\alpha\left(\frac{1}{\omega}+1\right)\,r^{-\left(\frac{1}{\omega}+3\right)},
\end{equation}
and
\begin{equation}
\label{28ch1}
\rho+P_t=\frac{1}{2}\,c\,\alpha\left(\frac{1}{\omega}-1\right)\,r^{-\left(\frac{1}{\omega}+3\right)}.
\end{equation}
Also, the NEC at the throat is given by
\begin{equation}
\label{29ch1}
\rho+P_r\mid_{r_0}=c\,\alpha\left(\frac{1}{\omega}+1\right)\,r_{0}^{-\left(\frac{1}{\omega}+3\right)}.
\end{equation}
In this case, it is clear that $\omega$ should not be equal to $-1$, i.e., $\omega\neq{-1}$. So, we consider $\omega<-1$ to imply the violation of NEC at the throat, i.e., the throat of the wormhole needs to open with phantom energy. 
\begin{figure}[h]
   \centering
   \includegraphics[width=8.5 cm]{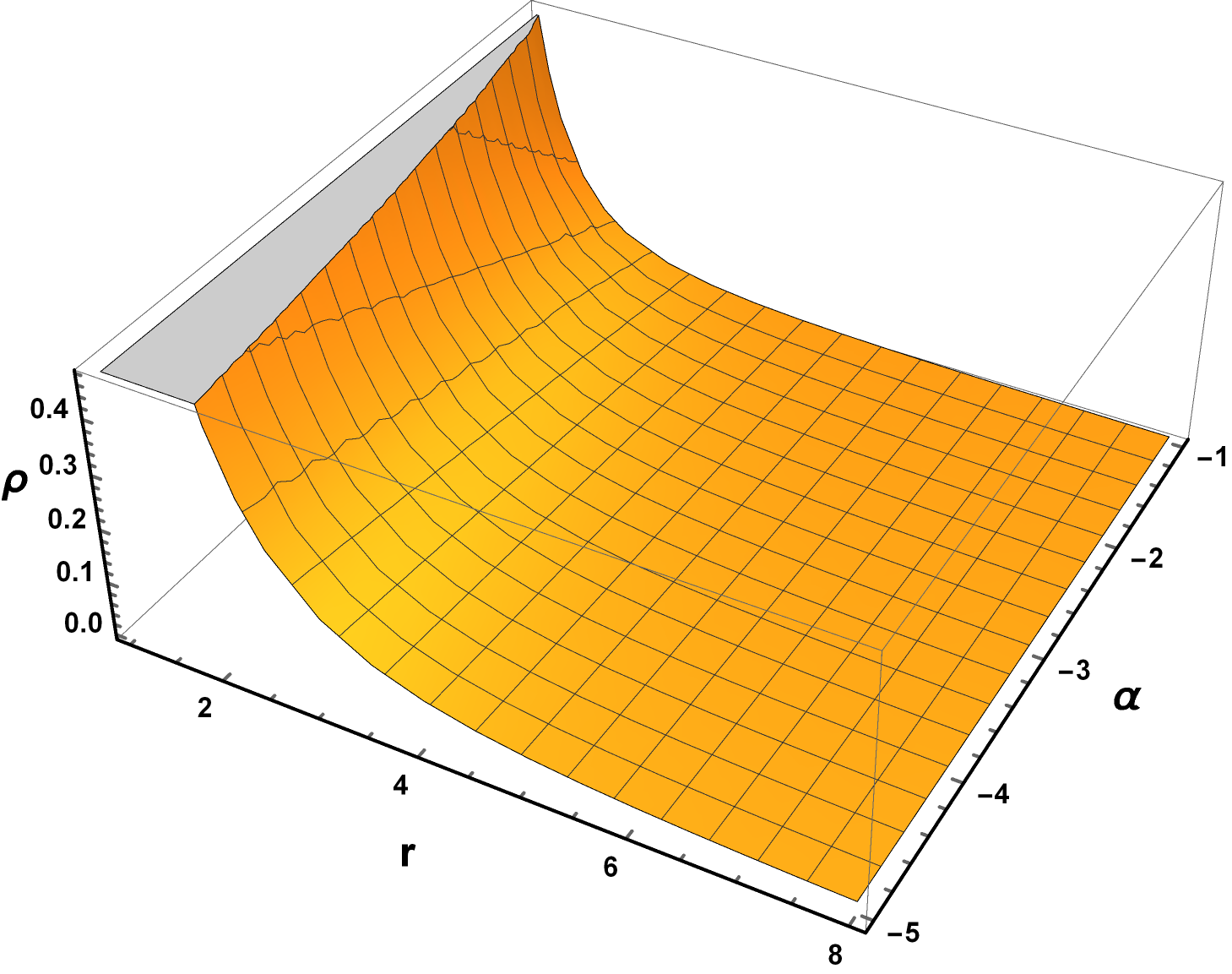}
   \caption{The profile of energy density,$\rho(r)$ w.r.t. $r$ and $\alpha$ with $\omega=-2,\,c=1$ for WH2.}
   \label{1chf5}
\end{figure}
We draw the graphical behavior of energy conditions depicted in Figs. \ref{1chf5} and \ref{1chf6}. For this model, we have considered a matter content that is related to the phantom equation of state, i.e., $\omega <-1$. In Fig. \ref{1chf5}, we have shown the behavior of energy density, and it takes positive values all over the range. It can be seen from the Fig. \ref{1chf6} that DEC is satisfied, i.e., $\rho-P_r\geq0$ and $\rho-P_t\geq0$, but we observe that NEC is violated due to $\rho+P_r<0$. The violation of NEC is proof of the presence of the exotic matter, which might be needed for the wormhole geometry.
\begin{figure}[h]
\centering
  \includegraphics[width=6.5 cm]{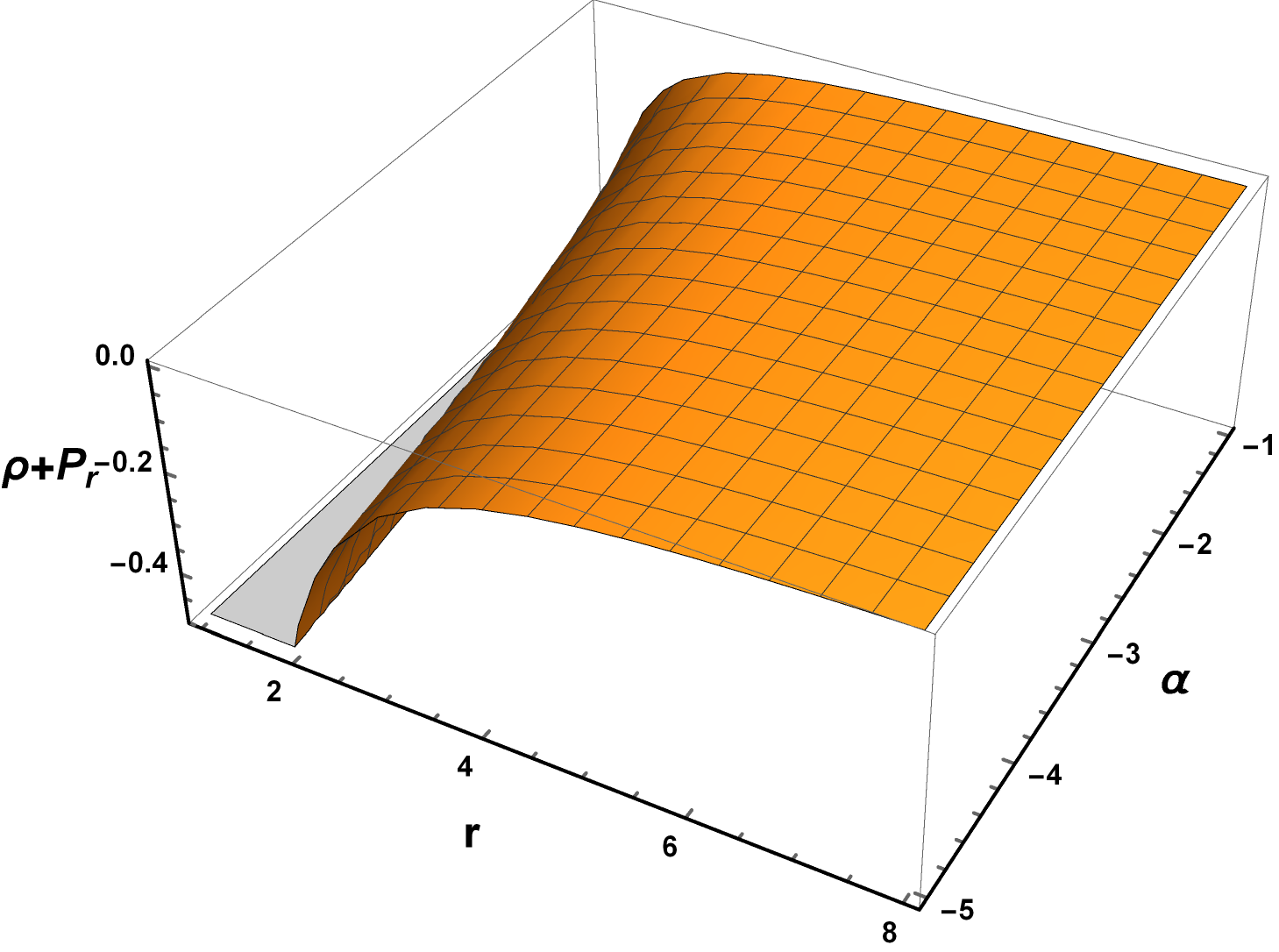}
  \includegraphics[width=6.5 cm]{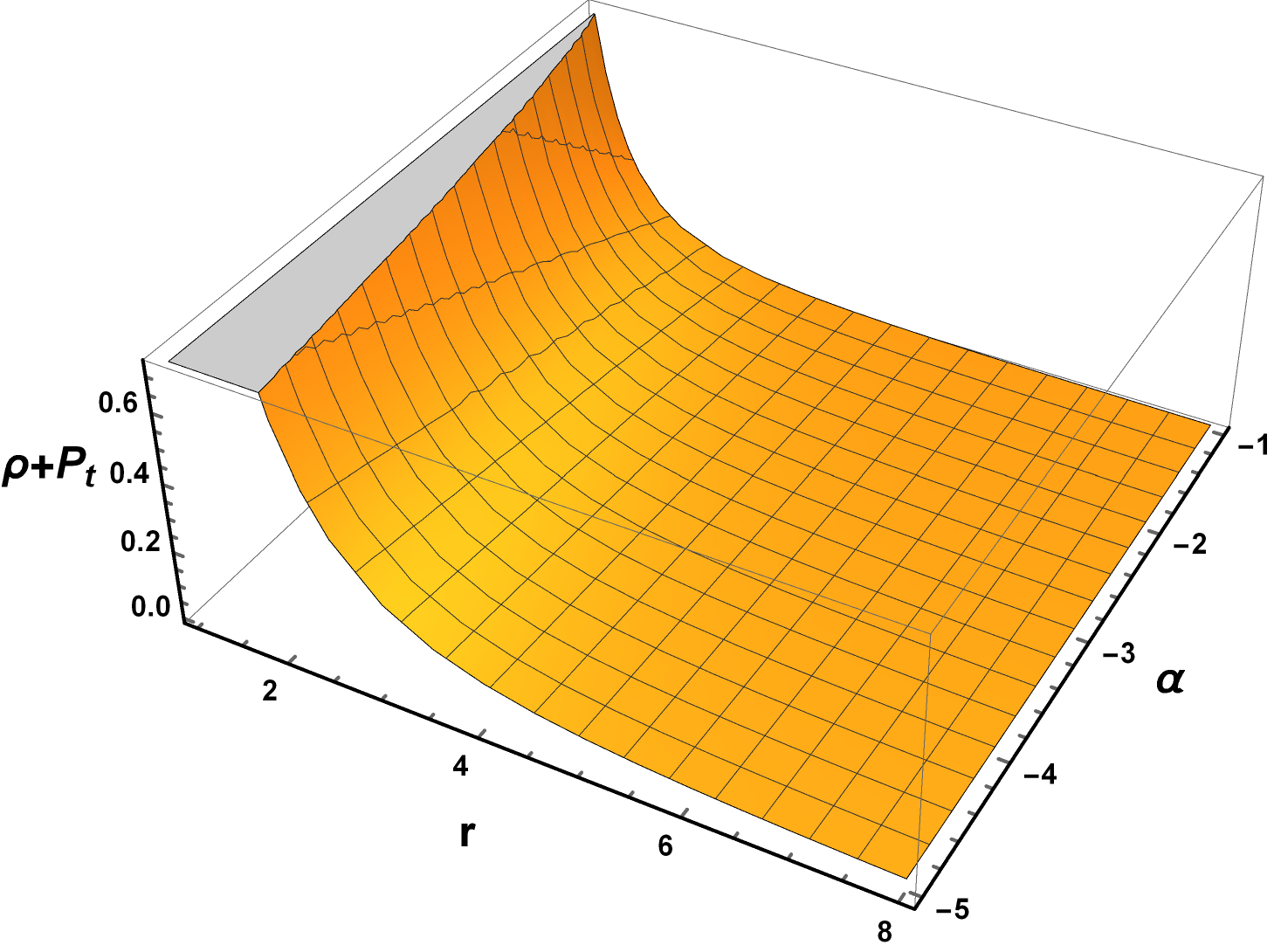}
  \includegraphics[width=6.5 cm]{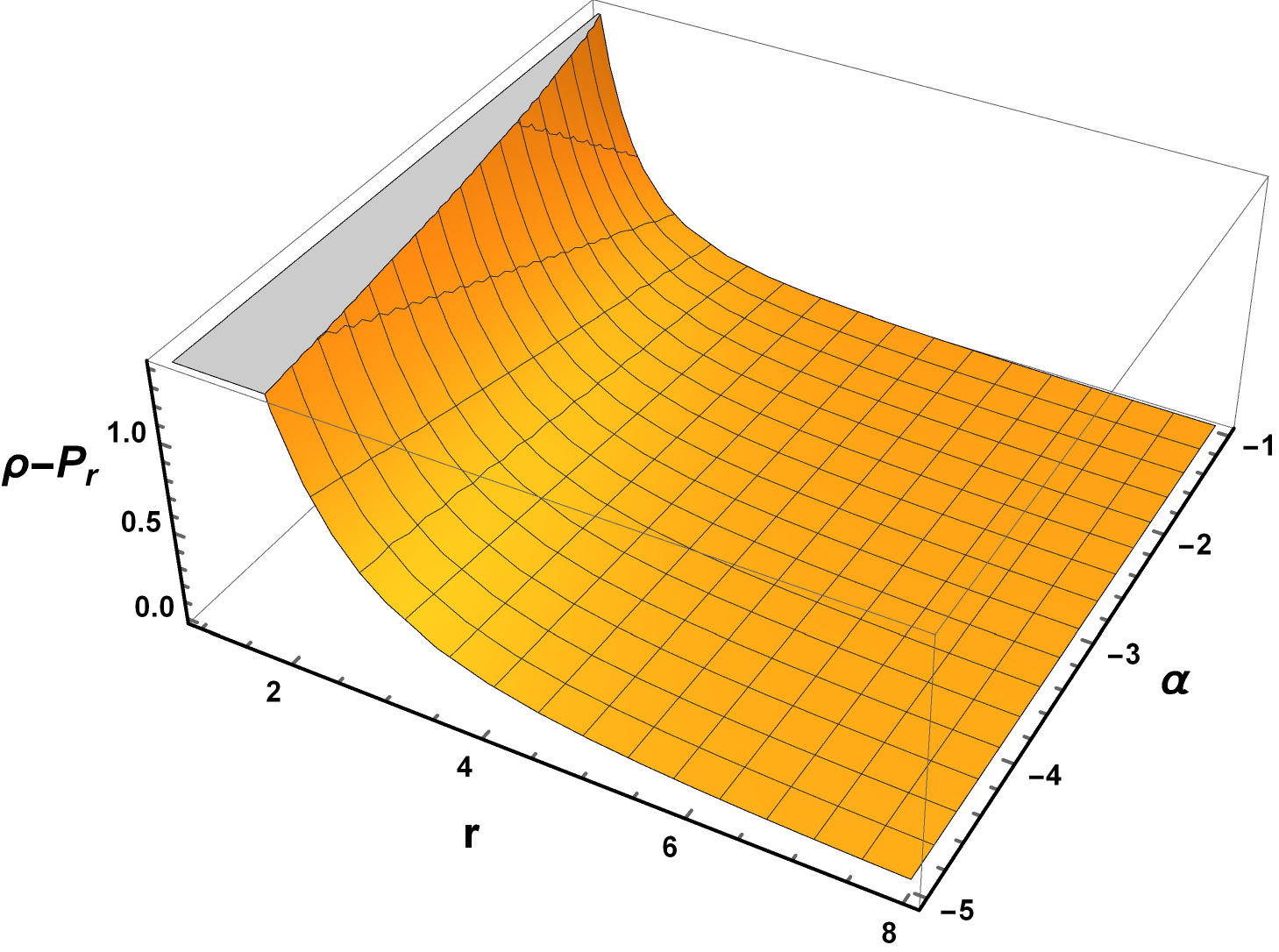}
  \includegraphics[width=6.5 cm]{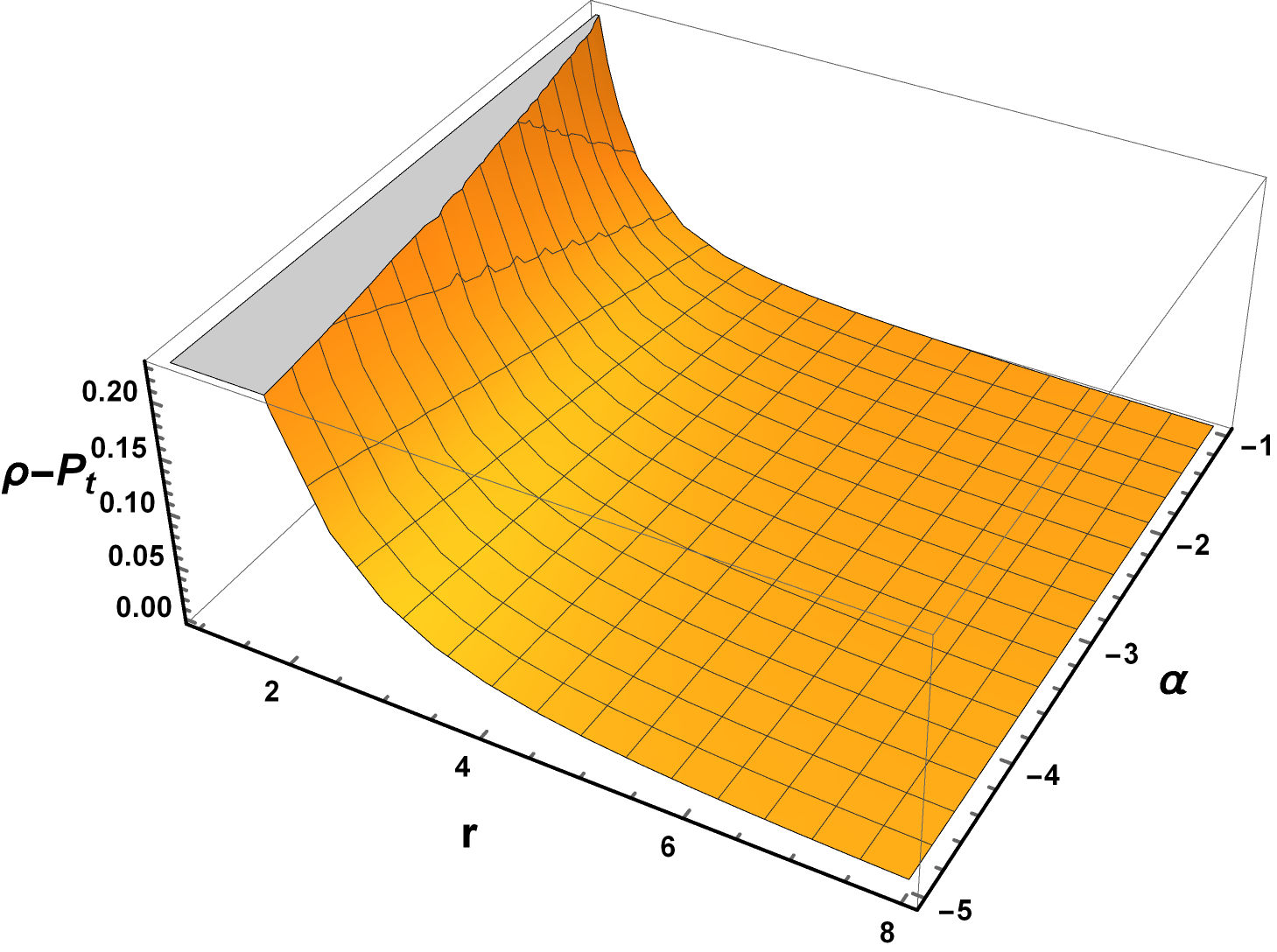}
\caption{The plots of WEC, NEC and DEC w.r.t. $r$  and $\alpha$ with $\omega=-2, c=1$ for WH2.}
\label{1chf6}
\end{figure}

\subsubsection{Wormhole (WH3) solution with $b(r)=r_0\left(\frac{r}{r_0}\right)^n$}
\begin{figure}[h]
\centering
	\includegraphics[width=6.5 cm]{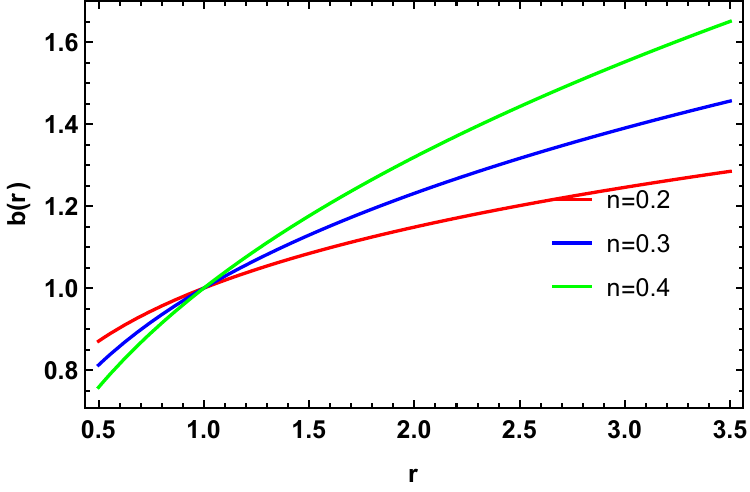}
	\includegraphics[width=6.5 cm]{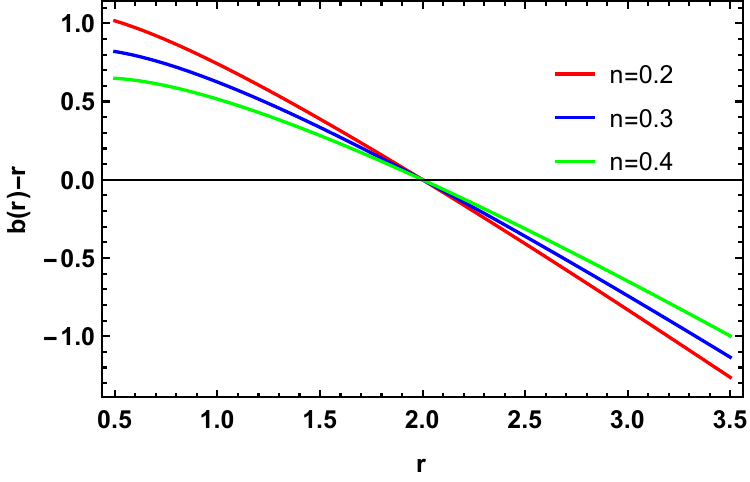}
	\includegraphics[width=6.5 cm]{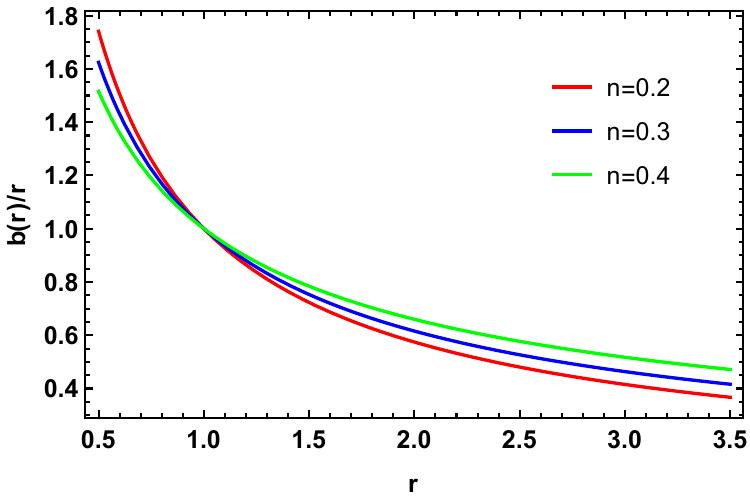}
	\includegraphics[width=6.5 cm]{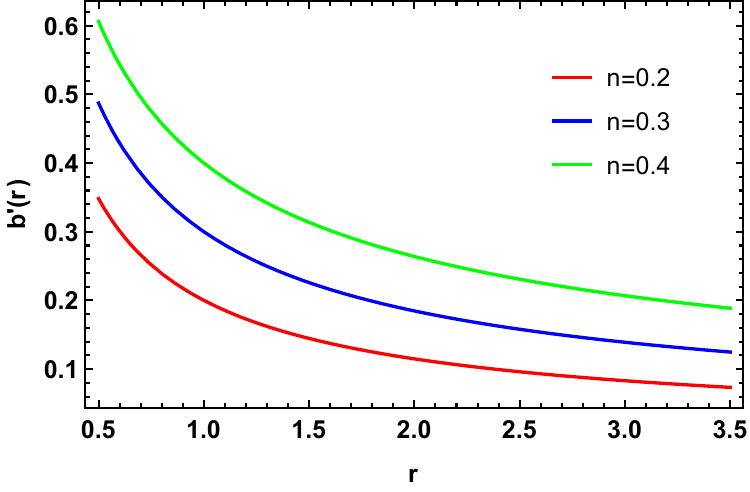}
	\caption{The plots of the shape functions for WH3.}
	\label{1chf7}
\end{figure}

\begin{figure}[h]
\centering
\includegraphics[width=8.5 cm]{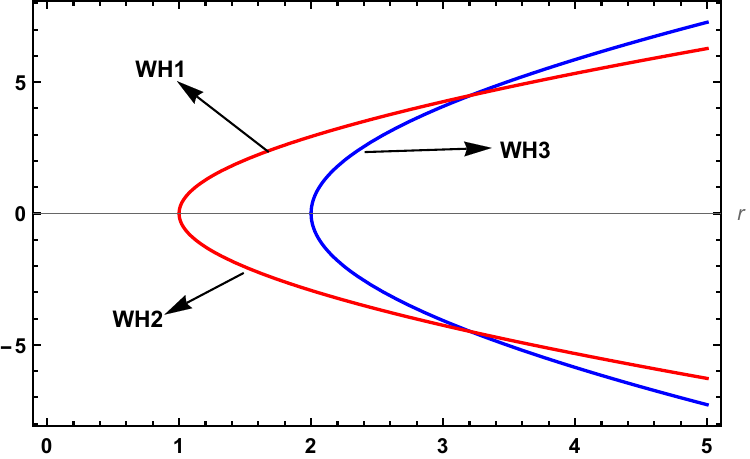}
    \caption{The plot shows two-dimensional embedding diagrams for WH1, WH2, and WH3 (here, WH1 and WH2 overlap each other).}
    \label{1chf8}
\end{figure}

\begin{figure}[h]
\centering
     \includegraphics[width=8.5 cm]{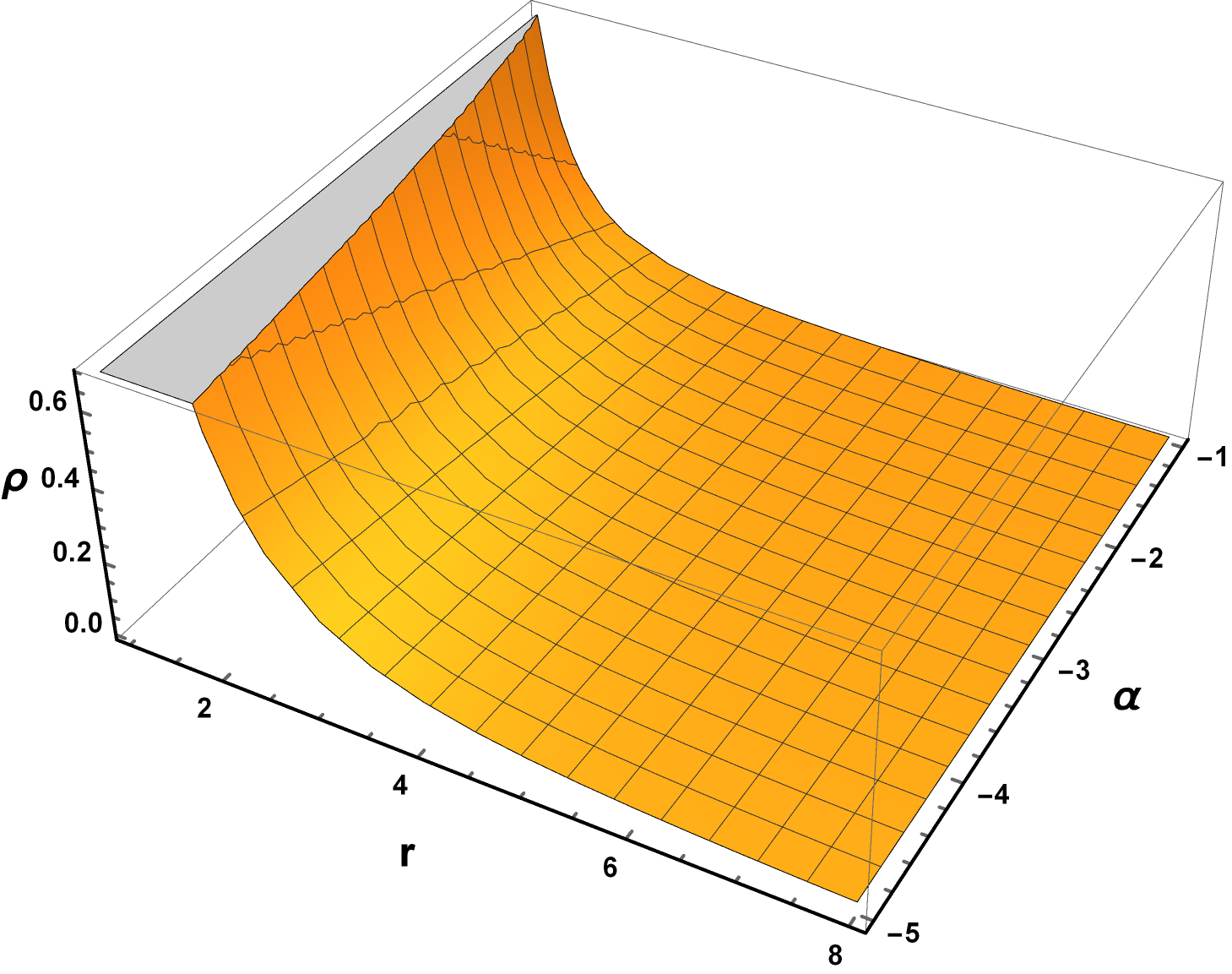}
     \caption{The plot of energy density, $\rho(r)\geq0$ with $n=-0.5$, $r_0=2$ for WH3.}
     \label{1chf9}
\end{figure}

\begin{figure}[h]
\centering
  \includegraphics[width=6.5 cm]{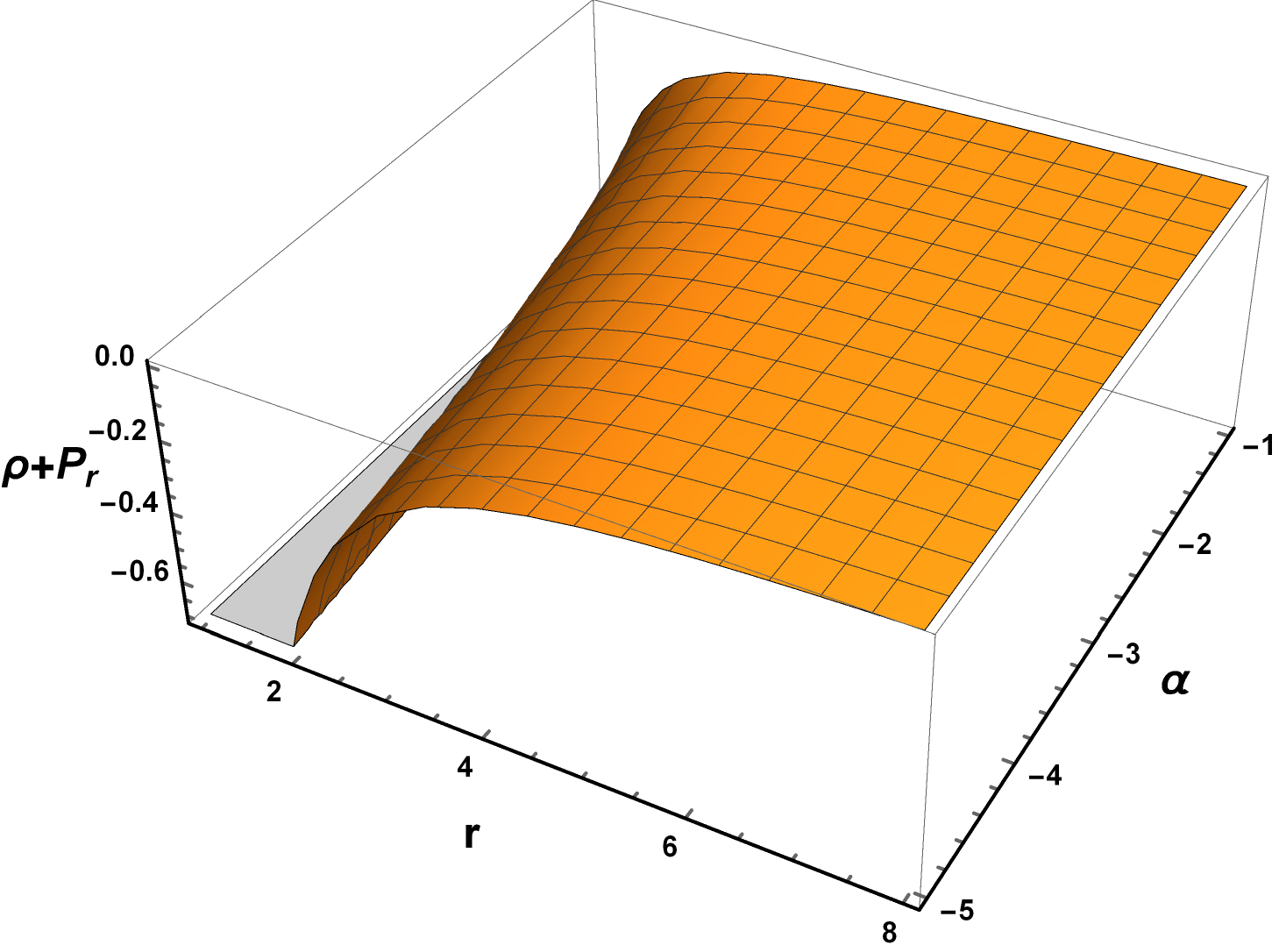}
  \includegraphics[width=6.5 cm]{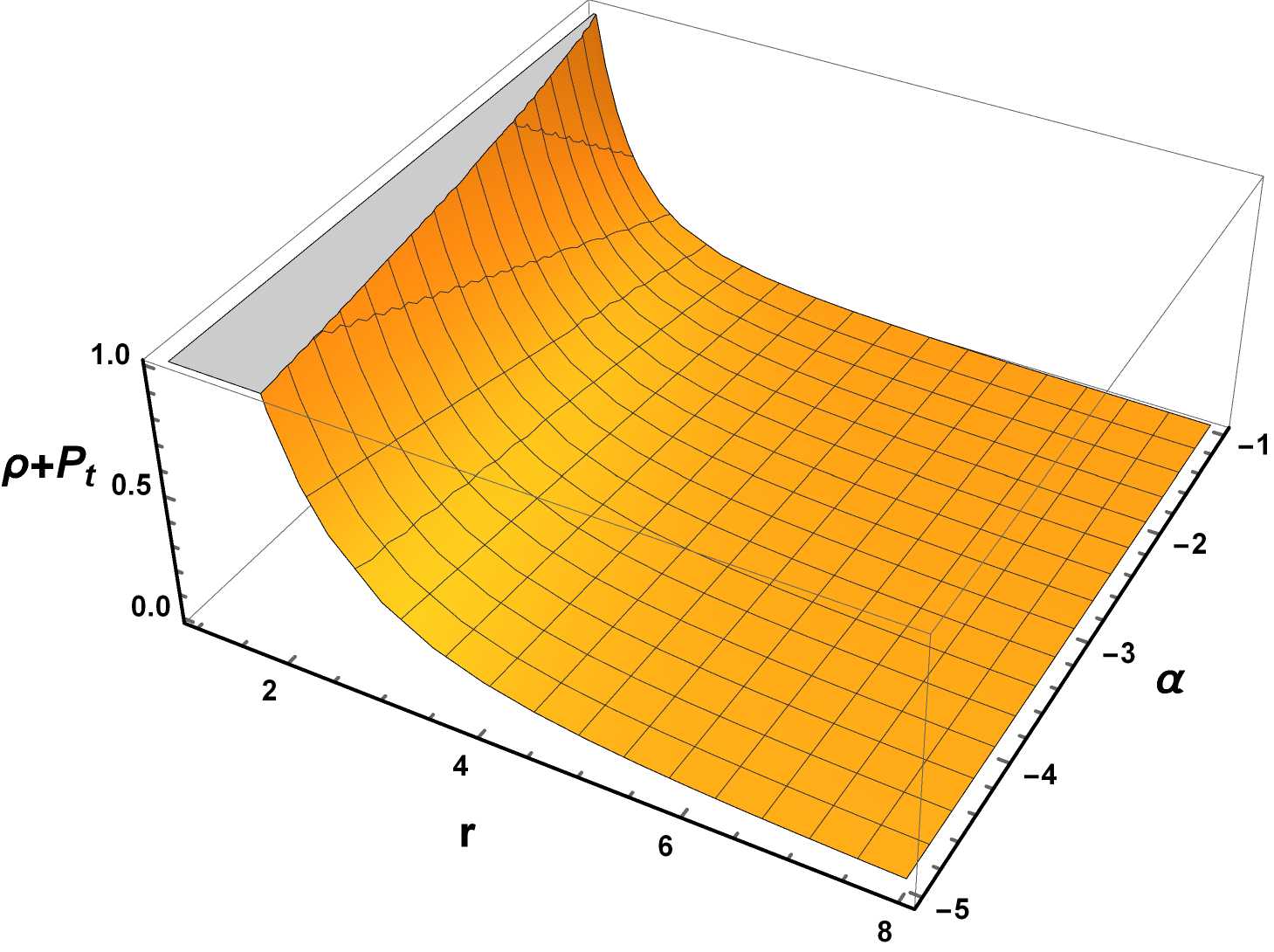}
  \includegraphics[width=6.5 cm]{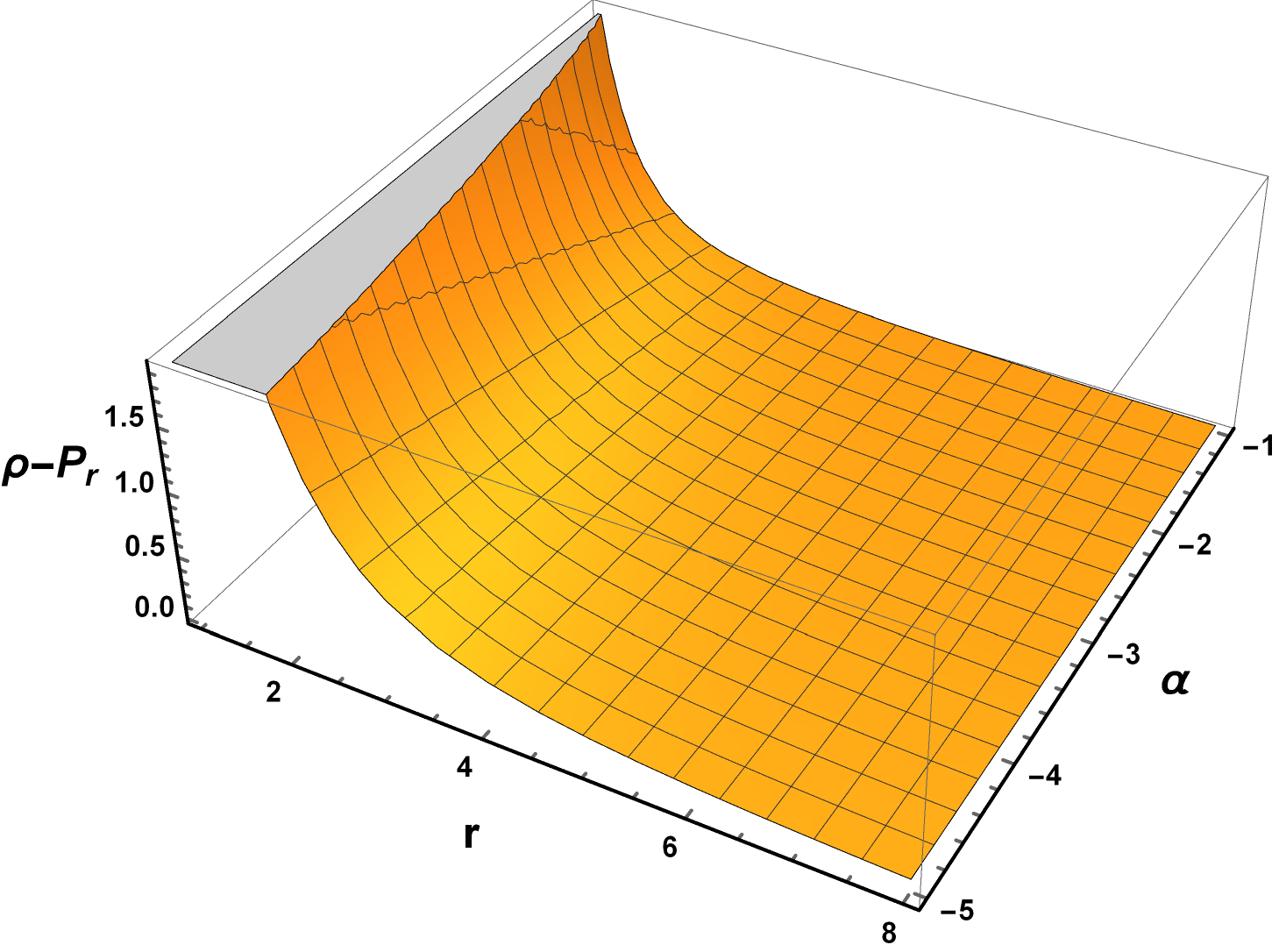}
  \includegraphics[width=6.5 cm]{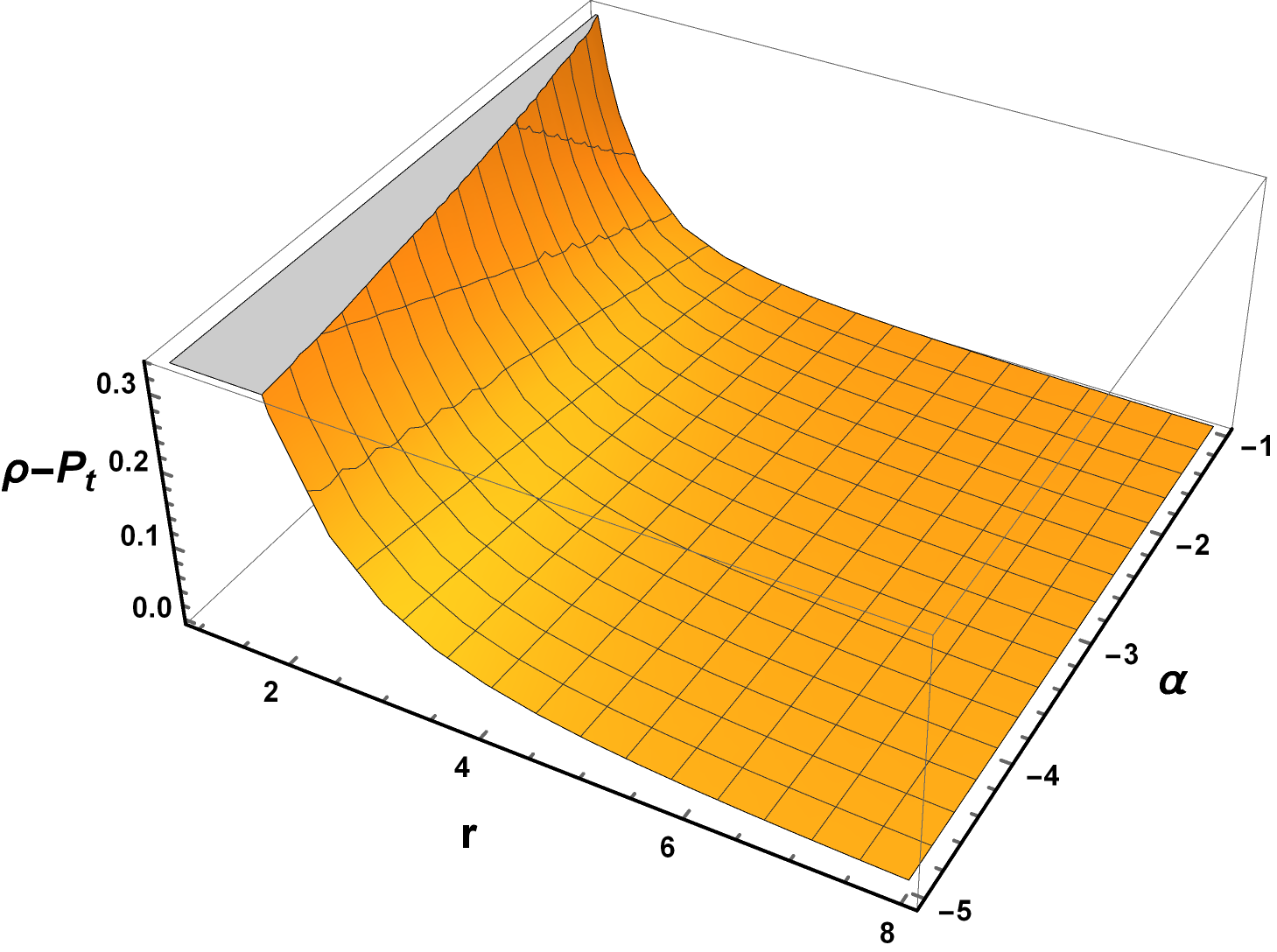}
\caption{The plots of WEC, NEC and DEC w.r.t. $r$  and $\alpha$ with $n=-0.5$, $r_0=2$ for WH3.}
\label{1chf10}
\end{figure}

In this subsection, we have considered the specific shape function, $b(r)=r_0\left(\frac{r}{r_0}\right)^n$, and for this choice, the expressions for energy conditions from Eqs. \eqref{9ch1}-\eqref{11ch1} are obtained as
\begin{equation}
\label{32ch1}
\rho=-n\,\alpha\,\frac{r^{n-3}}{r_0^{n-1}},
\end{equation}
\begin{equation}
\label{33ch1}
\rho+P_r=\left(1-n\right)\,\alpha\,\frac{r^{n-3}}{r_0^{n-1}},
\end{equation}
and
\begin{equation}
\label{34ch1}
\rho+P_t=-\frac{1}{2}\left(n+1\right)\alpha\,\frac{r^{n-3}}{r_0^{n-1}}.
\end{equation}
Now, at the throat i.e. at $r=r_0$, Eq. \eqref{33ch1} reduce to
\begin{equation}
\label{35ch1}
\rho+P_r\mid_{r_0}=\frac{\left(1-n\right)\alpha}{r_0^2}.
\end{equation}
Taking into account that the condition $\frac{b(r)}{r}\rightarrow0$  as  $r\rightarrow\infty$ will satisfy when $n<1$. In Fig. \ref{1chf7}, we have shown the behavior of $b(r),\,b(r)-r,\,\frac{b(r)}{r}$ and $b'(r)$ respectively and in this case $b(r)-r$ cuts the $r$-axis at $r_0$= 2, which is the throat radius of wormhole (see Fig. \ref{1chf8}). From Fig. \ref{1chf7}, one can see that $b(r)$ satisfies all the necessary conditions for a traversable wormhole.
Moreover, from Fig. \ref{1chf9}, we see that energy density, $\rho$, is always positive throughout the space-time. From Fig. \ref{1chf10}, we can observe that DEC is satisfied within the range of $\alpha$; however, NEC is violated as $\rho+P_r< 0$.

\section{Wormhole models with non-linear $f(Q)$ model}\label{ch1sec7}
In this section, by presuming a polynomial form of $f(Q)=a\,Q^2+B$, we continue trying to discuss the fundamental equations for the anisotropic fluid. For the choice of $P_t=mP_r$ and $P_r=w \rho$, the differential Eqs. \eqref{12ch1}-\eqref{14ch1} become highly non-linear and very complicated to solve and analyze these types of systems (see Ref. \cite{Newton/2020}). Therefore, we consider two different specific shape functions in finding wormhole solutions.
\subsubsection{Wormhole (WH4) solution with $b(r)=r_0\left(\frac{r}{r_0}\right)^n$}
\begin{figure}[h]
\centering
	\includegraphics[width=6.5 cm]{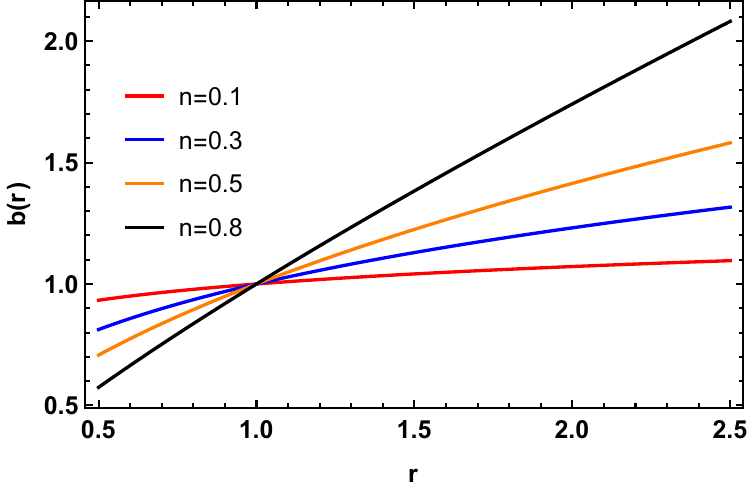}
	\includegraphics[width=6.5 cm]{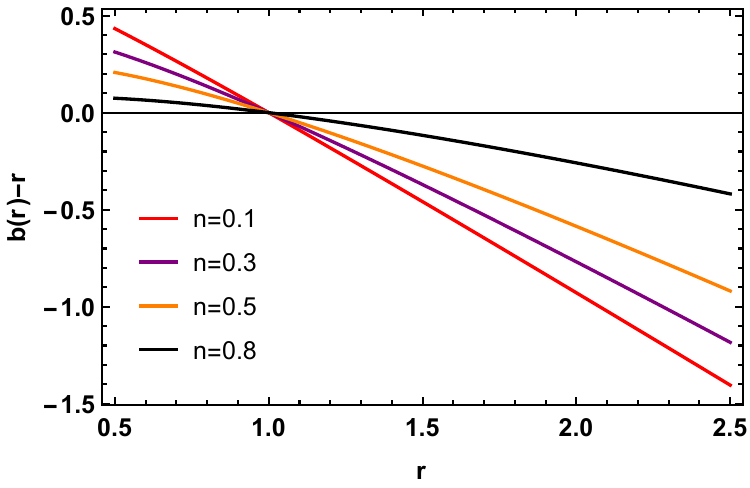}
	\includegraphics[width=6.5 cm]{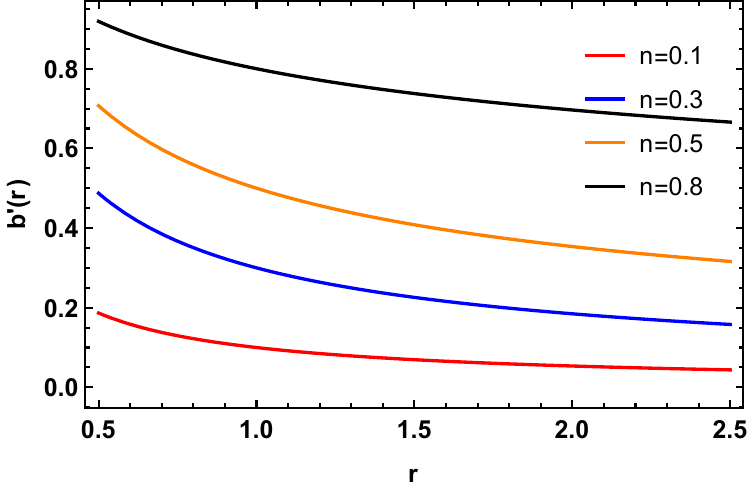}
	\includegraphics[width=6.5 cm]{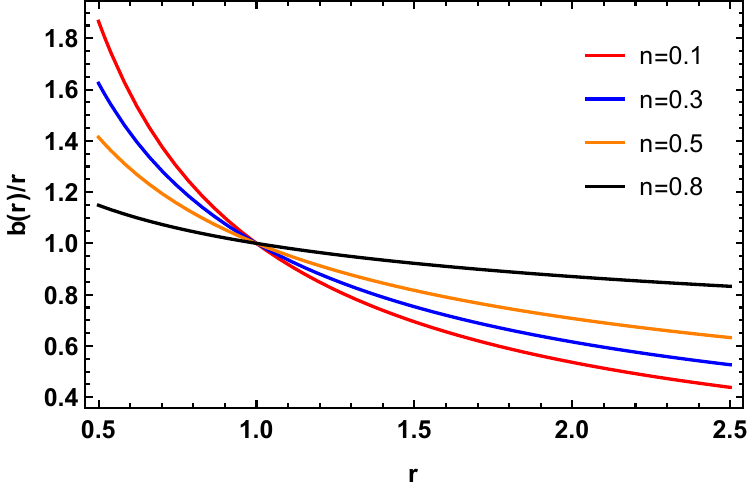}
	\caption{The plot of the shape functions for WH4}
	\label{1chf11}
\end{figure}

Here, in this subsection, we have assumed the specific shape function given by $b(r)=r_0\left(\frac{r}{r_0}\right)^n$ and for this choice the corresponding stress energy components from Eqs. \eqref{12ch1}-\eqref{14ch1} are obtained as follows
\begin{equation}
\rho=\frac{2 a \left((6 n-11) r_0 \left(\frac{r}{r_0}\right){}^n+7 r\right) \left(r-r_0 \left(\frac{r}{r_0}\right){}^n\right)}{r^6}+\frac{B}{2},
\end{equation}
\begin{equation}
\rho+P_r= \frac{4 a \left(r-r_0 \left(\frac{r}{r_0}\right){}^n\right) \left((3 n-7) r_0 \left(\frac{r}{r_0}\right){}^n+4 r\right)}{r^6},
\end{equation}
and
\begin{equation}
\rho+P_t= \frac{2 a \left(r-r_0 \left(\frac{r}{r_0}\right){}^n\right) \left((3 n-5) r_0 \left(\frac{r}{r_0}\right){}^n+4 r\right)}{r^6}.
\end{equation}
The behavior of the shape function is depicted in Fig. \ref{1chf11}. It can be seen from Fig. \ref{1chf11} that the shape function, $b(r)$, shows increasing behavior for different values of $n$. Furthermore, to check the asymptotically flatness condition we plot $\frac{b(r)}{r}$ verses $r$ which gives $\frac{b(r)}{r}\rightarrow 0$ as $r\rightarrow \infty$. Here $b(r)-r$ cuts the $r$-axis at $r_0=1$, which is the throat radius for this WH4 when $n<1$. All of the above results show the traversability of WH4.\\
\begin{figure}[h]
\centering
     \includegraphics[width=6.5 cm]{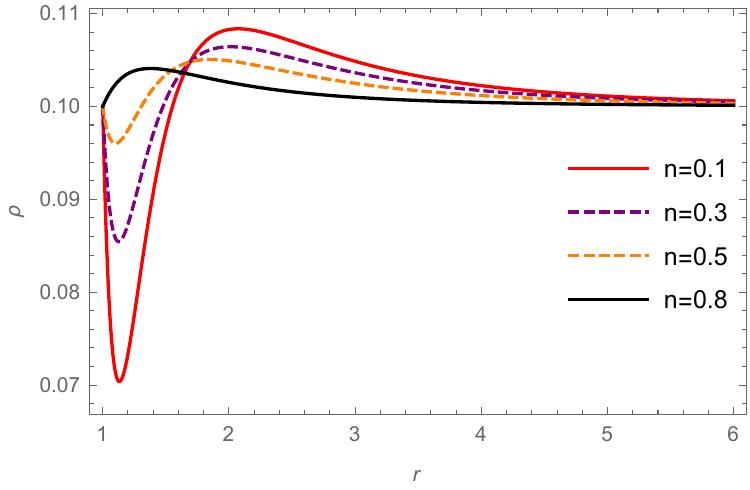}
     \caption{The plot of $\rho$ versus $r$ with $n=0.1$ (red), $n=0.3$ (purple), $n=0.5$ (orange), $n=0.8$ (black), with $a=0.1$, $B=0.2$ and $r_0=1$ for WH4.}
     \label{1chf12}
\end{figure}
\begin{figure}[h]
\centering
  \includegraphics[width=6.5 cm]{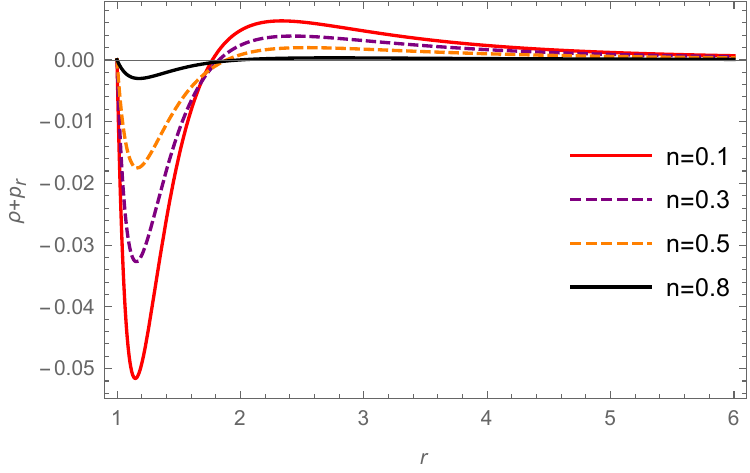}
  \includegraphics[width=6.5 cm]{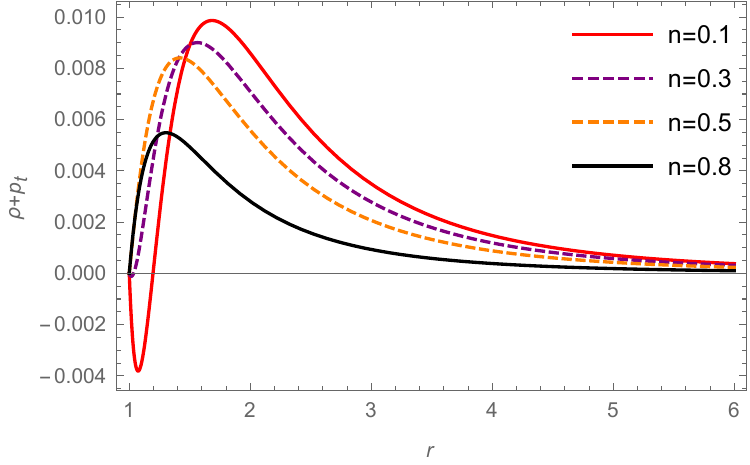}
\caption{The plots of $\rho+P_r$ and $\rho+P_t$ versus $r$ with $n=0.1$ (red), $n=0.3$ (purple), $n=0.5$ (orange), $n=0.8$ (black), with $a=0.1$, $B=0.2$ and $r_0=1$ for WH4.}
\label{1chf13}
\end{figure}
\begin{figure}[h]
\centering
  \includegraphics[width=6.5 cm]{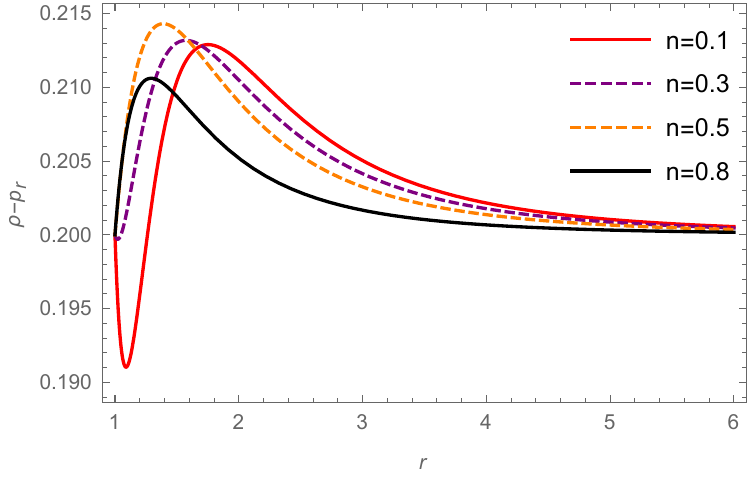}
  \includegraphics[width=6.5 cm]{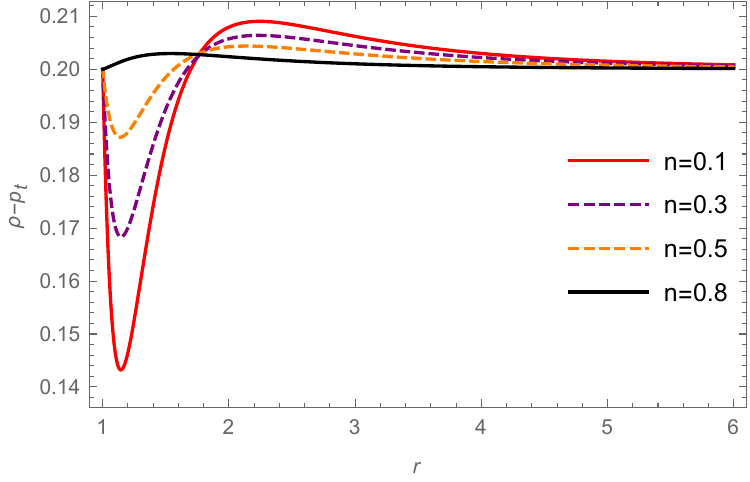}
\caption{The plots of $\rho-P_r$ and $\rho-P_t$ versus $r$ with $n=0.1$ (red), $n=0.3$ (purple), $n=0.5$ (orange), $n=0.8$ (black), with $a=0.1$, $B=0.2$ and $r_0=1$ for WH4.}
\label{1chf14}
\end{figure}
The behavior of energy conditions is shown in Figs. \ref{1chf12}-\ref{1chf14}. To verify the profiles of energy conditions, we have fixed the parameters as $a=0.1$, $B=0.2$, and $r_0=1$. As we know, energy density has to be positive everywhere, and its behaviors for WH4 are presented in Fig. \ref{1chf12} for different choices of $n$. Also, from Fig. \ref{1chf13}, we observed that $\rho+P_r<0$ and $\rho+P_t>0$, implying the violation of NEC at wormhole throat. Besides, we have shown the graph of DEC in Fig. \ref{1chf14}. The violation of NEC for different values of the parameter $n$ is proof of the presence of exotic matter at the wormhole throat, which may be required for the wormhole geometry.
\subsubsection{Wormhole (WH5) solution with $b(r)=\gamma\,r_0\,\left(1-\frac{r_0}{r}\right)+r_0$}
\begin{figure}[h]
\centering
	\includegraphics[width=6.5 cm]{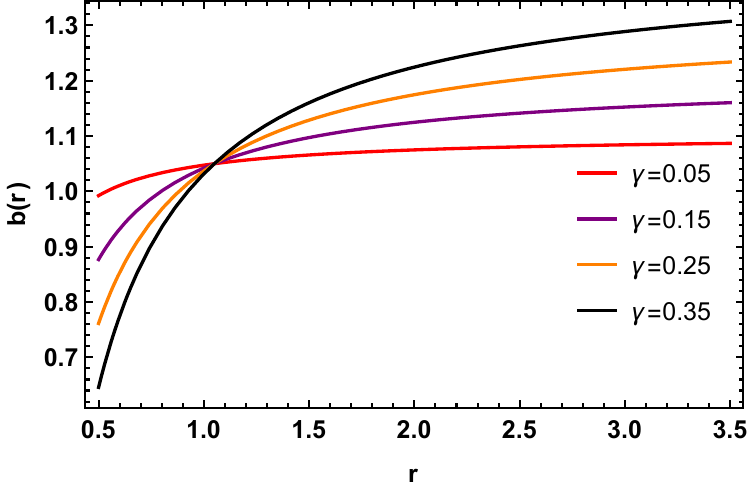}
	\includegraphics[width=6.5 cm]{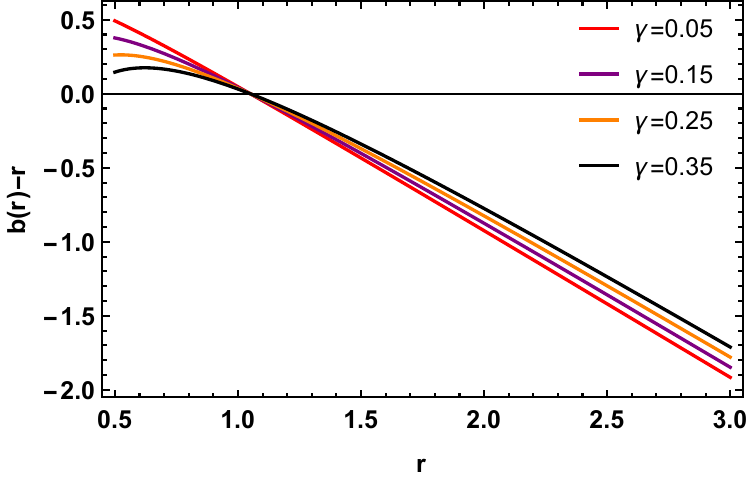}
	\includegraphics[width=6.5 cm]{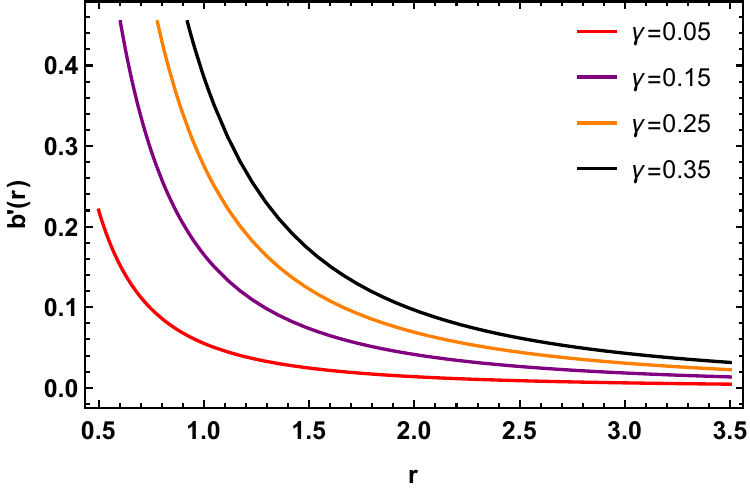}
	\includegraphics[width=6.5 cm]{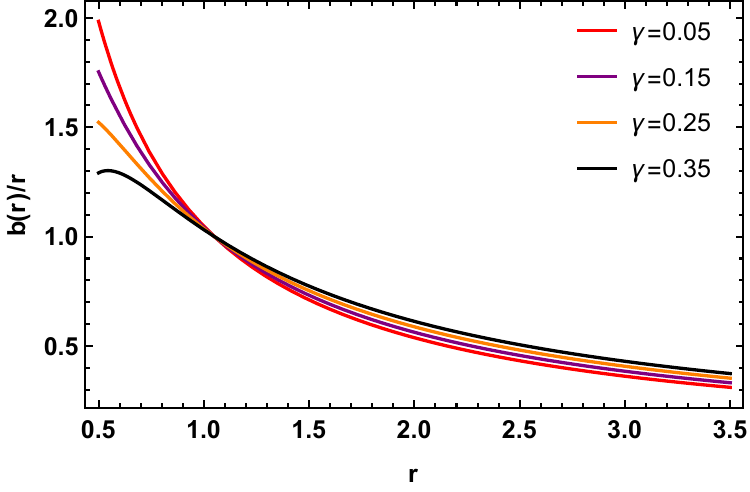}
	\caption{The plots of the shape functions for WH5}
	\label{1chf15}
\end{figure}

At last, we have considered another specific shape function given by $b(r)=\gamma\,r_0\,\left(1-\frac{r_0}{r}\right)+r_0$. We checked the behavior of this shape function which is depicted in Fig. \ref{1chf15}. It can be seen from Fig. \ref{1chf15} that the shape function, $b(r)$, shows increasing behavior for different values of $n$. We plot $\frac{b(r)}{r}$ with respect to $r$ to check the asymptotically flatness condition which gives $\frac{b(r)}{r}\rightarrow 0$ as $r\rightarrow \infty$. Here, in this case, $b(r)-r$ cuts the $r$-axis at $r_0=1.05$, which is the throat radius (see Fig. \ref{1chf15}) for this WH5. Flare out condition is also satisfied as $b^{'}(r)<1$ for $r>r_0$. All the above profiles for the shape function of WH5 satisfy the necessary properties for a traversable wormhole.
\begin{figure}[h]
\centering
     \includegraphics[width=6.5 cm]{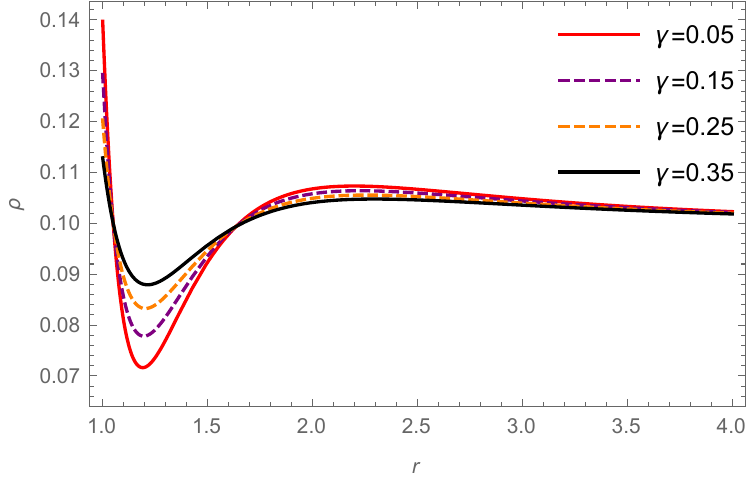}
     \caption{The plot of $\rho$ versus $r$ with $\gamma=0.05$ (red), $\gamma=0.15$ (purple), $\gamma=0.25$ (orange), $\gamma=0.35$ (black), with $a=0.1$, $B=0.2$ and $r_0=1.05$ for WH5.}
     \label{1chf16}
\end{figure}
\begin{figure}[h]
\centering
  \includegraphics[width=6.5 cm]{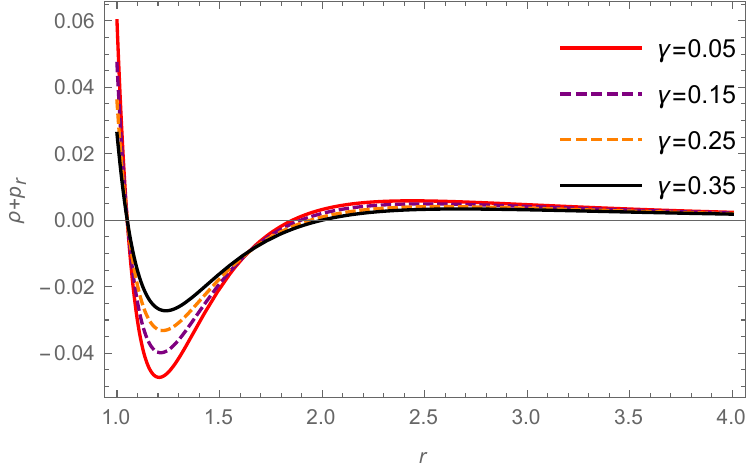}
  \includegraphics[width=6.5 cm]{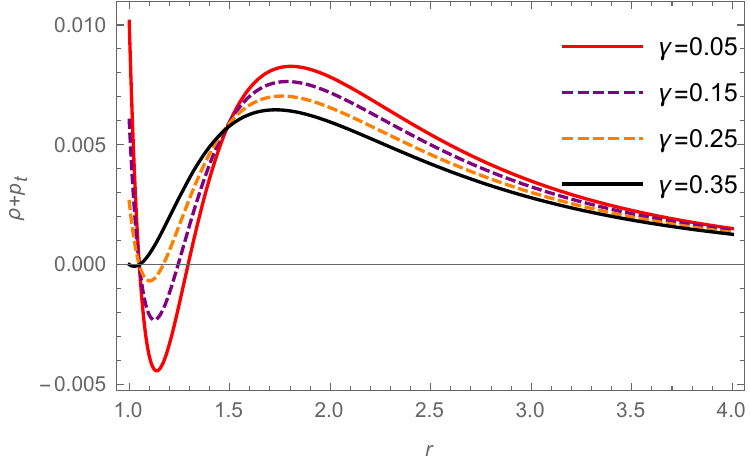}
\caption{The plots of $\rho+P_r$ and $\rho+P_t$ versus $r$ with $\gamma=0.05$ (red), $\gamma=0.15$ (purple), $\gamma=0.25$ (orange), $\gamma=0.35$ (black), with $a=0.1$, $B=0.2$ and $r_0=1.05$ for WH5.}
\label{1chf17}
\end{figure}
\begin{figure}[h]
\centering
  \includegraphics[width=6.5 cm]{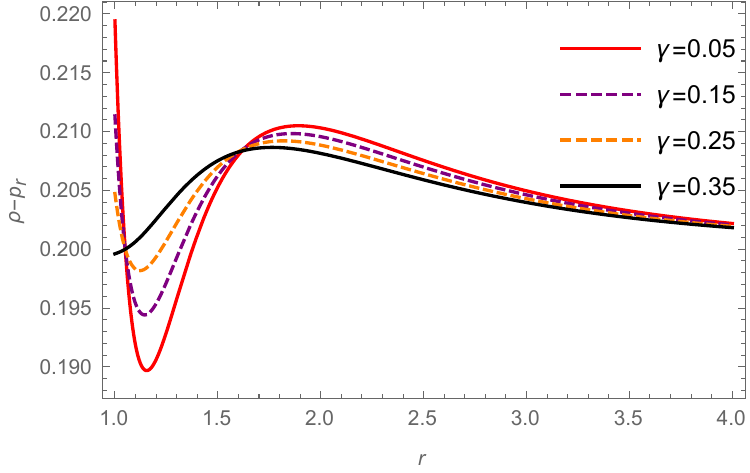}
  \includegraphics[width=6.5 cm]{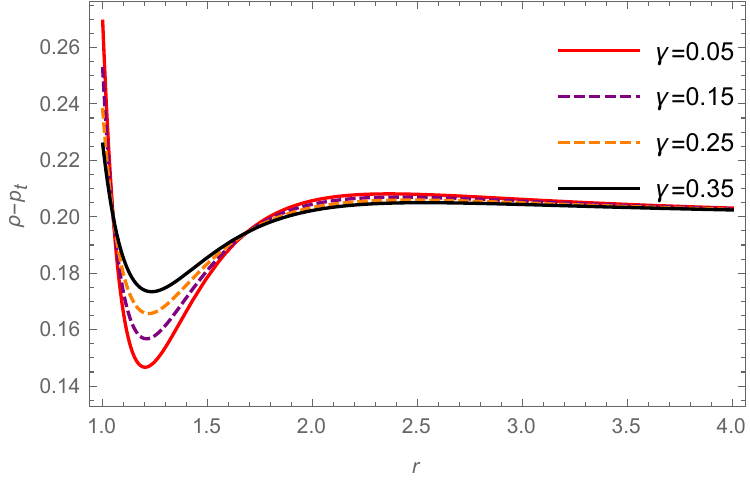}
\caption{The plots of $\rho+P_r$ and $\rho+P_t$ versus $r$ with $\gamma=0.05$ (red), $\gamma=0.15$ (purple), $\gamma=0.25$ (orange), $\gamma=0.35$ (black), with $a=0.1$, $B=0.2$ and $r_0=1.05$ for WH5.}
\label{1chf18}
\end{figure}
Now we check the energy conditions with this specific shape function, and for this choice, the components of energy-momentum tensors can be obtained from Eqs. (\ref{12ch1}-\ref{14ch1}) as
\begin{equation}
\rho=\frac{2 a \left(r-r_0\right) \left(r-\gamma  r_0\right) \left(7 r^2-11 (\gamma +1) r_0 r+17 \gamma  r_0^2\right)}{r^8}+\frac{B}{2}
\end{equation}
\begin{equation}
\rho+P_r= \frac{4 a \left(r-r_0\right) \left(r-\gamma  r_0\right) \left(4 r^2-7 (\gamma +1) r_0 r+10 \gamma  r_0^2\right)}{r^8}
\end{equation}
and
\begin{equation}
\rho+P_t= \frac{2 a \left(r-r_0\right) \left(r-\gamma  r_0\right) \left(4 r^2-5 (\gamma +1) r_0 r+8 \gamma  r_0^2\right)}{r^8}
\end{equation}
Figs. \ref{1chf16}-\ref{1chf18} shows the behavior of energy conditions for WH5. Here, we have chosen the parameters as $a=0.1$, $B=0.2$, and $r_0=1.05$. In Fig. \ref{1chf16}, we have shown the graph of energy density, $\rho$ versus $r$, which shows a positive behavior throughout space-time for different values of the parameter $\gamma$. Fig. \ref{1chf17} shows the behavior of NEC, and one can see from the figure that NEC is violated as $\rho+P_r<0$, which implies the violation of WEC. The violation of NEC is the primary requirement for a traversable wormhole. Moreover, we have shown the graph of DEC in Fig. \ref{1chf18}. These results tend to determine that the wormhole solutions that have been obtained are acceptable in this symmetric teleparallel gravity.
\section{Volume integral quantifier}\label{ch1sec9}
\begin{figure}[H]
\centering
	\includegraphics[width=6.5 cm]{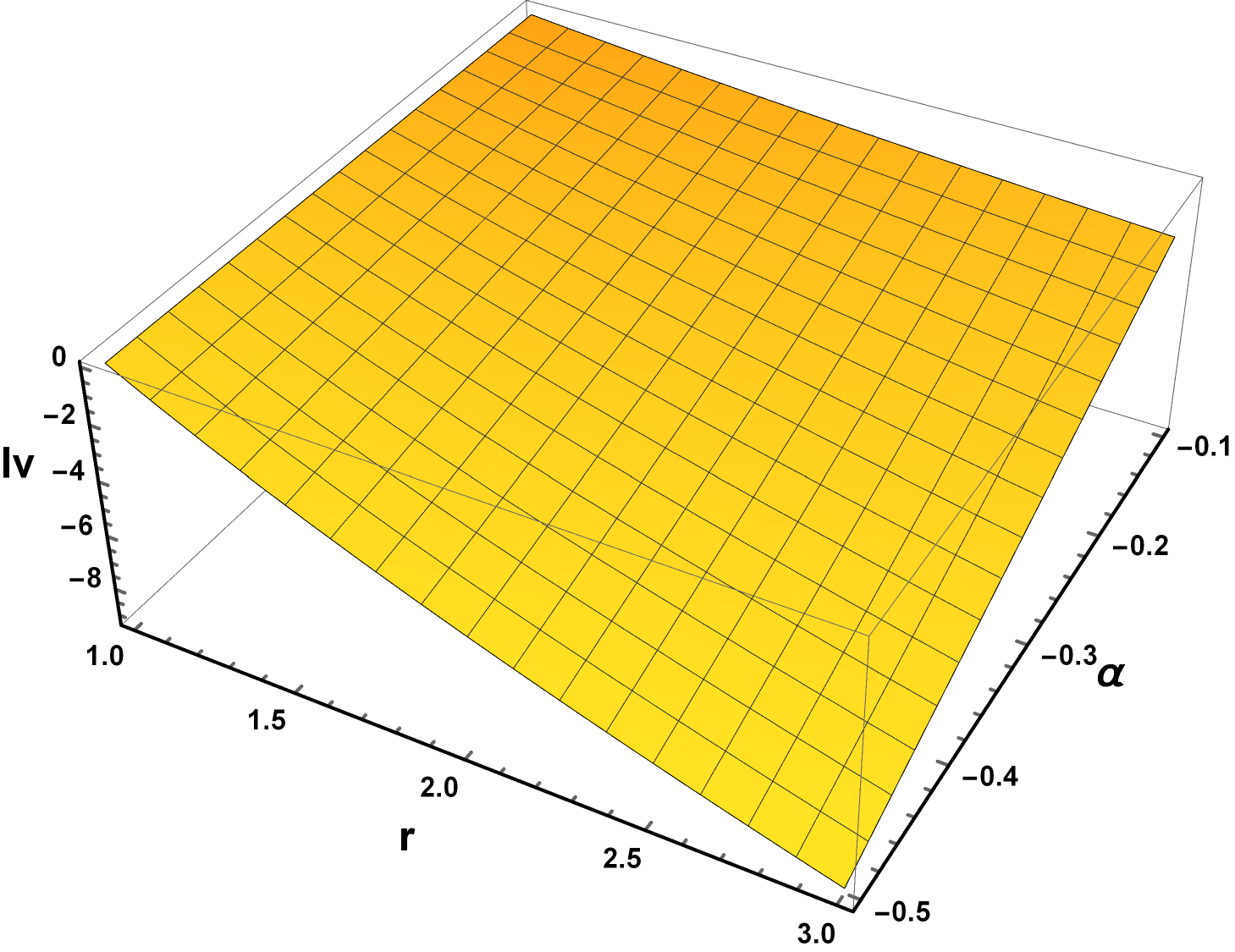}
	\caption{3D Plot of VIQ associated with model 1 and here it is clear that when $r_1\rightarrow r_0^+$ then $Iv\rightarrow0$, i.e., minimize, the violation of NEC would be possible. For this model, we consider $n=-0.25$, $r_0=1$, and $a=3$.}
	\label{1chf20}
	\end{figure}

\begin{figure}[H]
\centering
	\includegraphics[width=6.5 cm]{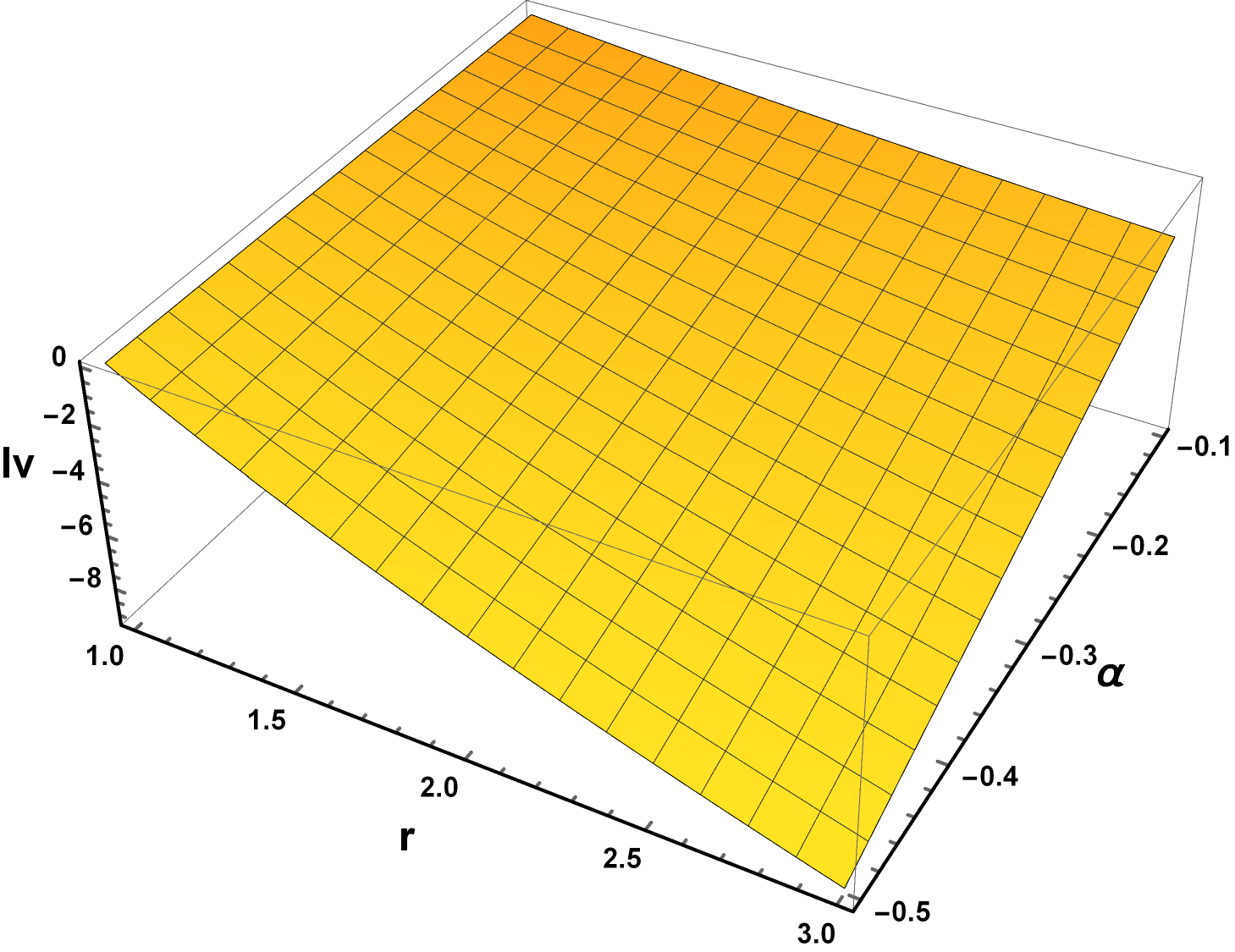}
	\caption{3D Plot of VIQ associated with model 2 and here it is clear that when $r_1\rightarrow r_0^+$ then $Iv\rightarrow0$. For this model, we consider $\omega=-2$, $r_0=1$ and $a=3$.}
	\label{1chf21}
	\end{figure}

\begin{figure}[H]
\centering
\includegraphics[width=6.5 cm]{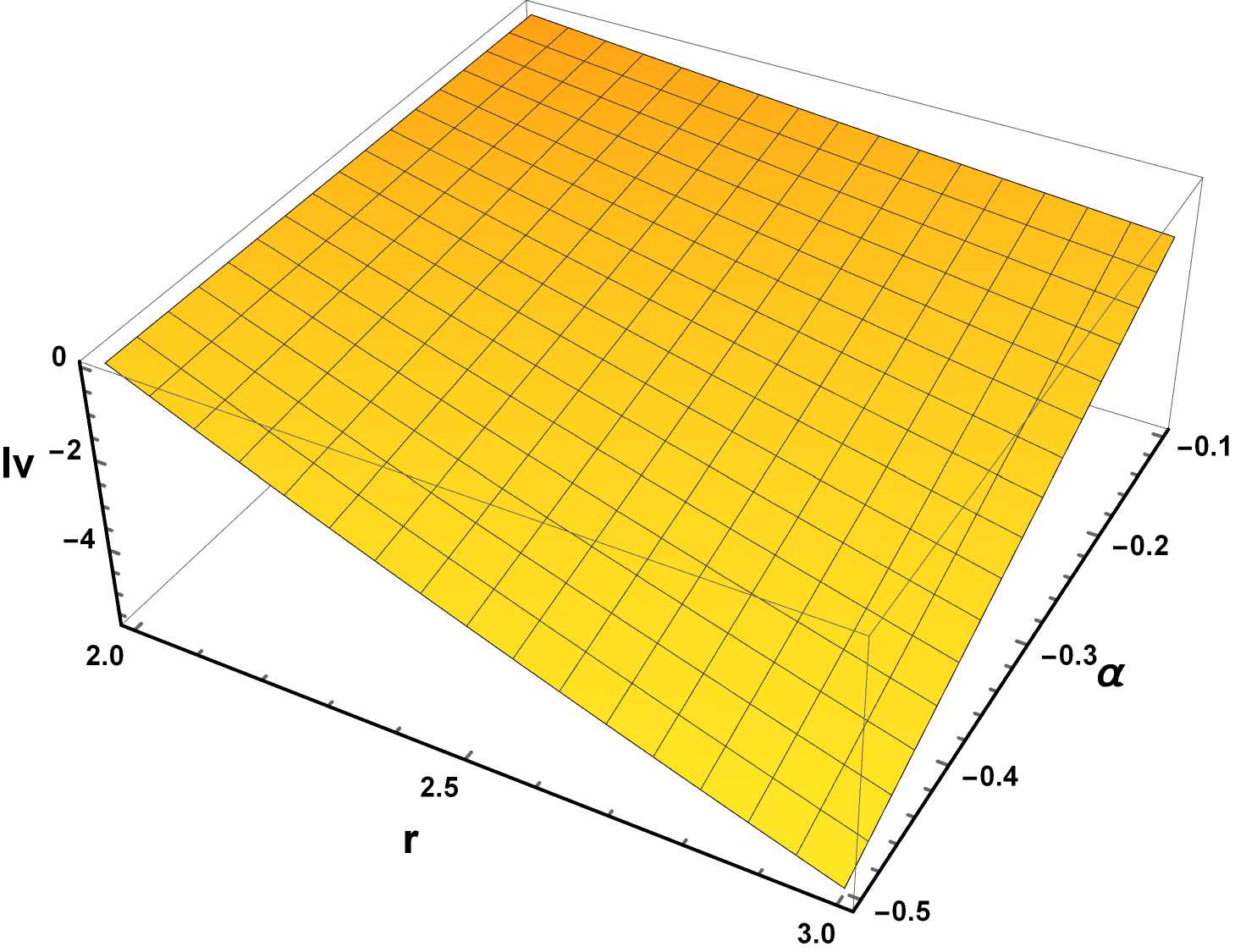}
\caption{3D Plot of VIQ associated with model 3 and here it is clear that when $r_1\rightarrow r_0^+$ then $Iv\rightarrow0$ and we consider $n=-0.5$, $r_0=2$ and $a=3$.}
	\label{1chf22}
	\end{figure}

\begin{figure}[H]
\centering
\includegraphics[width=6.5 cm]{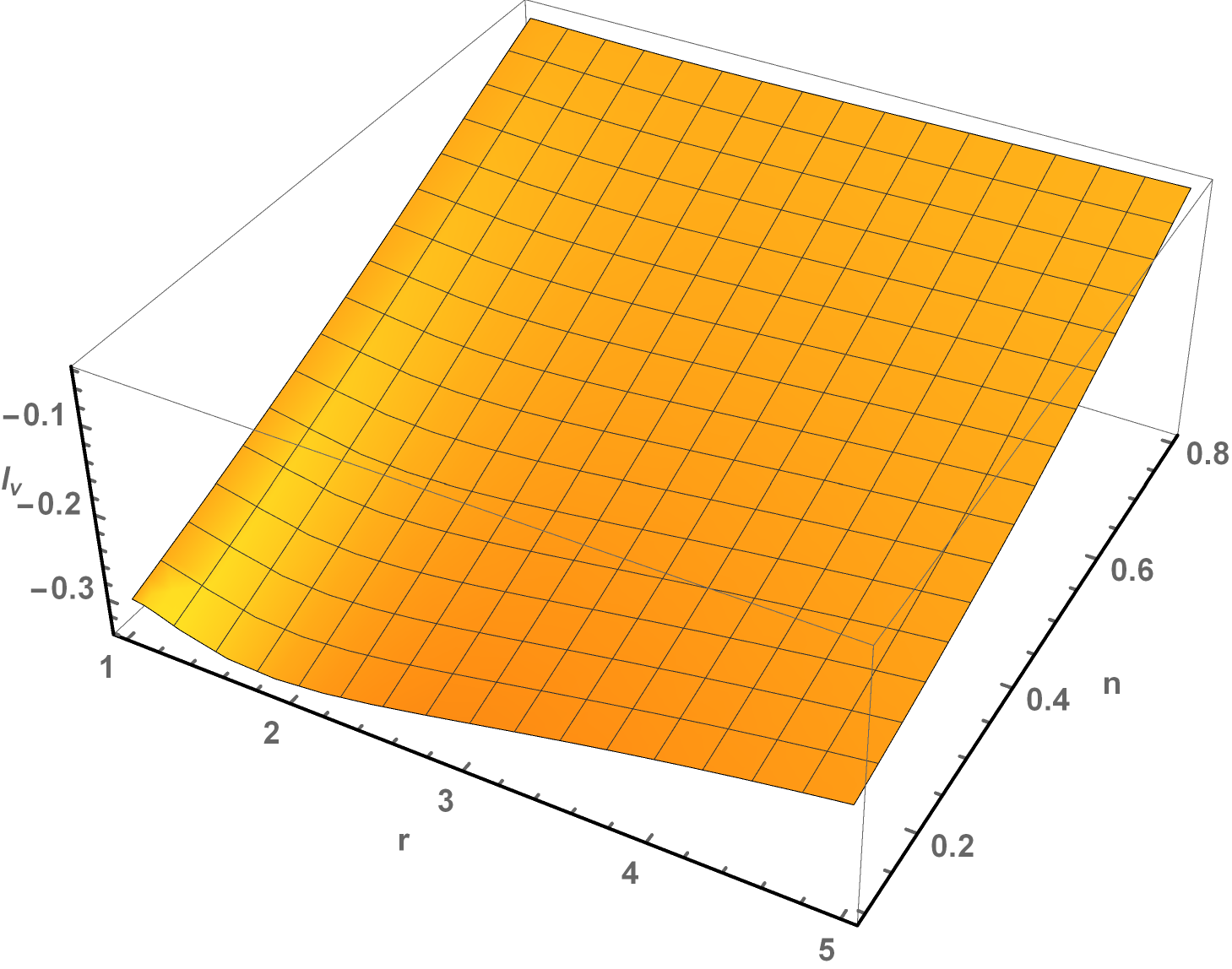}
\caption{3D Plot of VIQ associated with model 4 and here it is clear that when $r_1\rightarrow r_0^+$ then $Iv\rightarrow0$ and we consider $a=0.1$ and $B=0.2$.}
	\label{1chf23}
	\end{figure}

\begin{figure}[H]
\centering
\includegraphics[width=6.5 cm]{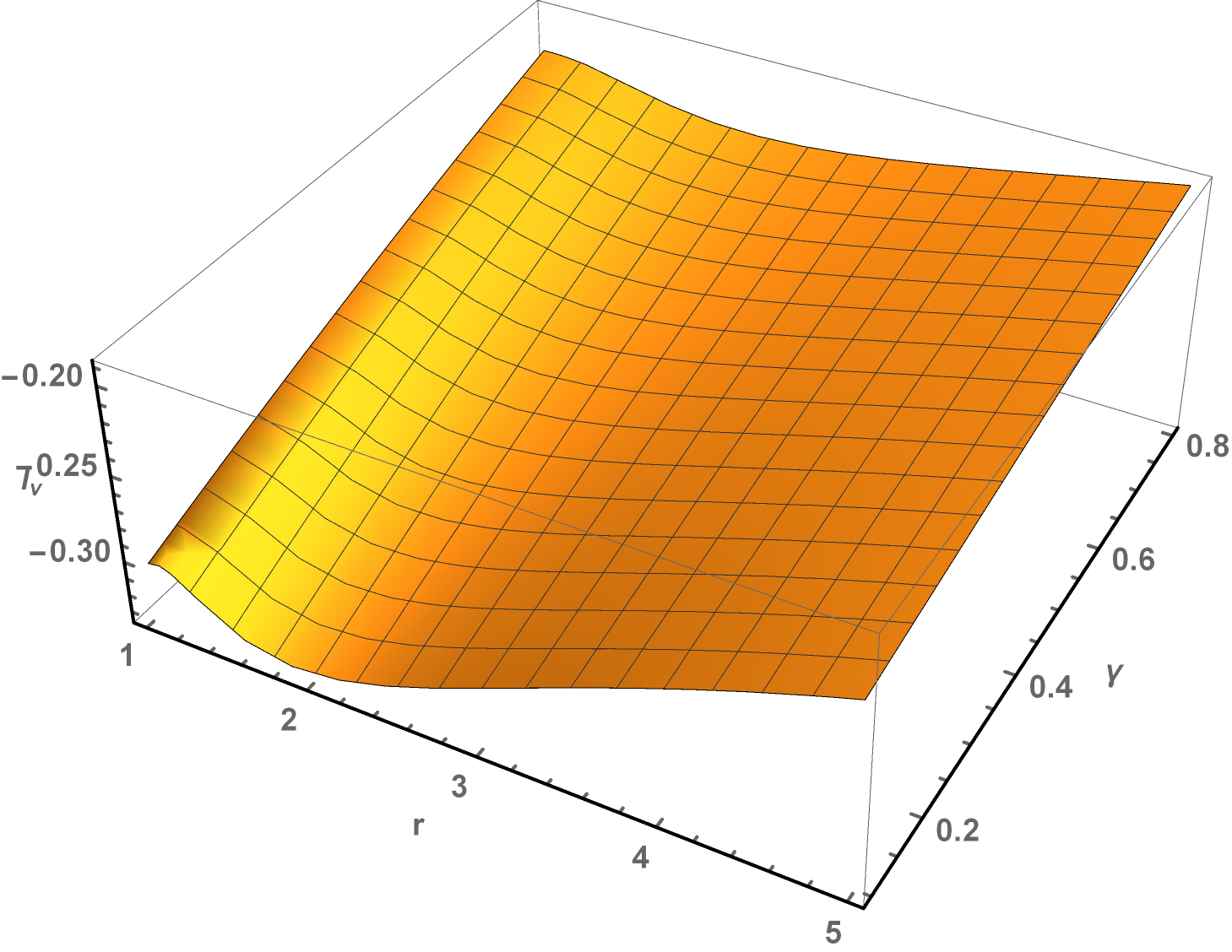}
\caption{3D Plot of VIQ associated with model 5 and here it is clear that when $r_1\rightarrow r_0^+$ then $Iv\rightarrow0$ and we consider $a=0.1$ and $B=0.2$.}
	\label{1chf24}
	\end{figure}

Volume Integral Quantifier (VIQ) provides information about the ``total amount of exotic matter" required for wormhole maintenance. To do this, one may compute the definite integral $\int T_{\mu\nu}u^\mu\,u^\nu$ and $\int T_{\mu\nu}k^\mu\,k^\nu$, where $u^\mu$ is the four-velocity \cite{Visser/2003}. For spherical symmetry and average null energy condition (ANEC), violating matter related to the radial component is defined as
\begin{equation}
Iv=\oint[\rho+P_r]dV,
\end{equation}
where $dV=r^2\sin\theta\,dr\,d\theta\,d\varphi$.
It can also be written as
\begin{equation}
Iv=8\pi\int_{r_0}^{\infty}\,(\rho+P_r)r^2\,dr.
\end{equation}
Now suppose that the wormhole enlarges from the throat, $r_0$, with a cutoff of the stress-energy tensor at a certain radius $r_1$, then it reduces to
\begin{equation}\label{viq1}
Iv=8\pi\int_{r_0}^{r_1}\,(\rho+P_r)r^2\,dr,
\end{equation}
where $r_0$ is the throat of the wormhole, which is the minimum value of $r$. The main point of this discussion is when limiting $r_1\rightarrow r_0^+$; one can verify that $Iv\rightarrow 0$. From Figs. (\ref{1chf20}-\ref{1chf24}), we found that for each wormhole solution, one may construct wormhole solutions with small quantities of exotic matter that need to open the wormhole throat.

\section{Conclusions}\label{ch1sec10}
In this chapter, we have discussed Morris-Throne wormholes, i.e., static and spherically symmetric traversable wormholes in the framework of symmetric teleparallel gravity (i.e., $f(Q)$ gravity), where the gravitational interaction is described by the non-metricity term $Q$. 
We explored three wormhole solutions by considering a linear functional form of $f(Q)$ and two wormhole solutions for the non-linear form of $f(Q)$. Also, we considered a constant redshift function, i.e., $\phi'(r)=0$ throughout our calculation, which simplifies our calculations and provides some exciting wormhole solutions.\\
For the first case, we have considered a relation between the radial and lateral pressure. We calculated the solution for shape function $b(r)$ in the power-law form. The parameter $'n'$ has to be negative to satisfy the asymptotically flatness condition. Keeping this in mind, we have tested all the necessary requirements of a shape function and the energy conditions. We found that the null energy condition is violated. For the second wormhole solution, we have considered the phantom energy EoS, which also violates the NEC. We have explored the wormhole solution by taking the range of $\omega$ as $ \omega< -1$. It is interesting to note here that the energy density $\rho >0$ throughout the space-time and the phantom energy may support traversable wormholes. In the third model, we have considered a specific shape function for $b(r)$. For this, we have discussed its' stability through the energy conditions and in this case, also NEC is violated. Thus, the violation of NEC for each model defines the possibility of the presence of exotic matter at the wormhole's throat.\\
Moreover, we have discussed two wormhole geometries by considering a quadratic Lagrangian $f(Q)$ with two specific shape functions. To test the traversability and stability of these two models, we have tested the flaring out condition and energy conditions. Both the models successfully passed all the tests. The advantage of these types of models is that they minimize the usage of the unknown form of matter called ``exotic matter" to have a traversable wormhole in comparison to the models for linear $f(Q)$ case (as it mimics the fundamental interaction of gravity, i.e., GR). Furthermore, in the case of the non-linear form of $f(Q)$, the profiles of NEC are violated near the throat of the wormholes and then satisfied (see Figs. \ref{1chf13} and \ref{1chf17}). Whereas, for linear $f(Q)$ case, it violated throughout the evolution of $r$ (see Figs. \ref{1chf3}, \ref{1chf6}, and \ref{1chf10}). Thus, it is worthy of exploring the wormhole geometries in the non-linear $f(Q)$ gravity.\\
Finally, we have discussed the volume integral quantifier to measure the exotic matter required for a traversable wormhole. We estimate that a small amount of exotic matter is required to have a traversable wormhole for our solutions.\\
Therefore, it is safe to conclude that in $f(Q)$ gravity, we found suitable geometries for traversable wormholes that violate the NEC at its throat. It will be interesting to explore the wormhole geometries in a more generalized $f(Q)$ form in the future, which may help to construct the traversable wormhole without the presence of exotic matter.\\
In the next chapter, we will study the effect of the GUP, a concept of quantum mechanics, on Casimir wormholes in the framework of $f(Q)$ gravity under different forms of redshift functions.

\chapter{GUP corrected Casimir wormholes in $f(Q)$ gravity} 
\label{Chapter3} 

\lhead{Chapter 3. \emph GUP Corrected Casimir Wormholes in $f(Q)$ Gravity}
\blfootnote{*The work in this chapter is covered by the following publication:\\
\textit{GUP Corrected Casimir Wormholes in $f(Q)$ Gravity}, General Relativity and Gravitation, \textbf{55}, 90 (2023).}
 
This chapter presents a detailed discussion of Casimir wormholes in $f(Q)$ gravity with GUP corrections. The detailed study of the work is outlined as follows:
\begin{itemize}
    \item We systematically presented the effect of the GUP in Casimir wormhole space-time in $f(Q)$ gravity.
    \item Two GUP models are considered, such as the KMM and DGS models, in this study.
    \item We assume two models namely linear $f(Q)=\alpha Q+\beta$ and the quadratic $f(Q)=Q+\gamma Q^2$, and studied wormhole solutions under different redshift functions.
    \item We analyzed the obtained wormhole solutions with energy conditions, especially NEC at the wormhole's throat, and encountered that some arbitrary quantity disrespects the classical energy conditions at the wormhole throat of radius $r_0$.
    \item The volume integral quantifier is also discussed to calculate the amount of exotic matter required near the wormhole throat.
\end{itemize}
\section{Introduction}\label{sec1}
It is well known that in Einstein's GR, for the wormhole to be traversable, we need to violate the NEC, which confirms the presence of exotic matter at the wormhole throat. One practical example of such matters can be found in the Casimir effect. The Casimir effect appears if we put two parallel conducting plates in a vacuum. They attract themselves as the zero modes of the quantum field theory give rise to the energy between the plates. It was first discovered in \cite{casimir} and later discussed in \cite{lifshitz} in a different way. The experimental evidence of the Casimir effect is also known and has been shown in \cite{experiment,bressi}. In \cite{Garattini}, the author presented a wormhole model probing the negative energy density because of the Casimir effect and explored the consequences of quantum weak energy conditions on the traversability of the wormhole.\\
The idea of the GUP comes from the fact that in the quantum gravity theory, there is usually a fundamental length scale beyond which the resolution is not possible, such as in string theory, and the length of string, etc. It may be shown via renormalization group theory methods that such an elementary length scale is inhabitable, as demonstrated in \cite{rainbowsmolin}. There are many other phenomenological implications in quantum gravity theory if we allow a minimum length scale, which has been discussed in \cite{sabinegup}.\\
The effect of GUP in the Casimir wormhole has been widely studied by Garattini in \cite{Garattini}. Further, Samart et al. \cite{Channuie} investigated the charged wormhole with and without GUP corrected Casimir wormhole for classical GR. These researches motivate us to study the effect of GUP-corrected Casimir wormholes in $f(Q)$ gravity. In particular, we consider two famous GUP relations, such as the KMM and DGS models, and investigate a class of asymptotically flat wormhole solutions in the background of the effect of GUP-corrected Casimir energy.\\
This chapter is organized as follows: the covariant formulation of $f(Q)$ gravity is discussed in section \ref{ch3sec2}. In section \ref{ch3sec3}, we have discussed wormhole field equations in $f(Q)$. A brief review of the Casimir effect under the generalized uncertainty principle is presented in section \ref{ch3sec4}. In section \ref{ch3sec5}, we construct the shape function and investigate the traversability conditions with different redshift functions under linear $f(Q)$ gravity. Further, a discussion on the energy conditions for both models under the linear $f(Q)$ form is placed in section \ref{ch3sec6}. Also, in \ref{ch3sec7}, we used a numerical approach to study the GUP Casimir wormhole for the quadratic case. Furthermore, we investigated the amount of exotic matter necessary for wormhole maintenance in \ref{ch3sec9}. Finally, we conclude our results in the last section.
\section{The covariant formulation of $f(Q)$ gravity}\label{ch3sec2}
As outlined in the Introduction, within this framework, the geometry's connection is treated as both curvature-free and devoid of torsion. Typically, the full connection $\tilde{\Gamma}$ cannot be uniquely determined from the conditions $R^a_{\;\:b}(\tilde{\Gamma})=0$ and $T^a(\tilde{\Gamma}) = 0$. Therefore, a method must be devised to specify $\tilde{\Gamma}$ or its components in any chosen reference frame before advancing to the spherically symmetric scenarios considered in this work. \\
According to Ref. \cite{g1}, it is always feasible to select $\tilde{\Gamma}^a_{\;\:b} = 0$ so that, alongside the nullification of $K^a_{\;\:b}$, both $R^a_{\;\:b}(\tilde{\Gamma}) = 0$ and $T^a(\tilde{\Gamma}) = 0$ are satisfied. This same gauge choice has been widely adopted in several studies of STEGR \cite{sp5,g3}, where it is commonly referred to as the coincident gauge \cite{fq1}. It is important to note that while $K^a_{\;\:b}$ is a tensor and its components remain zero across any frame, $\tilde{\Gamma}^a_{\;\:b}$ is not a tensor. Therefore, $\tilde{\Gamma}^a_{\;\:b} = 0$ does not guarantee that its components $\tilde{\Gamma}^\alpha_{\;\:\beta\gamma}$ in Eq. \eqref{gt1} will vanish in every frame. Consequently, $\tilde{\Gamma}^\alpha_{\beta\gamma} = 0$, and thus $L^\alpha_{\;\:\beta\gamma} = -{\Gamma}^{\alpha}_{\;\:\beta\gamma}$, only holds true in a specific set of frames. For any diffeomorphism or frame transformation $\chi(x)$, the components of $\tilde{\Gamma}$ can be determined by \cite{fq1}
\begin{equation}\label{cv1}
\tilde{\Gamma}^\alpha_{\;\:\beta\gamma} =\frac{\partial x^\alpha}{\partial \chi^\rho}\partial_\beta\partial_\gamma\chi^\rho.
\end{equation}
This transformation highlights the central role of $\tilde{\Gamma}^\alpha_{\;\:\beta\gamma}$ in the covariant formulation of $f(Q)$ gravity, which ensures Lorentz covariance within the nonmetricity framework. This leads to the fundamental question of identifying the frame in which $\tilde{\Gamma}^\alpha_{\;\:\beta\gamma} = 0$ is valid. Insights can be drawn from the $f(T)$ counterpart \cite{g4}. For local Lorentz transformations $\chi^\mu = \Lambda^\mu_\nu x^\nu$, Eq. \eqref{cv1} becomes
\begin{equation}\label{cv2}
\tilde{\Gamma}^\alpha_{\;\:\beta\gamma} = \Lambda^\alpha_\rho \partial_\beta(\Lambda^{-1})^\rho_\gamma,
\end{equation}
implying that the overall connection $\tilde{\Gamma}$ is purely inertial. Indeed, in the non-metricity framework, the tensor $Q_{\alpha\beta\gamma}$ is considered to encode all and only the information of gravity. It follows that the components of $Q_{\alpha\beta\gamma}$ should vanish when gravity is canceled, represented by $Q_{\alpha\beta\gamma}|_{G=0} = 0$. Consequently,
\begin{equation}
  \left. L^\alpha_{\beta\gamma}\right|_{G=0}=\left.\left( \frac12Q^\alpha_{\beta\gamma}-Q_{(\beta\gamma)}^{\alpha} \right)\right|_{G=0}=0,
\end{equation}
leading to
\begin{equation}
\left. \tilde{\Gamma}^\alpha_{\;\:\beta\gamma}\right|_{G=0}={\Gamma}^{\alpha}_{\;\:\beta\gamma}|_{G=0}+\left. L^\alpha_{\beta\gamma}\right|_{G=0}={\Gamma}^{\alpha}_{\;\:\beta\gamma}|_{G=0}.
\end{equation}
Furthermore, given that $R^a_{\;\:b}(\tilde{\Gamma}) = 0$, $T^a(\tilde{\Gamma}) = 0$, and all gravitational effects are contained in $Q_{\alpha\beta\gamma}$, the purely inertial connection $\tilde{\Gamma}^\alpha_{\beta\gamma}$ remains unchanged whether gravity is canceled or not, i.e.,
\begin{equation}
 \tilde{\Gamma}^\alpha_{\;\:\beta\gamma}=\left. \tilde{\Gamma}^\alpha_{\;\:\beta\gamma}\right|_{G=0}={\Gamma}^{\alpha}_{\;\:\beta\gamma}|_{G=0}.
\end{equation}
This provides a method to determine the components of $\tilde{\Gamma}$. For example, in the Cartesian frame, $\tilde{\Gamma}^\alpha_{\;\:\beta\gamma} = {\Gamma}^{\alpha}_{\;\:\beta\gamma}|_{G=0}=0$, which is commonly observed in the spatially flat cases of cosmology. In any other frame, the components of the connection $\tilde{\Gamma}^\alpha_{\beta\gamma}$ can either be derived from Eq. \eqref{cv1} or \eqref{cv2} relative to the Cartesian frame or directly calculated using ${\Gamma}^{\alpha}_{\;\:\beta\gamma}|_{G=0}$.
\section{Field equations in $f(Q)$ gravity}\label{ch3sec3}
In this study, we follow the procedure of \cite{rk} to calculate field equations in $f(Q)$ gravity. To make the consistency with the article \cite{rk}, we consider the metric of signature of the form $(+,-,-,-)$ and hence the wormhole metric can be read as
\begin{equation}\label{1ch3}
ds^2=U(r)dt^2-V(r)dr^2-r^2d\Omega^2,
\end{equation}
where, $U(r)=e^{\phi(r)}$ and $V(r)=\left(1-\frac{b(r)}{r}\right)^{-1}$.\\
For the above line element \eqref{1ch3}, we are able to find the following non-metricity scalar from Eq. \eqref{ga3}
\begin{equation}\label{11ch3}
Q=-\frac{b}{r^2}\left[\frac{r b'-b}{r (r-b)}+\phi^{'}\right].
\end{equation}
The field equations for the wormhole metric \eqref{1ch3} can be obtained from Eq. \eqref{ga4} under anisotropic matter source as follows
\begin{equation}
\label{12ch3}
\rho =\frac{(r-b)}{2 r^3}\left[2\,r\,f_{QQ}Q^{'}\frac{b}{r-b}
+f_Q \left(\frac{(2 r-b) \left(r b^{'}-b\right)}{(r-b)^2}+\frac{b \left(r \phi^{'}+2\right)}{r-b}\right)+\frac{f r^3}{r-b}\right],
\end{equation}
\begin{equation}
\label{13ch3}
P_r=-\frac{\left(r-b\right)}{2 r^3}\left[2\,r\,f_{QQ}Q^{'}\frac{b}{r-b}
+f_Q \left(\frac{b}{r-b}\left(\frac{r b{'}-b}{r-b}+r \phi^{'}+2\right)-2 r \phi^{'}\right)+\frac{f r^3}{r-b}\right],
\end{equation}
\begin{multline}
\label{14ch3}
P_t=-\frac{\left(r-b\right)}{4 r^2}\left[-2\,r\,\phi^{'} f_{QQ} Q^{'}
+ f_Q \left(\frac{\left(r b^{'}-b\right)}{r (r-b)}\left(\frac{2 r}{r-b}+r \phi^{'}\right)+\frac{2 (2 b-r) \phi^{'}}{r-b}\right.\right.\\\left.\left.
-r \left(\phi^{'}\right)^2-2 r \phi^{''}\right)+\frac{2 f r^2}{r-b}\right].
\end{multline}
\section{The Casimir effect under the GUP}
\label{ch3sec4}
\subsection{Casimir effect}
One of the natural sources of exotic matter which naturally comes for quantization of field is Casimir energy. It is well known that if we keep two parallel conducting plates in close proximity, they get attracted, and the energy is given by Eq. \eqref{cf1}.
This equation was first derived in \cite{casimir} and independently in \cite{lifshitz}, and later experimentally verified in \cite{experiment,bressi}.\\
One can show that the expression of Eq. \eqref{cf1} comes from summing over the normal modes of the field and adequately regularizing the sum. One can do the regularization in two ways: first, by introducing a cutoff limit \cite{zee}, and second via an analytic continuation of the Riemann Zeta function \cite{paddy}, both of which lead to the same answer.\\
It is known that the stability of a wormhole requires the NEC violation, and the exotic matter is necessary for the stability of the wormhole. Casimir energy can be used for such exotic matter sources as it has been studied extensively in various ``Casimir wormhole" articles (See Refs. \cite{Bezerra,Fuenmayor,Khabibullin}). However, on such a small quantum scale, it is not just necessary for the vacuum fluctuation but also for the fundamental length scale that gets important. So we need to use the fact that there is a natural length scale associated with the fundamental theory underlying quantum gravity. In the next sections, we will see that such a fundamental length scale will give rise to the GUP. In fact, the order of magnitude of such correction in the Casimir magnitude effect is quite comparable with the general Casimir force. 
\subsection{GUP corrected energy density}
In the introduction, we have discussed in detail the theory of GUP. In this section, we developed the energy density of Casimir force with the modifications of GUP models.\\
In \cite{gupcasimir}, the authors have employed the concept of minimal length and GUP to obtain the finite energy between the plane plates. They have derived the Hamiltonian and the corrections to the Casimir energy due to the minimal length. Up to a first-order correction term in the minimal uncertainty parameter $\lambda$, the Casimir energy for the two different cases of construction of maximally localized states are obtained as
\begin{equation}\label{18ch3}
E=-\frac{\pi^2 S}{720}\frac{\hbar}{a^3}\left[1+\Lambda_i\left(\frac{\hbar \sqrt{\lambda}}{a}\right)^2\right],
\end{equation}
where $S$ is the surface area of the plates, and $a$ is the separation between them. $\Lambda$ is a constant where $i=1,\,2$. In particular, we have the following two cases:
$$\Lambda_1=\pi^2\left(\frac{28+3\sqrt{10}}{14}\right)\quad (\text{for KMM model}),$$
$$\Lambda_2=\left(\frac{4\pi^2(3+\pi^2)}{21}\right) \quad (\text{for DGS model}).$$
Then the force can be obtained with the computation of
\begin{equation}\label{19ch3}
F=-\frac{dE}{da}=-\frac{3\pi^2 S}{720}\frac{\hbar}{a^4}\left[1+\frac{5}{3}\Lambda_i\left(\frac{\hbar \sqrt{\lambda}}{a}\right)^2\right].
\end{equation}
Thus, we get the formula for pressure
\begin{equation}\label{20ch3}
P=\frac{F}{S}=-\frac{3\pi^2}{720}\frac{\hbar}{a^4}\left[1+\frac{5}{3}\Lambda_i\left(\frac{\hbar \sqrt{\lambda}}{a}\right)^2\right]=\omega\rho.
\end{equation}
From the above equation, we can see that EoS can be defined by putting $\omega=3$. Now we can see that in natural units, the GUP-corrected energy density becomes
\begin{equation}\label{21ch3}
\rho=-\frac{\pi^2}{720}\frac{1}{a^4}\left[1+\frac{5}{3}\Lambda_i\left(\frac{\lambda}{a^2}\right)\right].
\end{equation}
Setting $\lambda=0$, we obtain the usual Casimir result.



\section{GUP corrected Casimir wormholes for the linear $f(Q)=\alpha\,Q+\beta$ model}\label{ch3sec5}
\begin{figure*}[h]
    \centering
    \includegraphics[scale=0.6]{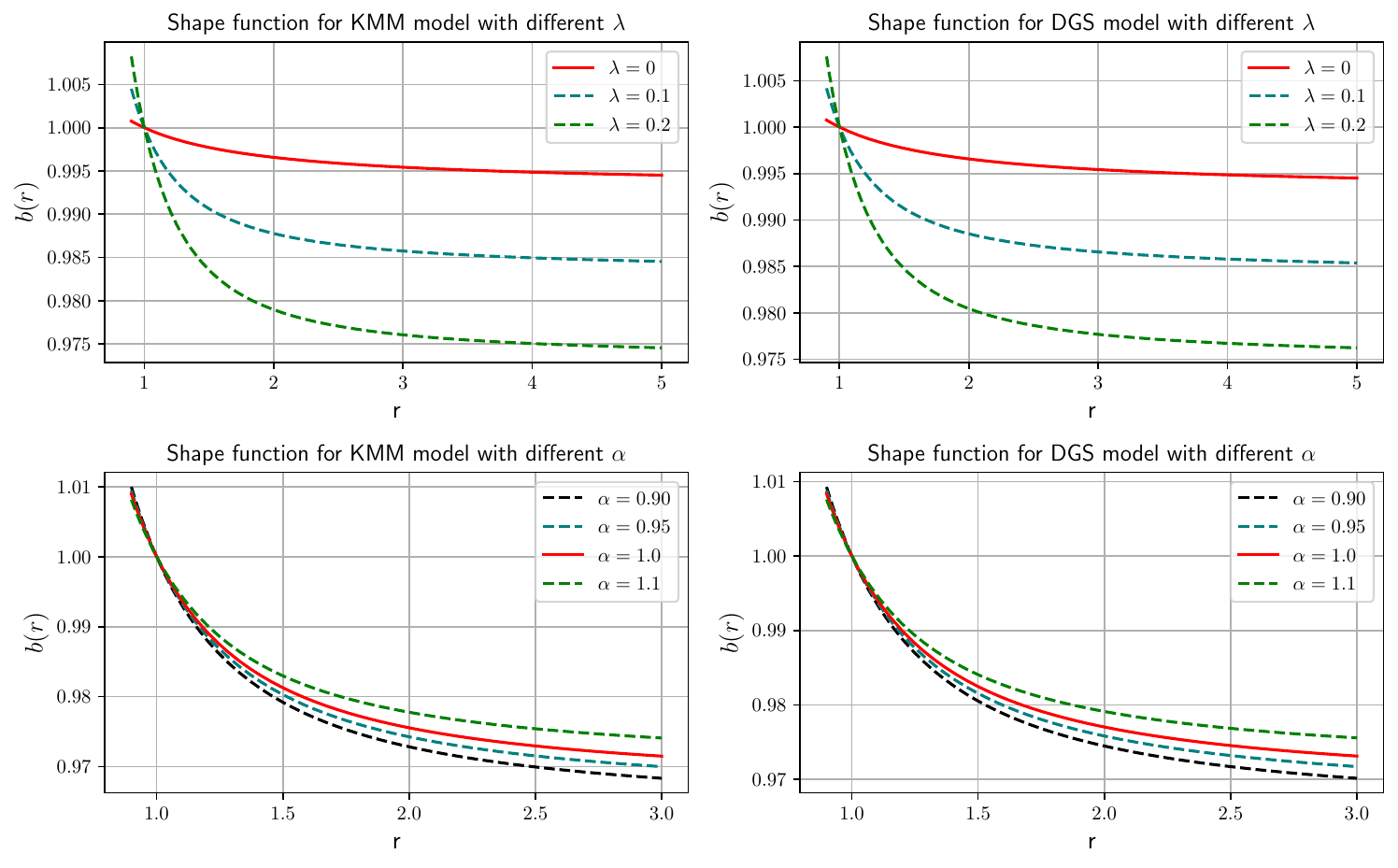}
    \caption{The plots of shape functions for KMM and DGS models against radial distance $r$ with $r_0=1$.  We fix $\alpha=2$ for the upper half and $\lambda=0.1$ for the lower half in the figures. Also, note that on the upper half figures $\lambda=0$ corresponds to the usual Casimir wormhole, and on the lower half figures $\alpha=1$ corresponds to the GR case.}
    \label{fig:g1}
\end{figure*}

\begin{figure*}[h]
    \centering
    \includegraphics[scale=0.6]{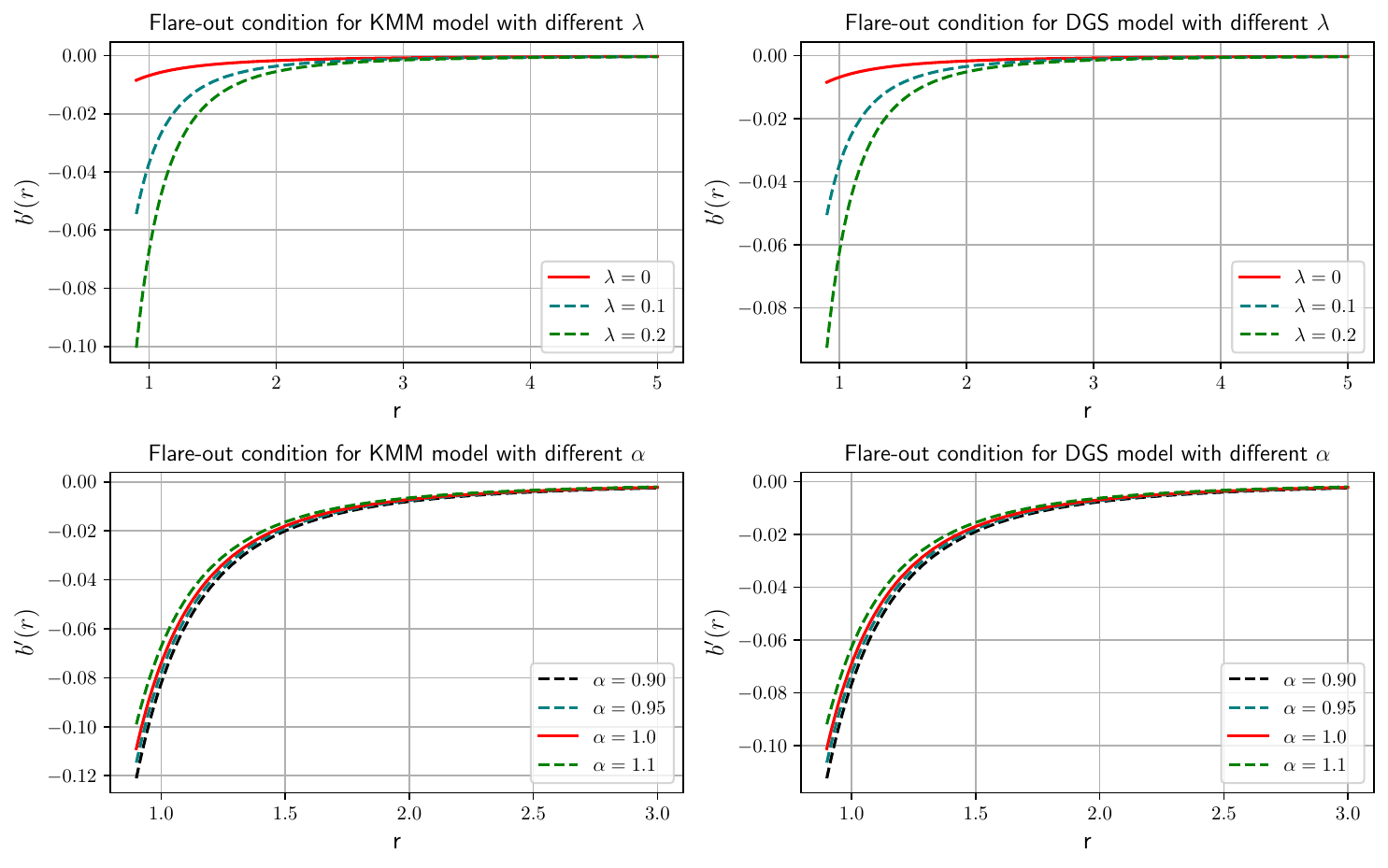}
    \caption{The plots of flare-out conditions for KMM and DGS models against radial distance $r$ with $r_0=1$.  We fix $\alpha=2$ for the upper half and $\lambda=0.1$ for the lower half in the figures. Also, note that on the upper half figures $\lambda=0$ corresponds to the usual Casimir wormhole, and on the lower half figures $\alpha=1$ corresponds to the GR case.}
    \label{fig:g2}
\end{figure*}

In this section, we assume the simplest linear functional form of $f(Q)$ gravity, such as $f(Q)=\alpha Q+\beta$, where $\alpha$ (the bound has been motivated from \cite{Solanki}) and $\beta$ are model parameters. Note that the above model can be reduced to GR if we consider $\alpha=1$ and $\beta=0$. This particular form is derived from the most general power law form $f(Q)=\alpha Q^{n+1}+\beta$ \cite{Parbati}. With this specific linear model, Solanki et al. \cite{Solanki} have investigated the late-time cosmic acceleration without invoking any dark energy component in the matter part. For this linear model, our general field equations (\ref{12ch3}-\ref{14ch3}) reduces to\\
\begin{equation}
\label{22ch3}
\rho=\frac{\alpha  b'}{r^2}+\frac{\beta }{2},
\end{equation}
\begin{equation}
\label{23ch3}
P_r=\frac{1}{r^3}\left[2 \alpha  r \left(r-b\right) \phi '-\alpha  b\right]-\frac{\beta }{2},
\end{equation}
\begin{equation}
\label{24ch3}
P_t=\frac{1}{2 r^3}\left[\alpha  \left(r \phi '+1\right) \left(-r b'+2 r (r-b) \phi '+b\right)\right]
+\frac{\alpha  (r-b) \phi ''}{r}-\frac{\beta }{2}.
\end{equation}
Further, to obtain the shape function of the GUP corrected energy density, we replace the plate separation distance $a$ by the radial distance $r$ in Eq. \eqref{21ch3}. In that case, we can rewrite the energy density from Eq. \eqref{21ch3} as
\begin{equation}\label{25ch3}
\rho=-\frac{\pi^2}{720}\frac{1}{r^4}\left[1+\frac{5}{3}\Lambda_i\left(\frac{\lambda}{r^2}\right)\right].
\end{equation}
Now comparing Eqs. \eqref{22ch3} and \eqref{25ch3}, and solving the differential equation for shape function $b(r)$, we obtain
\begin{equation}\label{26ch3}
b(r)=-\frac{1}{2160 \alpha }\left[-\frac{5 \pi ^2 \lambda\Lambda_i}{3 r^3}+360 \beta  r^3-\frac{3 \pi ^2}{r}\right]+c_1,
\end{equation}
where $c_1$ is the integrating constant, and to calculate it, we apply throat condition $b(r_0)=r_0$ in the above equation, we get
\begin{equation}\label{27ch3}
c_1=r_0+\frac{1}{2160 \alpha }\left[-\frac{5 \pi ^2 \lambda\Lambda_i}{3 r_0^3}+360 \beta  r_0^3-\frac{3 \pi ^2}{r_0}\right].
\end{equation}
Inserting Eq. \eqref{27ch3} into Eq. \eqref{26ch3}, we obtain the final version of shape function $b(r)$ as follows
\begin{equation}\label{28ch3}
b(r)=r_0+\frac{\xi_1}{5}\left(\frac{1}{r}-\frac{1}{r_0}\right)+\frac{\xi_1\lambda\Lambda_i}{9}\left(\frac{1}{r^3}-\frac{1}{r_0^3}\right)
+\frac{\beta}{6\alpha}\left(r_0^3-r^3\right),
\end{equation}
where 
\begin{equation}\label{29ch3}
\xi_1=\frac{\pi^2}{144\,\alpha}.
\end{equation}
It may be observed that the above equation is not asymptotically flat, that is, for $r\rightarrow \infty$, $\frac{b(r)}{r}\nrightarrow 0$. It happens because of the fourth term of the above equation. For $\beta \rightarrow 0$, it will satisfy the flatness condition. From now on, we consider $\beta=0$ in this work. The last equation reduces to
\begin{equation}\label{30ch3}
b(r)=r_0+\frac{\xi_1}{5}\left(\frac{1}{r}-\frac{1}{r_0}\right)+\frac{\xi_1\lambda\Lambda_i}{9}\left(\frac{1}{r^3}-\frac{1}{r_0^3}\right).
\end{equation}
The GUP correction term is proportional to the minimal uncertainty parameter $\lambda$. Clearly, in the limit $\lambda \rightarrow 0$, the shape function reduces to that of the Casimir wormhole \cite{Kazuharu}.\\
In Figs. \ref{fig:g1} and \ref{fig:g2}, we have depicted the behavior of shape functions for both models. It can be observed from Fig. \ref{fig:1} that for the increasing values of $\lambda$, the shape function $b(r)$ shows positively decreasing behavior, whereas, for the increase of $\alpha$, it shows increasing behavior. Also, from Fig. \ref{fig:2}, it is confirmed that the flaring-out condition is satisfied in the vicinity of the wormhole throat under the asymptotic background.\\
For the GUP-corrected Casimir wormhole, the field equations (\ref{22ch3}-\ref{24ch3}) can be read as
\begin{equation}\label{31ch3}
\rho(r)=-\frac{\pi ^2 \left(5 \Lambda_i \lambda +3 r^2\right)}{\mathcal{D}_1},
\end{equation}
\begin{equation}\label{32ch3}
P_r(r)=\frac{1}{3 \mathcal{D}_1 r_0^3}\left[r (r-r_0) \phi ' \mathcal{D}_2+\pi ^2 \left(5 \Lambda_i \lambda  \left(r^3-r_0^3\right)
+9 r^2 r_0^2 (r-r_0)\right)-\mathcal{D}_3\right],
\end{equation}
\begin{multline}\label{33ch3}
P_t(r)=\frac{1}{12 \mathcal{D}_1 r_0^3}\left[r \left(\phi ' \left(r (r-r_0) \phi '\mathcal{D}_2+5 \pi ^2 \Lambda_i \lambda  \left(r^3+2 r_0^3\right)+\mathcal{D}_4\right)+2 r (r-r_0) \phi '' \mathcal{D}_2\right) \right.\\\left.
+2 \pi ^2 \left(-5 \Lambda_i \lambda  \left(r^3-4 r_0^3\right)-9 r^2 r_0^2 (r-2 r_0)\right)+2\mathcal{D}_3\right],
\end{multline}
where,
\begin{equation}
\mathcal{D}_1=2160 r^6,
\end{equation}
\begin{equation}\label{34ch3}
\mathcal{D}_2=5 \pi ^2 \Lambda_i \lambda  \left(r^2+r r_0+r_0^2\right)+9 r^2 r_0^2 \left(720 \alpha  r r_0+\pi ^2\right),
\end{equation}
\begin{equation}\label{35ch3}
\mathcal{D}_3=6480 \alpha  r^3 r_0^4,
\end{equation}
\begin{equation}\label{36ch3}
\mathcal{D}_4=9 r^3 r_0^2 \left(720 \alpha  r_0 (2 r-r_0)+\pi ^2\right).
\end{equation}
Now, we shall present our study in the following subsections with the above components of generalized field equations.

\subsection{Case-I: $\phi(r)=k$}
In this subsection, we consider $\phi(r)=k$, where $k$ is any constant and hence $\phi^{'}(r)=0$.\\
For this case, the wormhole metric can be read as
\begin{equation}
ds^2=e^{k}dt^2-\frac{dr^2}{1-\frac{r_0}{r}+\frac{\xi_1}{5r}\left(\frac{1}{r}-\frac{1}{r_0}\right)+\frac{\xi_1\lambda\Lambda_i}{9r}\left(\frac{1}{r^3}-\frac{1}{r_0^3}\right)}
-r^2\,d\theta^2-r^2\text{sin}^2\theta\,d\Phi^2,
\end{equation}
where, $\xi_1$ is defined in Eq. \eqref{29ch3}.\\
Now, we derive the EoS for the radial pressure defined by
\begin{equation}\label{37ch3}
    P_r(r)=\omega_r(r)\rho(r),
\end{equation}
where $\omega$ is the EoS parameter, which is a function of $r$.\\
Considering Eqs. (\ref{22ch3}-\ref{24ch3}) with shape function \eqref{30ch3} under constant redshift function (zero tidal force), we obtain
\begin{equation}
\omega_r=\frac{\pi ^2 \left(5 \Lambda_i \lambda  \left(r_0^3-r^3\right)+9 r^2 r_0^2 (r_0-r)\right)+\mathcal{D}_3}{\mathcal{K}_1},
\end{equation}
where
\begin{equation}\label{38ch3}
\mathcal{K}_1=3 \pi ^2 r_0^3 \left(5 \Lambda_i \lambda +3 r^2\right).
\end{equation}
\begin{figure}[H]
    \centering
    \includegraphics[width=16cm,height=5cm]{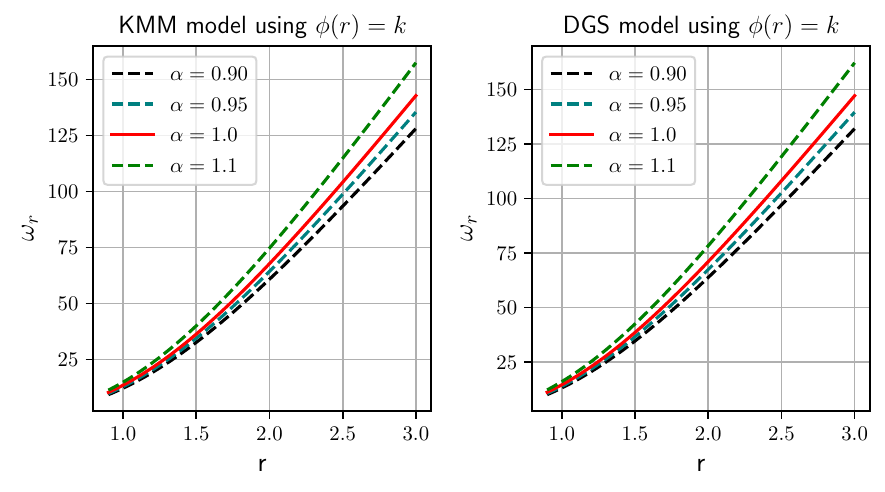}
    \caption{The plots of EoS parameter $\omega_r$ for KMM (left) and DGS (right) model using $\phi(r)=k$ for different $\alpha$ with $r_0=1$ and $\lambda=0.1$. In the figure, $\alpha=1$ corresponds to the GR case.}
    \label{fig:g3}
\end{figure}
The behavior of radial EoS parameter $\omega_r$ for both KMM and DGS models has been illustrated in Fig. \ref{fig:g3}. It is observed that radial EoS parameter $\omega_r$ increases positively with the increased values of $\alpha$ and radial distance $r$.

\subsection{Case-II: $\phi(r)=\frac{k}{r}$}
\begin{figure}[h]
    \centering
    \includegraphics[width=16cm,height=5cm]{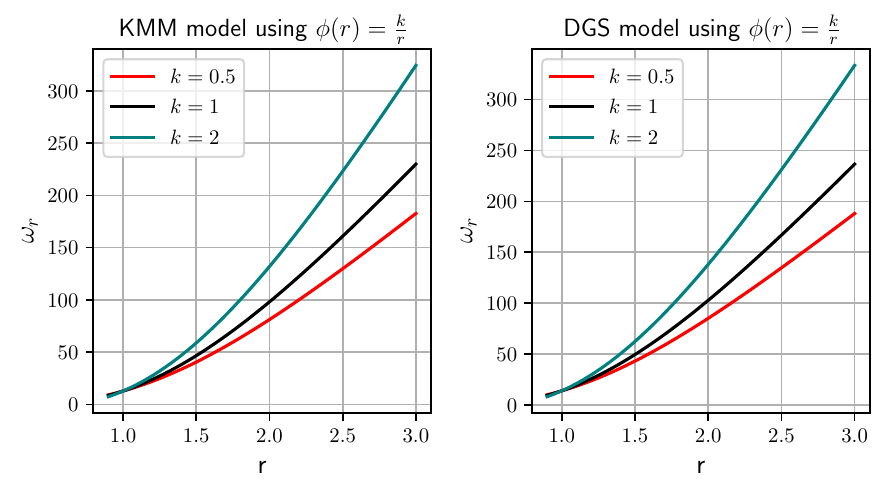}
    \caption{The plots of EoS parameter $\omega_r$ for KMM (left) and DGS (right) model using $\phi(r)=\frac{k}{r}$ for three different values of $k$ with $r_0=1$, $\lambda=0.1$ and $\alpha=0.95$.}
    \label{fig:g4}
\end{figure}

We shall start our investigation by considering the radial EoS relation \eqref{37ch3}. From the field equations \eqref{22ch3} and \eqref{23ch3}, we can determine the redshift function
\begin{equation}\label{39ch3}
\phi^{'}(r)=\frac{\omega_rb^{'}r+b}{r(r-b)},
\end{equation}
for the redshift function $\phi(r)=\frac{k}{r}$, we are able to find the EoS parameter
\begin{equation}
\omega_r(r)= -\frac{k(r-b)+br}{b^{'}r^2}.
\end{equation}
with shape function \eqref{30ch3}
\begin{equation}
\omega_r=\frac{1}{r\mathcal{K}_1}\left[5 \pi ^2 \Lambda_i \lambda  (k-r) \left(r^3-r_0^3\right)+9 r^2 r_0^2 \left(720 \alpha  r r_0 (k (r-r_0)+r r_0)+\pi ^2 (k-r) (r-r_0)\right)\right].
\end{equation}
at wormhole throat, the last equation reduces to
\begin{equation}
\omega_r\mid_{r=r_0}=\frac{2160 \alpha  r_0^4}{\pi ^2 \left(5 \Lambda_i \lambda +3 r_0^2\right)}.
\end{equation}
The behavior of the EoS parameter for both models is shown in Fig. \ref{fig:g4}.

\subsection{Case-III: $\phi(r)=\text{log} \left(\frac{\sqrt{\eta^2+r^2}}{r}\right)$}
\begin{figure}[h]
    \centering
    \includegraphics[width=16cm,height=5cm]{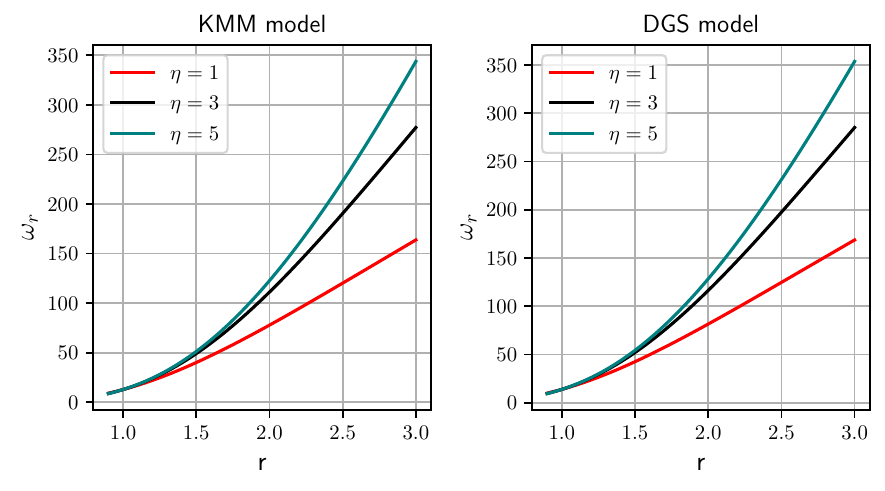}
    \caption{The plots of EoS parameter $\omega_r$ for KMM (left) and DGS (right) model using $\phi(r)=\frac{1}{2}\text{log}\left(1+\frac{\eta^2}{r^2}\right)$ for different $\eta$ with $r_0=1$, $\lambda=0.1$ and $\alpha=0.95$.}
    \label{fig:g5a}
\end{figure}

\begin{figure}[h]
    \centering
    \includegraphics[width=16cm,height=5cm]{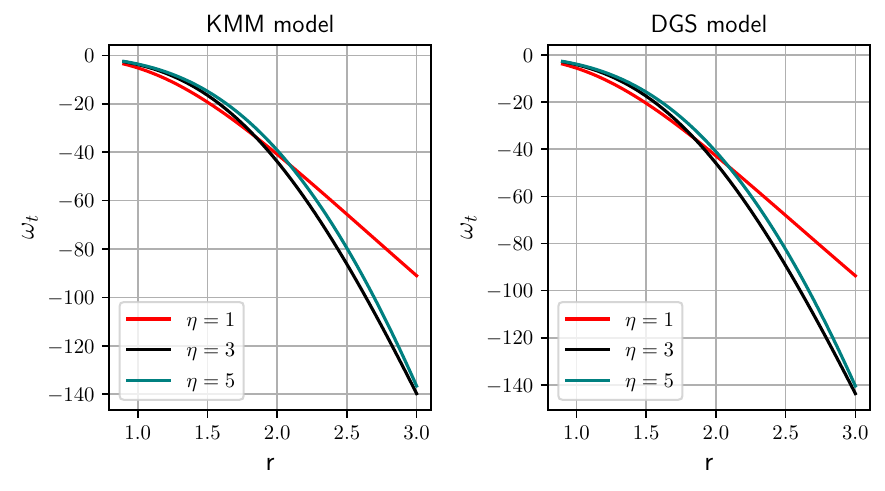}
    \caption{The plots of EoS parameter $\omega_t$ for KMM (left) and DGS (right) model using $\phi(r)=\frac{1}{2}\text{log}\left(1+\frac{\eta^2}{r^2}\right)$ for different $\eta$ with $r_0=1$, $\lambda=0.1$ and $\alpha=0.95$.}
    \label{fig:g8z}
\end{figure}

For this specific redshift function, our wormhole metric can be read as
\begin{equation}
 ds^2=\left(1+\frac{\eta^2}{r^2}\right)^{\frac{1}{2}}dt^2-V(r)dr^2-r^2\,d\Omega^2,
\end{equation}
where $\eta$ is any positive parameter. For this non-constant redshift function, from Eq. \eqref{37ch3},  we could find the EoS parameter $\omega_r$ as
\begin{equation}
\omega_r(r)=-\frac{r b+\eta ^2}{\left(\eta ^2+r^2\right) b'},
\end{equation}
and for the shape function \eqref{30ch3}, the above equation reduces to
\begin{equation}
\omega_r(r)=\frac{1}{\mathcal{K}_2}\left[5 \pi ^2 \Lambda_i \lambda  r^2 \left(r_0^3-r^3\right)+9 r^4 r_0^2 \left(720 \alpha  r_0 \left(\eta ^2+r r_0\right)+\pi ^2 (r_0-r)\right)\right],
\end{equation}
where $\mathcal{K}_2=\mathcal{K}_1\left(\eta ^2+r^2\right)$ and $\mathcal{K}_1$ is defined in Eq. \eqref{38ch3}. The graphical behavior of EoS parameter $\omega_r$ for both models has been depicted in Fig. \ref{fig:g5a}.\\
Now we consider another form of EoS, such as
\begin{equation}
P_t(r)=\omega_t(r)\rho(r),   
\end{equation}
where $\omega_t(r)$ is the EoS parameter which is a function of the radial coordinate $r$. With this form of EoS, we get the following differential equation
\begin{equation}
\alpha  \left[ b' \left(r \phi'+4 \omega +2\right)+b \left(r \phi'^2+\phi '+2 r \phi ''-\frac{2}{r}\right)
-r \left(\phi ' \left(r \phi '+2\right)+2 r \phi ''\right)\right]=0.
\end{equation}
Inserting the shape function \eqref{30ch3} with the redshift function $\phi(r)=\frac{1}{2}\text{log}\left(1+\frac{\eta^2}{r^2}\right)$ in the last equation, we could obtain the tangential EoS parameter
\begin{multline}\label{40ch3}
\omega_t(r)=\frac{1}{4 \mathcal{K}_3}\left[\pi ^2 \left(5 \Lambda_i \lambda  \left(-\eta ^2 r^2 \left(r^3+8 r_0^3\right)+2 r^4 \left(r^3-4 r_0^3\right)-3 \eta ^4 r_0^3\right)+9 r^2 r_0^2 \left(-\eta ^2 r^2 (r+2 r_0)\right.\right.\right.\\\left.\left.\left.
+2 r^4 (r-2 r_0)-\eta ^4 r_0\right)\right)-\mathcal{K}_5\right],
\end{multline}
where
\begin{equation}
\mathcal{K}_3=\mathcal{K}_2 \left(\eta ^2+r^2\right),
\end{equation}
\begin{equation}
\mathcal{K}_4=\frac{r}{r_0}\left(\eta ^4+2 r^3 r_0+\eta ^2 r (4 r-r_0)\right),
\end{equation}
\begin{equation}
\mathcal{K}_5=\mathcal{D}_3 \mathcal{K}_4,
\end{equation}
and $\mathcal{D}_3$ is defined in Eq. \eqref{35ch3}.\\
Also, we could find the radial EoS parameter $\omega_r$ from the expression \eqref{37ch3}
\begin{equation}\label{41ch3}
\omega_r(r)=\frac{1}{\mathcal{K}_6}\left[5 \pi ^2 \Lambda_i \lambda  r^2 \left(r_0^3-r^3\right)+9 r^4 r_0^2 \left(720 \alpha  r_0 \left(\eta ^2+r r_0\right)+\pi ^2 (r_0-r)\right)\right],
\end{equation}
where $\mathcal{K}_6=3 \pi ^2 r_0^3 \left(\eta ^2+r^2\right) \left(5 \Lambda_i \lambda +3 r^2\right)$.\\
At throat, the expressions \eqref{40ch3} and \eqref{41ch3} reduce to
\begin{equation}\label{42ch3}
\omega_r\mid_{r=r_0}=\frac{2160 \alpha  r_0^4}{\pi ^2 \left(5 \Lambda_i \lambda +3 r_0^2\right)},
\end{equation}
\begin{equation}\label{43ch3}
\omega_t\mid_{r=r_0}=-\frac{\left(\eta ^2+2 r_0^2\right) \left(\pi ^2 \left(5 \Lambda_i \lambda +3 r_0^2\right)+2160 \alpha  r_0^4\right)}{4 \pi ^2 \left(\eta ^2+r_0^2\right) \left(5 \Lambda_i \lambda +3 r_0^2\right)}.
\end{equation}
It is evident that the right-hand side of Eq. \eqref{42ch3} is a positive quantity, whereas that of Eq. \eqref{43ch3} is a negative quantity. Thus it turns out that the radial EoS parameter $\omega_r$ increases and tangential $\omega_t$ decreases with the increase of the radial distance.
We have depicted the graphical behavior of the tangential EoS parameter in Fig. \ref{fig:g8z}.

\section{Energy conditions}\label{ch3sec6}
\begin{figure}[h]
    \centering
    \includegraphics[scale=0.56]{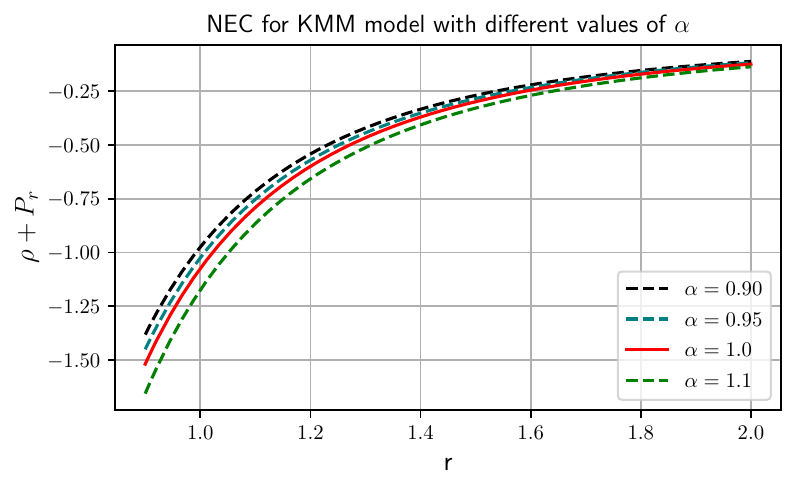}
    \includegraphics[scale=0.56]{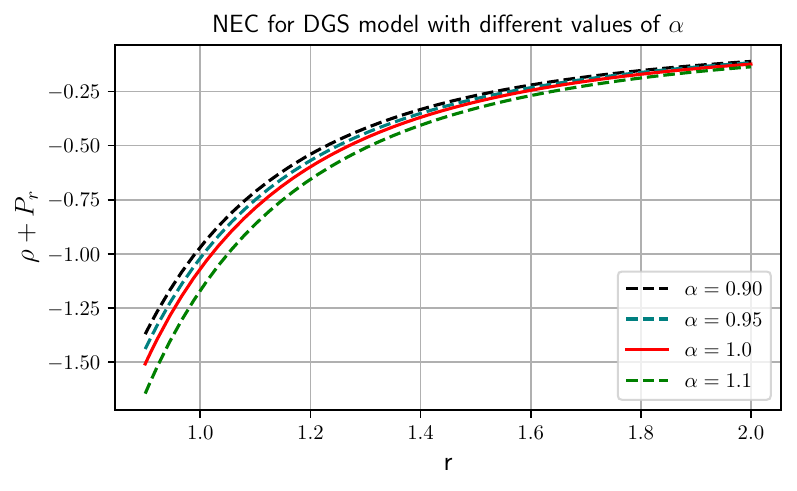}
    \caption{The plots of NEC for KMM and DGS models against radial distance $r$ using $\phi(r)=k$ with $r_0=1$ and GUP parameter $\lambda=0.1$. In the figure, $\alpha=1$ corresponds to the GR case.}
    \label{fig:g5}
\end{figure}

\begin{figure}[h]
    \centering
    \includegraphics[scale=0.56]{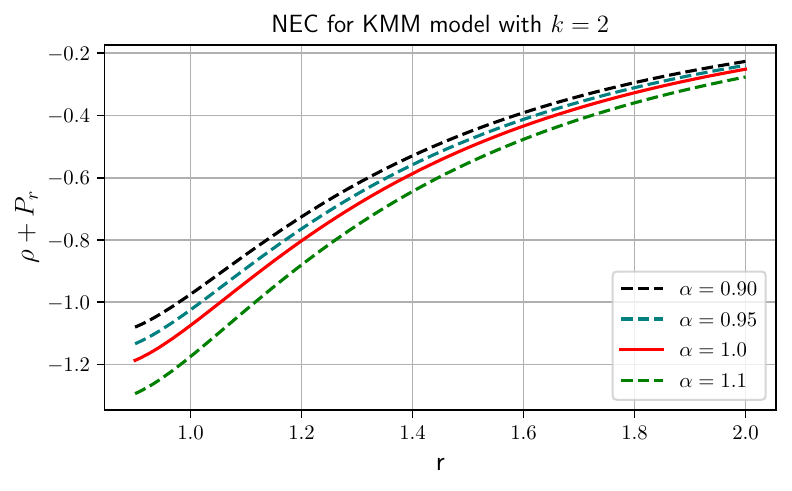}
    \caption{The plots of NEC for KMM model against radial distance $r$ using $\phi(r)=\frac{k}{r}$ with $r_0=1$ and GUP parameter $\lambda=0.1$. In the figure, $\alpha=1$ corresponds to the GR case.}
    \label{fig:g6}
\end{figure}

\begin{figure}[h]
    \centering
    \includegraphics[scale=0.56]{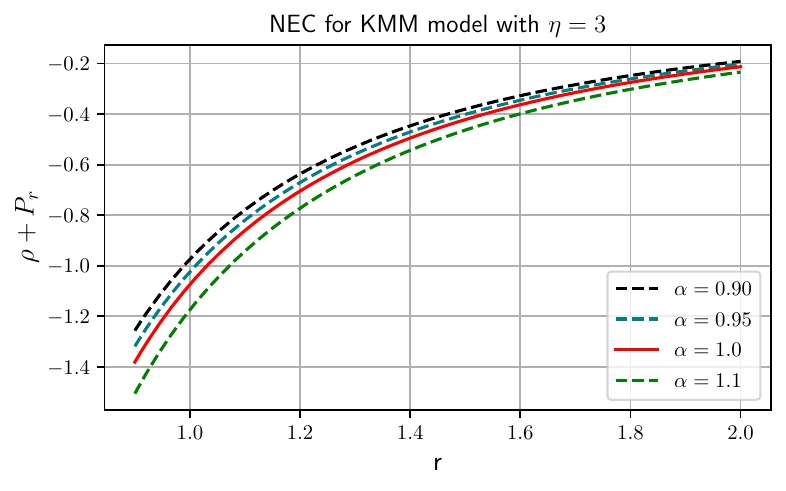}
    \caption{The plots of NEC for KMM model against radial distance $r$ using $\phi(r)=\frac{1}{2}\text{log}\left(1+\frac{\eta^2}{r^2}\right)$ with $r_0=1$ and GUP parameter $\lambda=0.1$. In the figure, $\alpha=1$ corresponds to the GR case.}
    \label{fig:g7}
\end{figure}

In this section, we shall discuss the classical energy conditions developed from the Raychaudhuri equations. In GR, the wormhole solutions are maintained by exotic matter involving a stress-energy tensor that disrespects the NEC (indeed, it disobeys all the energy conditions \cite{Visser}).\\
Now, using Eqs. (\ref{31ch3}-\ref{33ch3}), the NEC for the GUP-corrected Casimir wormhole can be written from the above expression as
\begin{equation}\label{44ch3}
\rho+P_r=\frac{\alpha r_0}{r^3 \mathcal{D}_3}\left[r (r-r_0) \phi ' \mathcal{D}_2+\pi ^2 \left(5 \Lambda_i \lambda  \left(r^3-4 r_0^3\right)
+9 r^2 r_0^2 (r-2 r_0)\right)-\mathcal{D}_3\right],
\end{equation}
\begin{multline}\label{45ch3}
\rho+P_t=\frac{\alpha r_0}{4 r^3 \mathcal{D}_3}\left[r \left(\phi ' \left(r (r-r_0) \phi ' \mathcal{D}_2+5 \pi ^2 \Lambda_i \lambda \left(r^3+2 r_0^3\right)+\mathcal{D}_4\right)+2 r (r-r_0) \phi '' \mathcal{D}_2\right)-2 \pi ^2 \right.\\\left.
\left(5 \Lambda_i \lambda  \left(r^3+2 r_0^3\right)+9 r^3 r_0^2\right)+\mathcal{D}_3\right],
\end{multline}
where, $\mathcal{D}_2$, $\mathcal{D}_3$ and $\mathcal{D}_4$ are already defined in Eqs. (\ref{34ch3}-\ref{36ch3}).
Here, the GUP correction term is proportional to the uncertainty parameter $\lambda$. In the limit $\lambda \rightarrow 0$, the expressions \eqref{44ch3} and \eqref{45ch3} reduce to usual Casimir wormhole's NEC (see Eqs. (28) and (29) of Ref. \cite{Kazuharu}). One can notice that the right-hand side of Eq. \eqref{44ch3} is a negative quantity for a radial distance $r \leq r_0$; hence, NEC is violated. Also, we observe that the contribution becomes more negative with the increase of GUP parameter $\lambda$ and model parameter $\alpha$.\\
At the throat of the wormhole, the above equations are reduced to
\begin{equation}\label{46ch3}
\rho+P_r\mid_{r=r_0}=-\left[\frac{\pi ^2 \left(5 \Lambda_i \lambda +3 r_0^2\right)}{2160 r_0^6}+\frac{\alpha }{r_0^2}\right],
\end{equation}
\begin{equation}\label{47ch3}
\rho+P_t\mid_{r=r_0}=\frac{1}{8640 r_0^6}\left[\phi ' \left(\pi ^2 \left(5 \Lambda_i \lambda  r_0+3 r_0^3\right)
+2160 \alpha  r_0^5\right)-2 \pi ^2 \left(5 \Lambda_i \lambda +3 r_0^2\right)+4320 \alpha  r_0^4\right].
\end{equation}
It is transparent that the right-hand side of the Eq. \eqref{46ch3} is a negative quantity for any positive $\alpha$. Thus, we could conclude that NEC is violated by the GUP-corrected Casimir wormhole at the throat. In Figs. \ref{fig:g5}-\ref{fig:g7}, we have plotted the graphs for NEC for both models with different redshift functions.

\section{GUP corrected Casimir wormholes for the quadratic $f(Q)=Q+\gamma\,Q^2$ model}\label{ch3sec7}
\begin{figure*}[h]
    \centering
    \includegraphics[scale=0.6]{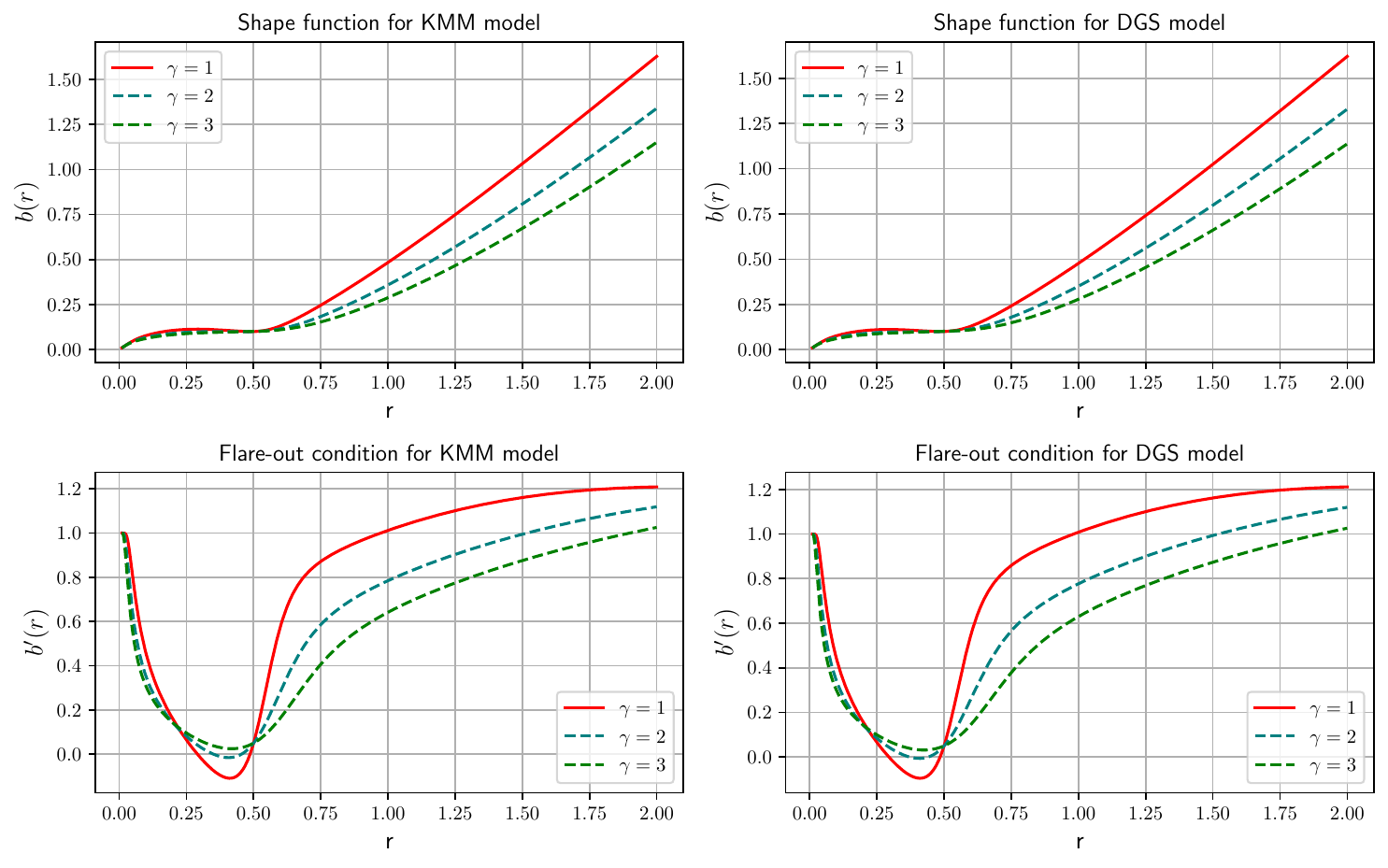}
    \caption{The plots of shape function and flare-out condition for KMM and DGS models for $\phi(r)=k$ under quadratic $f(Q)$ form. We fix the GUP parameter $\lambda=0.1$.}
    \label{fig:g8a}
\end{figure*}

\begin{figure*}[h]
    \centering
    \includegraphics[scale=0.6]{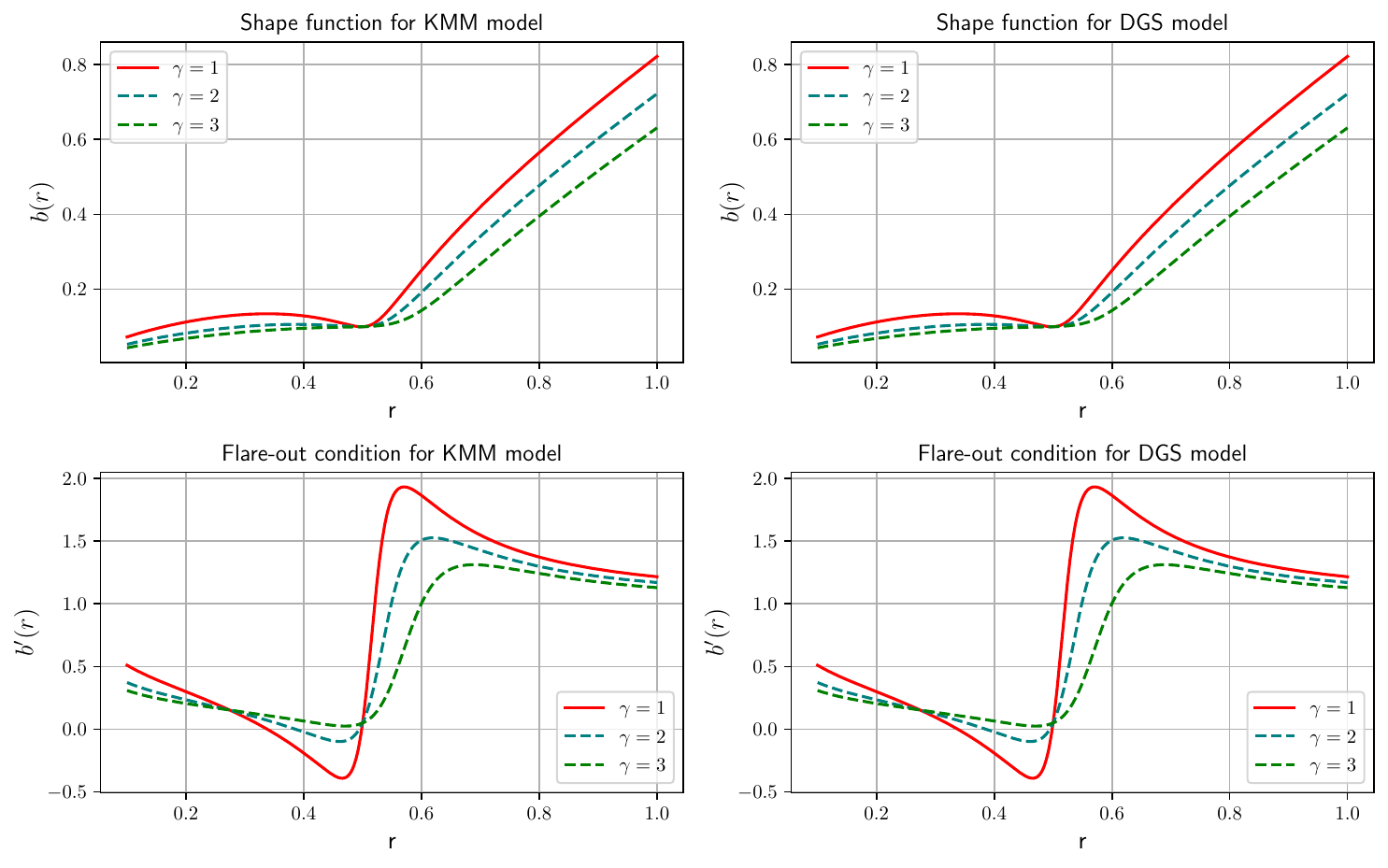}
    \caption{The plots of shape function and flare-out condition for KMM and DGS models for $\phi(r)=\frac{k}{r}$ under quadratic $f(Q)$ form. We fix the GUP parameter $\lambda=1$ and $k=1$.}
    \label{fig:g8b}
\end{figure*}

\begin{figure}[h]
    \centering
    \includegraphics[scale=0.56]{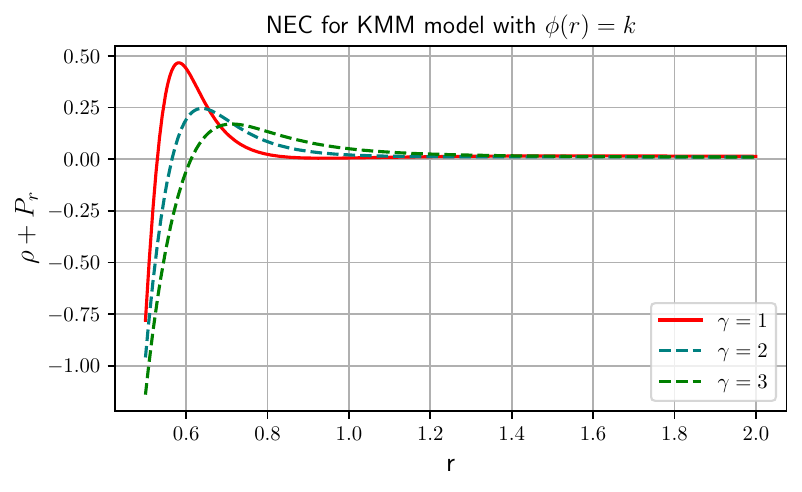}
    \includegraphics[scale=0.56]{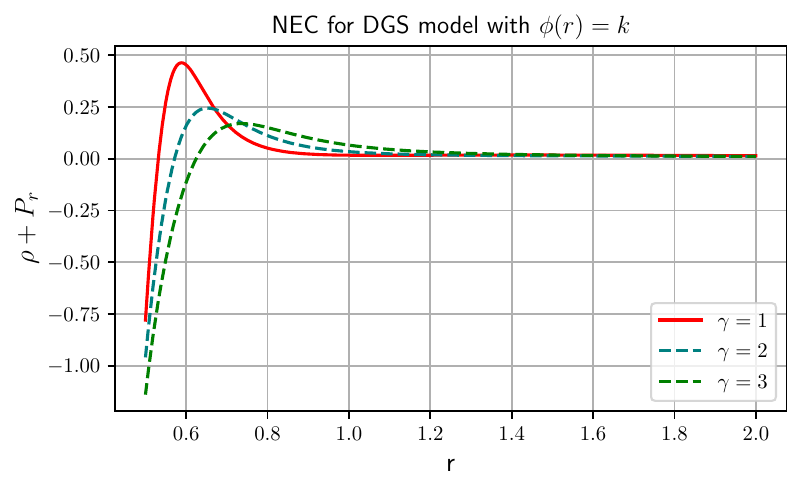}
    \caption{The plots of NEC for KMM and DGS models against radial distance $r$ for $f(Q)=Q+\gamma Q^2$ case with GUP parameter $\lambda=0.1$. In the figure, $\gamma=1$ corresponds to the GR case.}
    \label{fig:g8c}
\end{figure}

\begin{figure}[h]
    \centering
    \includegraphics[scale=0.56]{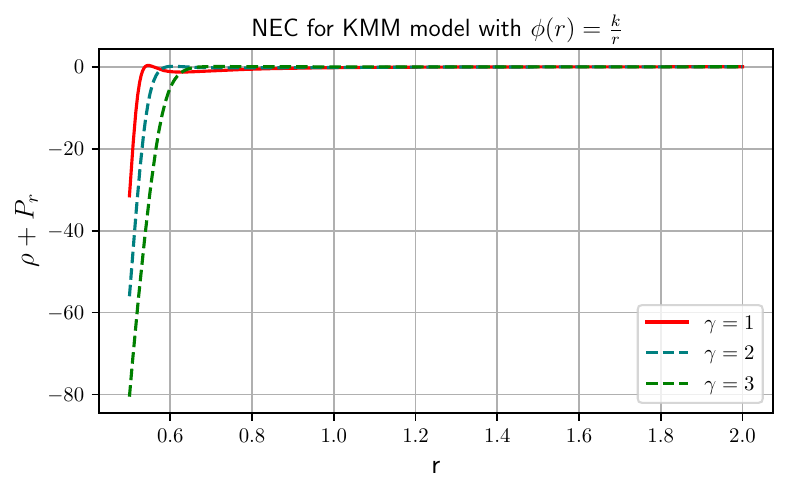}
    \includegraphics[scale=0.56]{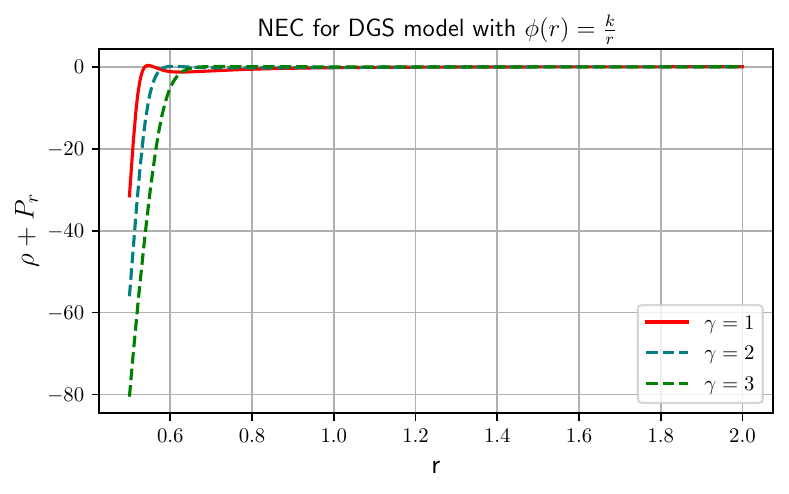}
    \caption{The plots of NEC for KMM and DGS models against radial distance $r$ for $f(Q)=Q+\gamma Q^2$ case with GUP parameter $\lambda=1$ and $k=1$. In the figure, $\gamma=1$ corresponds to the GR case.}
    \label{fig:g8d}
\end{figure}

Here, we consider a quadratic form of $f(Q)$ model such as $f(Q)=Q+\gamma Q^2$ where $\gamma$ is the model parameter. One can note that, for $\gamma=0$, the above model will be equivalent to the GR case. This model has been used for stellar structure with polytropic EoS \cite{rk}. With the same model, Banerjee et al. \cite{Pradhan} discussed wormhole solutions for different shape functions. For the quadratic model, the generalized field equations (\ref{12ch3}-\ref{14ch3}) can be read as
\begin{multline}\label{48ch3}
\rho=\frac{1}{2 r^6 (r-b(r))^2}\left[r^2 b^2 \left(2 b' \left(12 \gamma +r^2+7 \gamma  r \phi'\right)
-\gamma  r \left(4 b''+\phi ' \left(r \phi'-8\right)+4 r \phi''\right)+3 \gamma  b'^2\right)\right.\\\left.
-4 r^3 b b'\left(2 \gamma  \left(b'(r)+r \phi'\right)+r^2\right)+2 r^6 b'+2 \gamma  r b^3 \left(r \left(2 b''+\phi' \left(r \phi'-7\right)+4 r \phi''\right)-b' \left(3 r \phi'+7\right)\right.\right.\\\left.\left.
-8\right)+\gamma  b^4 \left(11-r \left(\phi' \left(r \phi'-6\right)+4 r \phi''\right)\right)\right],
\end{multline}
\begin{multline}\label{49ch3}
P_r=\frac{1}{2 r^6 (r-b)^2}\left[2 r^7 \phi'+r^2 b^2 \left(r \left(4 \left(\gamma  b''+\gamma  r \phi''+r\right)+6 r^2 \phi'+13 \gamma  r \phi'^2\right)+2 \gamma  b' \left(r \phi'-8\right)\right.\right.\\\left.\left.
+\gamma  b'^2\right)-2 r b^3\left(-6 \gamma +r \left(2 \gamma  \left(b''+2 r \phi''\right)+\left(\gamma +r^2\right) \phi '+7 \gamma  r \phi'^2\right)+\gamma  b' \left(r \phi '-3\right)+r^2\right)\right.\\\left.
-2 b \left(r^5 \left(2 \gamma  \phi'^2+3 r \phi '+1\right)
-2 \gamma  r^3 b'^2\right)+\gamma  b^4 \left(r \left(\phi ' \left(5 r \phi '+2\right)+4 r \phi ''\right)-7\right)\right],
\end{multline}
and
\begin{multline}\label{50ch3}
P_t=-\frac{1}{4 r^5 (r-b)^3}\left[\left(1-\frac{b}{r}\right) \left(-r \left(2 \gamma  b \left(-r b'+r (b-r) \phi '+b\right)+r^3 (r-b)\right) \left(\left(b-r b'\right)\right.\right.\right.\\\left.\left.\left.
\times \left((r-b) \phi '+2\right)+r (r-b)^2 \phi '^2+2 (r-2 b) (r-b) \phi '+2 r (r-b)^2 \phi ''\right)+4 \gamma r (r-b)\phi ' \right.\right.\\\left.\left.
\times \left(r^2 b \left(r \left(b''-2 \phi '+r \phi ''\right)-b' \left(2 r \phi '+5\right) \right)+r^3 b' \left(b'+r \phi '\right)+r b^2 \left(-r \left(b''-4 \phi '+2 r \phi ''\right)\right.\right.\right.\right.\\\left.\left.\left.\left.
+b'\left(r \phi '+3\right)+4\right)+b^3 \left(r^2 \phi ''-2 r \phi '-3\right)\right)+2 b \left(-r b'+r(b-r)\phi '+b\right) \left(b \left(\gamma  \left(-r b'+r (b-r) \phi '\right.\right.\right.\right.\right.\\\left.\left.\left.\left.\left.
+b\right)-r^3\right)+r^4\right)\right)\right].
\end{multline}
A comparison of Eqs. \eqref{25ch3} and \eqref{48ch3} yields the following non-linear differential equation:
\begin{multline}
\frac{1}{2 r^6 (r-b^2}\left[r^2 b^2 \left(2 b' \left(12 \gamma +r^2+7 \gamma  r \phi'\right)
-\gamma  r \left(4 b''+\phi ' \left(r \phi'-8\right)+4 r \phi''\right)+3 \gamma  b'^2\right)\right.\\\left.
-4 r^3 b b'\left(2 \gamma  \left(b'(r)+r \phi'\right)+r^2\right)+2 r^6 b'+2 \gamma  r b^3 \left(r \left(2 b''+\phi' \left(r \phi'-7\right)+4 r \phi''\right)-b' \left(3 r \phi'+7\right)\right.\right.\\\left.\left.
-8\right)+\gamma  b^4 \left(11-r \left(\phi' \left(r \phi'-6\right)+4 r \phi''\right)\right)\right]=-\frac{\pi^2}{720}\frac{1}{r^4}\left[1+\frac{5}{3}\Lambda_i\left(\frac{\lambda}{r^2}\right)\right].
\end{multline}
whose analytic solution is also not possible. Thus, we numerically evaluate the shape function's possible form by solving the above equation. Now, we shall examine the behavior of the shape function acquired by the numerical technique and their corresponding essential properties for the existence of wormhole structures for the GUP-corrected Casimir energy density. For this purpose, we use  Mathematica numerical ODE solver \textit{NDSolve} with the initial conditions $b(0.5)=0.1$ and $b'(0.5)=0.05$. We have depicted the behavior of shape function and flaring out condition for different redshift functions in Figs. \ref{fig:g8a} and \ref{fig:g8b}. It can be observed that shape function $b(r)$ is showing increasing behavior in the entire space-time, but for increases in the value of the model parameter $\gamma$, it is decreasing monotonically. During the numerical plot, we noticed that the asymptotic flatness condition $\frac{b(r)}{r}$ is validated for a small radius, the reason being the non-linearity of the Lagrangian. It is known that the role of GUP is to correct the Casimir energy, and hence the non-linearity of the Lagrangian is inevitable due to quantum correction. Due to such small-scale quantum correction, we note that the asymptotic flatness condition might be satisfied far from the throat as the GUP approximation to Casimir energy is not valid far from the throat. Also, we located the wormhole throat at $r_0\approx 0.005$. Moreover, we checked the flaring out near the throat and found that very near the throat, it was satisfied. However, far from the throat flare-out condition will not be validated for both redshift functions.\\
Further, we have studied the energy conditions, especially NEC, near the wormhole throat, which are given in Figs. \ref{fig:g8c} and \ref{fig:g8d}. We observed that NEC is disrespected near the throat for both KMM and DGS models under both redshift functions. Also, violation of NEC becomes more if we increase the value of $\gamma$. However, NEC will be satisfied for large $r$, or far from the throat. Thus there exists a possibility of having a micro or tiny wormhole.

\section{Volume integral quantifier}\label{ch3sec9}
Similar to the previous chapter, we will check the amount of exotic matter necessary for maintaining a wormhole using the VIQ technique. A brief discussion on this topic is already discussed in chapter-\ref{Chapter2}. In this chapter, we shall consider the appropriate equations and study the exotic matter requirement mathematically.
Considering the Eqs. \eqref{31ch3} and \eqref{32ch3}, integrating the expression \eqref{viq1} for the redshift function $\phi(r)=k,$ we obtain
\begin{equation}
Iv=\frac{1}{3\mathcal{F}_1}\left[3 r_1^3 \log \frac{r_1}{r_0} \left(\pi ^2 \left(5 \Lambda_i \lambda +9 r_0^2\right)-6480 \alpha  r_0^4\right)
+2 \pi ^2 \left(10 \Lambda_i \lambda  \left(r_0^3-r_1^3\right)+27 r_1^2 r_0^2 (r_0-r_1)\right)\right],
\end{equation}
where $\mathcal{F}_1=6480 r_1^3 r_0^3.$
Also, for $\phi(r)=\frac{k}{r},$ we obtain from the integral \eqref{viq1}
\begin{multline}
Iv=\frac{1}{12 r_1 r_0 \mathcal{F}_1}\left[(r_1-r_0) \left(54 r_1^2 r_0^2 \left(1440 r_1 \alpha  k r_0^2+\pi ^2 (k r_0-r_1 (k+4 r_0))\right)-5 \pi ^2 \Lambda_i \lambda \right.\right.\\\left.\left.
\times \left(r_1^3 (9 k+16 r_0)
+r_1^2 r_0 (16 r_0-3 k)+r_1 r_0^2 (16 r_0-3 k)-3 k r_0^3\right)\right)\right.\\\left.
+12 r_1^4 r_0 \log \frac{r_1}{r_0} \left(9 r_0^2 \left(\pi ^2-720 \alpha  r_0 (k+r_0)\right)+5 \pi ^2 \Lambda_i \lambda \right)\right].
\end{multline}
Moreover, we could find the volume integral $Iv$ for the redshift function $\phi(r)=\frac{1}{2}\text{log}\left(1+\frac{\eta^2}{r^2}\right)$ as
\begin{multline}
Iv=\frac{1}{2\mathcal{F}_1 \eta ^3}\left[r_1^3 \left(\mathcal{F}_2 \eta ^3  \left(\log \frac{r_1^2+\eta ^2}{\eta ^2+r_0^2}\right)+\mathcal{F}_3
\times \left(6480 \alpha  \eta ^4+\pi ^2 \left(9 \eta ^2-5 \Lambda_i \lambda \right)\right)\right)-2 \pi ^2 \eta  (r_1-r_0) \right.\\\left.
\times \left(5 \Lambda_i \lambda  \left(\eta ^2 \left(r_1^2+r_1 r_0+r_0^2\right)+r_1^2 r_0^2\right)+9 r_1^2 \eta ^2 r_0^2\right)\right],
\end{multline}
where $\mathcal{F}_2=\pi ^2 \left(5 \Lambda_i \lambda +9 r_0^2\right)-6480 \alpha  r_0^4\,,$ and
$\mathcal{F}_3=2 r_0^3\left[ \tan ^{-1}\left(\frac{r_0}{\eta }\right)- \tan ^{-1}\left(\frac{r_1}{\eta }\right)\right]\,.$\\
\begin{figure}[h]
    \centering
    \includegraphics[scale=0.63]{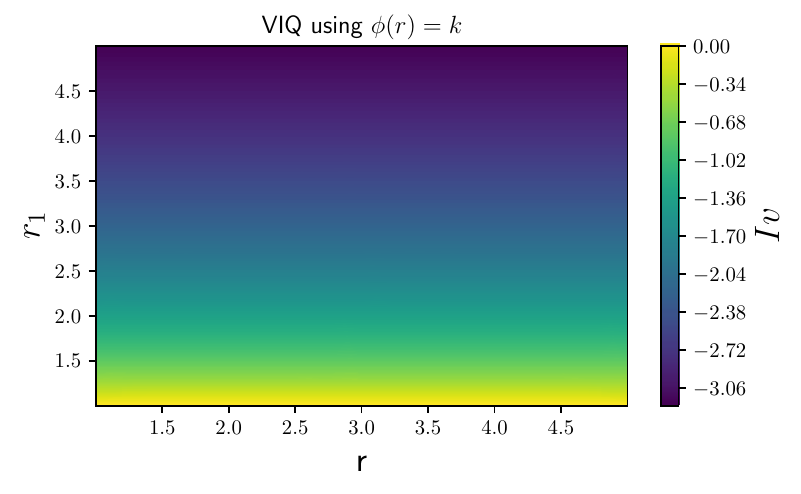}
    \caption{The plot of $Iv$ against $r$ and $r_1$ for KMM model. We consider GUP parameter $\lambda=0.1$, $r_0=1$ and $\alpha=0.95$.}
    \label{fig:g8}
\end{figure}
\begin{figure}[h]
    \centering
    \includegraphics[scale=0.63]{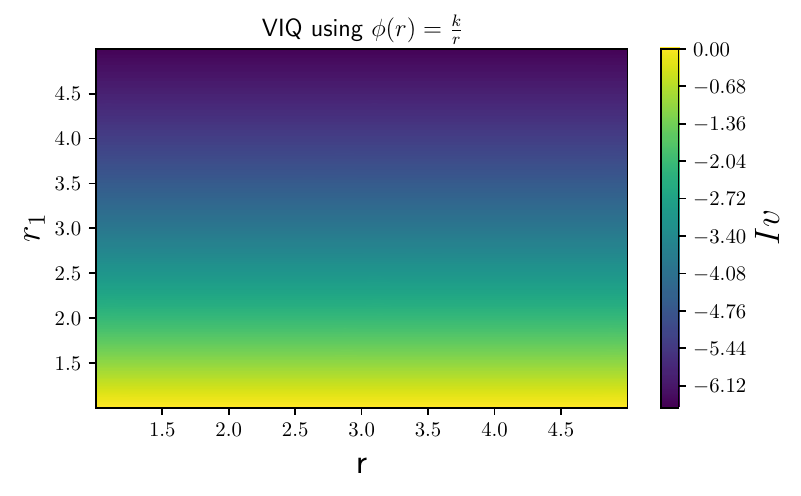}
    \caption{The plot of $Iv$ against $r$ and $r_1$ for DGS model. We consider GUP parameter $\lambda=0.1$, $r_0=1$, $k=2$ and $\alpha=0.95$.}
    \label{fig:g9}
\end{figure}
\begin{figure}[h]
    \centering
    \includegraphics[scale=0.63]{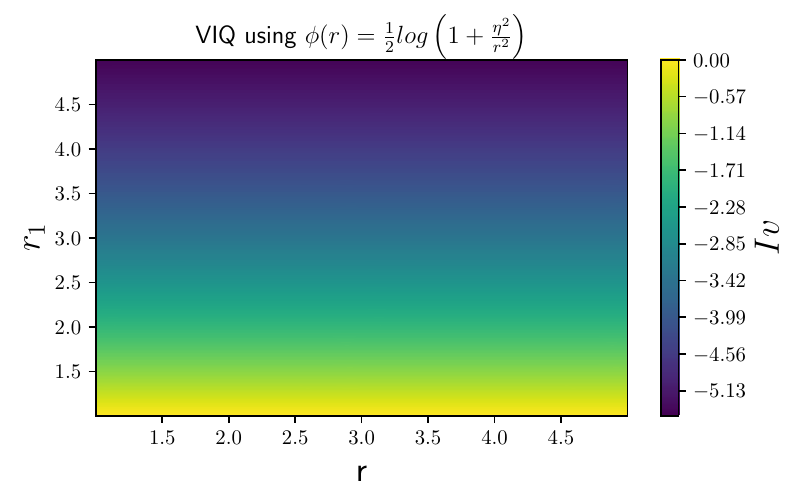}
    \caption{The plot of $Iv$ against $r$ and $r_1$ for KMM model. We consider GUP parameter $\lambda=0.1$, $r_0=1$, $\eta=3$ and $\alpha=0.95$.}
    \label{fig:g10}
\end{figure}
In Figs. (\ref{fig:g8}-\ref{fig:g10}), we have depicted the nature of volume integral $Iv$. Note that for $r_1\rightarrow r_0$, we should find $Iv\rightarrow 0$. One may observe from the figures that our obtained solutions satisfy the condition. Thus this reveals the existence of space-time geometries containing traversable wormholes sustained by arbitrarily small amounts of exotic matter. In fact, the total amount of ANEC-violating matter can be reduced by considering suitable wormhole geometry. Readers may check the Refs. \cite{Baransky,Channuie1} for a detailed discussion on this interesting topic.
\section{Conclusions}\label{ch3sec10}
In this chapter, we have investigated the effect of the GUP on the Casimir wormhole space-time in modified symmetric teleparallel gravity. The Casimir effect that occurs is attributable to the distorted quantized field of the vacuum between two parallel plane plates associated with exotic energy and pressure, which may be possible in the laboratory. Such types of exotic matter disrespect the energy conditions. Since in GR, wormhole material content must be exotic, and it should disobey some energy conditions and even present a negative mass. Hence, the quantum nature of the Casimir effect might help model these exotic objects. Here, we studied the MT wormhole field equations for $f(Q)$ gravity, describing the Casimir wormhole with the effect of GUP correction. However, GUP is not sufficient as a fundamental probe to the minimal length scale needed for quantum gravity as discussed in \cite{rainbowsmolin,sabinegup}, but it is a way to probe the length scales where the quantum gravity effects get non-negligible. Also, much work has been studied on the field-theoretic aspects of GUP, like showing how the ultraviolet divergence behaves. By calculating the first-order loop diagrams, \cite{regula}, it has been shown that the field theory is renormalizable and that there is no ultraviolet divergence.\\
In this study, we have employed two GUP relations: the KMM model and the DGS model, to show the effect of GUP. In \cite{Channuie1}, the authors have discussed the effect of GUP in Casimir wormholes by invoking the above GUP relations in GR. Later, Tripathy \cite{Tripathy} has investigated the GUP effect in $f(R,T)$ gravity. Here, we have utilized the mentioned GUP models to check the effect of GUP in the Casimir wormhole in this modified gravity. For this study, we have considered two $f(Q)$ models, such as linear $f(Q)=\alpha Q+\beta$ and quadratic $f(Q)=Q+\gamma\,Q^2$ models, where $\alpha$, $\beta$ and $\gamma$ are model parameters. Also, we have considered three different constant and non-constant redshift functions to acquire asymptotically flat wormhole solutions under GUP-affected Casimir density. In order to obtain the EoS parameters $\omega_r(r)$ and $\omega_t(r)$, we did use two famous EoS relations defined by $P_r=\omega_r(r)\rho$ and $P_t=\omega_t(r)\rho$, respectively. Our main theoretical observations are discussed below.\\
For the linear model, we have compared the GUP-corrected Casimir energy density with the energy density of the modified gravity and integrated it to obtain the shape function of the wormhole space-time metric. The resulting shape function respects the flare-out condition under the asymptotic background. Graphically we have shown the effect of the GUP parameter and modified gravity in shape functions. One may notice an increase in the GUP parameter $\lambda$ when the radial distance far from the throat decreases the shape function, whereas an increase in the model parameter $\alpha$ results in an increase in it. Nevertheless, in the throat, this effect is not substantial.\\
We have also investigated the behavior of EoS parameters for radial and tangential pressure to the Casimir energy density under different redshift functions. We have observed that the radial EoS parameter increases with the increase of radial distance $r$ and suitable parameters, whereas the tangential EoS parameter shows the opposite behavior. We can see the effect of modified gravity in the EoS parameter to a large extent, at least at distances away from the throat.\\
Again, we have studied the energy conditions, especially NEC, at the wormhole's throat with radius $r_0$ for both models. For each redshift function, we have noticed that NEC is violated in a small neighborhood of the throat. The violation contribution becomes more negative for an increase in $\alpha$. Thus this has demonstrated that some arbitrary amount of small quantity disrespects the classical energy condition at the wormhole's throat.\\
Further, for the quadratic model, We have used numerical techniques by setting some initial conditions and studying the graphical behavior of shape functions and energy conditions. We noticed that the shape function showed positively decreasing behavior as the values of $\gamma$  increased for both KMM and DGS models. Also, the flaring-out condition is satisfied near the throat, whereas, for large $r$, this condition will no longer be validated. Moreover, we have investigated the energy conditions and confirmed that NEC is violated for both models near the throat. Banerjee et al. \cite{Pradhan} discussed wormhole solutions for different shape functions and confirmed that solutions might not exist for considered shape functions under this quadratic model. But, from this study, it is worth mentioning that wormhole solutions could be possible numerically using appropriate initial conditions. However, this analysis shows the possibility of the existence of a macro or tiny wormhole.
Furthermore, we have examined the VIQ to study the amount of exotic matter required at the throat for a traversable wormhole. Our analysis found that a small amount of exotic matter is necessary for a traversable wormhole. 
Further, one can calculate the corrections of Casimir energy up to the next leading order using GUP-corrected quantum electrodynamics, as done in \cite{qedgup1,qedgup2} and can explore the significance in wormhole solutions.\\
In the next chapter, we shall explore the wormhole geometry with some popular dark matter models in the framework of $f(Q)$ gravity. We shall study how light deflects near a wormhole's throat due to different dark matter models.


\chapter{Deflection of light by wormholes and its shadow due to dark matter within modified symmetric teleparallel gravity formalism} 
\label{Chapter4} 

\lhead{Chapter 4. \emph{Deflection of light by wormholes and its shadow due to dark matter within modified symmetric teleparallel gravity formalism}} 
\blfootnote{*The work in this chapter is covered by the following publication:\\
\textit{Deflection of light by wormholes and its shadow due to dark matter within modified symmetric teleparallel gravity formalism}, Classical Quantum gravity \textbf{41}, 235001 (2024).}

This chapter presents a detailed discussion of wormhole shadow and the deflection angle of wormholes due to dark matter in the $f(Q)$ gravity framework. The detailed study of the work is outlined as follows:
\begin{itemize}
    \item We obtain the exact wormhole solutions based on the Bose-Einstein condensate, Navarro-Frenk-White, and Pseudo-Isothermal matter density profiles in $f(Q)$ gravity.
    \item We use the concept of rotational velocity to calculate the redshift functions and shape functions of the wormholes.
    \item We investigate the shadow of wormholes in the presence of various dark matter models.
    \item Further, the deflection of light by wormholes is also discussed.
\end{itemize}

\section{Introduction}\label{sec1}
The notion of dark matter, a mysterious form of matter containing approximately 25\% of the Universe's total matter content, emerges from observational predictions. Various candidates from particle physics and supersymmetric string theory, such as axions and weakly interacting massive particles, are considered compelling nominees for dark matter despite the absence of direct experimental confirmation. Nevertheless, indications of its presence are observed in phenomena such as galactic rotation curves \cite{Rubin1}, galaxy cluster dynamics \cite{Rubin2}, and cosmological observations of anisotropies in the cosmic microwave background as measured by PLANCK \cite{Rubin3}. The literature \cite{Rahaman1,Rahaman2} explores considerations of traversable wormholes within dark matter halos and galaxy formation regions, typically based on the NFW profile \cite{Rahaman3} of matter distribution. Rahaman et al. \cite{Rahaman4} initially proposed the existence of potential wormholes in the outer regions of galactic halos based on the NFW density profile, extending their analysis to utilize the URC dark matter model to derive analogous results within the central portion of the halo \cite{Rahaman2}. Also, dark matter, considered a non-relativistic matter describable by NFW and King profiles, is employed to construct wormholes \cite{Rahaman6}. Discrepancies between NFW halo velocity profiles and the observed dynamics of spiral galaxies remain unresolved, leading to modifications in the original NFW halo profiles within the $\Lambda$CDM scenario to align with observational data \cite{Rahaman5}. Additionally, it is shown that the presence of TWs in nature could be inferred through the study of scalar wave scattering \cite{Rahaman1}.\\
\indent In a recent study, Jusufi et al. \cite{Jusufi1} highlighted the potential formation of traversable wormholes through the presence of a BEC dark matter halo. This BEC dark matter model presents a more reasonable framework, particularly concerning the smaller scales of galaxies when compared to the Cold Dark Matter (CDM) model \cite{Jusufi2}. Notably, within the inner regions of galaxies, the interactions among dark matter particles are significantly vital, resulting in a deviation from cold dark matter behavior and rendering the density profile unsuitable. Consequently, the BEC dark matter model indicates considerably lower dark matter densities in the central regions of galaxies compared to those projected by the NFW profile. Additionally, an alternative category of dark matter characterized by a PI profile, alongside the CDM and BEC dark matter model, is associated with modifications to gravity, such as MOND \cite{Jusufi3}. MOND \cite{Milgrom1,Milgrom2,Milgrom3} indicates that the discrepancies in mass within galactic systems arise not from dark matter but from deviations from standard dynamics at lower accelerations. In \cite{Milgrom4}, Paul investigated the existence of traversable wormholes in the presence of MOND with or without a scalar field.\\
The exploration of shadows cast by compact objects in astrophysics has become a pivotal research focus, providing valuable insights into the intrinsic properties of these objects and the fabric of space-time \cite{sd1,sd2,sd3}. Observations of the shadows of compact objects, particularly those at the core of $SgrA^*$ and $M87$, have opened new routes for testing theories of gravity and examining various astrophysical models \cite{sd5,sd6}. Among these, the shadow of a wormhole is particularly fascinating, offering a theoretical means to explore the Universe's structure in novel ways. Notably, the shadow of a wormhole could offer indirect evidence of these hypothetical space-time tunnels, distinguishing itself from other compact objects. Analyzing the wormhole's shadow could yield valuable data on its structure, including throat size, spin, and potential accretion processes. The analysis of shadows from various objects, including black holes and wormholes, across different gravity models, with and without the presence of plasma, are extensively covered in \cite{sd7,sd8,sd9,sd10,sd11}.\\
Gravitational lensing stands as an early useful exploration within the realm of general relativity, which was initially delved into by Einstein \cite{Einstein10}. This phenomenon unfolds when a significantly massive celestial body bends incoming light, much like a lens, offering observers enhanced insights into the originating source. The interest in this area surged following the observed validation of light's deflection, as indicated theoretically \cite{Dyson,Eddington}. Its scope extends beyond celestial bodies, opening avenues to probe into exoplanets, dark matter, and dark energy. A notable milestone was achieved with the first successful measurement of a white dwarf's mass, Stein 2051 B, through astrometric microlensing \cite{Sahu}. An intriguing aspect of gravitational lensing is its potential to cause light to bend infinitely in unstable light rings, creating numerous relativistic images under strong lensing conditions \cite{Bozza1,Virbhadra,Bozza22}. This phenomenon, in both strong and weak forms, serves as a potent analytical tool for examining gravitational fields near various cosmic entities, including black holes and wormholes. Theoretical and astrophysical investigations have extensively applied gravitational lensing to study wormholes, reflecting its significance in contemporary research \cite{A1,A2,A3,A4}.\\
Motivated by the above discussions, we explore wormhole solutions under different dark matter halo models within $f(Q)$ gravity. The chapter is structured as follows: in section \ref{ch4sec2}, we explain the wormhole field equations in $f(Q)$ gravity. Section \ref{ch4sec3} discusses the dark matter profiles and wormhole solutions, while section \ref{ch4sec4} presents the energy conditions analytically and graphically. Wormhole shadows and deflection angles are examined in section \ref{ch4sec5} and \ref{ch4sec6}, respectively. The chapter concludes with final discussions in section \ref{ch4sec8}.
\section{Wormhole field equations}
\label{ch4sec2}
This chapter is the continuation of chapter-\ref{Chapter2} from the field equations (\ref{5ch1}-\ref{8ch1}).
One can note that the non-zero off-diagonal metric components \eqref{8ch1} obtained by the specific gauge choice for the field equations in the setting of $f(T)$ theory of gravity \cite{dz1} impose some constraints on the functional form of $f(T)$. As a consequence, one can put the same constraints on the functional form of the $f(Q)$ theory of gravity. In this regard, within the scope of anisotropic matter distribution, Wang et al. \cite{Wang2} developed the potential functional forms for $f(Q)$ gravity in the framework of the static and spherically symmetric space-time. Interestingly, they have proved that the exact form of Schwarzschild solution can be derived only when $f_{QQ}(Q)=0$, while the other related solutions obtained by taking nonmetricity term $Q^{\prime}=0$ or $Q=Q_0$, where $Q_0$ is constant, provides the deviation from the exact Schwarzschild solution. To solve the system of field equations for $f(Q)$ gravity theory and obtain self-gravitating solutions, we derive the functional form of $f(Q)$ by setting $f_{QQ}$ to zero as
\begin{eqnarray}
f_{QQ}(Q)=0~\Rightarrow~f_{Q}(Q)=c_0~\Rightarrow~f(Q)=c_0+c_1 Q,~~~~  \label{eq26}
\end{eqnarray}
where $c_0$ and $c_1$ are constants. We would like to emphasize that, at this point, the compatibility of a static spherically symmetric space-time with the coincident gauge can be achieved if one assumes the affine connection to be zero and $f(Q)$ gravity theory has vacuum solutions (i.e., $T_{\mu\nu}=0$). In this scenario, the off-diagonal component can be expressed as 
\begin{eqnarray}
\frac{\text{cot}\,\theta}{2}\,Q^\prime\,f_{QQ}=0, \label{eq27}
\end{eqnarray}
where $Q$ has been already provided in Eq. \eqref{ga3}. As a result of Eq. \eqref{eq26}, it is evident that $f(Q)$ must be linear, leading to the correct transformation of the equations of motion to $f_{QQ}=0$ \cite{dz3}. Therefore, in order to achieve a more generalized form of the spherically symmetric metric within a fixed coincident gauge, it is necessary to ensure compatibility with an affine connection $\Tilde{\Gamma}^\alpha_{\,\,\,\beta\gamma}=0$ \cite{dz3}. Hence, in this study, we have chosen a linear functional form by setting $f_{QQ}=0$ to derive the equations of motion, making the spherically symmetric coordinate system compatible with the considered affine connection. Therefore, we get the revised field equations as follows\\
\begin{equation}
\label{2b9}
\rho=\frac{c_0 }{2}-\frac{c_1  b'}{r^2},
\end{equation}
\begin{equation}
\label{2b10}
P_r=\frac{c_1  \left(2 r (b-r) \phi'+b\right)}{r^3}-\frac{c_0 }{2},
\end{equation}
\begin{equation}
\label{2b11}
P_t=\frac{c_1  \left(r \phi '+1\right) \left(r b'+2 r (b-r) \phi '-b\right)}{2 r^3}+\frac{c_1  (b-r) \phi ''}{r}-\frac{c_0 }{2}.
\end{equation}
Now, in the following sections, we shall study wormhole geometry under the effect of different dark matter models.
\section{Wormhole solutions due to dark matters}\label{ch4sec3}
In this section, we shall try to find the wormhole shape function as well as the redshift function and discuss the necessary properties of a traversable wormhole under the effect of dark matter. For this study, we will consider three well-known dark matter models: Bose-Einstein condensate, pseudo-isothermal, and Navarro-Frenk-White. A brief review of these profiles is already discussed in chapter-\ref{Chapter1}.

\subsection{BEC profile}
\begin{figure}[h]
    \centering
    \includegraphics[width=0.3\linewidth, height=1.5in]{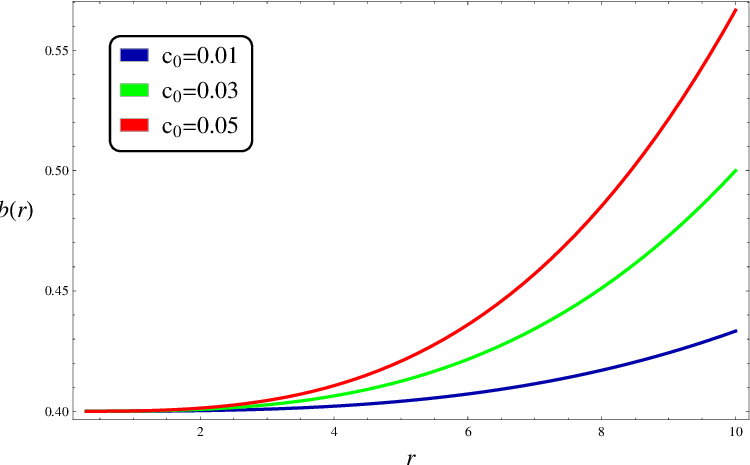}
    \includegraphics[width=0.3\linewidth, height=1.5in]{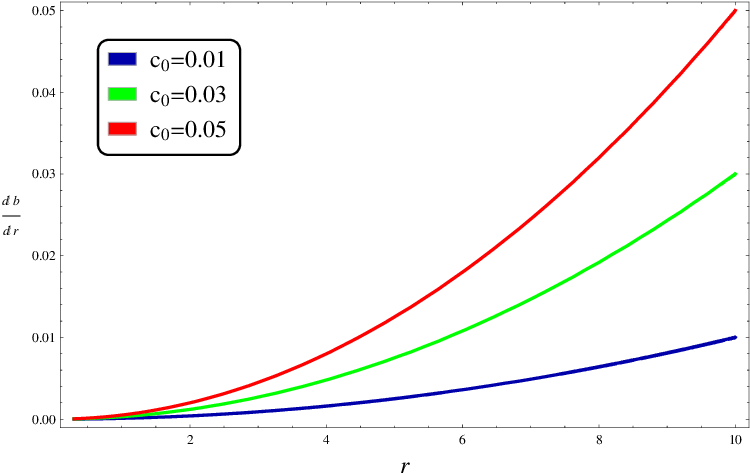}
    \includegraphics[width=0.3\linewidth, height=1.5in]{Figures/ch8/Figs1.eps}
    \caption{\label{F1} The plots of shape functions properties against the radial coordinate $r$ for the BEC model with $c_{1}=0.9$, $r_{0}=0.2$, $\rho_s=0.9$, and $R=1.94$.}
\end{figure}

The mass profile of the dark matter galactic halo can be read as
\begin{equation}\label{34}
    M(r)=4\pi \int_0^r \rho(r) r^2 dr.
\end{equation}
Now, for the density of BEC, given in Eq. \eqref{becd1}, the solution of the mass profile can be obtained as
\begin{equation}
\label{35}
M(r)=\frac{4\pi \rho_s}{k^2}r\left(\frac{\sin{kr}}{kr}-\cos{kr}\right).
\end{equation}
From the above Eq. \eqref{35}, we can obtain the tangential velocity of a test particle moving in the dark halo from the following relation
\begin{equation}\label{ab1}
v_{t}^2(r)=G\,M(r)/r,
\end{equation}
and hence
\begin{equation}\label{36}
v_{t}^2(r)=\frac{4\pi G \rho_s}{k^2}\left(\frac{\sin{kr}}{kr}-\cos{kr}\right),
\end{equation}
where $k=\pi/R$. For $r\rightarrow 0$, we have $v_{t}^2(r)\rightarrow 0$.\\
Note that within the equatorial plane, the rotational velocity of a test particle in spherically symmetric space-time is defined by \cite{dm67}
\begin{equation}\label{37}
v_{t}^2(r)=r\,\phi^{'}(r).
\end{equation}
Inserting the Eq. \eqref{36} in the above relation \eqref{37}, we have
\begin{equation}\label{38}
\frac{4 R^2 \rho_s}{\pi}\left[\frac{\sin\left({\frac{\pi r}{R}}\right)}{\frac{\pi r}{R}}-\cos\left({\frac{\pi r}{R}}\right)\right]=r\,\phi^{'}(r).
\end{equation}
On solving, we find
\begin{equation}\label{39}
\phi(r)=D_1-\frac{4 \rho_s R^3 \sin \left(\frac{\pi  r}{R}\right)}{\pi ^2 r},
\end{equation}
where $D_1$ is an integrating constant.\\
Now, we equate the density of the wormhole matter with the density of the BEC using Eqs. \eqref{2b9} and \eqref{becd1}, it follows that


\begin{equation}\label{40}
\frac{c_0 }{2}-\frac{c_1  b'(r)}{r^2}=\frac{\rho_s R}{\pi r}\sin\left({\frac{\pi r}{R}}\right).
\end{equation}
Solving for $b(r)$, we get
\begin{equation}\label{41}
b(r)=\frac{\frac{1}{3} \pi  c_0  r^3-\frac{2 \rho_s R^3 \sin \left(\frac{\pi  r}{R}\right)}{\pi ^2}+\frac{2 \rho_s r R^2 \cos \left(\frac{\pi  r}{R}\right)}{\pi }}{2 \pi  c_1 }+c_3.
\end{equation}
Now to find the integrating constant $c_3$, we impose the throat condition $b(r_0)=r_0$
\begin{equation}\label{42}
c_3=r_0-\frac{1}{2 \pi c_1 } \left[\frac{1}{3} \pi  c_0  r_0^3-\frac{2 \rho_s R^3 \sin \left(\frac{\pi  r_0}{R}\right)}{\pi ^2}+\frac{2 \rho_s R^2 r_0 \cos \left(\frac{\pi  r_0}{R}\right)}{\pi }\right],
\end{equation}
and hence the shape function $b(r)$ becomes
\begin{equation}\label{43}
    b(r)=\frac{1}{6 c_1 }\left[c_0 r^3-c_0  r_0^3+6 c_1  r_0 
    +\frac{1}{\pi ^3} \left(6 \rho_s R^2 \left(-R\Lambda_1
    +\pi \Lambda_2\right)\right)\right],
\end{equation}
where
\begin{equation}\label{44}
\Lambda_1=\left(\sin \left(\frac{\pi  r}{R}\right)-\sin \left(\frac{\pi  r_0}{R}\right)\right),
\end{equation}
and
\begin{equation}\label{45}
\Lambda_2=r \cos \left(\frac{\pi  r}{R}\right)-r_0 \cos \left(\frac{\pi  r_0}{R}\right).
\end{equation}
To sustain the wormhole's structure, the ``flare-out" condition must be met to ensure the wormhole mouth is open. The following relation gives this condition at the wormhole throat region
\begin{equation}\label{46}
b^{'}(r_0)=\frac{r_0 \left(\pi  c_0  r_0-2 \rho_s R \sin \left(\frac{\pi  r_0}{R}\right)\right)}{2 \pi  c_1 }<1.
\end{equation}
In Fig. \ref{F1}, we have depicted the flare-out condition for different values of $c_0$, which satisfies the condition around the throat under an asymptotic background. Also, we have checked the behavior of shape function $b(r)$ and noticed that as we increase the value of $c_0$, the shape function increases. Thus, our obtained shape function satisfies all the necessary properties of the shape function. Here, we considered the throat radius $r_0=0.2$.
\subsection{PI profile}
\begin{figure}[h]
    \centering
    \includegraphics[width=0.3\linewidth, height=1.5in]{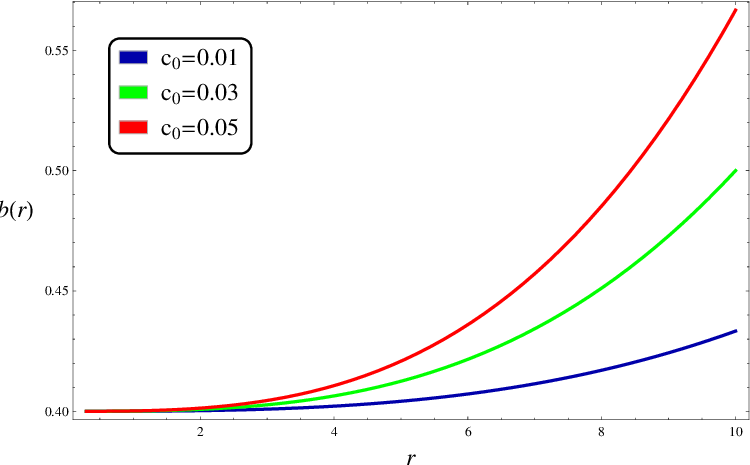}
    \includegraphics[width=0.3\linewidth, height=1.5in]{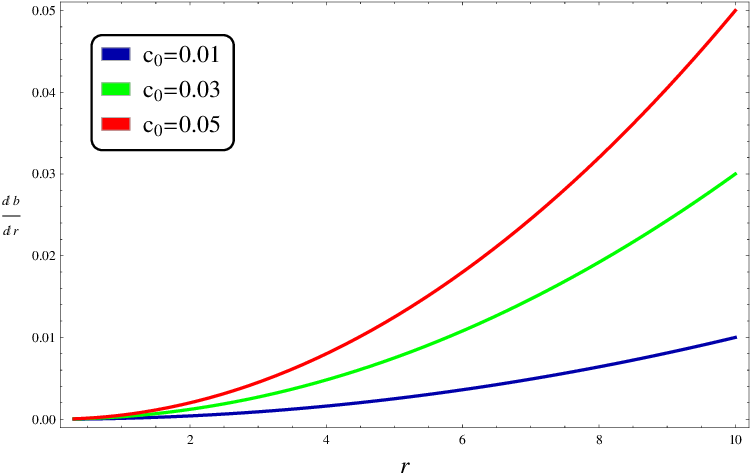}
    \includegraphics[width=0.3\linewidth, height=1.5in]{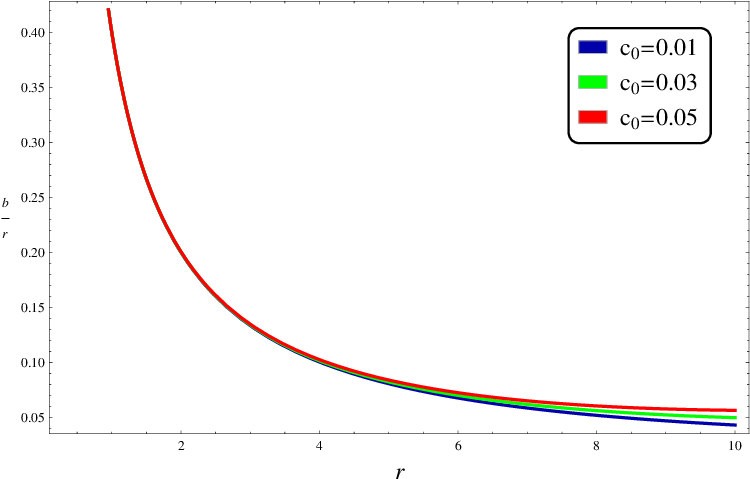}
    \caption{\label{F11} The plots of shape functions properties against the radial coordinate $r$ for the PI model with $c_{1}=0.9$, $r_{0}=0.2$, $\rho_s=0.9$, and $r_{s}=1.94$.}
\end{figure}

The solution of mass profile for the PI galactic halo, given in Eq. \eqref{4aa1}, can be obtained using Eq. \eqref{34}
\begin{equation}\label{4a2}
M(r)=4 \pi \rho_s r_s^2 \left(r-r_s \tan ^{-1}\left(\frac{r}{r_s}\right)\right).
\end{equation}
The tangential velocity $v_t^2(r)$ of a test particle moving in the dark halo can be obtained from Eq. \eqref{ab1}
\begin{equation}\label{4a3}
v_t^2(r)=\frac{4 \pi \rho_s r_s^2}{r} \left(r-r_s \tan ^{-1}\left(\frac{r}{r_s}\right)\right).
\end{equation}
Now, with the relation given in Eq. \eqref{37}, we can find the redshift function
\begin{equation}\label{4a4}
\phi(r)=2 \pi  \rho_s r_s^2 \log \left(r^2+r_s^2\right)+\frac{4 \pi  \rho_s r_s^3 \tan ^{-1}\left(\frac{r}{r_s}\right)}{r}+D_2,
\end{equation}
where $D_2$ is the integrating constant. Now, we shall try to calculate the shape function of the wormhole under the PI dark matter model by comparing the density of the wormhole with the density of PI dark matter. From Eqs. \eqref{2b9} and \eqref{4aa1}, one can get the relation
\begin{equation}\label{47}
\frac{c_0 }{2}-\frac{c_1  b'(r)}{r^2}=\frac{\rho_s}{1+\left(\frac{r}{r_s}\right)^2}.
\end{equation}
On solving
\begin{equation}\label{48}
b(r)=\frac{\frac{c_0  r^3}{3}+2 \rho_s r_s^3 \tan ^{-1}\left(\frac{r}{r_s}\right)-2 \rho_s r r_s^2}{2 c_1 }+c_5,
\end{equation}
where $c_5$ is the integrating constant. By imposing the throat condition on the above equation, we can obtain the final shape function
\begin{equation}\label{49}
b(r)=\frac{1}{6 c_1 }\left[c_0  r^3+6 \rho_s r_s^3 \Lambda_4
+6 \rho_s r_s^2 (r_0-r)
-c_0 r_0^3+6 c_1  r_0\right],
\end{equation}
where
\begin{equation}\label{50}
\Lambda_4=\left(\tan ^{-1}\left(\frac{r}{r_s}\right)-\tan ^{-1}\left(\frac{r_0}{r_s}\right)\right).
\end{equation}
Now, we check the flare-out condition at the throat, i.e., 
\begin{equation}
b'(r_0)= \frac{r_s^2 r_0^2 (c_0 -2 \rho_s)+c_0  r_0^4}{2 c_1  \left(r_s^2+r_0^2\right)},
\end{equation}
which is obviously satisfied. For more information, one can check Fig \ref{F11}.
\subsection{NFW profile}
\begin{figure}[h]
    \centering
    \includegraphics[width=0.3\linewidth, height=1.5in]{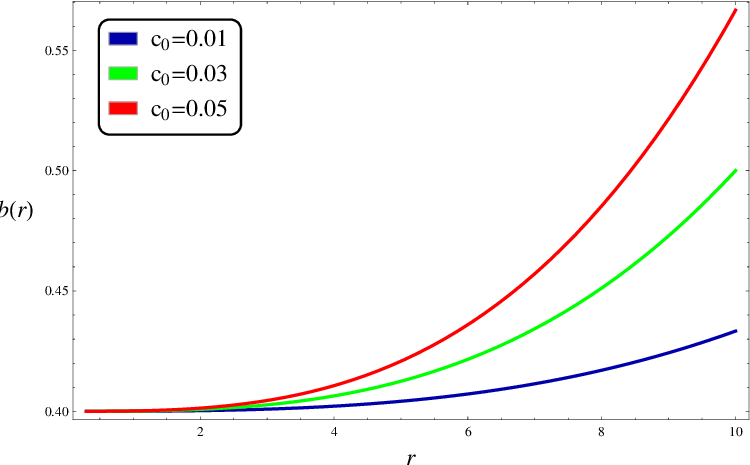}
    \includegraphics[width=0.3\linewidth, height=1.5in]{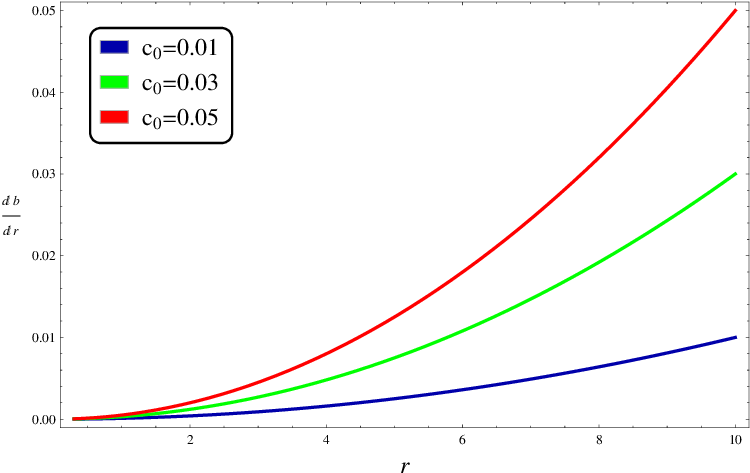}
    \includegraphics[width=0.3\linewidth, height=1.5in]{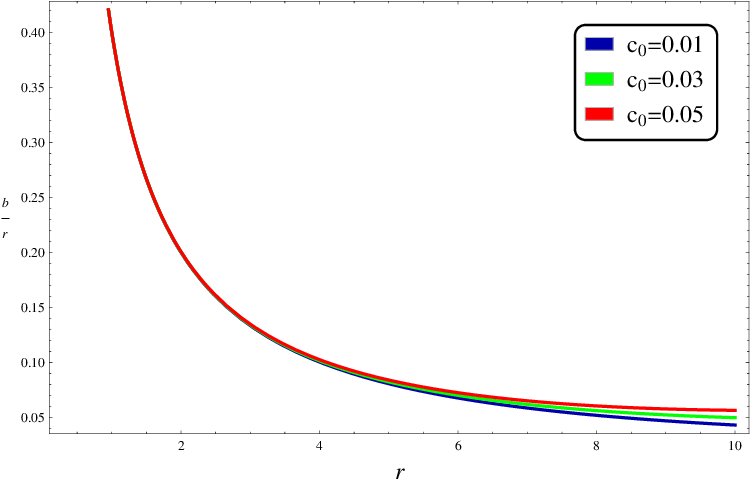}
    \caption{\label{F111} The plots of shape functions properties against the radial coordinate $r$ for the NFW model with $c_{1}=0.9$, $r_{0}=0.2$, $\rho_s=0.9$, and $r_{s}=1.94$.}
\end{figure}

For this NFW case, the mass function can be read as
\begin{equation}\label{4a2}
M(r)=4 \pi \rho_s r_s^3 \left(\frac{r_s}{r+r_s}+\log (r+r_s)-\log (r_s)-1\right),
\end{equation}
and consequently, the tangential velocity $v_t^2(r)$ is
\begin{equation}\label{52}
v_t^2(r)=\frac{4 \pi \rho_s r_s^3}{r} \left(\frac{r_s}{r+r_s}+\log (r+r_s)-\log (r_s)-1\right).
\end{equation}
Now from Eq. \eqref{37}, we can find the redshift function
\begin{equation}\label{53}
\phi(r)=\frac{4 \pi \rho_s r_s^3}{r}\left(\log (r_s)-\log (r+r_s)\right)+D_3,
\end{equation}
where $D_3$ is the integrating constant.
Again, from Eqs. \eqref{2b9} and \eqref{4aaa1}, it follows that
\begin{equation}\label{54}
\frac{c_0 }{2}-\frac{c_1  b'(r)}{r^2}=\frac{\rho_s}{(r/r_s)(1+r/r_s)^2}
\end{equation}
On solving the above differential equation, we obtain
\begin{equation}\label{55}
b(r)=\frac{1}{2 c_1 }\left[-\frac{2 \rho_s r_s^4}{r+r_s}-2 \rho_s r_s^3 \log (r+r_s)+c_0  r_s^2 (r+r_s)
-c_0  r_s (r+r_s)^2+\frac{1}{3} c_0  (r+r_s)^3\right]+c_7,
\end{equation}
where $c_7$ is the integrating constant. Now, we impose the condition $b(r_0)=r_0$ on the above equation and obtain the shape function
\begin{equation}\label{56}
b(r)=\frac{1}{6 c_1 }\left[c_0  r^3+\Lambda_6+6 \rho_s r_s^3 \Lambda_7-c_0  r_0^3+6 c_1  r_0\right],
\end{equation}
where,
\begin{equation}\label{57}
\Lambda_6=\frac{6 \rho_s r_s^4 (r-r_0)}{(r+r_s) (r_s+r_0)},   
\end{equation}
\begin{equation}\label{58}
\Lambda_7=(\log (r_s+r_0)-\log (r+r_s)).
\end{equation}
Now, we study the important condition, i.e., flare-out condition, with the obtained shape function \eqref{56} graphically. We noticed that with the appropriate choices of parameters, the flare-out condition is satisfied, i.e., $b(r)<1$ at $r=r_0$ (see Fig. \eqref{F111}).
\begin{figure}[h]
    \centering
    \includegraphics[width=.3\linewidth,
height=1.5in]{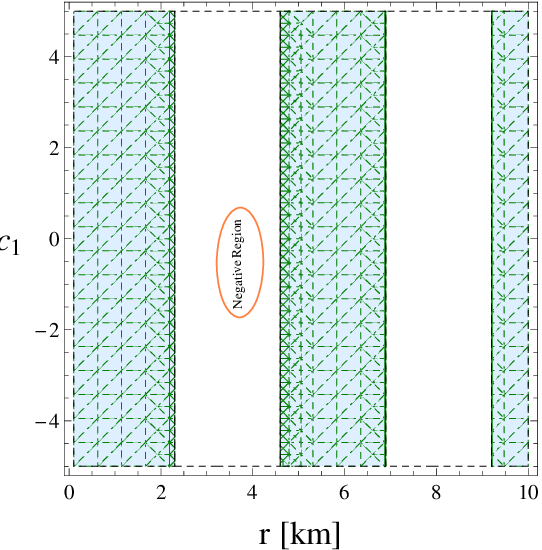}
\includegraphics[width=.3\linewidth,
height=1.5in]{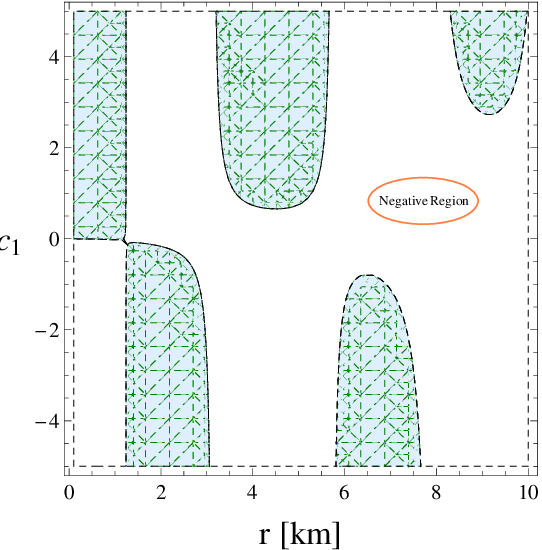}
\includegraphics[width=.3\linewidth,
height=1.5in]{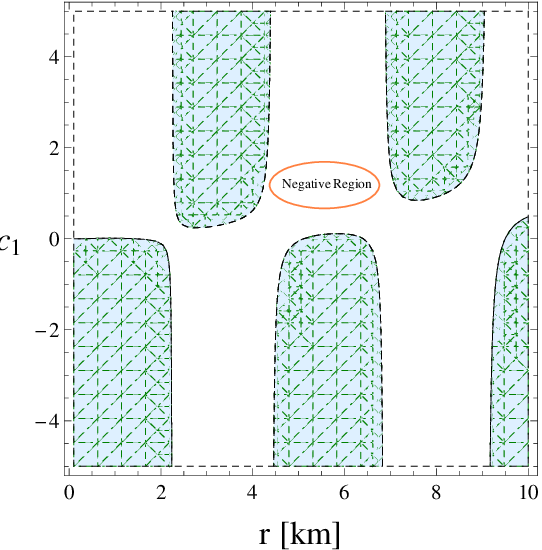}
\centering 
\includegraphics[width=.3\linewidth,
height=1.5in]{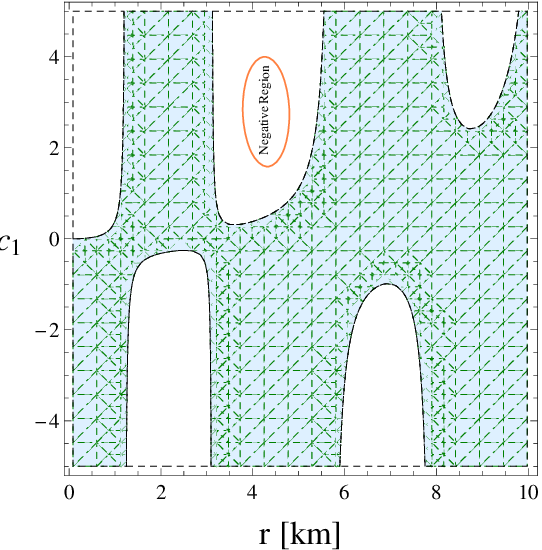}
\includegraphics[width=.3\linewidth,
height=1.5in]{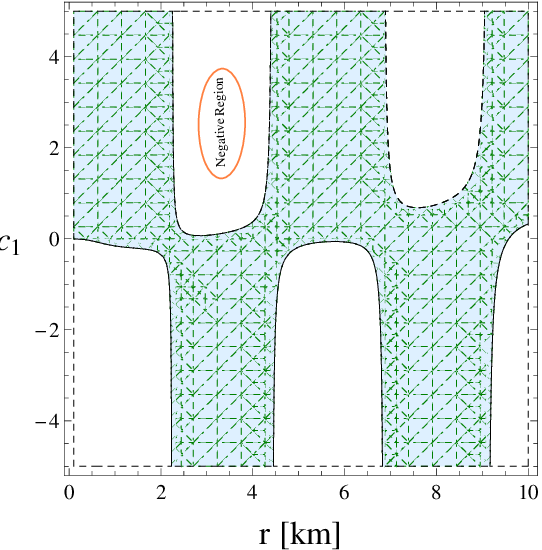}
\includegraphics[width=.3\linewidth,
height=1.5in]{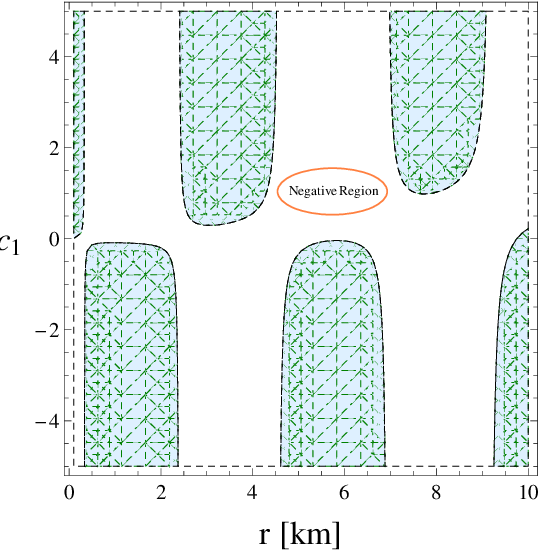}
 \caption{\label{F5} The plots represent the valid and negative regions of all the energy conditions for the BEC profile. In the first row $\rho$ (left), $\rho+P_r$ (middle), and $\rho+P_t$ (right) are presented. In second row, $\rho-|P_r|$ (left), $\rho-|P_t|$ (middle), and $\rho+P_r +2P_t$ (right) are shown with $c_{0}=0.05$, $r_{0}=0.2$, $\rho_{c}=0.9$, and $R=1.94$.}
\end{figure}
\begin{figure}[h]
    \centering
    \includegraphics[width=.3\linewidth,
height=1.5in]{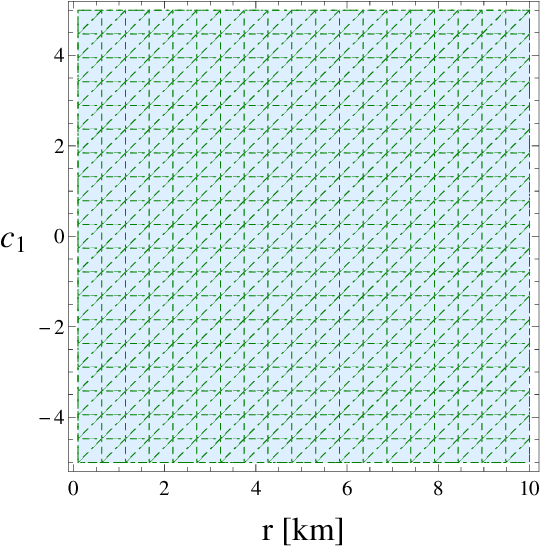}
\includegraphics[width=.3\linewidth,
height=1.5in]{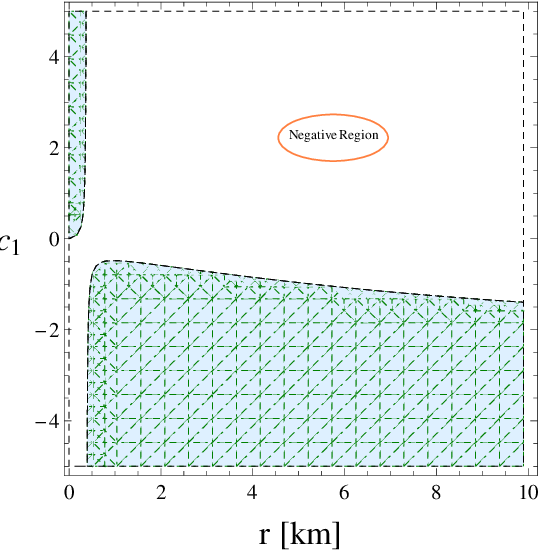}
\includegraphics[width=.3\linewidth,
height=1.5in]{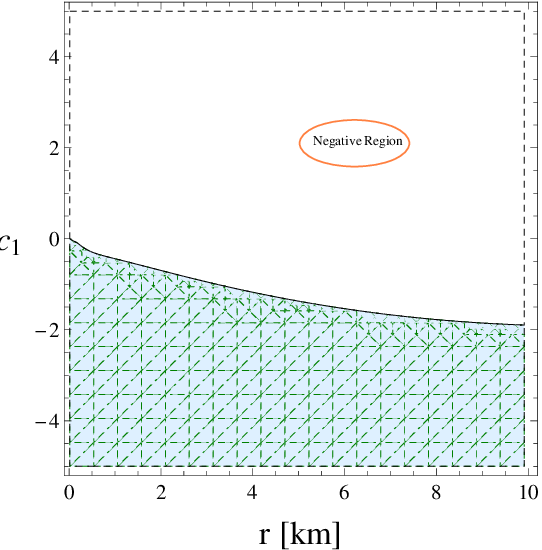}
\centering 
\includegraphics[width=.3\linewidth,
height=1.5in]{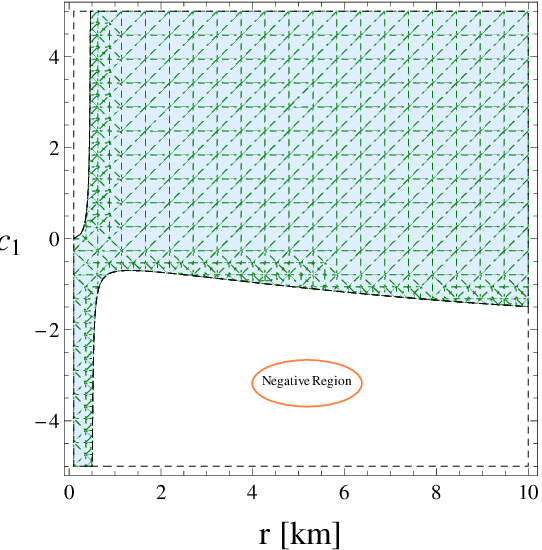}
\includegraphics[width=.3\linewidth,
height=1.5in]{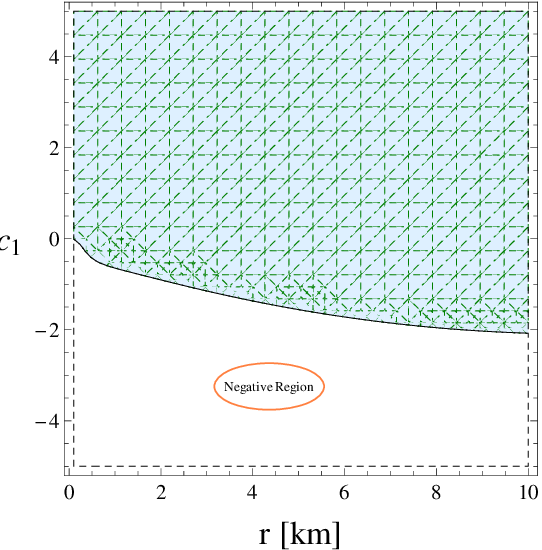}
\includegraphics[width=.3\linewidth,
height=1.5in]{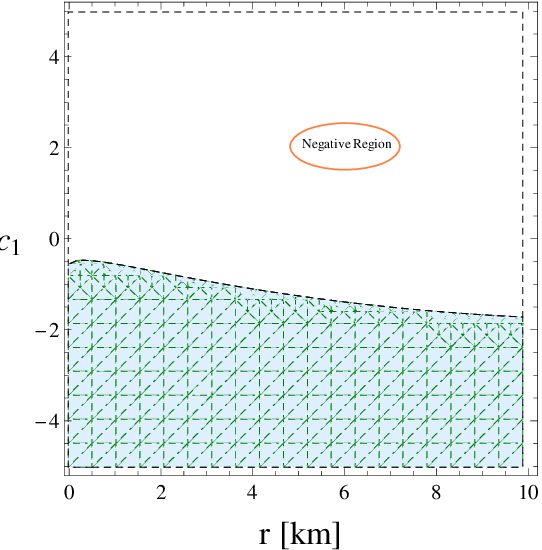}
 \caption{\label{F6} The plots represent the valid and negative regions of all the energy conditions for the PI profile. In the first row $\rho$ (left), $\rho+P_r$ (middle), and $\rho+P_t$ (right) are presented. In second row, $\rho-|P_r|$ (left), $\rho-|P_t|$ (middle), and $\rho+P_r +2P_t$ (right) are shown with $c_{0}=0.05$, $r_{0}=0.2$, $\rho_s=0.9$, and $r_{s}=1.94$.}
\end{figure}
\begin{figure}[h]
    \centering
    \includegraphics[width=.3\linewidth,
height=1.5in]{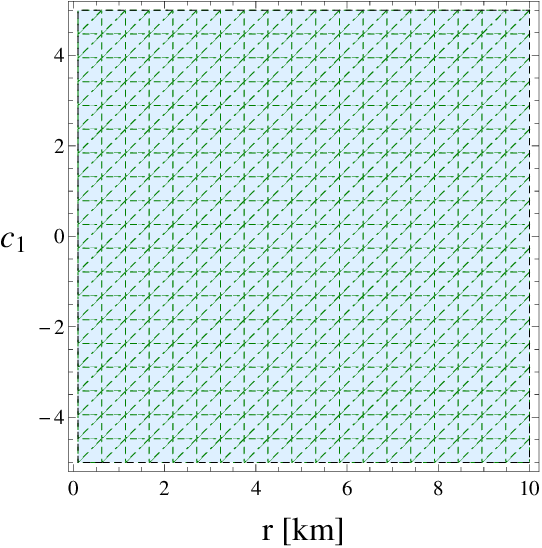}
\includegraphics[width=.3\linewidth,
height=1.5in]{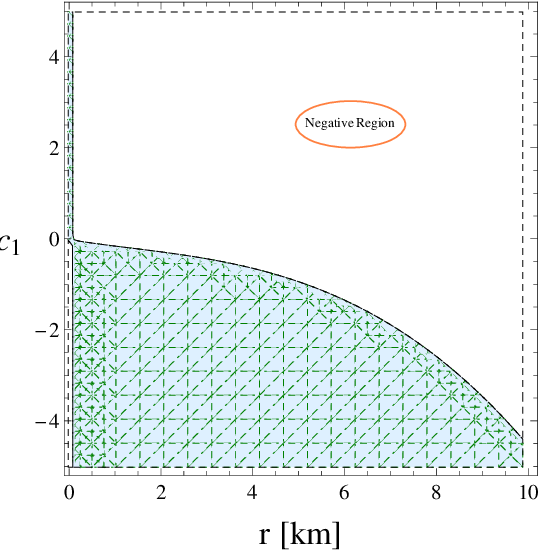}
\includegraphics[width=.3\linewidth,
height=1.5in]{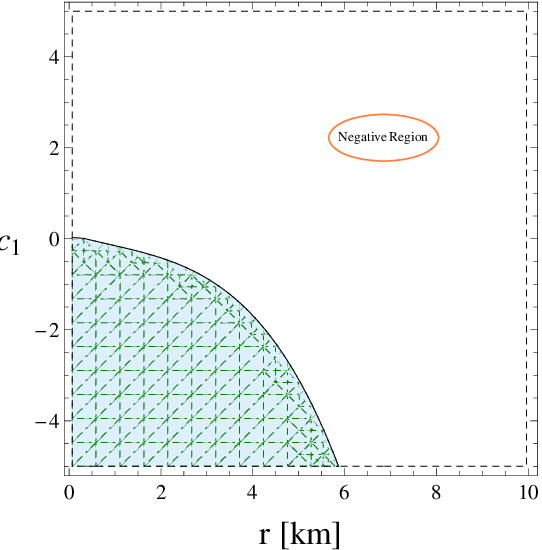}
\centering 
\includegraphics[width=.3\linewidth,
height=1.5in]{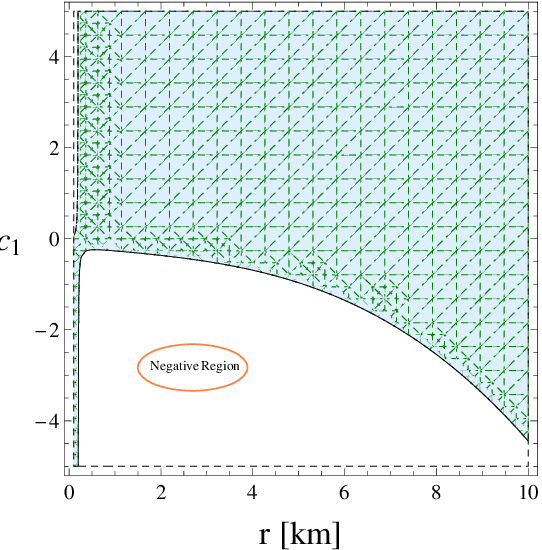}
\includegraphics[width=.3\linewidth,
height=1.5in]{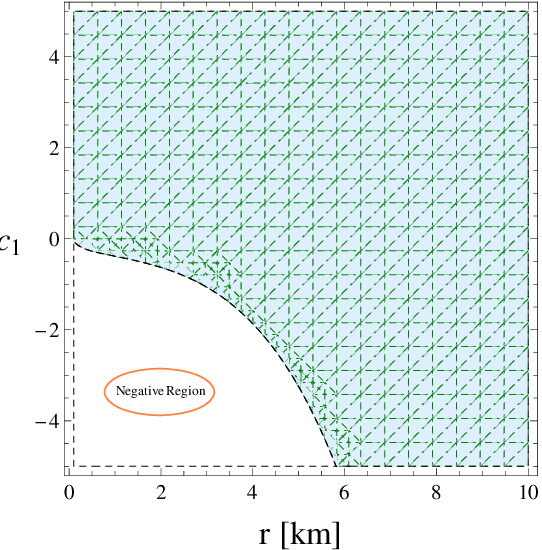}
\includegraphics[width=.3\linewidth,
height=1.5in]{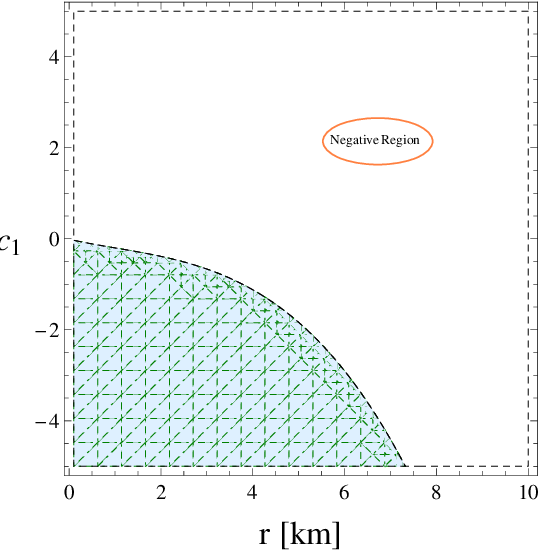}
 \caption{\label{F7} The plots represent the valid and negative regions of all the energy conditions for the NFW profile. In the first row $\rho$ (left), $\rho+P_r$ (middle), and $\rho+P_t$ (right) are presented. In second row, $\rho-|P_r|$ (left), $\rho-|P_t|$ (middle), and $\rho+P_r +2P_t$ (right) are shown with $c_{0}=0.05$, $r_{0}=0.2$, $\rho_s=0.9$, and $r_{s}=1.94$.}
\end{figure}
\section{Energy conditions}\label{ch4sec4}
In the study of modified gravitational theories, the validity of energy conditions of matter is often the key issue, and the energy conditions are the necessary conditions to explain the singularity theorem \cite{Hawking}. Moreover, the energy conditions aid in analyzing the entire space-time structure without precise solutions to Einstein's equations and play a crucial role in investigating wormhole solutions in the context of modified gravity theories. The common popular basic energy conditions (the null, weak, dominant, and strong energy conditions) originate from the Raychaudhuri equation \cite{Raychaudhuri}, which plays a crucial role in describing the attractive properties of gravity and positive energy density. These four popular energy conditions are considered here, and we shall discuss the behavior of wormholes using energy conditions.\\
The expression for radial NEC for the BEC dark matter profile can be read as
\begin{multline}\label{59}
\rho+P_r=\frac{1}{6} \left[-3 c_0 +\frac{1}{r^3}\left(c_0  r^3+\frac{1}{\pi ^3}\left(6 \rho_s R^2 \left(\pi \Lambda_2-R \Lambda_1\right)\right)-c_0  r_0^3+6 c_1  r_0\right)-\frac{1}{\pi ^5 r^4}\left(8 \rho_s R^2 \right.\right.\\\left.\left.
\times \Lambda_3 \left(\pi ^3 \left(-c_0  r^3+6 c_1  r +c_0  r_0^3-6 c_1  r_0\right)+6 \rho_s R^2 \left(R \Lambda_1 -\pi \Lambda_2\right)\right)\right)\right]+\frac{\rho_s R \sin \left(\frac{\pi  r}{R}\right)}{\pi  r},
\end{multline}
where $\Lambda_3=\left(R \sin \left(\frac{\pi  r}{R}\right)-\pi  r \cos \left(\frac{\pi  r}{R}\right)\right)$. $\Lambda_1$ and $\Lambda_2$ are defined in \eqref{44} and \eqref{45}.\\
Similarly, for the PI case, NEC can be obtained as follows
\begin{multline}\label{60}
\rho+P_r=\frac{\rho_s r_s^2}{r^2+r_s^2}+\frac{1}{6 r^4}\left[-3 c_0  r^4+r \left(c_0  r^3+6 \rho_s r_s^3 \Lambda_4+6 \rho_s r_s^2 (r_0-r)-c_0  r_0^3+6 c_1  r_0\right)\right.\\\left.
-8 \pi \rho_s r_s^2 \Lambda_5 \left(-c_0  r^3 +6 c_1  r -6 \rho_s r_s^3 \Lambda_4+6 \rho_s r_s^2 (r-r_0)+c_0  r_0^3-6 c_1  r_0\right)\right],
\end{multline}
where, $\Lambda_4$ is already defined in \eqref{50} and $\Lambda_5=\left(r-r_s \tan ^{-1}\left(\frac{r}{r_s}\right)\right)$.\\
At last, the expression for NEC for the NFW profile
\begin{multline}\label{61}
\rho+P_r=-\frac{c_0 }{2}+\frac{\Lambda_6+c_0  r^3+6 \Lambda_7 \rho_s r_s^3-c_0  r_0^3+6 c_1  r_0}{6 r^3}-\frac{1}{r^4 (r+r_s)}\left[8 \pi  c_1  \rho_s r_s^3 ((r+r_s) \Lambda_8+r) \right.\\\left. \times \left(\frac{1}{6 c_1 }\left(\Lambda_6+c_0  r^3
+6 \Lambda_7 \rho_s r_s^3-c_0  r_0^3+6 c_1 r_0\right)-r\right)\right]+\frac{\rho_s r_s^3}{r (r+r_s)^2},
\end{multline}
where $\Lambda_8=(\log (r_s)-\log (r+r_s))$. The expression of $\Lambda_6$ and $\Lambda_7$ are presented in \eqref{57} and \eqref{58}, respectively.\\
Now, at the throat ($r=r_0$), the NEC for each dark matter halo profile has been obtained and shown in Eq. \eqref{62}
\begin{equation}\label{62}
\rho+P_r\bigg\vert_{r=r_0}=
     \begin{cases}
      -\frac{c_0 }{2}+\frac{\rho_s R \sin \left(\frac{\pi  r_0}{R}\right)}{\pi  r_0}+\frac{c_1 }{r_0^2},  & \text{(BEC)}\\
      \\
      -\frac{c_0 }{2}+\frac{\rho_s r_s^2}{r_s^2+r_0^2}+\frac{c_1 }{r_0^2}, &  \text{(PI)}\\
      \\
       \frac{1}{r_0^2}\left[c_1 +\frac{\rho_s r_s^3 r_0}{(r_s+r_0)^2}\right]-\frac{c_0 }{2},   &  \text{(NFW)}
     \end{cases}
\end{equation}
\subsection{Detailed analysis of these dark matter models with graphical descriptions}
The primary focus of research has been on a method where the energy conditions are not violated by the actual matter itself but rather by an effective energy-momentum tensor that emerges within a modified gravitational theory framework. In this section, we will examine the energy conditions for the solutions we have explored. A comprehensive graphical analysis, including a regional investigation of the energy conditions in these dark matter models, is presented in Figs. (\ref{F5}-\ref{F7}). The energy density $\rho$ for the BEC model shows positive behavior in the vicinity of the throat within $-5 \leq c_1 \leq 5$ and $0 \leq r \leq 2$. Interestingly, for the PI and NFW cases, energy density shows positive behavior in the entire space-time. Next, we plotted the radial NEC $\rho+P_r$ graphs for each case. It was noticed that for the BEC case, $\rho+P_r$ is violated near the throat for $-5\leq c_1 \leq 0$ and $0\leq r \leq 1.7$, and satisfied for $0\leq c_1 \leq 5$. For the PI case, the valid region of $\rho+P_r$ is $-5\leq c_1 < -0.5$ and negative region is $-0.5\leq c_1 < 5$ against the radial co-ordinate $0.4 \leq r \leq10$. The energy condition $\rho+P_r$ for the NFW profile depicted violated for $0\leq c_1 \leq 5$ and obeyed for $-5\leq c_1 \leq 0$ within $0\leq r \leq 10$. Further, $\rho+P_t$ is depicted for each case and observed that it is disrespected for the BEC case within $0\leq c_1 \leq 5$ and satisfied within $-5 \leq c_1 \leq 0$ against $0\leq r \leq 2.2$. Also, for the PI case, $\rho+P_t$ shows negative behavior for $c_1\geq 0$ whereas satisfied within $c_1<0$. In addition, $\rho+P_t$ for the NFW case portrays a valid region for $-5\leq c_1 \leq 0$ and $0\leq r \leq 6$, and the remaining region shows the violation of $\rho+P_t$. Furthermore, we have thoroughly investigated the behavior of SEC, which can be found in the lower left plot of Figs. (\ref{F5}-\ref{F7}). We observed that the negative region of $\rho+P_r+2P_t$ for the BEC case is $0\leq c_1 \leq 5$, PI case is $c_1>-0.5$, and for the NFW case is $c_1\geq 0$. Finally, we checked the behavior of DEC for each profile, and interestingly, we noticed that DEC's behavior was completely opposite to NEC's behavior. One can check Figs. (\ref{F5}-\ref{F7}) for a complete overview. It is important to note that the model parameter, $c_1$, significantly influences all the energy conditions in the current analysis. In fact, within the range of parameters involved, all the energy conditions are violated in maximum regions, confirming the presence of exotic matter, which is necessary for creating the obtained wormhole solutions due to the violation of energy conditions, specifically the violation of NEC. This exotic matter is believed to contribute to the stability and traversability of wormholes by counteracting the gravitational collapse caused by ordinary matter. The study of energy violation in the context of wormholes sheds light on these structures. The energy conditions for each case (as shown in Figs. (\ref{F5}-\ref{F7})) support the existence of these wormhole solutions in the background of symmetric teleparallel gravity.\\
All the results mentioned above regarding the energy conditions for each dark matter model are also summarized in table-\ref{Table1}.

\begin{table*}[t]
    \centering
\begin{tabular}{ p{3cm}| p{12cm}}
 \hline
  \multicolumn{2}{|c|}{The behavior of the energy conditions around the throat} \\
 \hline
Physical expressions & \quad \quad \quad \quad \quad \quad \quad \quad \quad \quad \quad BEC profile\\
 \hline
$\rho$ & $\rho>0$ for $-5\leq c_1 \leq 5$\\
\hline
$\rho+P_r$ & $\rho+P_r<0$ for $-5\leq c_1 \leq 0$ and $\rho+P_r>0$ for $0\leq c_1 \leq 5$\\
\hline
$\rho+P_t$ & $\rho+P_t<0$ for $0\leq c_1 \leq 5$ and $\rho+P_t>0$ for $-5\leq c_1 \leq 0$\\
\hline
$\rho+P_r+2 P_t$ & $\rho+P_r+2 P_t<0$ for $0\leq c_1 \leq 5$ and $\rho+P_r+2 P_t>0$ for $-5\leq c_1 \leq 0$\\
\hline
$\rho-|P_r|$ & $\rho-|P_r|<0$ for $0\leq c_1 \leq 5$ and $\rho-|P_r|>0$ for $-5\leq c_1 \leq 0$\\
\hline
$\rho-|P_t|$ & $\rho-|P_t|<0$ for $-5\leq c_1 \leq 0$ and $\rho-|P_t|>0$ for $0\leq c_1 \leq 5$\\
\hline
\multicolumn{2}{|c|} {PI profile}\\
\hline
$\rho$ & $\rho>0$ for $-5\leq c_1 \leq 5$\\
\hline
$\rho+P_r$ & $\rho+P_r<0$ for $-0.5\leq c_1 < 5$ and $\rho+P_r>0$ for $-5\leq c_1<-0.5$\\
\hline
$\rho+P_t$ & $\rho+P_t<0$ for $c_1 \geq 0$ and $\rho+P_t>0$ for $c_1<0$\\
\hline
$\rho+P_r+2 P_t$ & $\rho+P_r+2 P_t<0$ for $c_1 >-0.5$ and $\rho+P_r+2P_t>0$ for $c_1 \leq -0.5$\\
\hline
$\rho-|P_r|$ & $\rho-|P_r|<0$ for $-5\leq c_1<-0.5$ and $\rho-|P_r|>0$ for $-0.5\leq c_1 < 5$\\
\hline
$\rho-|P_t|$ & $\rho-|P_t|<0$ for $c_1<0$ and $\rho-|P_t|>0$ for $c_1 \geq 0$\\
\hline
\multicolumn{2}{|c|} {NFW profile}\\
\hline
$\rho$ & $\rho>0$ for $-5\leq c_1 \leq 5$\\
\hline
$\rho+P_r$ & $\rho+P_r<0$ for $0\leq c_1 < 5$ and $\rho+P_r>0$ for $-5\leq c_1<0$\\
\hline
$\rho+P_t$ & $\rho+P_t<0$ for $c_1 \geq 0$ and $\rho+P_t>0$ for $-5\leq c_1<0$\\
\hline
$\rho+P_r+2 P_t$ & $\rho+P_r+2 P_t<0$ for $c_1 \geq 0$ and $\rho+P_r+2P_t>0$ for $0< c_1 \leq -5$\\
\hline
$\rho-|P_r|$ & $\rho-|P_r|<0$ for $-5\leq c_1<0$ and $\rho-|P_r|>0$ for $0\leq c_1 \geq 5$\\
\hline
$\rho-|P_t|$ & $\rho-|P_t|<0$ for $-5\leq c_1<0$ and $\rho-|P_t|>0$ for $c_1 \geq 0$\\
\hline
\end{tabular}
\caption{Table shows the summary of the energy conditions.}
\label{Table1}
\end{table*}
\section{Shadows of wormholes}\label{ch4sec5} 
In this section, we shall discuss wormhole shadows under the effect of three different kinds of dark matter models.
To study the impact of wormholes on light deflection, we need to calculate the movement of light rays. We consider the motion of a light beam using the null geodesic equation, which allows us to predict its trajectory. The equation can be found by applying the Euler-Lagrange equation
$\mathcal{L}=-\frac{1}{2}g_{\mu \nu}\dot{x}^{\mu}\dot{x}^{\nu}$.
Without loss of generality, we can consider the equatorial plane, i.e., $\theta=\frac{\pi}{2}$. The Lagrangian equation describing the motion of light rays around the wormhole geometry is given as
\begin{equation}\label{rrr2}
\begin{split}
 & \mathcal{L}=-\frac{1}{2}g_{\mu \nu}\dot{x}^{\mu}\dot{x}^{\nu}\\
& =- e^{2\phi(r)} \dot{t}^2+\frac{1}{1-\frac{b(r)}{r}} \dot{r}^2 +r^2 (\dot{\theta}^2 + \sin^2\theta \dot{\varphi}^2),\\
\end{split}
\end{equation}
where $\dot{x}^{\mu}$ denotes the four-velocity of the photon, and the dot represents the differentiation with respect to the affine parameter $\tau$. Now, by applying the Lagrangian equation of motion within the scope of wormhole geometry in Eq. (\ref{rrr2}), one can get the following relations
\begin{equation}\label{rrr3}
\dot{t}=\frac{dt}{d\tau}=\frac{E}{e^{2\phi(r)}},
\end{equation}
\begin{equation}\label{rrr4}
\dot{\varphi}= \frac{d\varphi}{d\tau}=\frac{L}{r^2 \sin^2(\theta)}.
\end{equation}
In the above relations, $E$ and $L$ represent the energy and angular momentum of the particle around the wormhole throat. To simplify the geodesics, we propose two dimensionless impact parameters: $\xi=\frac{L}{E}$ and $\eta=\frac{\kappa}{E^2}$, where $\kappa$ denotes the Carter constant. By considering the null geodesic ($\mathcal{L}=0$), which can be expressed in terms of kinetic and potential energy. Scale an affine parameter $\tau \rightarrow \frac{\tau}{L}$. The orbit equation of motion can be revised as
\begin{equation}\label{rrr6}
   K_{E}+V_{eff}=\frac{1}{\Tilde{b}^2},
\end{equation}
where  $\Tilde{b}=\frac{L}{E}$ is the  kinetic energy function $K_{E}$ and potential function $V_{eff}$ are described as
\begin{equation}\label{rrr7}
 K_{E}=  \frac{e^{2\phi(r)}}{1-\frac{b(r)}{r}}\dot{r}^2,
\end{equation}
and
\begin{equation}\label{rrr7}
 V_{eff}=   \frac{e^{2\phi(r)}}{r^2}.
\end{equation}
\par
\begin{figure}
    \centering
    \includegraphics[width=.45\linewidth,
height=2.2in]{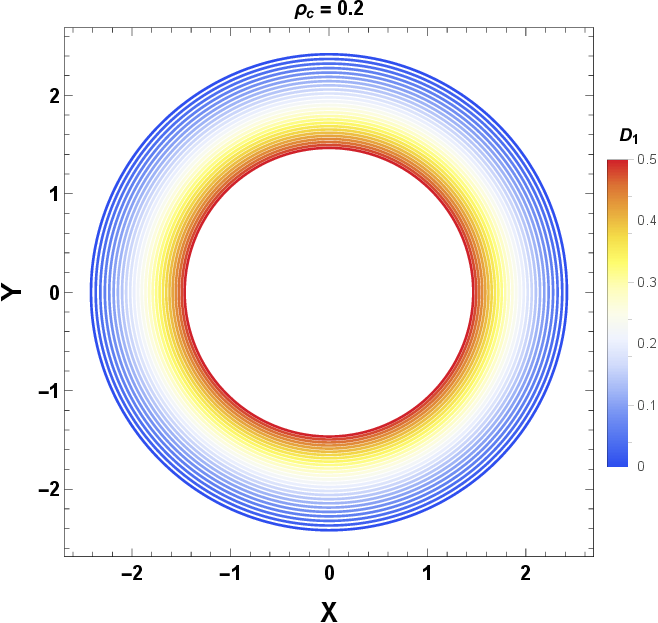}
\includegraphics[width=.45\linewidth,
height=2.2in]{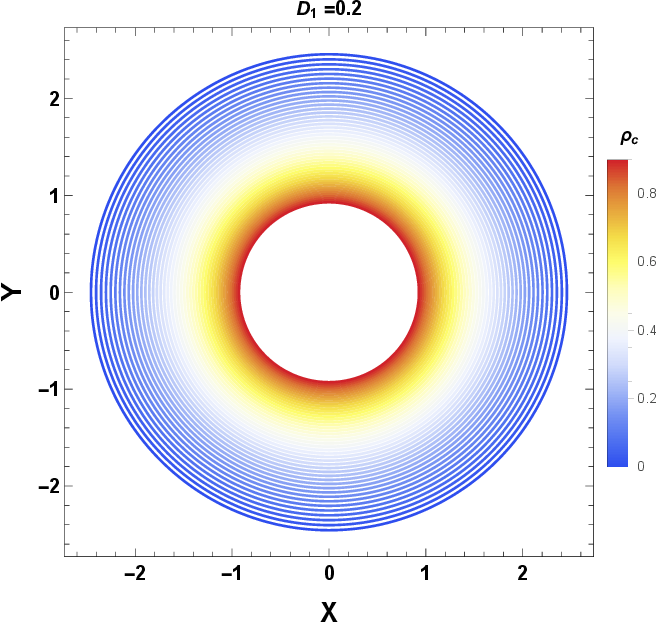}
\centering
\includegraphics[width=.45\linewidth,
height=2.2in]{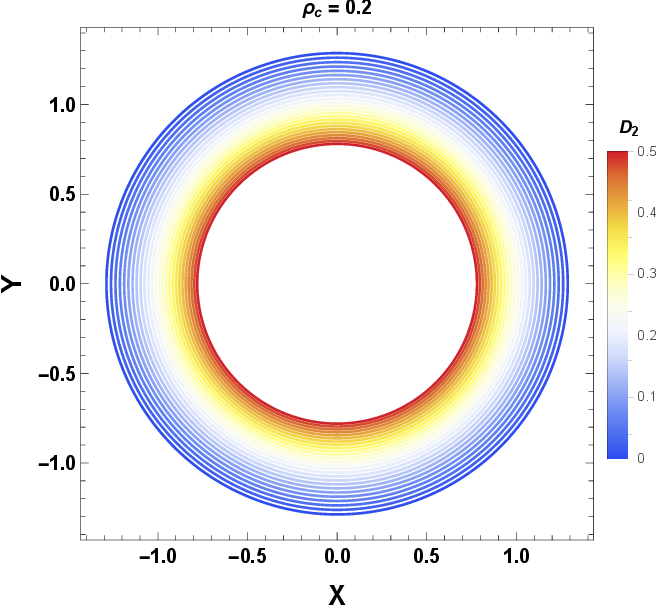}
\includegraphics[width=.45\linewidth,
height=2.2in]{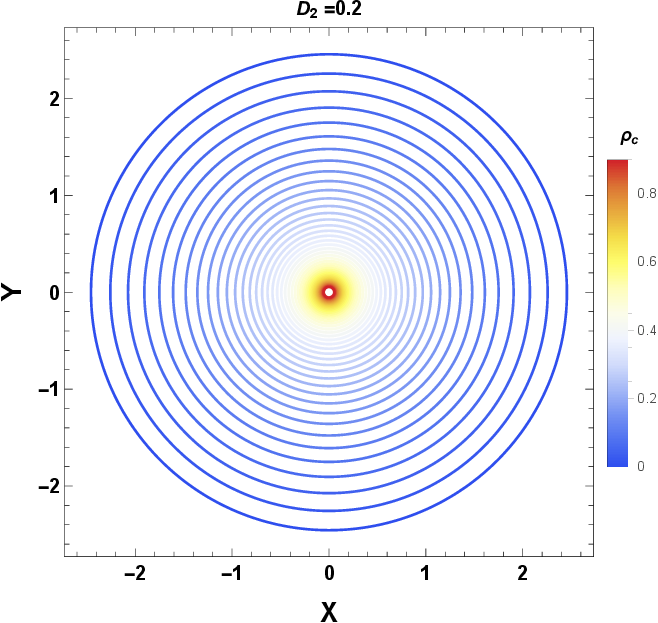}
\centering
\includegraphics[width=.45\linewidth,
height=2.2in]{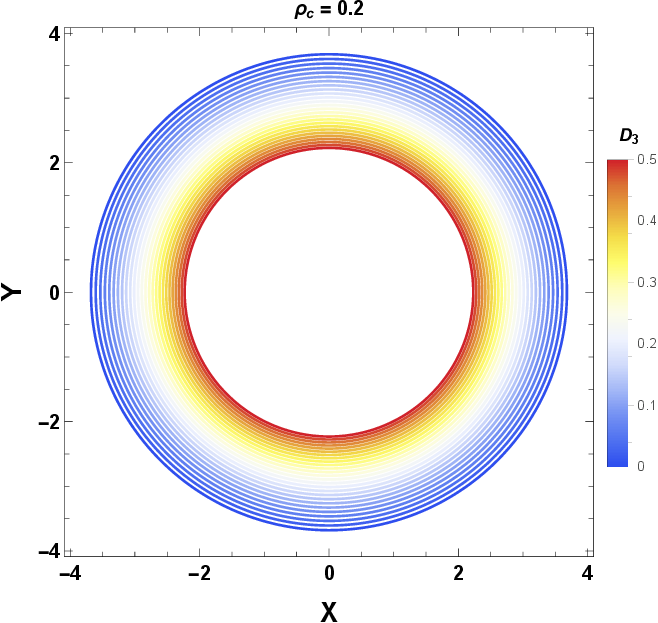}
\includegraphics[width=.45\linewidth,
height=2.2in]{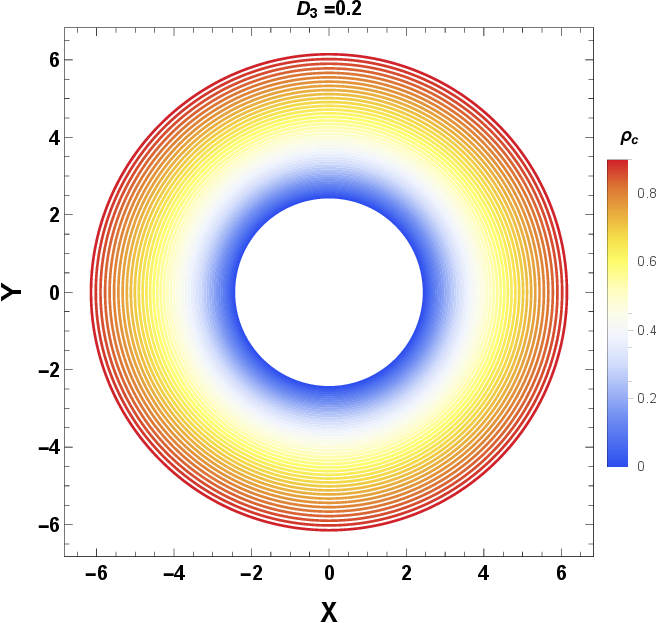}
 \caption{\label{rrF8} The plots represent the parametric plots for wormhole shadows with BEC profile (first row), PI profile (second row), and NFW profile (third row) with $r_{0}=0.2$, $R=1.94$, and $r_{s}=1.94$.}
\end{figure}

In order to describe the wormhole shadows around the wormhole throat, one can use celestial coordinates ($X,Y$), which are further defined as \cite{ws1,ws2,ws3}
\begin{equation}\label{19}
X=\lim_{r_0\rightarrow \infty}(-r_0^2  \sin(\theta_0))\frac{d\phi}{dr},
\quad
Y=\lim_{r_0\rightarrow \infty}(r_0^2  \frac{d\theta}{dr}),
\end{equation}
where $r_0$ is the wormhole throat. Further, $\theta_0$ is the inclination angle between the wormhole and the observer. After simplification, one can obtain the celestial coordinates for wormhole geometry as \cite{ws4}
\begin{equation}\label{rrr37}
X=-\frac{\xi}{\sin(\theta_0)},
\quad
Y=\sqrt{\eta-\frac{\xi^2}{\sin^2(\theta_0)}}.
\end{equation}
Assuming the static observer is at infinity, the radius of the wormhole shadow $R_s$ as seen from the equatorial plane, i.e., $\theta_0=\pi/2$ can be stated as
\begin{equation}\label{rrr35}
R_s=\sqrt{\xi^2 +\eta}=\frac{r_0}{e^{\phi(r_0)}}.
\end{equation}
The parametric plot for the Eqs. (\ref{rrr37}) and (\ref{rrr35}) in the $(X, Y)$ plane can cast a variety of wormhole shadows for the different ranges of involved parameters. The wormhole shadows for the wormhole geometry are depicted in Fig. (\ref{rrF8}) for three models of dark matter halos, including BEC. From the first row of Fig. (\ref{rrF8}), it is noticed that the BEC profile has an influence on the wormhole shadows. The larger values of $D_1$ and central density $\rho_c$ lead the wormhole shadow closer to the wormhole throat. For the smaller values of these mentioned parameters wormhole shadow radius is going away from the wormhole throat. The same behavior is also noticed for two other cases, the PI profile and NFW profile from the second row and third row of Fig. (\ref{rrF8}).
\section{Deflection angle}\label{ch4sec6}
\label{secDef}
\begin{figure}[h]
    \centering
    \includegraphics[width=.3\linewidth,
height=1.5in]{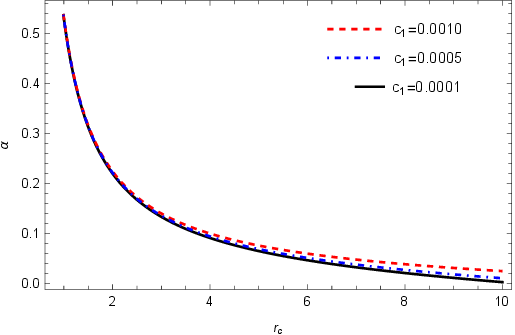}
\includegraphics[width=.3\linewidth,
height=1.5in]{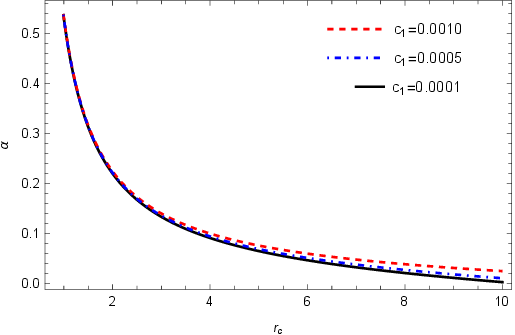}
\includegraphics[width=.3\linewidth,
height=1.5in]{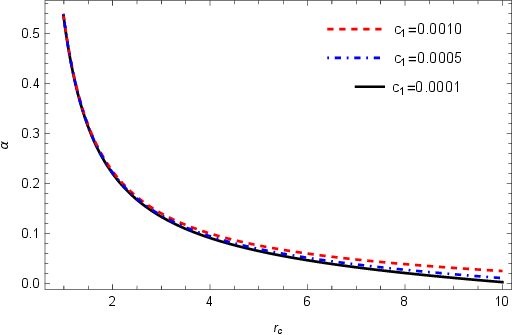}
\caption{\label{rrrF2} The plots represent the deflection angle for BEC (left), PI (middle), and NFW (right) profiles with $c_{0}=0.05$, $R=1.94$, $r_{0}=0.2$, $\rho_s=0.9$, and $r_{s}=1.94$.}
\end{figure}

This section deals with the deflection angle cast by wormhole geometry under the effect of three different kinds of dark matter halos. The mass and energy generate the curvature of space-time, which influences the speed of light. The curvature of space-time can cause light to bend near a big object, such as a black hole or wormhole. The fascination among researchers with gravitational lensing, especially its strong form, has seen a notable increase following the works by Virbhadra and colleagues \cite{Virbhadra1, Virbhadra3}. Additionally, Bozza, in Ref. \cite{Bozza22}, introduced an analytical approach to compute gravitational lensing in the strong field limit for any spherical symmetric space-time. This method has been applied in numerous subsequent studies, such as Refs. \cite{Quinet, Tejeiro}. This background encourages us to use this analytical technique in our current research. We adopt a numerical technique to study the deflection angle near the wormhole's throat to achieve this aim. For a detailed derivation of the deflection angle, one can refer to the Refs. \cite{g1,r4}.
The formula for deflection angle $\alpha$ for the MT wormhole geometry can be read as \cite{Bozza1}
\begin{equation}\label{def.angle}
    \alpha=-\pi+2\int_{r_c}^{\infty} \frac{e^{\phi(r)} dr}{r^2 \sqrt{(1-\frac{b(r)}{r})(\frac{1}{\beta^2}-\frac{e^{2\phi(r)}}{r^2})}},
\end{equation}
where $r_c$ is the closest path of light near the throat and $\beta$ is the impact parameter. For null geodesic, we have the relation between $\beta$ and $r_c$, defined by
\begin{equation}
\beta=r_{c} e^{-\phi(r_c)}.    
\end{equation}
We obtained redshift and shape functions in the previous sections for three dark matter models. Here, we shall use those redshift and shape functions to study the deflection angles around the throat.
We substitute the three different sets of redshift functions by Eq. (\ref{39}), Eq. (\ref{4a4}), and Eq. (\ref{53}) and shape functions by Eq. (\ref{41}), Eq. (\ref{48}), and Eq. (\ref{55}) into Eq.(\ref{def.angle}) and solve numerically, one can get deflection angle $\alpha$ with respect to $r_c$ for three different backgrounds. Fig. (\ref{rrrF2}) shows that the deflection angle tends to zero as the distance $r_c$ increases to infinity. In other words, as the light ray goes away from the wormhole throat, where the gravitational field of the wormhole is negligible, it does not deflect from the original path. Conversely, when the value of the distance $r_c$ is close to the radius of the wormhole throat, the deflection angle increases significantly and is positive. Also, in the wormhole throat, where the gravitational field is extremely strong, the deflection of light tends to infinity.
\section{Conclusions}\label{ch4sec8}
\indent In this chapter, we have uncovered a novel wormhole solution sustained by dark matter frameworks such as BEC, PI, and NFW within the $f(Q)$ gravity theory framework. Specifically, our approach involved utilizing the density profile equations of the dark matter frameworks in conjunction with the rotational velocity to determine both the redshift and shape functions of the wormholes. It is critical to highlight that the model's parameters significantly impact the investigation of the shape of the wormhole. Our findings demonstrate that choosing particular parameters, including the wormhole throat radius $r_0$, leads to a wormhole solution that meets the flare-out condition at the throat, maintaining this characteristic under an asymptotic background. Further, we have investigated the energy conditions using the same parameter involved in the shape functions. Mathematically, we have calculated the expressions of NEC for each model at the throat (see Eq. \eqref{62}). Also, we have depicted the positive and negative regions of all the energy conditions in Figs. (\ref{F5}-\ref{F7}) as well as summarized in table- \ref{Table1}.\\
\indent Furthermore, we have investigated the wormhole shadow under the effect of dark matter models. Our findings indicate that the BEC dark matter model influences the wormhole shadow. Specifically, larger values of $D_1$ and central density $\rho_s$ bring the wormhole shadow closer to the throat, while smaller values of these parameters push the shadow radius away from the throat. We have also observed similar behavior for the PI and NFW profiles.\\
In addition to the shadow, we have extensively analyzed gravitational lensing, specifically the strong gravitational lensing resulting from wormhole geometry under each dark matter model. The method used in this analysis was developed by Bozza et al. \cite{Bozza1} to explore black hole physics within the strong field regime. Following this, Bozza \cite{Bozza22} advanced an analytical approach to derive gravitational lensing for a general spherically symmetric metric under strong field conditions. This approach has been applied in various studies \cite{r1,r2,r3} to investigate the behavior of deflection angles around wormholes within both GR and modified gravity frameworks.
Consistent with this methodology, we explored the convergence of the deflection angle using the derived solutions for the shape and redshift functions in the context of $f(Q)$ gravity. Our findings reveal that the deflection angle of an outward light ray diverges precisely at the wormhole's throat, representing the surface's correspondence to the photon sphere. Detecting a photon sphere near the wormhole's throat would be highly significant due to its implications. It would confirm the presence of a strong gravitational field and support theoretical predictions related to wormholes. From an observational astronomy standpoint, such a phenomenon would offer a unique opportunity to directly study and observe these mysterious structures, thereby enhancing our understanding of gravity and the fundamental nature of wormholes.\\
In recent decades, numerous researchers have developed various models to explore the presence of wormholes within the framework of Einstein's gravity and modified gravitational theories in the galactic halos. Rahaman and his collaborators, for instance, investigated the potential existence of wormholes within the galactic halo by employing the NFW \cite{Rahaman4} and URC dark matter density profiles \cite{Rahaman2}, utilizing a redshift function derived from a flat rotational curve. Additionally, the Bose-Einstein dark matter density profile has been shown to support the presence of wormholes in the galactic halo \cite{Jusufi1}. Moreover, the geometry of wormholes has been explored with embedded class-I wormholes using URC, NFW, and SFDM dark matter models in Einstein cubic gravity \cite{c4}. In this study, we have employed the BEC, NFW, and PI dark matter models to generate new wormhole solutions within the context of $f(Q)$ gravity. We have identified more physically viable wormhole solutions by calculating redshift functions derived from flat rotational curves for each dark matter model. Furthermore, we have explored the deflection angles and shadows cast by these wormholes for each dark matter model, marking a novel investigation within the scope of modified gravitational theories.\\
In summary, this study presents a novel wormhole solution within the $f(Q)$ gravity framework, sustained by dark matter models such as BEC, PI, and NFW. Using the dark matter density profiles and rotational velocity, we determine the wormhole's redshift and shape functions, highlighting the significant influence of model parameters on the wormhole's shape, particularly the throat radius $r_0$. The energy conditions are evaluated at the throat, revealing regions where these conditions are met or violated. The study also explores the wormhole shadow, showing that larger values of specific parameters bring the shadow closer to the throat in the BEC dark matter model, a trend similarly observed in PI and NFW profiles. Additionally, gravitational lensing effects are examined, demonstrating that the deflection angle increases near the wormhole throat, where the gravitational field is strongest, and tends to zero as the distance from the throat increases, indicating minimal light deviation in weaker gravitational fields. Thus, it is safe to conclude that our findings suggest the possibility of the existence of wormholes in the galactic halos caused by BEC, PI, and NFW dark matter within the framework of $f(Q)$ gravity.\\
In the next chapter, we will examine these BEC and NFW models along with another dark matter model, such as the URC model, and discuss wormhole solutions through a different modified gravity, namely $4$D EGB gravity.


\chapter{Possibility of the traversable wormholes in the galactic halos within $4$D Einstein-Gauss-Bonnet gravity} 
\label{Chapter5} 

\lhead{Chapter 5. \emph{Possibility of the Traversable Wormholes in the Galactic Halos within $4$D Einstein-Gauss-Bonnet Gravity}} 
\blfootnote{*The work in this chapter is covered by the following publication:\\
\textit{Possibility of the Traversable Wormholes in the Galactic Halos within $4$D Einstein-Gauss-Bonnet Gravity}, Annalen der Physik \textbf{536}, 2400114 (2024).}

This chapter presents a detailed discussion of wormhole solutions in the galactic halos due to dark matter in the context of $4$D EGB gravity. The detailed study of the work is outlined as follows:
\begin{itemize}
    \item We investigate the wormhole solutions in the galactic halos based on three different choices of dark matter profiles, such as URC, NFW, and Scalar Field Dark Matter (SFDM), within the framework of $4$D EGB gravity.
    \item The Karmarkar condition was used to find the exact solutions for the shape functions under different non-constant redshift functions.
    \item We discussed the energy conditions for each dark matter profile and noticed the influence of the GB coefficient in violating energy conditions, especially null energy conditions.
    \item Further, some physical features of wormholes, viz. complexity factor, active gravitational mass, total gravitational energy, and embedding diagrams, have been explored.
\end{itemize}

\section{Introduction}
In string theories, particularly in their low-energy limits, effective field theories of gravity emerge. In these theories, the Lagrangian incorporates terms of quadratic and higher orders in the curvature alongside the conventional scalar curvature term \cite{Zwiebach1,Zwiebach3,Zwiebach4}. Notably, the gravitational action can undergo modification to encompass quadratic curvature correction terms while maintaining the equations of motion in the second order. This modification is permissible as long as the quadratic terms manifest in specific combinations corresponding to the GB invariants given in \eqref{gb7}.
Such a gravitational theory is designated as $D\geq 5$ dimensional EGB gravity theory, with $(D-4)$ extra dimensions playing a pivotal role. 
The EGB gravity constitutes a specific instance within the broader framework of the Lovelock theory of gravitation \cite{Lovelock1}. Significantly, the equations of motion associated with EGB gravity exhibit no more than two metric derivatives, ensuring the absence of ghost-related issues \cite{Lovelock2}. 
However, it is crucial to note that the GB term \eqref{gb7} behaves as a topological invariant in four dimensions, contributing proportionally to $(D-4)$ in all components of Einstein's equations. Consequently, this term does not contribute to the equations of motion, necessitating $ D\geq 5$ to manifest non-trivial gravitational dynamics.\\
\indent Nevertheless, Glavan and Lin \cite{Lin1} demonstrated a significant impact on gravitational dynamics in $D = 4$ could be achieved by re-scaling the GB coupling constant in the EGB gravity action.
A consistent theory of $D\rightarrow 4$ EGB gravity, maintaining invariance under spatial diffeomorphism while breaking the time diffeomorphism, has been discussed in \cite{Aoki1,Aoki2}. The 4D EGB gravity has garnered substantial attention and is undergoing extensive examination \cite{Lin2,Lin4,Lin5,Lin6}. 
A cloud of strings in 4D EGB was also explored in \cite{Lin8}. The exploration of non-static Vaidya-like spherical radiating black hole solutions in 4D EGB gravity has been reported \cite{Lin9}, and further black hole solutions can be found in \cite{Doneva1,Doneva2,Doneva3}. Furthermore, in Ref. \cite{Guo}, the authors discussed the photon geodesics and the influence of the GB coupling parameter on the shadow of the 4D EGB black hole. A detailed review of this topic has been discussed in \cite{Mulryne}. 
Wormhole solutions were also investigated using the same formalism in $4$D EGB gravity. Jusufi et al. \cite{Jusufi11} discussed wormhole solutions in $4$D EGB gravity for isotropic and anisotropic matter sources. Moreover, the stability of thin-shell wormholes \cite{Jusufi22} and Yukawa-Casimir wormholes \cite{Jusufi33} have been investigated. However, wormholes in the region of galactic dark matter halos have yet to be explored, which motivates us to study this particular gap.\\
Recent revelations about the composition of the Universe have brought to light the fact that a mere 5\% of the total matter and energy consists of the visible planets, stars, and ordinary matter, with the remainder being dominated by dark matter and dark energy of which 27\% is attributed to dark matter. Unlike regular baryonic matter, this dark matter does not interact with photons but exerts its influence solely through gravitational forces. In 1933, astronomer Zwicky was the first to ascertain the presence of dark matter within a distant galaxy cluster \cite{dm26,dm27}. Oort \cite{Oort1,Oort2} initially proposed the existence of dark matter in the Milky Way galaxy and subsequently confirmed through observational evidence by Diemand and Springel \cite{Diemand,Springel}. 
Dark matter plays a crucial role in the hypothetical construction of wormholes, with the density profile of dark matter proposed by researchers playing a pivotal role in the theoretical development of a fully stable and traversable wormhole. In Ref. \cite{Rahaman4}, the authors have provided evidence indicating the role of dark matter in sustaining the wormhole geometry within the outer reaches of the galactic halo. Sarkar et al. \cite{Sarkar2} have demonstrated the presence of wormholes within the isothermal galactic halo, supported by the presence of dark matter. Additionally, Kuhfitting \cite{r3} has examined the gravitational lensing phenomena associated with wormholes within the galactic halo region.\\
In this chapter, we consider three dark matter models such as URC, NFW, and SFDM profiles, and investigate the wormhole solutions through recently proposed $4$D EGB gravity. The structure of this chapter is arranged as follows: in section \ref{ch5sec2}, we have provided the wormhole field equations under this EGB gravity. The Karmarkar condition and the dark matter halo profiles have been briefly discussed in sections \ref{ch5sec3} and \ref{ch5sec4}, respectively. Moreover, section \ref{ch5sec5} discusses wormhole solutions for different redshift functions under each dark matter profile. Further, some physical features of wormholes have been studied in section \ref{ch5sec6}. Finally, we conclude our finding in section \ref{ch5sec7}.

\section{Field equations in $4$D EGB gravity}\label{ch5sec2}
The $4$D EGB theory is defined by considering the limit $D\rightarrow 4$ at the level of equations of motion rather than in the action. Consequently, in this framework, the GB term plays a non-trivial role in gravitational dynamics \cite{Lin1}.
Also, to generate wormhole solutions within the framework of $4$D EGB, we employ the regularization process outlined in \cite{Lin1}. Notably, the $4$D spherical solutions derived in \cite{Lin1} align with those obtained in various other regularized theories \cite{Hennigar1,Hennigar2,Hennigar3}.\\
The gravitational field equation \eqref{2a3} for the metric \eqref{1ch1} under anisotropic energy-momentum tensor \eqref{3ch1} and in the limit $D\rightarrow 4$ provides the following relations \cite{Jusufi11}
\begin{equation}\label{2c3}
8\pi\rho(r)=\frac{s b(r)}{r^6}\left(2rb'(r)-3b(r)\right)+\frac{b'(r)}{r^2},
\end{equation}
\begin{equation}\label{2c4}
8\pi P_r(r)=\frac{s b(r)}{r^6}\left[4 r\phi'(r)\left(r-b(r)\right)+b(r)\right]
+\frac{2\phi'(r)\left(r-b(r)\right)}{r^2}-\frac{b(r)}{r^3},
\end{equation}
\begin{multline}\label{3c5}
8\pi P_t(r)=\left(1-\frac{b(r)}{r}\right)\left[\left(\phi''(r)+(\phi'(r))^2\right)\left(1+\frac{4s b(r)}{r^3}\right)
+\frac{1}{r}\left(\phi'(r)-\frac{r b'(r)-b(r)}{2 r (r-b(r)}\right)\right.\\\left.
\times \left(1-\frac{2s b(r)}{r^3}\right)
-\left(\frac{\left(r b'(r)-b(r)\right)\phi'(r)}{2 r (r-b(r)}\right) \left(1-\frac{8s}{r^2}+\frac{12 s b(r)}{r^3}\right)\right]-\frac{2s b^2(r)}{r^6}.
\end{multline}
In this chapter, we will examine the energy conditions by defining\\
$NEC1=\rho+P_r$, \quad $NEC2=\rho+P_t$, \quad $DEC1=\rho-|P_r|$, \quad $DEC2=\rho-|P_t|$, \quad $SEC=\rho+P_r+2P_t$\\
for this investigation.

\section{Embedding class-1 space-time}\label{ch5sec3}
In this study, our main focus is to develop a wormhole shape function using the Karmarkar conditions that describe wormhole geometry. For this purpose, we assume static spherically symmetric space-time defined as
\begin{equation}\label{2b1}
ds^2=-e^{\xi(r)}dt^2+e^{\lambda(r)}dr^2+r^2 d\Omega^2,
\end{equation}
where $\xi$ and $\lambda$ are function of radial coordinate $r$ only. According to the above line element, the non-vanishing covariant Riemannian curvature components are
\begin{equation}
\nonumber
R_{1212}=R_{2121},\,\,\,R_{1221}=R_{2112},\,\,\,R_{1313}=R_{3131},
\end{equation}
\begin{equation}\nonumber
R_{1331}=R_{3113},\,\,\,R_{2323}=R_{3232},\,\,\,R_{2332}=R_{3223},
\end{equation}
\begin{equation}\nonumber
R_{1414}=R_{4141},\,\,\,R_{4114}=R_{1441},\,\,\,R_{4242}=R_{2424},
\end{equation}
\begin{equation}\label{rcb1}
R_{4224}=R_{2442},\,\,\,R_{4343}=R_{3434},\,\,\,R_{4334}=R_{3443}.
\end{equation}
The concept of embedding $n$-dimensional space $V_n$ within a pseudo-Euclidean space $E_n$ has been a subject of significant interest, as highlighted in the works of Eisland \cite{Eiesland} and Eisenhart \cite{Eisenhart1}. When an $n$-dimensional space $V_n$ can be isometrically immersed in an $(n + m)$-dimensional space, where $m$ represents the minimum number of additional dimensions, $V_n$ is considered to possess $m$-Class embedding, in that context, metric described in \eqref{2b1} typically yields a four-dimensional spherically symmetric space-time, placing it within Class two, denoting $m = 2$ and its embedding in a six-dimensional pseudo-Euclidean space. On a different note, it's worth mentioning that it is possible to devise a parametrization that allows the space-time outlined in equation \eqref{2b1} to be incorporated into a five-dimensional pseudo-Euclidean space, corresponding to Class $m = 1$, commonly referred to as embedding Class one \cite{Eiesland,Eisenhart1,Karmarkar1}. In order for a spherically symmetric space-time, whether static or non-static, to be classified as Class-one, the system must adhere to the following requisite and appropriate conditions
\begin{itemize}
\item The Gauss equation:
\begin{eqnarray}\label{eqcls1.1}
\mathcal{R}_{mnpq}=2\,\epsilon\,{b_{m\,[p}}{b_{q]n}}\,,
\end{eqnarray}
\item The Codazzi equation:
\begin{eqnarray}\label{eqcls1.2}
b_{m\left[n;p\right]}-{\Gamma}^q_{\left[n\,p\right]}\,b_{mq}+{{\Gamma}^q_{m}}\,{}_{[n}\,b_{p]q}=0.
\end{eqnarray}
\end{itemize}
Here, we consider the case where $\epsilon=\pm1$ and square brackets denote antisymmetrization. The coefficients of the second differential form are represented by $b_{mn}$. By utilizing Eqs.~(\ref{eqcls1.1}) and ~(\ref{eqcls1.2}) and applying the prescribed mathematical procedure, we can calculate the Karmarkar condition as follows
\begin{equation}\label{2b5}
R_{1414}=\frac{R_{1224}R_{1334}+R_{1212}R_{3434}}{R_{2323}},
\end{equation}
with $R_{2323}\neq0$ \cite{Pandey1}. The form of space-time satisfying the Karmarkar condition is known as embedding class-I. By substituting the non-zero
components of Riemann curvature \eqref{rcb1} in the Karmarkar relation \eqref{2b5}, we get the following differential equation
\begin{equation}\label{2b6}
\frac{\xi'(r) \lambda'(r)}{1-e^{\lambda(r)}}=\xi'(r)\lambda'(r)-\xi'^2(r)-2\xi^{''},
\end{equation}
where $e^{\lambda(r)}\neq1$.
On solving the above differential equation, one can find
\begin{equation}\label{2b7}
e^{\lambda(r)}=1+c_1e^{\xi(r)}\xi^{'2}(r),
\end{equation}
where $c_1$ is the constant of integration. Using the above expression \eqref{2b7}, we shall obtain wormhole shape functions for different redshift functions in section-\ref{ch5sec5}.

\section{Dark matter halos}\label{ch5sec4}
Several inquiries are highlighted herein in the dark matter halos context, particularly those involving galactic wormholes, primarily deduced from rotation curves \cite{r3,Rahaman2}. It is essential to acknowledge that the postulation of ``dark matter" arose from the review of Oort constants \cite{Oort4} and investigations into the masses of galaxy clusters \cite{dm26,dm27}, aiming to explain the observed flat rotation curves. To address this phenomenon, the Navarro-Frenk-White density profile function \cite{Rahaman3,Frenk3} and, alternatively, the Einasto profile was suggested, with the latter demonstrating improved alignment with specific dark matter halo simulations \cite{Frenk44,Frenk5}. Readers interested in galactic wormholes are encouraged to refer to some of the recent relevant literature \cite{Rizwan2,Rizwan4}. Generally, the wormhole throat is surrounded by varying types of dark matter halos. The observed deviation from expected energy conditions substantiates the presence of these dark halos. This study explores the representations associated with different kinds of dark matter halos, incorporating wormhole solutions within various dark matter halo profiles.\\
Three dark matter profiles are considered here in this chapter such as URC, NFW, and SFDM profiles. Note that the SFDM model contains two distinct density profiles, namely the BEC profile and the finite BEC profile \cite{Matos1,Matos2}. For the sake of simplicity, we concentrate on the BEC profile. This particular profile aligns with the static solution of the Klein-Gordon equation and involves a quadratic potential governing the scalar field.
\section{Wormhole solutions for different redshift functions}\label{ch5sec5}
In this chapter, we will explore wormhole solutions with non-zero tidal force solutions, where $\phi(r)\neq 0$. A non-constant redshift function can improve the stability of a wormhole and enhance its traversability. By appropriately choosing the redshift function, one can control the gravitational tidal forces experienced by travelers. For instance, a well-chosen redshift function can ensure that these tidal forces remain weak enough for safe passage through the wormhole, which is essential for the traversability condition. However, many researchers have focused on the zero tidal force (ZTF) model, where $\phi(r)$ is kept constant, to simplify the calculations of the energy conditions. Maintaining a constant redshift may not be realistic from a physics perspective. Therefore, removing this assumption and allowing for a non-constant redshift function is crucial. In \cite{Liempi1}, the authors confirm that the ZTF condition does not enable wormholes to exist with isotropic pressure. The physical properties of traversable wormholes under a phantom energy state were investigated in \cite{Liempi2}, and it was discussed that a non-constant redshift function could help minimize the amount of exotic matter needed to sustain the wormhole. Additionally, the impact of redshift functions on the WEC within the framework of $f(R)$ gravity has been discussed in \cite{Liempi3}. Thus, a non-constant redshift function offers numerous theoretical advantages in wormhole geometry, thereby improving their feasibility and versatility as solutions within GR and beyond. Keeping these in mind, we shall consider two different non-constant redshift functions to study the properties of wormhole solutions. Now, to compute the shape functions, we compare the coefficients of the metrics \eqref{1ch1} and \eqref{2b1}, and we obtain
\begin{equation}\label{41}
  \nu(r)=2\phi(r) \,\,\,\,\,\,\,\,\,\text{and}\,\,\,\,\,\,\,\,e^{\lambda(r)}=\left(1-\frac{b(r)}{r}\right)^{-1}.  
\end{equation}
Now, with the above relations, we could able to obtain the shape function from Eq. \eqref{2b7} under different redshift functions.
\subsection{$\phi(r)=-\frac{k}{r}$}
\begin{figure}[h]
    \centering
    \includegraphics[width=7.5cm,height=5cm]{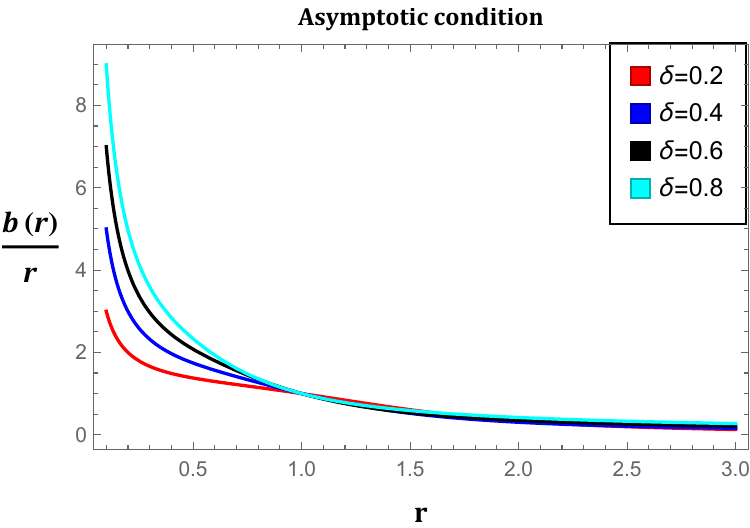}
    \caption{The plot of the flatness condition of the wormhole against the radial distance $r$ for the redshift $\phi(r)=-\frac{k}{r}$. We use $k=0.2$ and $r_0=1$.}
    \label{fig:1}
\end{figure}

\begin{figure}[h]
    \centering
    \includegraphics[width=7.5cm,height=5cm]{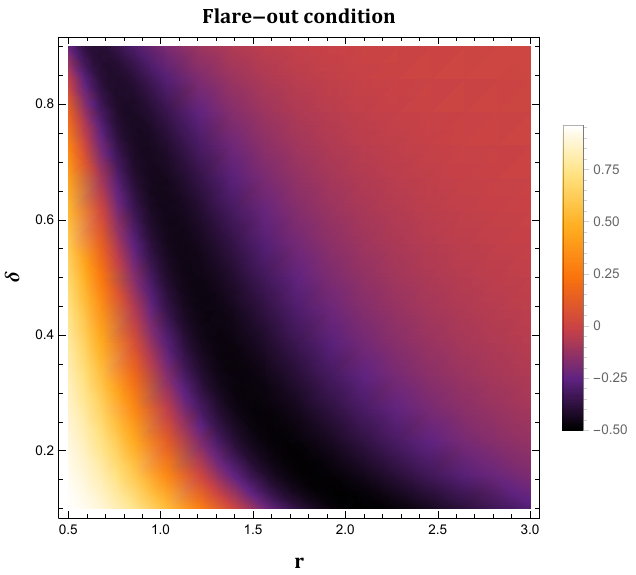}
    \caption{The plot of the flare-out condition of the wormhole against the radial distance $r$ for the redshift $\phi(r)=-\frac{k}{r}$. We use $k=0.2$ and $r_0=1$.}
    \label{fig:2}
\end{figure}

For the first case, we consider the redshift function defined by
\begin{equation}\label{4a1}
\phi(r)=-\frac{k}{r},
\end{equation}
where $k$ is any positive constant. Note that this form of redshift function satisfies the asymptotic flatness condition, i.e., $\phi(r)\rightarrow 0$ as $r\rightarrow \infty$. Kar and Sahdev \cite{Sahdev} proposed this specific redshift function and investigated the nature of matter (WEC) and the embedding of the space-like slices. They also discussed the human traversability of these space-times under this redshift function. Later, L. A. Anchordoqui et al. \cite{Anchordoqui} discussed evolving wormholes with this specific redshift function. They studied some issues related to the WEC violation and human traversability in these time-dependent geometries. Additionally, with this specific redshift function, wormhole solutions have been investigated in different modified theories of gravity, including $f(R)$ gravity \cite{Fayyaz1} and $f(R,\Phi)$ gravity \cite{Fayyaz2}.\\
Using Eqs. \eqref{2b7}, \eqref{41} and \eqref{4a1}, one can obtain the shape function
\begin{equation}\label{4a2}
b(r)=r-\frac{r^5}{r^4+4c_1k^2 e^{-2k/r}}.
\end{equation}
Note that when we impose the throat condition, i.e., $b(r_0)=r_0$, we get the trivial solution $r_0=0$. Thus, to avoid this issue, we add a free parameter $\delta$, and hence the shape function \eqref{4a2} becomes 
\begin{equation}\label{4a3}
b(r)=r-\frac{r^5}{r^4+4c_1k^2 e^{-2k/r}}+\delta.
\end{equation}
Now, to find the integrating constant $c_1$, we use the throat condition and obtain
\begin{equation}\label{4a4}
c_1=\frac{r_0^4(r_0-\delta)}{4\delta k^2 e^{-2k/r_0}}.
\end{equation}
Substituting the value of $c_1$ in Eq. \eqref{4a3}, one can find the final version of the shape function
\begin{equation}\label{4a5}
b(r)=r-\frac{\delta r^5}{\delta r^4+r_0^4(r_0-\delta) e^{-2k(\frac{1}{r}-\frac{1}{r_0})}}+\delta,
\end{equation}
where $0<\delta<r_0$.
Hence, in this case, the Morris-Throne wormhole metric \eqref{1ch1} can be read as
\begin{equation}\label{4a6}
ds^2=-e^{-\frac{2k}{r}}dt^2+\left(\frac{\delta r^4}{\delta r^4+r_0^4(r_0-\delta) e^{-2k(\frac{1}{r}-\frac{1}{r_0})}}-\frac{\delta}{r}\right)^{-1} dr^2+r^2d\Omega^2.
\end{equation}
Clearly, in the limit $r\rightarrow \infty$, we obtain
$$\lim \limits_{r\rightarrow \infty}\frac{b(r)}{r}\rightarrow 0,$$
which confirms that the asymptotically flatness condition is satisfied. Moreover, the flare-out condition is also satisfied at the throat for $0<\delta<r_0$ as well as for the appropriate choice of the other free parameters. The graphical behavior of the shape functions can be found in Figs. \ref{fig:1} and \ref{fig:2}.\\
For the redshift function \eqref{4a1}, the radial and tangential pressures can be read as
\begin{multline}\label{4a7}
P_r=\frac{1}{8 \pi}\left[\frac{1}{r^7}\left(-\delta  r \mathcal{L}_2+s  \delta ^2 (r-4 k)+s  r^3-r^5\right)+s  \delta ^2 r^3 (r-4 k) \mathcal{L}_1^2+\frac{\delta \mathcal{L}_1}{r^2} \right.\\\left.
\times \left(2 k \left(4 s  \delta +r^3+2 s  r\right)+r \left(-2 s  \delta +r^3-2 s  r\right)\right)\right],
\end{multline}
\begin{multline}\label{4a8}
P_t=\frac{1}{8 \pi}\left[\frac{\delta }{2 r^8}\left(-8 s  \delta  k^2+\mathcal{L}_1^2\left(2 \delta  r^9 \left(k^2 \left(12 s  \delta+r^3
+12 s  r\right)-k r \left(26 s  \delta +r^3+26 s  r\right)-2 r^2 \right.\right.\right.\right.\\\left.\left.\left.\left.
\times \left(-2 s  \delta+r^3-4 s  r\right)\right)\right)+\mathcal{L}_1\left(-8 s  \delta  k^2 r^5-6 s  \delta  r^7-8 s  r^8+4 r^{10}\right)-2 k^2 r^3 -8 s  k^2 r+24 s  k r^2 \right.\right.\\\left.\left.
-\mathcal{L}_1^3\left(4 s  \delta ^2 r^{14} (r-6 k) (2 r-k)\right)+3 k r^4 +32 s  \delta  k r-2 s  \delta  r^2-2 s  r^3+r^5\right)-\mathcal{L}_3\right],
\end{multline}
where, $\mathcal{L}_1=\frac{1}{r_0^4 (r_0-\delta ) e^{2 k \left(\frac{1}{r_0}-\frac{1}{r}\right)}+\delta  r^4}$,
$\mathcal{L}_2=\left(2 k \left(2 s +r^2\right)+r^3-2 s  r\right)$,
and $\mathcal{L}_3=\frac{2 s  \left(\delta -\mathcal{L}_1\delta  r^5+r\right)^2}{r^6}$.\\
Now, for the URC model, the NEC can be read at the throat
\begin{equation}\label{4a9}
NEC1\bigg\vert_{r=r_0}=\frac{1}{8 \pi  r_0^4}\left[s -\frac{r_0^2 \left(r_s^3 \left(1-8 \pi  \rho_s r_0^2\right)+\mathcal{L}_4\right)}{(r_s+r_0) \left(r_s^2+r_0^2\right)}\right],
\end{equation}
\begin{multline}\label{4a10}
NEC2\bigg\vert_{r=r_0}=\frac{\rho_s r_s^3}{(r_s+r_0) \left(r_s^2+r_0^2\right)}+\frac{1}{16 \pi  r_0^8}\left[2 \delta ^2 (k-2 r_0)\left(k \left(4 s +r_0^2\right)+r_0^3-2 s  r_0\right) \right.\\\left.
+\delta  r_0 (5 r_0-2 k) \left(k \left(4 s +r_0^2\right)+r_0^3-2 s  r_0\right)-4 s  r_0^4\right],
\end{multline}
where $\mathcal{L}_4=r_s^2 r_0+r_s r_0^2+r_0^3$.\\
Again, for the NFW model, the expression for NEC can read as
 \begin{multline}\label{4a11}
NEC1=\frac{\rho_s r_s^3}{r (r_s+r)^2}+\frac{1}{8 \pi}\left[\frac{1}{r^7}\left(-\delta  r \mathcal{L}_2+s  \delta ^2 (r-4 k)+s  r^3-r^5\right)+\mathcal{L}_1^2s  \delta ^2 r^3 (r-4 k) \right.\\\left.
+\frac{\delta \mathcal{L}_1}{r^2}\left(2 k \left(4 s  \delta +r^3 
+2 s  r\right)+r \left(-2 s  \delta +r^3-2 s  r\right)\right)\right].
\end{multline}
At wormhole throat $r=r_0$, the above expression reduces to
\begin{equation}\label{4a12}
NEC1\bigg\vert_{r=r_0}=\frac{1}{8 \pi  r_0^2} \left[ \frac{s}{r_0^2}-\frac{1}{(r_s+r_0)^2}\left(-8 \pi  \rho_s r_s^3 r_0+r_s^2
+2 r_s r_0+r_0^2\right)\right].
\end{equation}
Similarly, we could obtain the NEC at the throat for the SFDM model as
\begin{equation}\label{4a13}
NEC1\bigg\vert_{r=r_0}=\frac{s +8 \rho_s r_s r_0^3 \sin \left(\frac{\pi  r_0}{r_s}\right)-r_0^2}{8 \pi  r_0^4}.
\end{equation}
\begin{figure*}[]
\centering
\includegraphics[width=5cm,height=4cm]{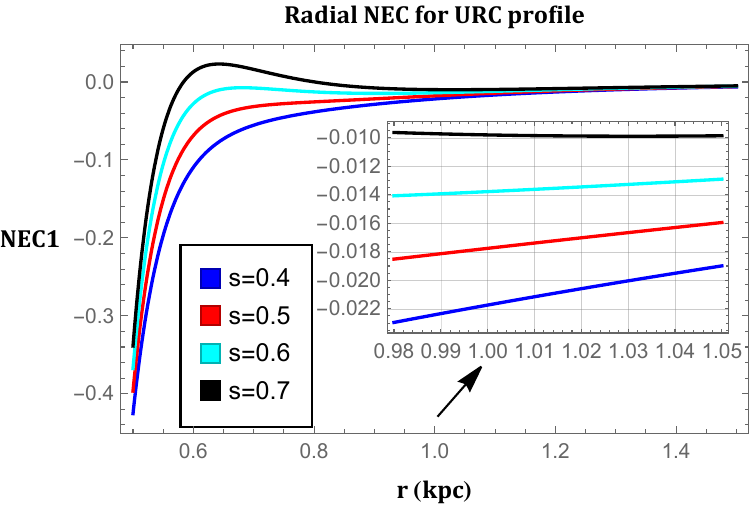}
\includegraphics[width=5cm,height=4cm]{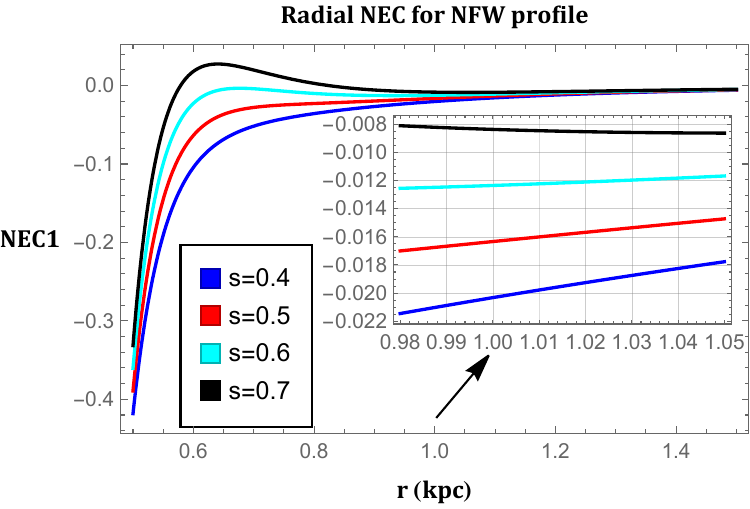}
\includegraphics[width=5cm,height=4cm]{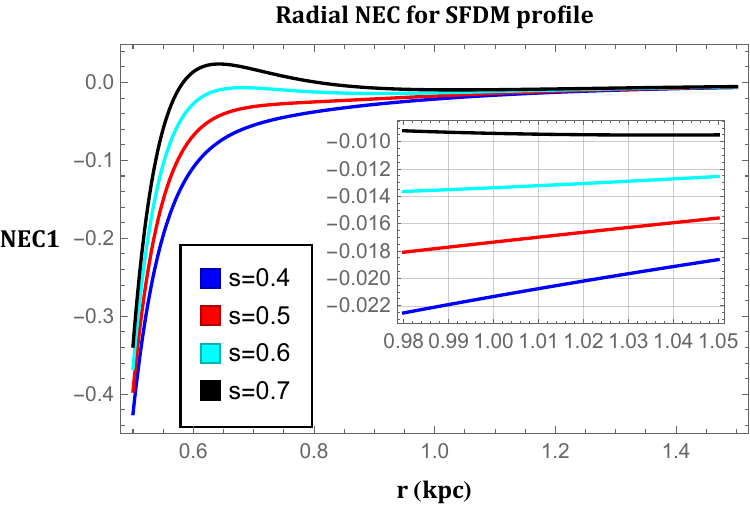}
\caption{The plots of $\rho+P_r$ against the radial distance $r$ for the redshift $\phi(r)=-\frac{k}{r}$. We use $\rho_s=0.004\, kpc^{-2}$, $r_s=2\,kpc$, $\delta=0.9$, $k=0.2$ and $r_0=1$.}
\label{fig:3}
\end{figure*}
\begin{figure*}[]
\centering
\includegraphics[width=5cm,height=4cm]{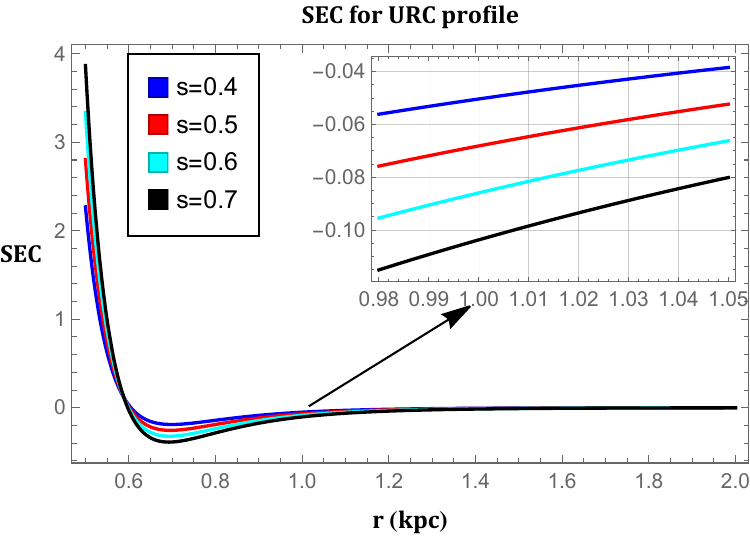}
\includegraphics[width=5cm,height=4cm]{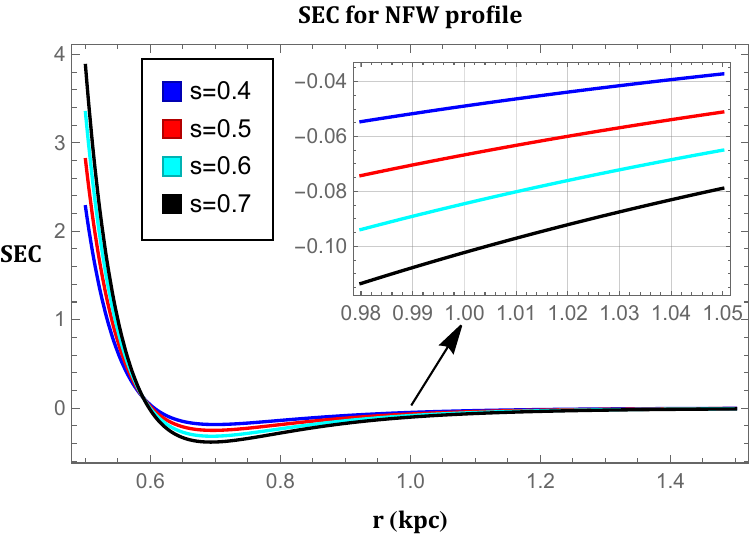}
\includegraphics[width=5cm,height=4cm]{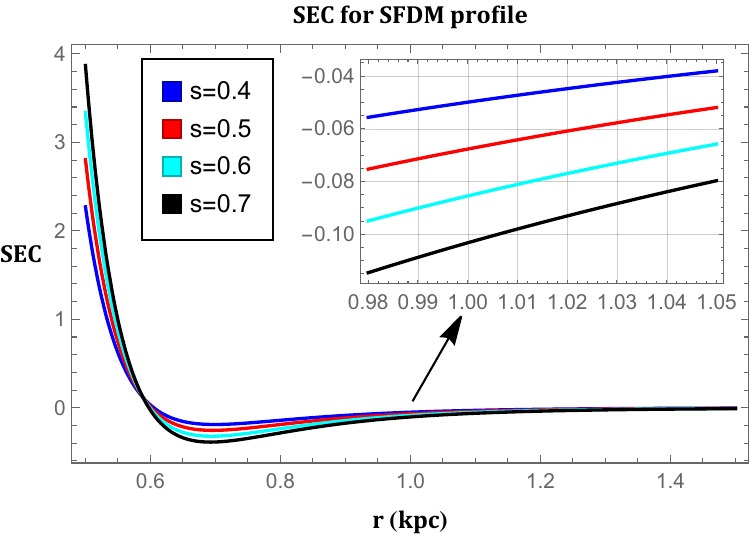}
\caption{The plots of $\rho+P_r+2 P_t$ against the radial distance $r$ for the redshift $\phi(r)=-\frac{k}{r}$. We use $\rho_s=0.004\, kpc^{-2}$, $r_s=2\,kpc$, $\delta=0.9$, $k=0.2$ and $r_0=1$.}
\label{fig:4}
\end{figure*}
\begin{figure*}[]
\centering
\includegraphics[width=5cm,height=4cm]{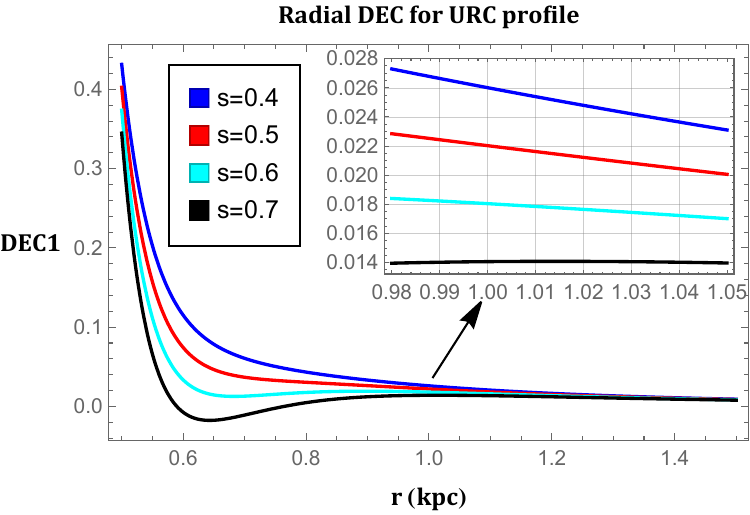}
\includegraphics[width=5cm,height=4cm]{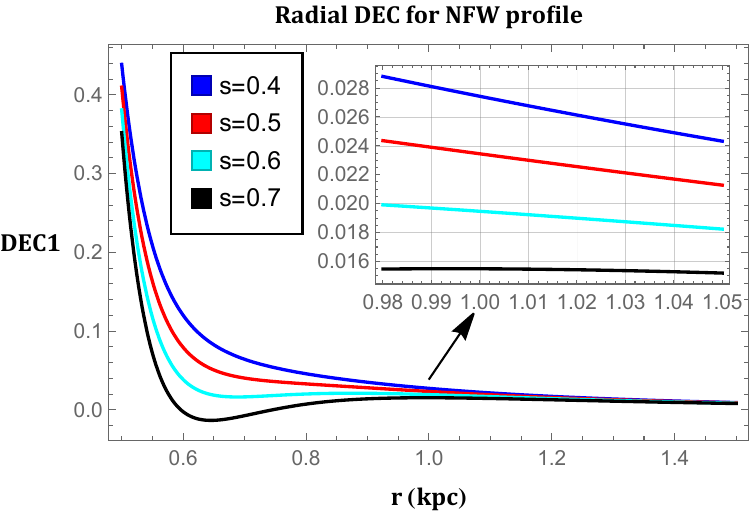}
\includegraphics[width=5cm,height=4cm]{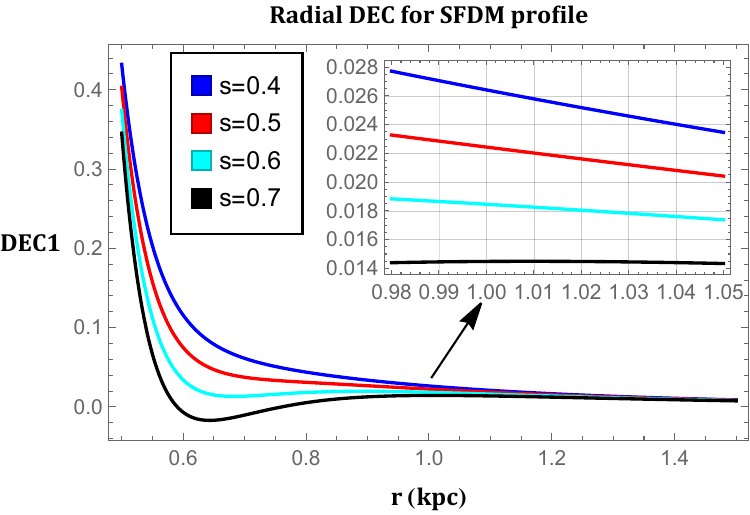}
\caption{The plots of $\rho-|P_r|$ against the radial distance $r$ for the redshift $\phi(r)=-\frac{k}{r}$. We use $\rho_s=0.004\, kpc^{-2}$, $r_s=2\,kpc$, $\delta=0.9$, $k=0.2$ and $r_0=1$.}
\label{fig:5}
\end{figure*}
Now, we will discuss the energy conditions for URC, NFW, and SFDM dark matter halo profiles. We consider some particular choice of free parameters $\rho_s=0.004\, kpc^{-2}$, $r_s=2\,kpc$ \cite{Sarkar}, throat radius $r_0=1$, $\delta=0.9$ (as $0<\delta<r_0$) and $k=0.2$. Mathematically, one can check the RHS of Eqs. \eqref{4a9}, \eqref{4a12}, and \eqref{4a13} by simply putting the values of the parameters and taking the GB coefficient $s>0$ provides a negative quantity. This confirms the violation of NEC at the wormhole throat. Graphically, we have presented the behavior of NEC for each dark matter profile in Fig. \ref{fig:3}. Moreover, we have studied other energy conditions, such as SEC and DEC in Figs. \ref{fig:4} and \ref{fig:5}. SEC is violated near the throat; however, far from the throat, SEC will be satisfied. Moreover, radial DEC is satisfied at the throat. But for $s> > 0$, DEC will disrespect the energy conditions.

\subsection{$\phi(r)=\frac{1}{2}\log(1+\frac{\eta^2}{r^2})$}
\begin{figure}[h]
    \centering
    \includegraphics[width=7.5cm,height=5cm]{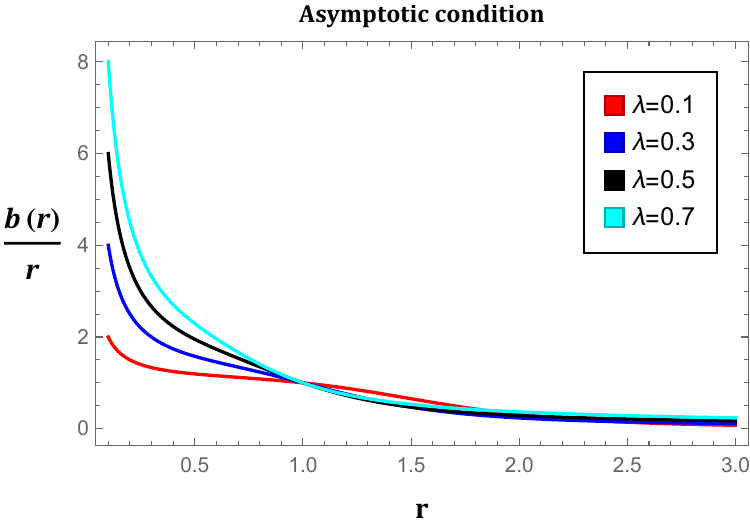}
    \caption{The plot of the flatness condition of the wormhole against the radial distance $r$ for the redshift $\phi(r)=\frac{1}{2}\log(1+\frac{\eta^2}{r^2})$. We use $\eta=2$ and $r_0=1$.}
    \label{fig:6}
\end{figure}

\begin{figure}[h]
    \centering
    \includegraphics[width=8cm,height=5cm]{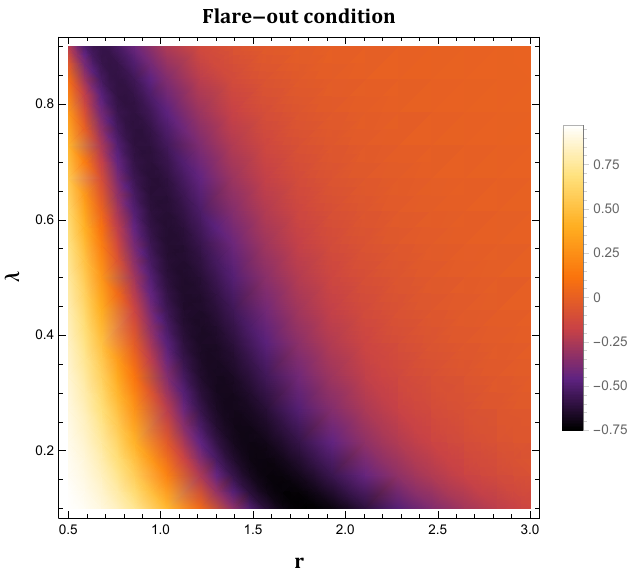}
    \caption{The plot of the flare-out condition of the wormhole against the radial distance $r$ for the redshift $\phi(r)=\frac{1}{2}\log(1+\frac{\eta^2}{r^2})$. We use $\eta=2$ and $r_0=1$.}
    \label{fig:7}
\end{figure}

\begin{figure*}[]
\centering
\includegraphics[width=5cm,height=4cm]{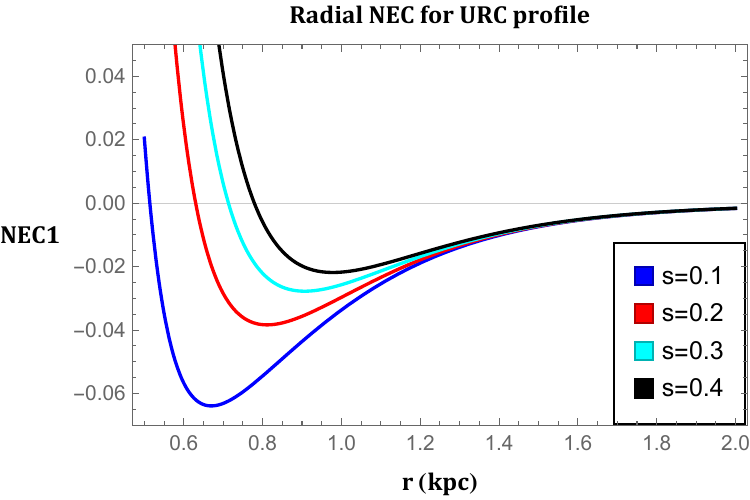}
\includegraphics[width=5cm,height=4cm]{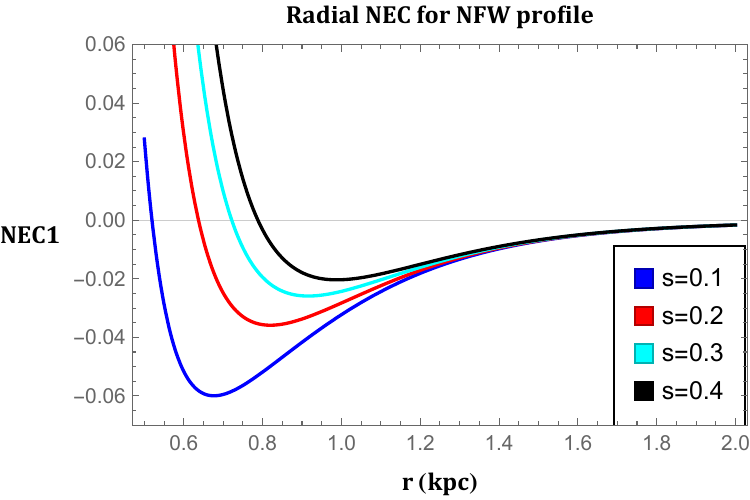}
\includegraphics[width=5cm,height=4cm]{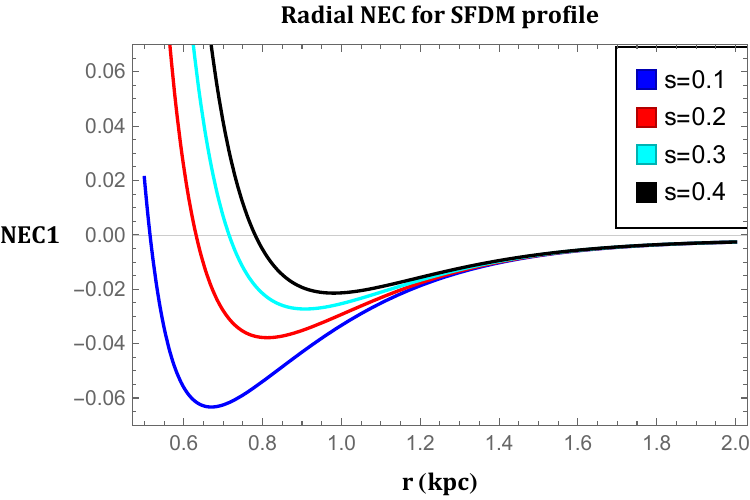}
\caption{The plots of $\rho+P_r$ against the radial distance $r$ for the redshift $\phi(r)=\frac{1}{2}\log(1+\frac{\eta^2}{r^2})$. We use $\rho_s=0.004\, kpc^{-2}$, $r_s=2\,kpc$, $\lambda=0.3$, $\eta=2$ and $r_0=1$.}
\label{fig:8}
\end{figure*}

\begin{figure*}[]
\centering
\includegraphics[width=5cm,height=4cm]{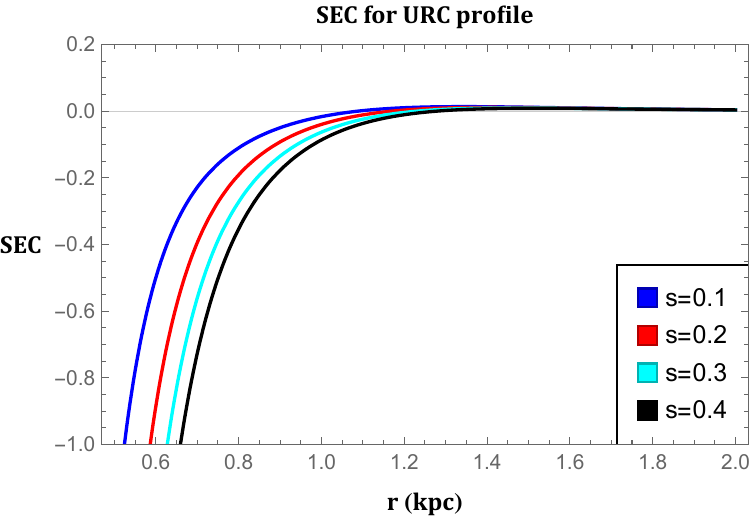}
\includegraphics[width=5cm,height=4cm]{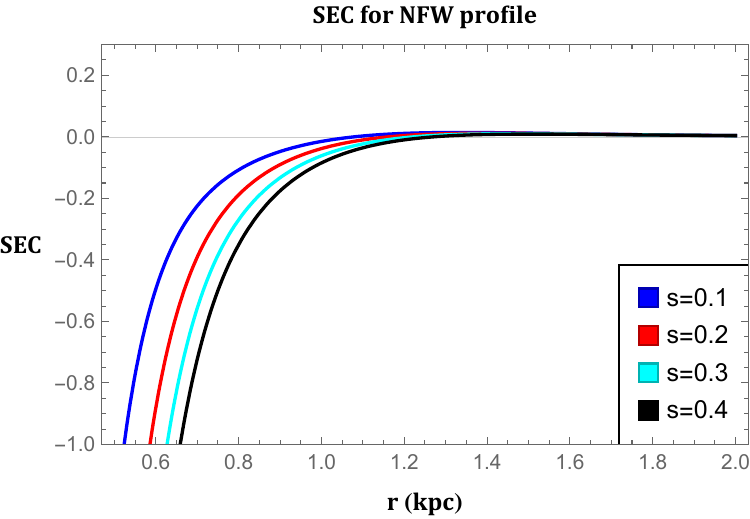}
\includegraphics[width=5cm,height=4cm]{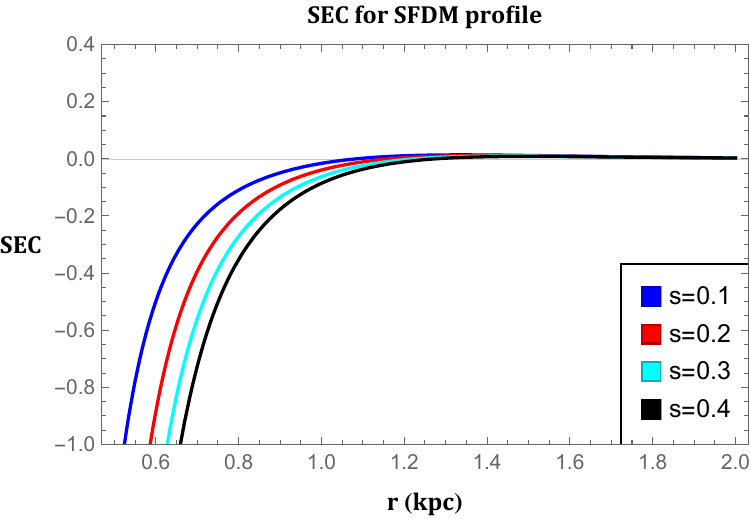}
\caption{The plots of $\rho+P_r+2 P_t$ against the radial distance $r$ for the redshift $\phi(r)=\frac{1}{2}\log(1+\frac{\eta^2}{r^2})$. We use $\rho_s=0.004\, kpc^{-2}$, $r_s=2\,kpc$, $\lambda=0.3$, $\eta=2$ and $r_0=1$.}
\label{fig:9}
\end{figure*}

\begin{figure*}[]
\centering
\includegraphics[width=5cm,height=4cm]{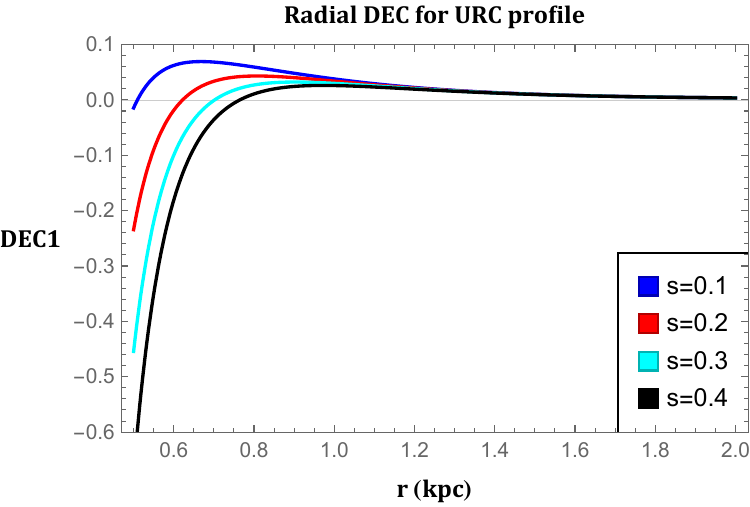}
\includegraphics[width=5cm,height=4cm]{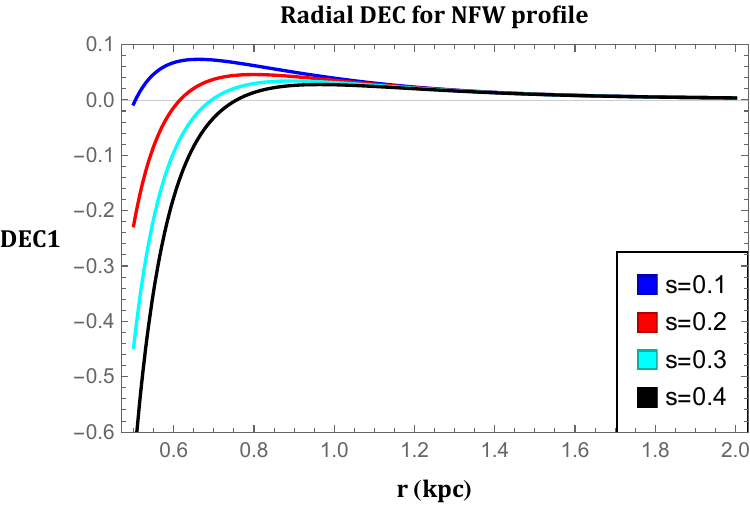}
\includegraphics[width=5cm,height=4cm]{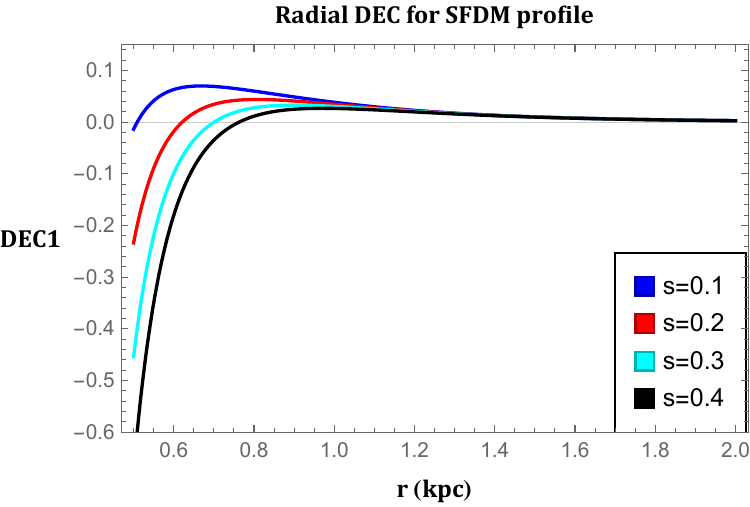}
\caption{The plots of $\rho-|P_r|$ against the radial distance $r$ for the redshift $\phi(r)=\frac{1}{2}\log(1+\frac{\eta^2}{r^2})$. We use $\rho_s=0.004\, kpc^{-2}$, $r_s=2\,kpc$, $\lambda=0.3$, $\eta=2$ and $r_0=1$.}
\label{fig:10}
\end{figure*}

In this case, we consider the redshift function of the form
\begin{equation}\label{4b1}
\phi(r)=\frac{1}{2}\log(1+\frac{\eta^2}{r^2}),
\end{equation}
where $\eta$ is any positive parameter. Note that this non-constant redshift function respects the asymptotic condition of a traversable wormhole. With the above choice of redshift function, the authors of \cite{Channuie} studied traversable wormholes supported by Casimir energy in general relativity. 
We consider this specific choice of redshift function to check the stability of wormhole solutions in $4$D EGB gravity.\\
Similar to the previous subsection, we could obtain the shape function by using Eqs. \eqref{2b7}, \eqref{41} and \eqref{4b1},
\begin{equation}\label{4b2}
b(r)=r-\frac{r^5\left(r^2+\eta^2\right)}{r^4\left(r^2+\eta^2\right)+4\eta^2 c_1}.
\end{equation}
It is clear that when we impose the throat condition on the above equation, we get the trivial solution $r_0=0$. Thus, we introduced a free parameter $\lambda$ to the above shape function, and hence it becomes
\begin{equation}\label{4b3}
b(r)=r-\frac{r^5\left(r^2+\eta^2\right)}{r^4\left(r^2+\eta^2\right)+4\eta^2 c_1}+\lambda.
\end{equation}
Now, we impose $b(r_0)=r_0$ to the above expression \eqref{4b3} to obtain $c_1$
\begin{equation}\label{4b4}
c_1=\frac{\mathcal{K}_1}{4\eta^2 \lambda},
\end{equation}
where $\mathcal{K}_1=r_0^4 \left(r_0^2+\eta^2\right)\left(r_0-\lambda\right)$.
Substituting the value of $c_1$ in Eq. \eqref{4b3}, we can obtain the shape function
\begin{equation}\label{4b5}
b(r)=r-\frac{r^5 \left(r^2+\eta^2\right) \lambda}{\lambda r^4 \left(r^2+\eta^2\right)+\mathcal{K}_1}+\lambda,
\end{equation}
where $0<\lambda<r_0$.
Hence, the Morris-Throne wormhole metric \eqref{1ch1} reduces to
\begin{equation}\label{4b6}
ds^2=-\left(1+\frac{\eta^2}{r^2}\right)^{-1}dt^2+
\left(\frac{r^4 \left(r^2+\eta^2\right) \lambda}{\lambda r^4 \left(r^2+\eta^2\right)+\mathcal{K}_1}-\frac{\lambda}{r}\right)^{-1} dr^2+r^2d\Omega^2,
\end{equation}
which is asymptotically flat space-time. In fact, one can smoothly verify that the case $0<\lambda<r_0$ gives the asymptotically flat solution. Also, we investigate the important criteria of a traversable wormhole is the flare-out condition, which is also satisfied within $0<\lambda<r_0$. In this case, we consider the throat radius $r_0=1$ and $\eta=2$. The graphical behavior of shape functions can be found in Figs. \ref{fig:6} and \ref{fig:7}.\\
For the given redshift function \eqref{4b1} and shape function \eqref{4b5}, we can compute the components of the energy-momentum tensor. In fact, the radial pressure can be read as follows
\begin{multline}\label{4b7}
P_r=\frac{1}{8 \pi  r^6}\left[s  \left(\lambda -\frac{4 \lambda ^3 \left(\mathcal{K}_2-1\right)}{r^2 \left(\frac{\lambda ^2}{r^2}+1\right)}-\lambda \mathcal{K}_2 +r\right) \left(\lambda- \lambda \mathcal{K}_2+r\right)
-\frac{2 \lambda ^3 r \left(\mathcal{K}_2-1\right)}{\frac{\lambda ^2}{r^2}+1}\right.\\\left.
-\left(r^3 \left(\lambda -\lambda  \mathcal{K}_2+r\right)\right)\right],
\end{multline}
where $\mathcal{K}_2=\frac{r^5 \left(\eta ^2+r^2\right)}{\lambda  r^4 \left(\eta ^2+r^2\right)+r_0^4 \left(\eta ^2+r_0^2\right) (r_0-\lambda )}$.
On the other hand, tangential pressure, in this case, can be read as
\begin{multline}\label{4b8}
P_t=\frac{1}{8 \pi } \left[\frac{1}{r} \left(\lambda  \left(\mathcal{K}_2-1\right) 
\left(\frac{\lambda ^2 \left(\mathcal{K}_3+1\right)}{2 r^4 \left(\frac{\lambda ^2}{r^2}+1\right) \left(1-\mathcal{K}_2\right)}\left(\frac{12 s  \lambda }{r^3}+\frac{4 s }{r^2}-\frac{12 s  \lambda \mathcal{K}_2}{r^3}+1\right) \right.\right.\right.\\\left.\left.\left.
+\frac{\left(2 \lambda ^4+3 \lambda ^2 r^2\right)}{r^2 \left(\lambda ^2+r^2\right)^2} \left(\frac{4 s  \left(\lambda -\lambda \mathcal{K}_2+r\right)}{r^3}+1\right)
+\frac{1}{r}\left(\left(-\frac{\lambda ^2}{r^3+\lambda ^2 r} 
+\frac{\mathcal{K}_3+1}{2 r \left(\mathcal{K}_2-1\right)}\right) \right.\right.\right.\right.\\\left.\left.\left.\left.
\times\left(1-\frac{2 s  \left(\lambda -\lambda  \mathcal{K}_2+r\right)}{r^3}\right)\right)\right)\right) -\frac{2 s  \left(\lambda -\lambda  \mathcal{K}_2+r\right)^2}{r^6}\right],
\end{multline}  
where $\mathcal{K}_3=\frac{2 r^5 r_0^4 \left(2 \eta ^2+3 r^2\right) \left(\eta ^2+r_0^2\right) (r_0-\lambda )}{\left(\lambda  r^4 \left(\eta ^2+r^2\right)-\eta ^2 \lambda  r_0^4+\eta ^2 r_0^5-\lambda  r_0^6+r_0^7\right)^2}$.
Now, with the above pressure components, we can examine the energy conditions for the URC, NFW, and SFDM dark matter halo profiles and try to generate plots to evaluate the validity of the energy conditions.\\
The NEC at wormhole throat ($r=r_0$) for each dark matter halo profile has been obtained and shown in Eq. \eqref{4b9}.
\begin{equation}\label{4b9}
NEC1\bigg\vert_{r=r_0}=
     \begin{cases}
      \frac{1}{8 \pi  r_0^4}\left[s -\frac{r_0^2 \left(r_s^3 \left(1-8 \pi  \rho_s r_0^2\right)+r_s^2 r_0+r_s r_0^2+r_0^3\right)}{(r_s+r_0) \left(r_s^2+r_0^2\right)}\right],  & \text{(URC profiles)}\\
      \\
      \frac{s  (r_s+r_0)^2-r_0^2 \left(-8 \pi  \rho_s r_s^3 r_0+r_s^2+2 r_s r_0+r_0^2\right)}{8 \pi  r_0^4 (r_s+r_0)^2},   &  \text{(NFW profiles)}\\
      \\
      \frac{s +8 \rho_s r_s r_0^3 \sin \left(\frac{\pi  r_0}{r_s}\right)-r_0^2}{8 \pi  r_0^4}, &  \text{(SFDM profiles).}
     \end{cases}
\end{equation}
Note that the above expression \eqref{4b9} is independent of the free parameters $\eta$ and $\lambda$, which confirms no influence of those parameters at the throat. However, outside the throat, we can see their influences. Also, in this case, we consider the appropriate choice of free parameters $\lambda=0.3$, $\eta=2$, $r_0=1$ along with the same values of $\rho_s$ and $r_s$, which we choose in the previous subsection. Numerically, one can check by putting the above values along with $s> 0$, which gives a negative quantity of NEC. Graphically, we have presented the behavior of NEC in Fig. \ref{fig:8}. Moreover, we studied SEC in Fig. \ref{fig:9} and found that SEC is violated near the throat for $s > 0$. We also noticed that as we increase the values of $s$, the contribution of violation becomes more. Further, we checked DEC and observed that DEC is satisfied in the vicinity of the throat (see Fig. \ref{fig:10}). A summary of the energy conditions for URC, NFW, and SFDM galactic dark matter profiles against both redshift functions have been calculated and shown in Table- \ref{Table0}.

\begin{table*}[t]
    \centering
\begin{tabular}{ p{3cm} p{5.8cm} p{5.8cm}}
 \hline
 \multicolumn{3}{|c|}{The behavior of the energy conditions around the throat} \\
 \hline
 Physical expressions & $\phi(r)=-\frac{k}{r}$ with $\rho_s=0.004\, kpc^{-2}$, $r_s=2\,kpc$, $\delta=0.9$, $k=0.2$ and $r_0=1$ & $\phi(r)=\frac{1}{2}\log(1+\frac{\eta^2}{r^2})$ with $\rho_s=0.004\, kpc^{-2}$, $r_s=2\,kpc$, $\lambda=0.3$, $\eta=2$ and $r_0=1$\\
 \hline
\multicolumn{3}{|c|} {Energy conditions for URC profile}\\
\hline
$\rho$ & $\rho>0$ & $\rho>0$\\
\hline
$\rho+P_r$ & $\rho+P_r<0$ for $s>0$ & $\rho+P_r<0$ for $s>0$\\
\hline
$\rho+P_t$ & $\rho+P_t<0$ for $s>0$ & $\rho+P_t<0$ for $s>0$\\
\hline
$\rho+P_r+2P_t$ & $\rho+P_r+2P_t<0$ for $s>0$ & $\rho+P_r+2P_t<0$ for $s>0$\\
\hline
$\rho-|P_r|$ & $\rho-|P_r|>0$ for $s>0$ & $\rho-|P_r|>0$ for $s>0$\\
\hline
$\rho-|P_t|$ & $\rho-|P_t|>0$ for $s>0$ & $\rho-|P_t|<0$ for $s>0$\\
\hline
\multicolumn{3}{|c|} {Energy conditions for NFW profile}\\
\hline
$\rho$ & $\rho>0$ & $\rho>0$\\
\hline
$\rho+P_r$ & $\rho+P_r<0$ for $s>0$ & $\rho+P_r<0$ for $s>0$\\
\hline
$\rho+P_t$ & $\rho+P_t<0$ for $s>0$ & $\rho+P_t<0$ for $s>0$\\
\hline
$\rho+P_r+2P_t$ & $\rho+P_r+2P_t<0$ for $s>0$ & $\rho+P_r+2P_t<0$ for $s>0$\\
\hline
$\rho-|P_r|$ & $\rho-|P_r|>0$ for $s>0$ & $\rho-|P_r|>0$ for $s>0$\\
\hline
$\rho-|P_t|$ & $\rho-|P_t|>0$ for $s>0$ & $\rho-|P_t|>0$ for $s>0$\\
\hline
\multicolumn{3}{|c|} {Energy conditions for SFDM profile}\\
\hline
$\rho$ & $\rho>0$ & $\rho>0$\\
\hline
$\rho+P_r$ & $\rho+P_r<0$ for $s>0$ & $\rho+P_r<0$ for $s>0$\\
\hline
$\rho+P_t$ & $\rho+P_t<0$ for $s>0$ & $\rho+P_t<0$ for $s>0$\\
\hline
$\rho+P_r+2P_t$ & $\rho+P_r+2P_t<0$ for $s>0$ & $\rho+P_r+2P_t<0$ for $s>0$\\
\hline
$\rho-|P_r|$ & $\rho-|P_r|>0$ for $s>0$ & $\rho-|P_r|>0$ for $s>0$\\
\hline
$\rho-|P_t|$ & $\rho-|P_t|>0$ for $s>0$ & $\rho-|P_t|>0$ for $s>0$\\
\hline
\end{tabular}
\caption{Table shows the summary of the energy conditions}
\label{Table0}
\end{table*}
\section{Some physical features of wormhole solutions}\label{ch5sec6}
In this particular section, we study some analysis of the obtained wormhole solutions, such as the complexity factor, active gravitational mass, total gravitational energy, and the embedding diagrams of the wormholes.
\subsection{Complexity factor}
A recent contribution in Ref. \cite{L. Herrera} proposed a novel complexity definition for self-gravitating fluid distributions. This definition centers on the intuitive concept that the gravitational system's least complexity should be represented by a homogeneous energy density distribution accompanied by isotropic pressure. The work in \cite{L. Herrera} reveals that in static spherically symmetric space-times, an associated scalar emerges from the orthogonal splitting of the Riemann tensor \cite{Gomez1,Gomez2}. This scalar, denoted as $\Upsilon_{TF}$, summarizes the core aspects of complexity, and its expression can be read as
\begin{equation}\label{6a1}
\Upsilon_{TF}=8\pi \Delta  -\dfrac{4 \pi}{r^{3}}\int _{0}^{r}\Tilde{r}^{3}\rho'(r) d\Tilde{r}\,,
\end{equation}
where $\Delta=P_r-P_t$. Moreover, the above expression \eqref{6a1} allows us to write the Tolman mass as
\begin{equation}
m_T=(m_T)_\Sigma \left(\frac{r}{r_\Sigma}\right)^3+r^3 \int_0^{r_\Sigma}\frac{e^{(\xi+\lambda)/2}}{\Tilde{r}}\Upsilon_{TF} d\Tilde{r},
\end{equation}
and this serves as a vital justification for defining the complexity factor using the aforementioned scalar, as it effectively incorporates all modifications arising from both energy density inhomogeneity and pressure anisotropy on the active gravitational mass. Notably, the condition of vanishing complexity $(\Upsilon_{TF}=0)$ can be fulfilled not solely in the simplest scenario of isotropic and homogeneous systems but also across all cases where
\begin{equation}
\Delta=\dfrac{1}{2r^{3}}\int _{0}^{r}\Tilde{r}^{3}\rho'(r) d\Tilde{r}.
\end{equation}
In this context, fulfilling the vanishing complexity condition results in a non-local equation of state, providing an additional condition to close the system of Einstein's field equations. This application has been presented in some recent papers, as seen in Refs. \cite{Casadio1,Casadio2,Casadio3}.\\
In this chapter, we are interested in studying the complexity factor for the wormhole solutions under galactic regions. It is known that wormholes are defined for $r_0\leq r <\infty$, in that case the complexity factor \eqref{6a1} should be modified and can be redefined as
\begin{equation}\label{6a2}
\Upsilon_{TF}=8\pi \Delta  -\dfrac{4 \pi}{r^{3}}\int _{r_0}^{r}\Tilde{r}^{3}\rho'(r) d\Tilde{r}.
\end{equation}
Note that the standard definition requires $r_0 = 0$, but we must discard this case to ensure a finite size of the wormhole throat.\\
\begin{figure*}[]
\centering
\includegraphics[width=5cm,height=4cm]{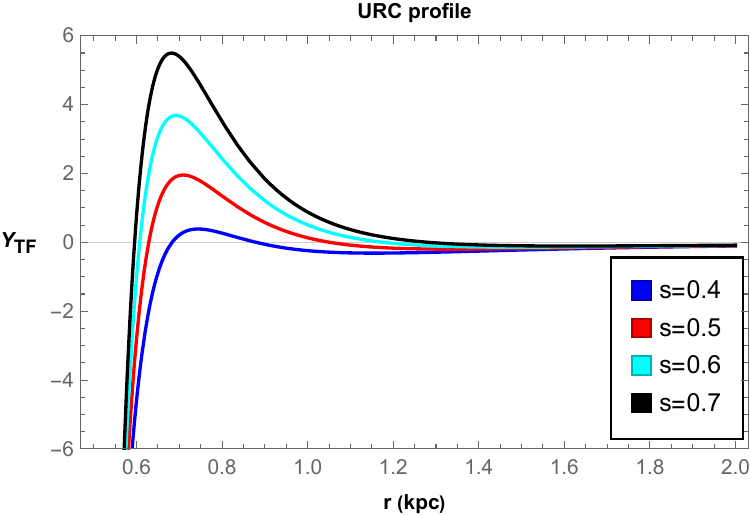}
\includegraphics[width=5cm,height=4cm]{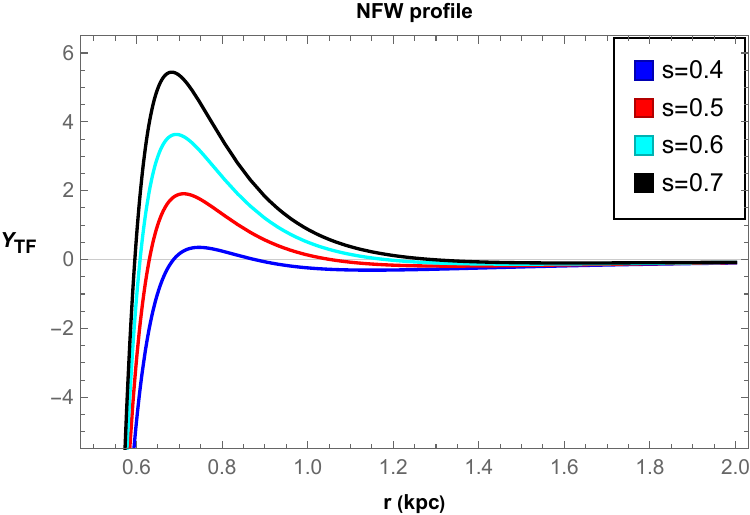}
\includegraphics[width=5cm,height=4cm]{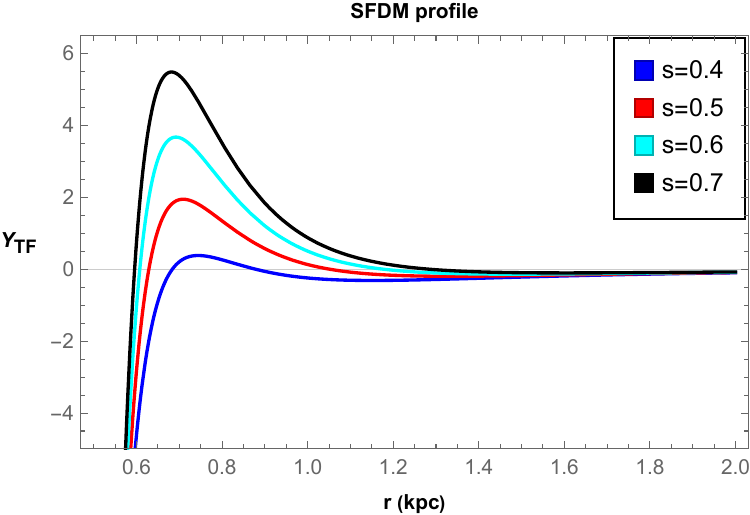}
\caption{The plots of $\Upsilon_{TF}$ against the radial distance $r$ for the redshift $\phi=\frac{-k}{r}$.  We use $\rho_s=0.004\, kpc^{-2}$, $r_s=2\,kpc$, $\delta=0.9$, $k=0.2$ and $r_0=1$.}
\label{fig:11}
\end{figure*}
\begin{figure*}[]
\centering
\includegraphics[width=5cm,height=4cm]{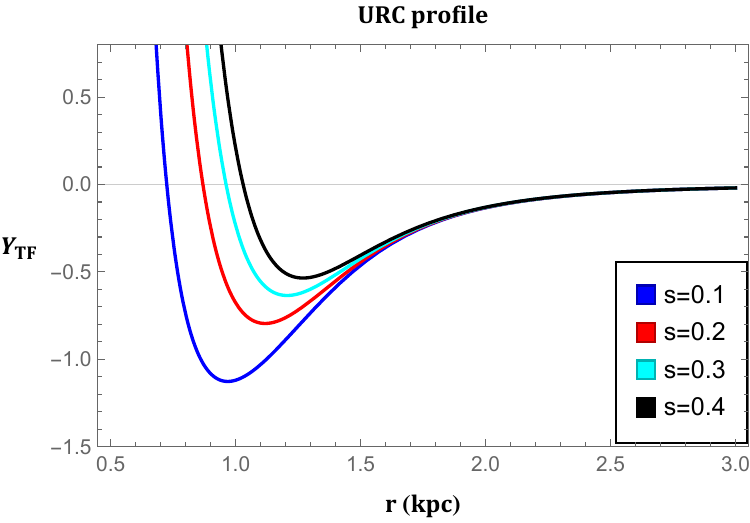}
\includegraphics[width=5cm,height=4cm]{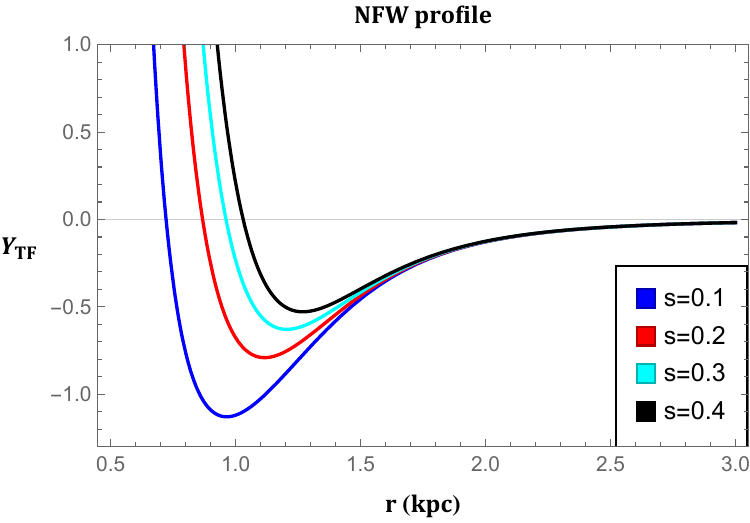}
\includegraphics[width=5cm,height=4cm]{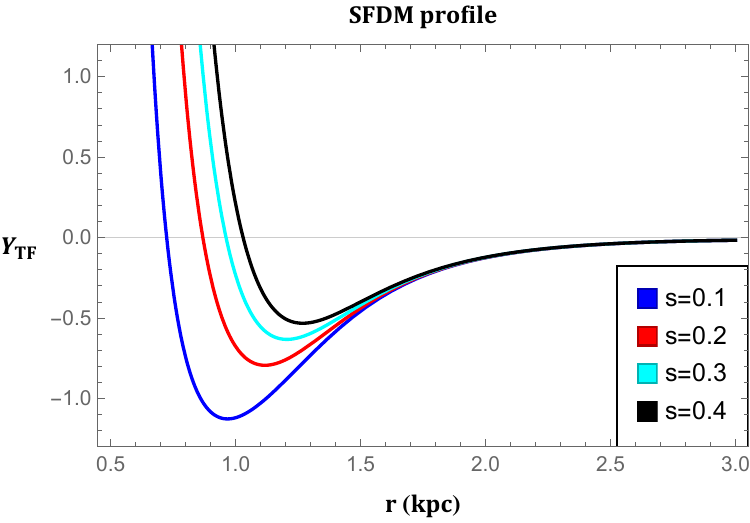}
\caption{The plots of $\Upsilon_{TF}$ against the radial distance $r$ for the redshift $\phi(r)=\frac{1}{2}\log(1+\frac{\eta^2}{r^2})$. We use $\rho_s=0.004\, kpc^{-2}$, $r_s=2\,kpc$, $\lambda=0.3$, $\eta=2$ and $r_0=1$.}
\label{fig:12}
\end{figure*}
Now, considering the relevant equations, we have studied the behavior of the complexity factor (using the above Eq. \eqref{6a2}) for three galactic halos wormholes with respect to the radial coordinate shown in Figs. \ref{fig:11} and \ref{fig:12}. We noticed that the complexity factor $\Upsilon_{TF}$ tends toward zero as the radial coordinate approaches infinity $r \rightarrow \infty$ or moves away from the wormhole throat. As explained in \cite{L. Herrera}, a minimal complexity factor, corresponds to a configuration characterized by homogeneous energy density and isotropic pressure. Additionally, a complexity factor of zero suggests the presence of inhomogeneous energy density and anisotropic pressure, provided these two effects counterbalance each other in the overall complexity factor. Consequently, in the vicinity of the wormhole throat, the complexity factor exhibits a monotonically increasing direction, and as we increase the range of the radial coordinate, $\Upsilon_{TF}$ approaches zero. Moreover, we noticed that the above holds for any values of $s>0$. Therefore, the complexity factor converges to zero for high radial coordinates and $s>0$ in the context of galactic halo wormholes in EGB gravity. Further, in the dynamics of the complexity factor, the role of pressure isotropy emerges as more crucial compared to the homogeneity of energy density. Furthermore, one can read some recent interesting papers on this topic given in Refs. \cite{Butt1,Butt2,Butt3}.

\subsection{Active gravitational mass}
The active mass function for our wormhole within the region from the wormhole throat $r_0$ up to the radius $R$ can be read as
\begin{equation}\label{6b1}
\mathcal{M}_{\mathcal{A}}=4 \pi \int_{r_0}^{R}\rho(r) r^2 dr,
\end{equation}
where $\mathcal{M}_{\mathcal{A}}$ is the active gravitational mass. Note that the positive
nature of the active gravitational mass indicates that the models are physically acceptable. Keeping this in mind, we shall investigate three different types of dark matter halo models and try to find some conditions for which one can find positive mass function. \\
For the URC profile, the expression for $\mathcal{M}_{\mathcal{A}}$ can be read as
\begin{equation}\label{6b2}
\mathcal{M}_{\mathcal{A}}=4 \pi  \rho_s r_s^3 \left[\frac{1}{4} \log \left(r_s^2+r^2\right)+\frac{1}{2} \log (r_s+r)
-\frac{1}{2} \tan ^{-1}\left(\frac{r}{r_s}\right)\right]_{r_0}^{R}.
\end{equation}
It is observed from the above expression \eqref{6b2} that the active gravitational mass $\mathcal{M}_{\mathcal{A}}$ of the wormhole for URC galactic is positive under the constraint $\frac{1}{4} \log \left(r_s^2+r^2\right)+\frac{1}{2} \log (r_s+r)> \frac{1}{2} \tan ^{-1}\left(\frac{r}{r_s}\right)$.\\
Again, for the NFW profile, we can obtain the active gravitational mass
\begin{equation}\label{6b3}
\mathcal{M}_{\mathcal{A}}=4 \pi  \rho_s r_s^3 \left[\frac{r_s}{r_s+r}+\log (r_s+r)\right]_{r_0}^{R}.
\end{equation}
Note that the RHS of the Eq. \eqref{6b3} is a positive quantity for any values of the parameters.\\
At last, for the SFDM profile, $\mathcal{M}_{\mathcal{A}}$ can be read as
\begin{equation}\label{6b4}
\mathcal{M}_{\mathcal{A}}=4 \rho_s r_s \left[\frac{r_s^2 \sin \left(\frac{\pi  r}{r_s}\right)}{\pi ^2}-\frac{r_s r \cos \left(\frac{\pi  r}{r_s}\right)}{\pi }\right]_{r_0}^{R}.
\end{equation}
In this case, the active gravitational mass is positive under the constraint $\frac{r_s^2 \sin \left(\frac{\pi  r}{r_s}\right)}{\pi ^2}>\frac{r_s r \cos \left(\frac{\pi  r}{r_s}\right)}{\pi }$.\\
Thus, we can conclude that the above dark matter models are physically acceptable under some restrictions.
\subsection{Total gravitational energy}
\begin{figure}[h]
    \centering
    \includegraphics[width=7.5cm,height=5cm]{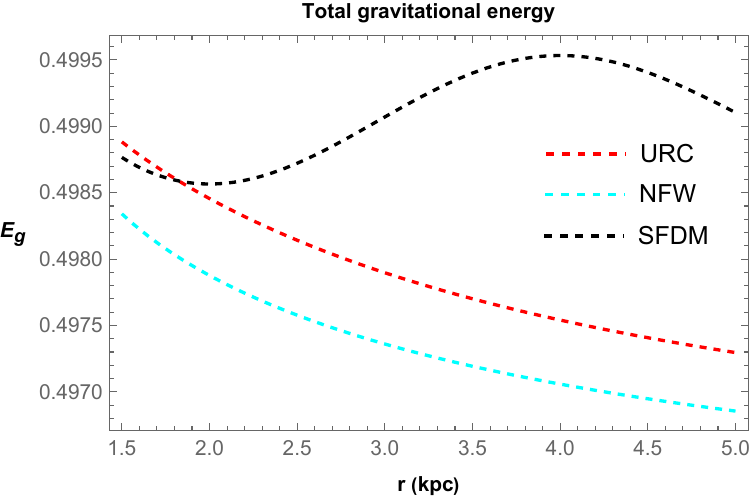}
    \caption{The plot of $\mathcal{E}_g$ under the shape function \eqref{4a5}. We use $\rho_s=0.004\, kpc^{-2}$, $r_s=2\,kpc$, $\delta=0.9$, $k=0.2$ and $r_0=1$.}
    \label{fig:13}
\end{figure}

\begin{figure}[h]
    \centering
    \includegraphics[width=7.5cm,height=5cm]{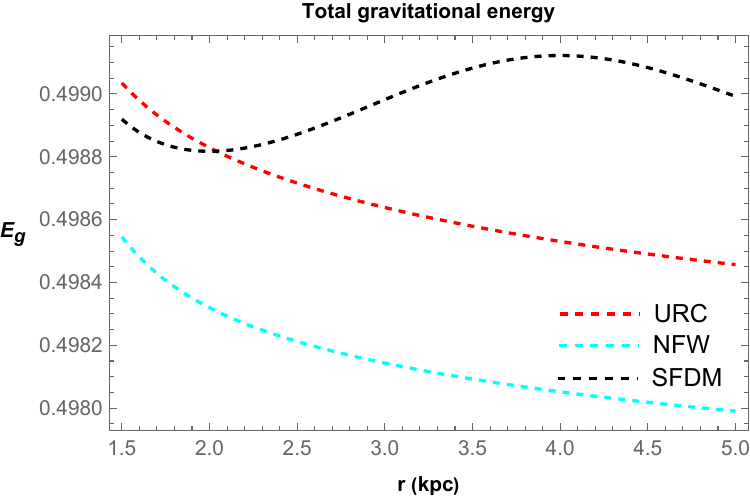}
    \caption{The plot of $\mathcal{E}_g$ under the shape function \eqref{4b5}. We use $\rho_s=0.004\, kpc^{-2}$, $r_s=2\,kpc$, $\lambda=0.3$, $\eta=2$ and $r_0=1$.}
    \label{fig:14}
\end{figure}

We have already noticed that the material forming the wormhole violates the NEC and, therefore, must be exotic rather than normal matter. The total gravitational energy of a structure composed of normal baryonic matter is negative. Therefore, it is crucial to check the nature of the gravitational energy in a wormhole background. Here, we follow the works of Lyndell-Bell et al. \cite{Katz1}, and Nandi et al. \cite{Katz2} and will try to find out the total gravitational energy of the dark matter galactic wormholes. The total gravitational energy $\mathcal{E}_g$ can be define as \cite{Katz2}
\begin{equation}\label{6c1}
\mathcal{E}_g=\mathcal{M}c^2-\mathcal{E}_M,
\end{equation}
where $\mathcal{M}c^2$ represents the total energy, and it can be expressed as
$$\mathcal{M}c^2=\frac{1}{2}\int_{r_0}^{r}T_0^0 r^2 dr+\frac{r_0}{2},$$
where the quantity $\frac{r_0}{2}$ is linked to the effective mass \cite{Katz2}.  $\mathcal{E}_M$ is the sum of other forms of energy like kinetic energy, rest energy, internal energy, etc., defined by
$$\mathcal{E}_M=\frac{1}{2}\int_{r_0}^{r} T_0^0 (g_{rr})^{\frac{1}{2}} r^2 dr,\,\,\,\text{with} \quad g_{rr}=\left(1-\frac{b(r)}{r}\right)^{-1}.$$
It can be claimed that since $(g_{rr})^{\frac{1}{2}}>1$ (by definition), then one can instantly deduce the criteria that $\mathcal{E}_g<0$ (attractive) if $T_0^0>0$ and $\mathcal{E}_g>0$ (repulsive) if $T_0^0<0$ \cite{A. Wheeler}. Since it is very complicated to find the exact solutions of the integral \eqref{6c1}, we solve it numerically by setting the integration range from the throat at $r_0$ to the embedded radial space of the wormhole geometry (refer to Tables \eqref{Table1} and \eqref{Table2}). As depicted in Figs. \ref{fig:13} and \ref{fig:14}, the results demonstrate that $\mathcal{E}_g>0$ signifies the repulsive behavior in the proximity of the throat. Note that we obtain $\mathcal{E}_g>0$ here despite $T_0^0>0$. This is because the matter distribution supporting the wormhole structure violates the NEC. Similar behavior of $\mathcal{E}_g$ can be found in \cite{Manna1,Manna2,Baransky}. This repulsive nature of $\mathcal{E}_g$ aligns with expectations for the formation of a physically viable wormhole.
\begin{table}[t]
    \centering
\begin{tabular}{ |p{1.5cm} p{1.5cm} p{1.5cm} p{1.5cm}| }
 \hline
 \multicolumn{4}{|c|}{The values of $\mathcal{E}_g$ for different $r$.} \\
 \hline
 $r$ & URC & NFW & SFDM\\
 \hline
 1.5 & 0.498883 & 0.498341 & 0.498766\\
 2.0 & 0.498456 & 0.497876 & 0.498566\\
 2.5 & 0.498142 & 0.497577 & 0.498722\\
 3.0 & 0.497897 & 0.49736  & 0.499071\\
 3.5 & 0.497701 & 0.497193 & 0.499401\\
 4.0 & 0.497541 & 0.497058 & 0.499534\\
 4.5 & 0.497408 & 0.496948 & 0.499406\\
 5.0 & 0.497296 & 0.496855 & 0.499104\\
 \hline
\end{tabular}
\caption{Table shows the values of $\mathcal{E}_g$ for different dark matter profiles under the shape function \eqref{4a5}. We use $\rho_s=0.004\, kpc^{-2}$, $r_s=2\,kpc$, $\delta=0.9$, $k=0.2$ and $r_0=1$.}
\label{Table1}
\end{table}

\begin{table}[t]
    \centering
\begin{tabular}{ |p{1.5cm} p{1.5cm} p{1.5cm} p{1.5cm}| }
 \hline
 \multicolumn{4}{|c|}{The values of $\mathcal{E}_g$ for different $r$.} \\
 \hline
 $r$ & URC & NFW & SFDM\\
 \hline
 1.5 & 0.499035 & 0.498545 & 0.49892\\
 2.0 & 0.498829 & 0.49832 & 0.498818\\
 2.5 & 0.498716 & 0.498212 & 0.498872\\
 3.0 & 0.498639 & 0.498143  & 0.498982\\
 3.5 & 0.498579 & 0.498092 & 0.499083\\
 4.0 & 0.498531 & 0.498052 & 0.499122\\
 4.5 & 0.498491 & 0.498019 & 0.499084\\
 5.0 & 0.498457 & 0.497991 & 0.498992\\
 \hline
\end{tabular}
\caption{Table shows the values of $\mathcal{E}_g$ for different dark matter profiles under the shape function \eqref{4b5}. We use $\rho_s=0.004\, kpc^{-2}$, $r_s=2\,kpc$, $\lambda=0.3$, $\eta=2$ and $r_0=1$.}
\label{Table2}
\end{table}
\subsection{Embedding diagram}
\begin{figure}[h]
    \centering
    \includegraphics[width=7.5cm,height=5cm]{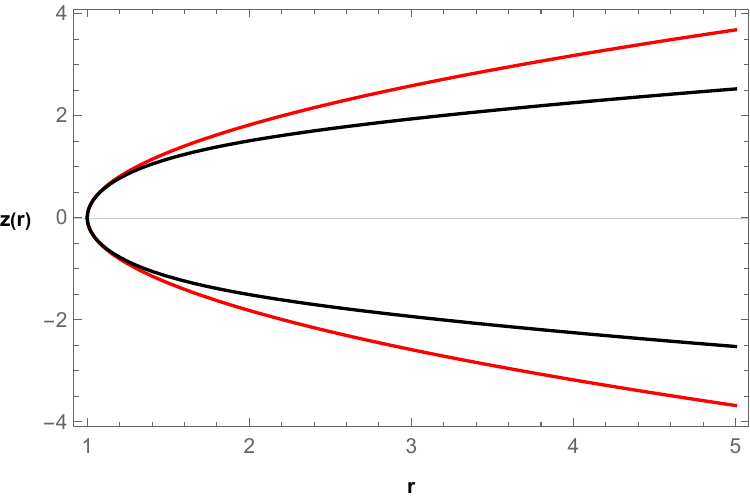}
    \caption{The plot of $2D$ embedding diagram for the shape functions \eqref{4a5} (\textit{Red}) and \eqref{4b5} (\textit{Black}). We use $\delta=0.9$, $k=0.2$, $\lambda=0.3$, $\eta=2$ and $r_0=1$.}
    \label{fig:15}
\end{figure}

\begin{figure}[h]
    \centering
    \includegraphics[width=3cm,height=5cm]{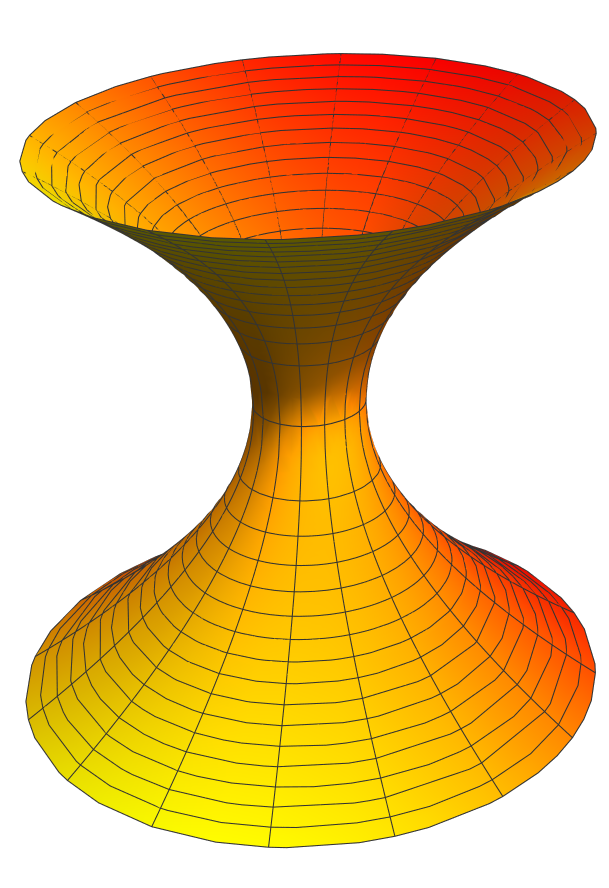}
    \includegraphics[width=5.5cm,height=5cm]{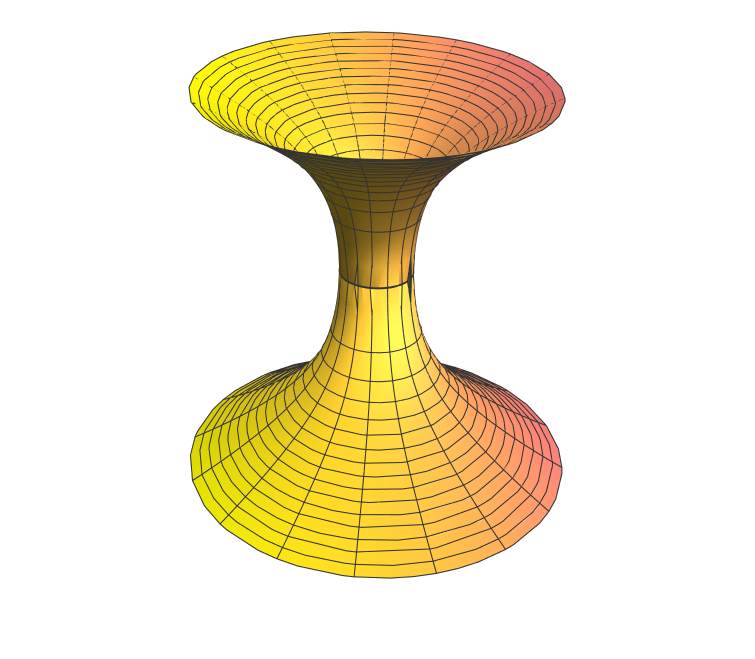}
    \caption{The plot of full visualization of the embedding diagram for the shape functions \eqref{4a5} (\textit{left}) and \eqref{4b5} (\textit{right}). Left panel: we use $r_0=1$, $\delta=0.9$, $k=0.2$. Right panel: we use $r_0=1$, $\lambda=0.3$, $\eta=2$.}
    \label{fig:16}
\end{figure}

This subsection will discuss the embedding diagram that helps us visualize wormhole space-time. As per the framework outlined by Morris and Thorne \cite{a17}, the embedding surface of the wormhole is represented by the function $z(r)$, which satisfies the given differential equation:
\begin{equation}\label{6d1}
\frac{dz}{dr}=\pm \frac{1}{\sqrt{\frac{r}{b(r)}-1}}.
\end{equation}
In this expression, it is observed that $\frac{dz}{dr}$ diverges at the wormhole's throat, implying that the embedding surface assumes a vertical orientation at the throat. Note that the above differential equation \eqref{6d1} provides the following relation
\begin{equation}\label{6d2}
z(r)=\pm \int_{r_0}^{r} \frac{dr}{\sqrt{\frac{r}{b(r)}-1}}.
\end{equation}
Further, the radial distance of the wormhole can be read as
\begin{equation}\label{6d3}
l(r)=\pm \int_{r_0}^{r} \frac{dr}{\sqrt{1-\frac{r}{b(r)}}}.
\end{equation}
Note that the above integral is given in Eqs. \eqref{6d2} cannot be solved analytically. Hence, we will solve it numerically by fixing some values of the free parameters and by changing the upper limit $r$. The numerical plot for embedding diagram $z(r)$ for the shape functions \eqref{4a5} and \eqref{4b5} is given in Fig. \ref{fig:15}. Moreover, one can check Fig. \ref{fig:16} for the full visualization of wormholes.

\section{Conclusions}\label{ch5sec7}
Indeed, exploring wormhole geometry has recently sparked considerable enthusiasm among theoretical researchers. Consequently, wormholes have been identified in the galactic halo region, supported by various dark matter density profiles. This chapter searches for wormhole existence in the galactic halo supported by three different dark matter profiles, such as URC, NFW, and SFDM models within recently proposed $4$D EGB gravity. Also, we employed the Karmarkar condition to find the shape function for two different non-constant redshift functions. The detailed outcomes of this work are as follows:\\
\indent Firstly, we obtained the shape function using an embedding class- I approach under the redshift function $\phi(r)=-\frac{k}{r}$. We noticed that the obtained shape function follows the flare-out condition under asymptotically flatness conditions within the range $0<\delta<r_0$. Later, we inserted the shape function into the pressure components of the $4$D EGB gravity. Further, we have investigated the energy conditions in the galactic halo with dark matter density profiles. We have observed that NEC is violated in the neighborhood of the throat. It was noticed the influence of GB coefficient $s$ in the violation of NEC. Mathematically, one can check the RHS of Eqs. \eqref{4a9}, \eqref{4a12}, and \eqref{4a13} by simply putting the values of the parameters and taking the GB coefficient $s>0$ provides a negative quantity. Also, we checked SEC for each dark matter profile and found that SEC was violated. Moreover, as we increase the value of $s$, the contribution of violation becomes more. Additionally, DEC for both pressures was found satisfying near the throat. Moreover, one can check the summary of the energy conditions near the throat of the wormhole in Table-\ref{Table0}.\\
\indent Similar to the previous case, we extracted the shape function under the redshift function $\phi(r)=\frac{1}{2}\log(1+\frac{\eta^2}{r^2})$. We investigated the necessary criteria for a traversable wormhole, i.e., the flare-out condition, which is satisfied in the range $0<\lambda<r_0$. Later, we checked the NEC, SEC, and DEC for obtained solutions in the galactic halo regions. NEC is disrespected at the throat for $s>0$. Moreover, one can check numerically from the expression given in Eq. \eqref{4b9}. DEC was investigated for each dark matter profile, and it was found that DEC was satisfied with both pressures in the galactic halos, except tangential DEC, which is violated in the URC dark matter profile. The calculated energy conditions are shown in Table-\ref{Table0}.\\ 
\indent Note that the confirming violated behavior of energy conditions supports the presence of the dark halos. We checked the behavior of energy conditions for each dark matter halo profile and noticed the violations of energy conditions, which means that traversable wormholes may exist in the galactic regions supported by dark matter in the context of $4$D EGB gravity.\\
\indent Further, some physical features of wormholes, such as complexity factor, active gravitational mass, total gravitational mass, and embedding diagrams, have been explored in this chapter. The complexity factor in the context of galactic dark matter halo wormholes in EGB gravity has been calculated, and it was noticed that the complexity factor converges to zero for high radial coordinates and GB coefficient $s>0$. Such a study has been done in Ref. \cite{Butt3} for Casimir wormholes in higher dimensional EGB gravity. Moreover, the active gravitational mass for each dark matter density profile has been performed, and it has been observed that these dark matter models are physically acceptable under some restrictions. Rahaman et al. \cite{Rahaman1} studied the active gravitational mass of the NFW profile; however, in this chapter, we have studied the active gravitational mass of URC and SFDM models along with the NFW model. Further, we have studied and numerically calculated the total gravitational energy for each dark matter profile under obtained shape functions. it was observed that $\mathcal{E}_g>0$, which signifies the presence of repulsion near the throat. This characteristic nature of $\mathcal{E}_g$ aligns with expectations for the formation of a physically viable wormhole.\\
In \cite{c4}, wormhole geometry in the galactic halos has been explored with two embedded wormhole-specific shape functions in the context of Einsteinian cubic gravity. They used observational data within the signature of the M87 galaxy and the Milky Way galaxy to check the effect of the dark matter halos. Further, in \cite{Milgrom4}, the author discussed wormhole solution with density profile obtained from MOND with or without a scalar field in GR. 
In this chapter, we have investigated wormhole solutions under URC, NFW, and SFDM profiles in $4$D EGB gravity with two different redshift functions. Our study confirms that the obtained wormhole solutions might exist in the galactic halos within $4$D EGB gravity. Also, traversable wormholes in the galactic halos with observational data sets in higher dimensional gravity would be an interesting problem that is being actively considered.

\chapter{Concluding remarks and future perspectives} 

\label{Chapter7} 

\lhead{Chapter 6. \emph{Concluding Remarks and Future Perspectives}} 

Let us take a moment to summarise the findings brought to light by this thesis. The primary objective of this thesis is to investigate the wormhole geometry in modified theories of gravity. This thesis is a review of the most important concepts in wormhole physics and of the most recent recipes to construct these extreme objects. Let us discuss the results obtained in four concrete works from the previous chapters \ref{Chapter1}-\ref{Chapter5}.\\
We start chapter-\ref{Chapter1} by briefly discussing the motivations for this thesis, the history of wormhole geometry, and energy conditions. Further, we have discussed some important concepts like Casimir force, GUP implications, and dark matter. We further discussed fundamental elements, mathematical concepts, and fundamental theories of gravity. Besides this, it is well-known that the fundamental theory of gravity, like general relativity, fails to address certain issues, such as fine-tuning and the flatness problem. Therefore, its modifications and generalizations are more effective in addressing these concerns, and the chapter concludes with a summary of the modified gravity theories.\\
In chapter-\ref{Chapter2}, we first developed the field equations for the Morris-Throne wormhole metric in the generic $f(Q)$ gravity. Then, we have considered two specific $f(Q)$ forms, such as $f(Q)=\alpha Q+\beta$ and $f(Q)=a Q^2+B$. We then obtained two wormhole models by imposing EoS relations, namely $P_r=\omega \rho$ and $P_t=m P_r$ under linear $f(Q)$ form. We note that in order to satisfy the asymptotically flatness condition, $\omega<-1$, which shows the phantom region of the Universe. Also, for the $P_t=m P_r$ case, the parameter $n$ has to be negative to satisfy the asymptotically flatness condition. Considering all these restrictions, we have tested all the necessary requirements of a shape function and the energy conditions. Interestingly, we noticed that the NEC is violated for each case. Thus, the violation of NEC for each model defines the possibility of the presence of exotic matter at the wormhole’s throat. Further, for the quadratic $f(Q)$ model, we have considered two specific shape functions. To test the traversability and stability of these two models, we have tested the flaring out condition and energy conditions. Both the models successfully passed all the tests. The advantage of these types of models is that they minimize the usage of the unknown form of matter called ``exotic matter” to have a traversable wormhole in comparison to the models for linear $f(Q)$ case (as it mimics the fundamental interaction of gravity, i.e., GR). In non-linear $f(Q)$, the NEC was violated near the throat but later satisfied, while linear $f(Q)$ violated NEC throughout. This highlights the potential of exploring wormholes in non-linear $f(Q)$ gravity. Finally, we used the volume integral quantifier to estimate the amount of exotic matter required for traversability. We found that only a small amount of exotic matter was needed for each solution.\\
We further check the strength of $f(Q)$ gravity with Casimir force, discussed elaborately in chapter-\ref{Chapter3}. By utilizing the KMM and DGS GUP models, we derived solutions for the Morris-Thorne wormholes for both linear and quadratic $f(Q)$ forms under different redshift functions. For the linear $f(Q)=\alpha Q+\beta$ model, our analysis demonstrates that the GUP parameter $\lambda$ reduces the shape function as radial distance increases, while the model parameter $\alpha$ enhances it. We examined the asymptotic flatness condition and the flare-out condition of the Casimir wormhole, and both were found to be satisfied in proximity to the throat. We also analyzed the EoS parameters, observing that the radial EoS parameter increases with radial distance while the tangential EoS parameter exhibits the opposite behavior. We can see the effect of modified gravity in the EoS parameter to a large extent, at least at distances away from the throat. Further, violation of NEC near the throat was confirmed for both KMM and DGS models under each redshift function. The violation contribution becomes more negative for an increase in $\alpha$. Thus this has demonstrated that some arbitrary amount of small quantity disrespects the classical energy condition at the wormhole's throat. Furthermore, numerical methods applied to the quadratic $f(Q)=Q+\gamma Q^2$ model indicate the possibility of traversable wormhole solutions contingent on suitable initial conditions. We noticed that the shape function shows positively decreasing behavior as the values of $\gamma$ increased for both KMM and DGS models. Also, the flaring-out condition is satisfied near the throat, whereas, for large $r$, this condition will no longer be validated. Moreover, we have investigated the energy conditions and confirmed that NEC is violated for both models near the throat. Thus, it is worth mentioning that wormhole solutions could be possible numerically using appropriate initial conditions. However, this analysis shows the possibility of the existence of a macro or tiny wormhole.\\
In chapter-\ref{Chapter4}, we furthermore extended our work focusing on galactic halo dark matter models in the $f(Q)$ gravity framework. In this study, we derived novel wormhole solutions supported by dark matter models, including BEC, PI, and NFW, within the $f(Q)$ gravity framework. By applying dark matter density profiles and rotational velocity, we analytically obtained the redshift functions and shape functions of the wormholes. Our results show that the wormhole throat radius $r_0$ and other model parameters play a crucial role in satisfying the flare-out condition under asymptotic flatness. Energy condition analysis revealed significant violations, including the NEC, across most parameter maximum regions, confirming the presence of exotic matter required for sustaining the wormhole. We also explored the wormhole shadow under dark matter influence, noting that larger values of parameters like $D_1$ and central density $\rho_s$ shift the shadow closer to the throat, a behavior consistent across the PI and NFW profiles. Additionally, we investigated strong gravitational lensing due to wormhole geometry, and our findings reveal that the deflection angle of an outward light ray diverges precisely at the wormhole's throat, corresponding to the photon sphere. Detecting a photon sphere near the throat would not only verify strong gravitational fields around the wormhole but also offer observational opportunities to study these structures, providing deeper insights into gravity and the nature of wormholes.\\
In chapter-\ref{Chapter5}, we focused on the extension of the newly proposed geometrical framework of GR, named the $4$D EGB gravity. Here, we explored wormhole existence in the galactic halo using three dark matter profiles, namely URC, NFW, and SFDM, within $4$D EGB gravity. We Applied the Karmarkar condition and derived the shape function for two non-constant redshift functions. For the redshift $\phi(r)=-\frac{k}{r}$, the shape function satisfied the flare-out condition within $0<\delta<r_0$, and energy condition analysis revealed NEC violation near the throat, influenced by the GB coefficient $\alpha$. Increasing $\alpha$ enhanced the violations of NEC and SEC, while DEC was satisfied near the throat. For the redshift $\phi(r)=\frac{1}{2}\log(1+\frac{\eta^2}{r^2})$, we confirmed the flare-out condition within $0<\lambda<r_0$. NEC was again violated for $\alpha>0$, while DEC was satisfied except for the tangential DEC in the URC profile. These violations indicate the possible existence of traversable wormholes in galactic halos supported by dark matter in $4$D EGB gravity. Additionally, we investigated wormhole properties such as complexity factor, active gravitational mass, and total gravitational energy. The complexity factor converged to zero for large radial distances and $\alpha>0$. The active gravitational mass was physically acceptable under certain conditions, and the total gravitational energy, $\mathcal{E}_g>0$, indicated a repulsive effect near the throat, consistent with the formation of a viable wormhole.\\
In terms of the future outlook for these works, it is important to explore more general forms of action, different variational formalisms, and more general modified theories. Additionally, there are many opportunities to expand on the work presented in this thesis. For example, in the third chapter, the GUP correction up to the first order on the minimal length scale has been applied. This approach can also be used to calculate the corrections of Casimir energy up to the next leading order and investigate its significance on the wormhole geometry. Furthermore, in the last two chapters, we have assumed the parameter values of dark matter models. Therefore, these works can be extended using astrophysical observational data from the $M87$ galaxy and Milky Way galaxy datasets. The construction of traversable wormholes is a field of physics that, despite the recent progress and ongoing development, still has many unanswered questions. It is an exciting time for physicists working in this area, and one can only hope that the coming decades will bring even more wonderful revelations.






\addtocontents{toc}{\vspace{2em}} 

\backmatter


\label{References}
\lhead{\emph{References}}

\cleardoublepage
\pagestyle{fancy}

\label{Publications}
\lhead{\emph{List of Publications}}

\chapter{List of Publications}
\section*{Thesis Publications}
\begin{enumerate}

\item \textbf{Zinnat Hassan}, S. Mandal, P.K. Sahoo, \textit{Traversable Wormhole Geometries in $f(Q)$ Gravity}, \textcolor{blue}{Fortschritte der Physik} \textbf{69}, 2100023 (2021).

\item \textbf{Zinnat Hassan}, S. Ghosh, P.K. Sahoo, \textit{GUP corrected Casimir wormholes in $f(Q)$ gravity}, \textcolor{blue}{General Relativity and Gravitation} \textbf{55}, 90 (2023).

\item G. Mustafa, \textbf{Zinnat Hassan}, P.K. Sahoo, \textit{Deflection of light by wormholes and its shadow due to dark matter within modified symmetric teleparallel gravity}, \textcolor{blue}{Classical Quantum Gravity} \textbf{41}, 235001 (2024).

\item \textbf{Zinnat Hassan}, P.K. Sahoo \textit{Possibility of the Traversable Wormholes in the Galactic Halos within 4D Einstein-Gauss-Bonnet Gravity}, \textcolor{blue}{Annalen der Physik} \textbf{536}, 2400114 (2024).

\end{enumerate}
\section*{Other Publications}
\begin{enumerate}
\item M. M. Rizwan, \textbf{Zinnat Hassan}, P.K. Sahoo, A. Ovgun, \textit{Influence of GUP corrected Casimir energy on zero tidal force wormholes in modified teleparallel gravity with matter coupling}, \textcolor{blue}{The European Physical Journal C} \textbf{84}, 1132 (2024).
\item S. Pradhan, \textbf{Zinnat Hassan}, P.K. Sahoo, \textit{Wormhole geometries supported by strange quark matter and phantom-like generalized Chaplygin gas within $f(Q)$ gravity}, \textcolor{blue}{Physics of the Dark Universe} \textbf{46}, 101620 (2024).
\item M. Tayde, \textbf{Zinnat Hassan}, P.K. Sahoo, \textit{Conformally symmetric wormhole solutions supported by non-commutative geometries in the context of $f(Q,T)$ gravity}, \textcolor{blue}{Chinese Journal of Physics} \textbf{89}, 195-209 (2024).
\item M. Tayde, \textbf{Zinnat Hassan}, P.K. Sahoo, \textit{Impact of dark matter galactic halo models on wormhole geometry within $f(Q,T)$ gravity}, \textcolor{blue}{Nuclear Physics B} \textbf{1000}, 116478 (2024).
\item M. Tayde, \textbf{Zinnat Hassan}, P.K. Sahoo, \textit{Existence of wormhole solutions in $f (Q,T)$ gravity under non-commutative geometries}, \textcolor{blue}{Physics of the Dark Universe} \textbf{42}, 101288 (2023).

\item R. Solanki, \textbf{Zinnat Hassan}, P.K. Sahoo, \textit{Wormhole solutions in $f(R,\,Lm)$ gravity}, \textcolor{blue}{Chinese Journal of Physics} \textbf{85}, 74-88 (2023).

\item \textbf{Zinnat Hassan}, Sayantan Ghosh, P.K. Sahoo, Kazuharu Bamba, \textit{Casimir wormholes in modified symmetric teleparallel gravity}, \textcolor{blue}{The European Physical Journal C} \textbf{82}, 1116 (2022).

\item \textbf{Zinnat Hassan}, G. Mustafa, Joao RL Santos, P.K. Sahoo, \textit{Embedding procedure and wormhole solutions in $f(Q)$ gravity}, 
\textcolor{blue}{Europhysics Letters} \textbf{139}, 39001 (2022).

\item O. Sokoliuk, \textbf{Zinnat Hassan}, P.K. Sahoo, A. Baransky, \textit{Traversable wormholes with charge and non-commutative geometry in the $f(Q)$ gravity}, \textcolor{blue}{Annals of Physics} \textbf{443}, 168968 (2022).

\item M. Tayde, \textbf{Zinnat Hassan}, P.K. Sahoo, S. Gutti, \textit{Static spherically symmetric wormholes in $f(Q,T)$ gravity}, \textcolor{blue}{Chinese Physics C} \textbf{46}, 115101 (2022). 

\item S. Mandal, G. Mustafa, \textbf{Zinnat Hassan}, P.K. Sahoo, \textit{A study of anisotropic spheres in $f(Q)$ gravity with quintessence field}, \textcolor{blue}{Physics of the Dark Universe} \textbf{35}, 100934 (2022).

\item G. Mustafa, \textbf{Zinnat Hassan}, P.K. Sahoo, \textit{Traversable wormhole inspired by non-commutative geometries in $f(Q)$ gravity with conformal symmetry}, \textcolor{blue}{Annals of Physics} \textbf{437}, 168751 (2022).

\item G. Mustafa, \textbf{Zinnat Hassan}, P.H.R.S. Moraes, P.K. Sahoo, \textit{Wormhole solutions in symmetric teleparallel gravity}, \textcolor{blue}{Physics Letters B} \textbf{821}, 136612 (2021).

\item \textbf{Zinnat Hassan}, G. Mustafa, P.K. Sahoo, \textit{Wormhole Solutions in Symmetric Teleparallel Gravity with Noncommutative Geometry}, \textcolor{blue}{Symmetry} \textbf{13}, 1260 (2021).

\end{enumerate}

\chapter{Conferences/workshops attended}
\begin{enumerate}
\item Participated in a national workshop on ``\textit{Contemporary Issues in Astronomy and Astrophysics}” organized by the \textbf{Department of Physics, Shivaji University, Kolhapur, Maharashtra} in collaboration with \textbf{IUCAA, Pune} during \textcolor{blue}{13th - 15th Sept. 2024}.

\item Participated in an international workshop on ``\textit{Astronomy Data Analysis with Python}” jointly organized by \textbf{the Department of Physics and the Department of Mathematics, Maulana Azad National Urdu
University (MANUU), Hyderabad} and sponsored by \textbf{IUCAA, Pune} during \textcolor{blue}{5th – 8th Sept. 2023}.

\item Participated in a national workshop on ``\textit{General Relativity and Cosmology}” organized by \textbf{Centre for Cosmology, Astrophysics and Space Science GLA University, Mathura,} and sponsored by \textbf{IUCAA, Pune} during \textcolor{blue}{24th – 26th Nov. 2022}.

\item Presented a research paper entitled ``\textit{Traversable wormhole inspired by non-commutative geometries in $f(Q)$ gravity with conformal symmetry}” at the international conference “\textbf{International Conference on Mathematical Sciences and its Applications}” organized by the \textbf{Department of Mathematics, SRTM University, Nanded} during \textcolor{blue}{28th - 30th July 2022}.

\item Presented a research paper entitled ``\textit{Traversable wormholes with charge and non-commutative geometry in the $f(Q)$ gravity}” at the international conference “\textbf{Differential Geometry and its Applications}” jointly organized by the \textbf{Department of Mathematics, Kuvempu University, Shivamogga, Karnataka} and \textbf{The Tensor Society, Lucknow} during \textcolor{blue}{4th - 5th March 2022}.

\item Presented research paper entitled “\textit{Wormhole solutions in symmetric Teleparallel gravity}" at the international conference “\textbf{International Academy of Physical Sciences on Advances in Relativity and Cosmology}” organized by the \textbf{Department of Mathematics, BITS-Pilani, Hyderabad Campus} during \textcolor{blue}{26th - 28th Oct 2021}.
\end{enumerate}
\cleardoublepage
\pagestyle{fancy}
\lhead{\emph{Biography}}

\chapter{Biography}

\section*{Brief Biography of the Candidate:}
\textbf{Mr. Zinnat Hassan} earned his Bachelor's degree in Mathematics from Gauhati University in 2016, followed by a Master's degree in Mathematics from the University of Science and Technology, Meghalaya, in 2019. His academic record is exemplary, complemented by impressive research achievements and outstanding performance as a research scholar. He was awarded the gold medal in Mathematics during his master's program, and he received the prestigious DST-INSPIRE Fellowship from the Government of India, supporting his Ph.D. studies from 2021 to 2024. Over his four-year research career, Mr. Hassan has published 18 research articles in highly reputed international journals and has presented some of his works at various national and international conferences.


\section*{Brief Biography of the Supervisor:}
\textbf{Prof. Pradyumn Kumar Sahoo} has over 24 years of immense research experience in Applied Mathematics, Cosmology, Astrophysical Objects, General Theory of Relativity, and Modified Theories of Gravity. He obtained his Ph.D. from Sambalpur University, Odisha, India, in 2004. In 2009, he joined the Department of Mathematics at BITS Pilani, Hyderabad Campus, as an Assistant Professor and is currently a Professor. He is also an Associate Member of IUCAA, Pune. In 2022, he received the ``Prof. S. Venkateswaran Faculty Excellence Award" from BITS Pilani. He has been awarded a visiting professor fellowship at Transilvania University of Brașov, Romania. According to a survey by researchers from Stanford University, he has been ranked among the top 2\% of scientists worldwide in the field of Nuclear and Particle Physics in the last five years. Throughout his career, he has published more than 250 research articles in various renowned national and international journals. As a visiting scientist, he had the opportunity to visit the European Organization for Nuclear Research (CERN) in Geneva, Switzerland, a renowned center for scientific research. He has participated in numerous national and international conferences, often presenting his work as an invited speaker. Prof. Sahoo has engaged in various research collaborations at both the national and international levels. He has contributed to BITS through five sponsored research projects: University Grants Commission (UGC 2012-2014), DAAD Research Internships in Science and Engineering (RISE) Worldwide (2018, 2019, 2023, and 2024), Council of Scientific and Industrial Research (CSIR 2019-2022), National Board for Higher Mathematics (NBHM 2022-2025), and Science and Engineering Research Board (SERB), Department of Science and Technology (DST 2023-2026). He also serves as an expert reviewer for Physical Science Projects for SERB, DST (Government of India), and UGC research schemes. Additionally, he is an editorial board member for various reputable journals, contributing to the research community.

\end{document}